\documentclass[pdflatex,sn-basic]{sn-jnl}

\usepackage{fancyhdr}

\jyear{2023}%

\newcommand{\sixj}[6]{\Bigl\{\mbox{\small$\hspace{0.01cm}\begin{array}{ccc} #1 \hspace{0.01cm} & \hspace{0.01cm} #3 \hspace{0.01cm} & \hspace{0.01cm} #5  \hspace{0.01cm}\\[-0mm]  #2 \hspace{0.01cm}  & \hspace{0.01cm} #4 \hspace{0.01cm}  & \hspace{0.01cm} #6  \end{array}\hspace{0.005cm}$}\Bigr\}}

\newcommand{\threej}[6]{\left(\mbox{\small$\!\begin{array}{ccc} #1 \! & \!\! #3 \! & \!\! #5 \\[-1mm]  #2 \!  & \!\! #4 \!  & \!\! #6 \end{array}\!$}\!\right)}

\theoremstyle{thmstyleone}%
\theoremstyle{thmstyletwo}%

\theoremstyle{thmstylethree}%

\numberwithin{equation}{section}

\begin{document}

\title[Solvable models of quantum black holes: a review on JT gravity]{Solvable models of quantum black holes: a review on Jackiw--Teitelboim gravity}


\author[1]{\fnm{Thomas G.} \sur{Mertens}}\email{thomas.mertens@ugent.be}

\author[2,3]{\fnm{Gustavo J.} \sur{Turiaci}}\email{turiaci@uw.edu}

\affil[1]{\orgdiv{Department of Physics and Astronomy}, \orgname{Ghent University}, \orgaddress{\street{Krijgslaan, 281-S9, 9000 Gent}, \country{Belgium}}}

\affil[2]{\orgname{Institute for Advanced Study}, \orgaddress{\city{Princeton}, \state{NJ}, \country{USA}}}

\affil[3]{\orgdiv{Physics Department}, \orgname{University of Washington}, \orgaddress{ \city{Seattle}, \state{WA}, \country{USA}}}

\abstract{
We review recent developments in Jackiw--Teitelboim (JT) gravity. This is a simple solvable model of quantum gravity in two dimensions (that arises e.g. from the s-wave sector of higher dimensional gravity systems with spherical symmetry). Due to its solvability, it has proven to be a fruitful toy model to analyze important questions such as the relation between black holes and chaos, the role of wormholes in black hole physics and holography, and the way in which information that falls into a black hole can be recovered.}



\maketitle

\pagestyle{myheadings}
\markright{T. G. Mertens and G. J. Turiaci} 
\setcounter{tocdepth}{3}
\tableofcontents
\section{Introduction}\label{sec:Intro}

Understanding quantum gravity and quantum black hole physics are some of the most pressing open problems in contemporary theoretical physics. In particular, deep questions arise in the problem of black hole formation and subsequent evaporation, starting with Hawking's work in the 1970s \cite{Hawking:1975vcx,Hawking:1976ra}. 

The road towards quantum gravity, starting with the problem of non-renormalizability of pure Einstein--Hilbert gravity in 3+1d, has in the years led us through higher dimensions, string theory, compactification, branes .... And some significant successes have been made using this approach \cite{Strominger:1996sh}. However, it is safe to say that we do not have a fully satisfying understanding, and any alternative approaches that could shed new light on the problem would be most welcome. 

With this goal in mind, an alternative strategy toward quantum gravity would be to work in lower dimensions (2d or 3d), where gravitational models can make sense at the level of the Euclidean path integral. If we furthermore do this in the framework of holography, then we have a preferred anchoring point, and are guided by the major breakthroughs in that field throughout the years \cite{Maldacena:1997re}.

Even within this strategy, finding interesting lower-dimensional solvable models of gravity is an art on its own. Few such models exist, but the degree of solvability that we can obtain in 2d Jackiw--Teitelboim gravity is unprecedented as we aim to explain with this review. If we work in two spacetime dimensions, the simplest candidate model would be Einstein--Hilbert gravity
\begin{equation}
\label{eq:seh}
S =  \frac{1}{16 \pi G_N} \int d^2 x \sqrt{-g} R + \hdots.
\end{equation}
However, this model is topological since its Euclidean action is just the Euler characteristic, and the Einstein tensor vanishes identically. This model does play an important role as the 2d gravity on the string worldsheet, weighing different topologies. Additionally coupling this model to a matter action $S_m$, the gravitational equations lead to $T_{\mu\nu}^m =0$ and no energy flows exist. Hence when using it as a classical toy model for black hole formation and evaporation, this model has little value.\footnote{Quantum mechanically, gauge fixing this model with conformal matter, leads to Liouville gravity which is an interesting model appearing in the context of the non-critical string \cite{Polyakov:1981rd,Distler:1988jt,David:1988hj}. We do not follow this route.}

To get something more interesting, the required adjustment in two dimensions is to introduce a direct coupling of the Ricci scalar to a new scalar field $\Phi$, called the dilaton field for historical (string-theoretic) reasons:
\begin{equation}
S =  \frac{1}{16 \pi G_N} \int d^2 x \sqrt{-g} \Phi R + \hdots
\end{equation}
Models of this kind have been introduced in 3+1d as scalar-tensor (Brans--Dicke type) models, and provide deformations of general relativity with spacetime-varying gravitational coupling. Here we view these dilaton gravity models as quantum mechanical solvable toy models of quantum gravity and black hole physics. Jackiw--Teitelboim (JT) gravity is one such particular model that we will specify below. Interestingly, this particular model also captures the dynamics close to the horizon of near-extremal black holes in higher dimensions, see the upcoming review by \cite{ROPreviewNEBH}.

Gravitational models of this kind have attracted widespread attention over the years (see e.g. the reviews by \cite{Grumiller:2002nm,Nojiri:2000ja}), but it has been only recently, sparked by developments in 2015 by A. Kitaev in solvable many-body models (the Sachdev--Ye--Kitaev or SYK models; \cite{Sachdev:1992fk,kitaevTalks1,kitaevTalks2,kitaevTalks3}), that we have reached a far deeper understanding of their quantum mechanical aspects. This work refined the original proposal relating SYK and AdS$_2$ gravity put forward by \cite{Sachdev:2010um}. 

Our aim here is to provide an in-depth review of these developments in JT gravity. This review is organized into four main sections. In Sect.~\ref{sec:JT} we introduce the model, and fully solve its classical equations of motion, crucially incorporating boundary conditions at the holographic boundary that allow us to map the dynamics to its boundary Schwarzian description. Sect.~\ref{sec:JTquantum} proceeds with the exact quantization of the model, by computing the Euclidean gravitational path integral in the Schwarzian language. In Sect.~\ref{sec:JTRMT} we furthermore include non-trivial topological corrections (or wormholes) to the quantum amplitudes, that modify the heavy quantum regime even further. Finally, Sect.~\ref{sec:Generalization} contains several applications of the exact solvability of JT gravity. We do not treat these in technical detail, but refer to the literature for more information. In particular, in the past few years significant progress has been made on Hawking's information loss paradox, made possible to a large extent due to the solvability of JT gravity. We discuss some of these developments, but will not do justice to this topic in this work. We refer e.g. to the excellent recent review by \cite{Almheiri:2020cfm} for details. 

A few reviews have been written before on the connection between JT gravity and SYK. We refer the interested reader to the excellent review on this topic by \cite{Chowdhury:2021qpy} that focuses mostly on the SYK side. Other earlier reviews are \cite{Sarosi:2017ykf} and \cite{Rosenhaus:2018dtp}. However, no comprehensive review exists that combines these earlier developments with the current state-of-the-art of JT gravity specifically and lower-dimensional gravitational models more generally, something we hope to address with this review.

Finally, a few comments on conventions: in Lorentzian signature our metric signature convention is $(-,+,\hdots +)$. We denote Lorentzian actions as $S$ and Euclidean actions as $I$.

\section{Classical Jackiw--Teitelboim gravity}\label{sec:JT}
We begin by introducing and motivating the JT model and its coupling to matter. In this section, we study the classical solution of this model including gravitational backreaction. Our endeavors will ultimately lead us to a description in terms of a boundary Schwarzian model that will be the starting point for a quantum mechanical solution in the next Sect.~\ref{sec:JTquantum}.

\subsection{Dilaton gravity models}\label{sec:DilGravMod}
To start, it is instructive to consider what is the most general theory of dilaton gravity in two dimensions with a two-derivative action \cite{Banks:1990mk,Louis-Martinez:1993bge,Ikeda:1993fh,Grumiller:2002nm}. Working in Euclidean signature, any such theory can be written as 
\begin{eqnarray}
\label{eq:gendil}
    I = - \frac{1}{16 \pi G_N} \int_\mathcal{M} d^2 x \sqrt{g} \Big( U_1(\tilde{\Phi}) R + U_2(\tilde{\Phi}) g^{\mu \nu} \partial_\mu \tilde{\Phi} \partial_\nu \tilde{\Phi} + U_3 (\tilde{\Phi}) \Big),
\end{eqnarray}
where $g_{\mu\nu}$ is the two-dimensional metric and $\tilde{\Phi}(x)$ is the dilaton field. The 2d Newton's constant $G_N$ is dimensionless, and hence unlike in higher dimensions does not set the scale of physics.\footnote{Later on, around Eq.~\eqref{eq:scale} in JT gravity, we will see an effective scale emerge nonetheless.} This theory seems to be parametrized by three functions $U_1(\tilde{\Phi})$, $U_2(\tilde{\Phi})$ and $U_3(\tilde{\Phi})$ but two of them are redundant. To see this, first perform a field redefinition on the dilaton $\tilde{\Phi} \to \Phi = U_1(\tilde{\Phi})$. We assume there is no value of $\tilde{\Phi}$ such that $U_1'(\tilde{\Phi})=0$ so the field redefinition is invertible, otherwise the resulting kinetic term in $\Phi$ will be ill-defined. 

We can further set $U_2(\Phi)=0$ by making a Weyl transformation on the metric as follows. Under a local rescaling, the 2d Ricci scalar transforms as
\begin{equation}
    g'_{\mu\nu} = e^{2\omega}g_{\mu\nu},~~~\rightarrow g'^{1/2}R' = g^{1/2}(R-2\nabla^2\omega).
\end{equation}
It is then a simple calculation to show that the choice
\begin{equation}
\omega(x) = \frac{1}{2}\int^{\Phi(x)} U_2(\Phi')d\Phi',
\end{equation}
will cancel the term in \eqref{eq:gendil} proportional to $U_2$.
In dimensions greater than two, this Weyl field redefinition $g'_{\mu\nu} = e^{2\omega}g_{\mu\nu}$, with a suitable choice of $\omega$, allows us to redefine $\Phi R \to R'$, a procedure that is well-known in string theory to transfer between the so-called string frame and the Einstein frame. In 2d however, the above Weyl field redefinition allows us instead to set the kinetic term of the dilaton to zero, and it is impossible to get rid of the $\Phi R$ term: dilaton gravity is (in this sense) an invariant notion in 2d. This observation also implies that calling $\Phi$ a dilaton is not a good nomenclature in 2d since it is not related to rescalings of the metric. Regardless of this point, to avoid confusion we will follow tradition and continue calling it the dilaton field. 

This leaves us with a very general class of \textbf{2d dilaton gravity models}:
\begin{eqnarray}
\label{gendilgrav}
    I[g,\Phi] = - \frac{1}{16 \pi G_N}\int_\mathcal{M} d^2 x \sqrt{g} \left( \Phi R +  U(\Phi)\right),
\end{eqnarray}
parametrized by a single function $U(\Phi)$ called the \textbf{dilaton potential}. Let us first make some comments on this class of models.

Notice that the above field transformations from \eqref{eq:gendil} to \eqref{gendilgrav} are done at the classical level. In the full quantum theory, care has to be taken when performing these steps, in particular with regard to the thermodynamics of the resulting models.

We started with a two-derivative action \eqref{eq:gendil}. One can relax this assumption and include higher-derivative terms. The resulting model becomes power-counting non-renormalizable, but since there are no local degrees of freedom anyway, this may not be a problem. See \cite{Grumiller:2021cwg} for investigations of this generalization of Eq.~\eqref{eq:gendil}.
    
In some works, partially motivated by string theory, the dilaton field is denoted instead as $\Phi \to \Phi^2$ or $\Phi \to \exp(-2\Phi)$, emphasizing its positivity throughout spacetime. The dilaton coupling parametrizes a spacetime-dependent effective Newton's constant
    \begin{equation}
    \label{Geff}
        G_{\text{eff}}(x) = \frac{G_N}{\Phi(x)},
    \end{equation}
    and to interpret a classical solution physically, one wants this effective gravitational coupling to remain positive everywhere\footnote{This will be revisited when considering the integration contour for the dilaton in the quantum theory in Sect.~\ref{sec:JTquantum}.}. This will also lead to another more appropriate interpretation for the dilaton: as a measure of entropy.

\textbf{Jackiw--Teitelboim (or JT) gravity} corresponds to a specific linear choice of dilaton potential in \eqref{gendilgrav}, where $U(\Phi)=-\Lambda\Phi$ \cite{Jackiw:1984je,Teitelboim:1983ux}:
\begin{align}
I_{\text{JT}}^\Lambda[g,\Phi] = - \frac{1}{16 \pi G_N}\int_\mathcal{M}  \sqrt{g} \Phi\left(  R - \Lambda\right).
\end{align}
 The quantity $\Lambda$ (of dimension energy squared) is the cosmological constant of the model.
To focus on the Anti-de Sitter space (AdS) case, we choose negative $\Lambda$ and set $\Lambda = -2/L^2$. Working in units where the AdS length $L=1$, we can write down our JT model: 
\begin{align}
\label{JTaction}
I_{\text{JT}}[g,\Phi] = - \frac{1}{16 \pi G_N}\int_\mathcal{M}  \sqrt{g} \Phi\left(  R +2\right)- \frac{1}{8 \pi G_N}\oint_{\partial \mathcal{M}}\sqrt{h} \Phi (K-1).
\end{align}
For manifolds that have a boundary $\partial \mathcal{M}$, we wrote explicitly the Gibbons-Hawking-York (GHY) boundary term \cite{Gibbons:1976ue, York:1972sj}, and a holographic counterterm (the ``$-1$'') which is needed a posteriori to make sense of the model at an asymptotically AdS boundary as will become clear later on.
It will be important later to add the topological Einstein--Hilbert term as well, proportional to a parameter of the model $S_0$. The final action we will study throughout this review is hence
\begin{align}
\label{JTactiontotal}
    I[g,\Phi] = - S_0 \chi + I_{\text{JT}}[g,\Phi],
\end{align}
where we introduced the Euler characteristic of the manifold
\begin{equation}
\chi = \frac{1}{4\pi} \int_\mathcal{M} \sqrt{g} R + \frac{1}{2\pi} \oint_{\partial \mathcal{M}} \sqrt{h} K.
\end{equation}
The topological term will be important in Sect.~\ref{sec:JTRMT}, but in this section it will only produce an overall shift of the action. This term can be thought of as being produced by a shift in the dilaton field by a constant $\Phi(x) \to \Phi_0 + \Phi(x)$ producing the term \eqref{JTactiontotal} with $S_0 = \frac{\Phi_0}{4 G_N}$.

It is possible to study JT gravity in asymptotically de Sitter (dS) by choosing the same dilaton potential with positive $\Lambda = + 2/L^2$ and we will consider this theory briefly in Sect.~\ref{sec:JTdS}. Finally, the asymptotically flat case where $U(\Phi)={\rm constant}$, also has interesting applications which we will briefly mention there as well. For the body of this work, we will restrict to AdS. 

\subsubsection{First-order formulation}
\label{s:firstorder}
Just as any gravitational model, 2d dilaton gravity can be written in the first-order formulation. Unlike gravity in four or higher spacetime dimensions, the resulting model is a topological gauge theory: the Poisson-sigma model \cite{Schaller:1994es}. For the specific case of JT gravity, this gauge theory further simplifies into the topological BF model \cite{Horowitz:1989ng}.  Here we review this connection for closed manifolds $\mathcal{M}$. We start with the Euclidean dilaton gravity model:
\begin{align}
\label{secondorder}
I &= -\frac{1}{16\pi G_N}\int_{\mathcal{M}} d^2 x\sqrt{g}\, (\Phi R + U(\Phi)).
\end{align}
Introducing the frame field $e_\mu^a$ as:
\begin{equation}
g_{\mu\nu} = e_\mu^a e_\nu^b\delta_{ab},
\end{equation}
where $a, b\in \{0, 1\}$, and the torsion-free spin connection $\omega^{ab} = \smash{\omega_\mu^{[ab]}}\, dx^\mu$, determined by $de^a + \omega^a{}_b\wedge e^b = 0$, we have the following relations in 2d: 
\begin{equation}
\omega^{ab} = \epsilon^{ab}\omega, \qquad d^2 x\, \sqrt{g} = e^0\wedge e^1, \qquad d^2 x\, \sqrt{g}R = 2\, d\omega.
\end{equation}
We can then write the gravitational action \eqref{secondorder} in the first-order form:
\begin{align}
\label{firstor}
I &= -\frac{1}{8\pi G_N}\int_{\mathcal{M}} \left[\Phi\, d\omega + \frac{1}{4}U(\Phi)\epsilon^{ab} e_a \wedge e_b + X^a (de_a + \epsilon_a{}^b \omega \wedge e_b)\right].
\end{align}
Indeed, integrating over the Lagrange multipliers $X^{0,1}$, we produce the torsion-free conditions and the remaining action reduces back to the second-order form \eqref{secondorder}.
This model can be identified with a topological Poisson sigma model, with 3d target space \cite{Ikeda:1993aj,Ikeda:1993fh}. To see this, introduce a connection $A_i = (e_0, e_1, \omega)$ and a 3d space parametrized by scalars $X^i = (X^0, X^1, X^2\equiv\Phi)$. Poisson manifolds are characterized by a (possibly degenerate) bracket $\{ X^i, X^j \}_{\text{PB}} \equiv P^{ij}(X)$. The antisymmetric tensor $P^{ij}$ satisfies the Jacobi identity $\partial_l P^{[ij\vert} P^{l\vert k]} = 0$. In terms of this data the Poisson sigma model action is given by
\begin{equation}
\label{PSMbos}
I_{\text{PSM}} ~=~ -\frac{1}{8\pi G_N} \int_{\mathcal{M}} \left( A_i \wedge dX^i  + \frac{1}{2} P^{ij}(X) A_i \wedge A_j \right), \quad i=0,1,2,
\end{equation}
where the brackets needed to reproduce \eqref{firstor} can be read off as
\begin{equation}
\label{bospoi}
 P^{01} =\left\{X^0, X^1\right\}_{\text{PB}}= \frac{U(X^2)}{2}, \qquad  P^{a2} =\left\{X^a, X^2\right\}_{\text{PB}} = \epsilon^a{}_b X^b.
\end{equation}
Upon redefining the generators as $E^\pm \equiv -X^0\pm X^1$ and $H \equiv X^2$, this non-linear Poisson algebra becomes:
\begin{align}
\label{bosalg}
\left\{H,E^\pm\right\}_{\text{PB}} = \pm E^\pm, \qquad \left\{E^+,E^-\right\}_{\text{PB}} = U(H).
\end{align}
For the special case of JT gravity where $U(\Phi)=2\Phi$, this algebra becomes the linear $\mathfrak{sl}(2,\mathbb{R})$ Lie algebra. Starting with \eqref{PSMbos}, one can perform an integration by parts in the first term and, since the Poisson tensor $P^{ij}(X)$ is linear in $X$ in this case, rewrite the action \eqref{firstor} as a BF model \cite{Fukuyama:1985gg,Isler:1989hq,Chamseddine:1989yz,Jackiw:1992bw}:
\begin{equation}
\label{BFbos}
I ~=~ -\frac{1}{8\pi G_N} \int_{\mathcal{M}} X^i F_i ~=~ -\frac{1}{4\pi G_N} \int_{\mathcal{M}} \text{Tr}~X F, ~~~ F = dA + A \wedge A,
\end{equation}
where $X= X^i \lambda_i$ (and similarly for $F$) with $\{\lambda_1,\lambda_2,\lambda_3\}$ the generators of $\mathfrak{sl}(2,\mathbb{R})$, normalized such that $\text{Tr} \lambda_i \lambda^j = \delta_{i}{}^{j}/2$.

This discussion was valid for manifolds without boundaries. We will come back to the case with boundaries later on in Sect.~\ref{s:BF}. The $\mathfrak{sl}(2,\mathbb{R})$ BF description of JT gravity is extremely useful to obtain explicit solutions, but care has to be taken: the BF description is only \emph{locally} equivalent with JT gravity, and the relation with the global group-theoretic description is a bit more subtle, and contains several layers of complications:
\begin{itemize}
    \item A good gauge connection $A_\mu$ can be singular in Euclidean gravity. An example is $A_\mu=0$, leading to $g_{\mu\nu} = 0$. These singular metrics are to be excluded from the gravitational path integral. 
    \item Gravity contains large diffeomorphisms (such as the SL$(2,\mathbb{Z})$ modular group of the torus; more generally in 2d this is the mapping class group) that are not contained within the gauge transformations of the BF description.
    \item Gravity contains a sum over topologies that is not included in the gauge theory description.
\end{itemize}
All of these items are present for any model of dilaton gravity, but they can be dealt with very explicitly even in quantum JT gravity, for which we refer to the recent works \cite{Blommaert:2018iqz,Fan:2021wsb,Fan:2021bwt}, and to some of the older literature \cite{Schaller:1993np}. These complications parallel the difficulties in writing 3d gravity as a Chern--Simons gauge theory, explained nicely in \cite{Witten:2007kt}.

\subsubsection{Motivation: near-extremal black holes}
\label{sec:NEBH}
One of the attractive features of JT gravity is the fact that it universally describes the physics near the horizon of near-extremal black holes in higher dimensions. The simplest realization of this is a near-extremal charged black hole in 4d Einstein--Maxwell theory. The 4d theory has a $U(1)$ gauge field beside the metric, and the Lorentzian action is:
\begin{eqnarray}
\label{EM}
    S =  \frac{1}{16 \pi G_N^{(4)}} \int d^4 x \sqrt{-g} \left( R - F^2 \right) + S_{\rm bdy}.
\end{eqnarray}
$S_{\rm bdy}$ includes the necessary boundary terms and counterterms to make the variational problem well-defined, such as the Gibbons-Hawking-York one. The Reissner-Nordstr\"om black hole solution with charge $Q$ is given by 
\begin{eqnarray}\label{eq:RN4d}
    ds^2 = -f(r) dt^2 + \frac{dr^2}{f(r)} + r^2 d\Omega_2^2,~~~f(r)=1-\frac{2 M}{r} + \frac{Q^2}{r^2},
\end{eqnarray}
where we absorbed $G_N^{(4)}$ into $M$ for brevity of notation. The only non-zero component of the electric field is $F_{rt} \propto Q/r^2$, and $d\Omega_2^2$ is the metric element of the transverse 2-sphere. The horizons are at $r_\pm = M \pm \sqrt{M^2-Q^2}$ and the Hawking temperature of the black hole is $T=\vert f'(r_+)\vert /4\pi$. The near-extremal limit corresponds to low temperatures such that $M\sim \vert Q \vert $ and hence $r_+\sim r_- \sim \vert Q \vert$. From now on we take $Q>0$.

The first step to see how JT gravity arises from this 4d theory, is to identify the near-horizon near-extremal geometry. We consider the scaling limit of small $\Delta M \equiv M-Q$, where $r = Q + Q^2\tilde{r}$, and the radial coordinate scales as $\tilde{r}\sim \sqrt{\Delta M/Q^3}$. In this region, the spacetime metric \eqref{eq:RN4d} becomes:
\begin{equation}
    ds^2 \approx -Q^2\left(\tilde{r}^2-\frac{2\Delta M}{Q^3}\right) dt^2 + \frac{Q^2d\tilde{r}^2}{\tilde{r}^2-\frac{2\Delta M}{Q^3}} + Q^2 d\Omega_2^2.
\end{equation}
This is the near-horizon AdS$_2$ $\times$ S${^2}$ geometry, with AdS length and 2-sphere radius both $Q$.
Rescaling lengths by $Q^{-1}$, we can rewrite this in the canonical form:
\begin{equation}\label{eq:RNNH}
    Q^{-2} ds^2 \approx -\left(\tilde{r}^2-r_h^2\right) dt^2 + \frac{d\tilde{r}^2}{\tilde{r}^2-r_h^2} + d\Omega_2^2,~~~r_h = \sqrt{\frac{2 \Delta M}{Q^3}},
\end{equation}
with the AdS length and sphere radius equal to $1$ in these units. The exactly extremal case is found by setting $\Delta M =0$ which implies $r_h=0$. Otherwise, we have a black hole in AdS$_2$ with a horizon at a finite distance $r_h$. Within this AdS$_2$ region, $\tilde{r}-r_h \ll r_h$ corresponds to the Rindler region whereas $\tilde{r}  \to \infty$ corresponds to the boundary of AdS$_2$. 

\begin{figure}[t!]
    \centering
\includegraphics[scale=0.22]{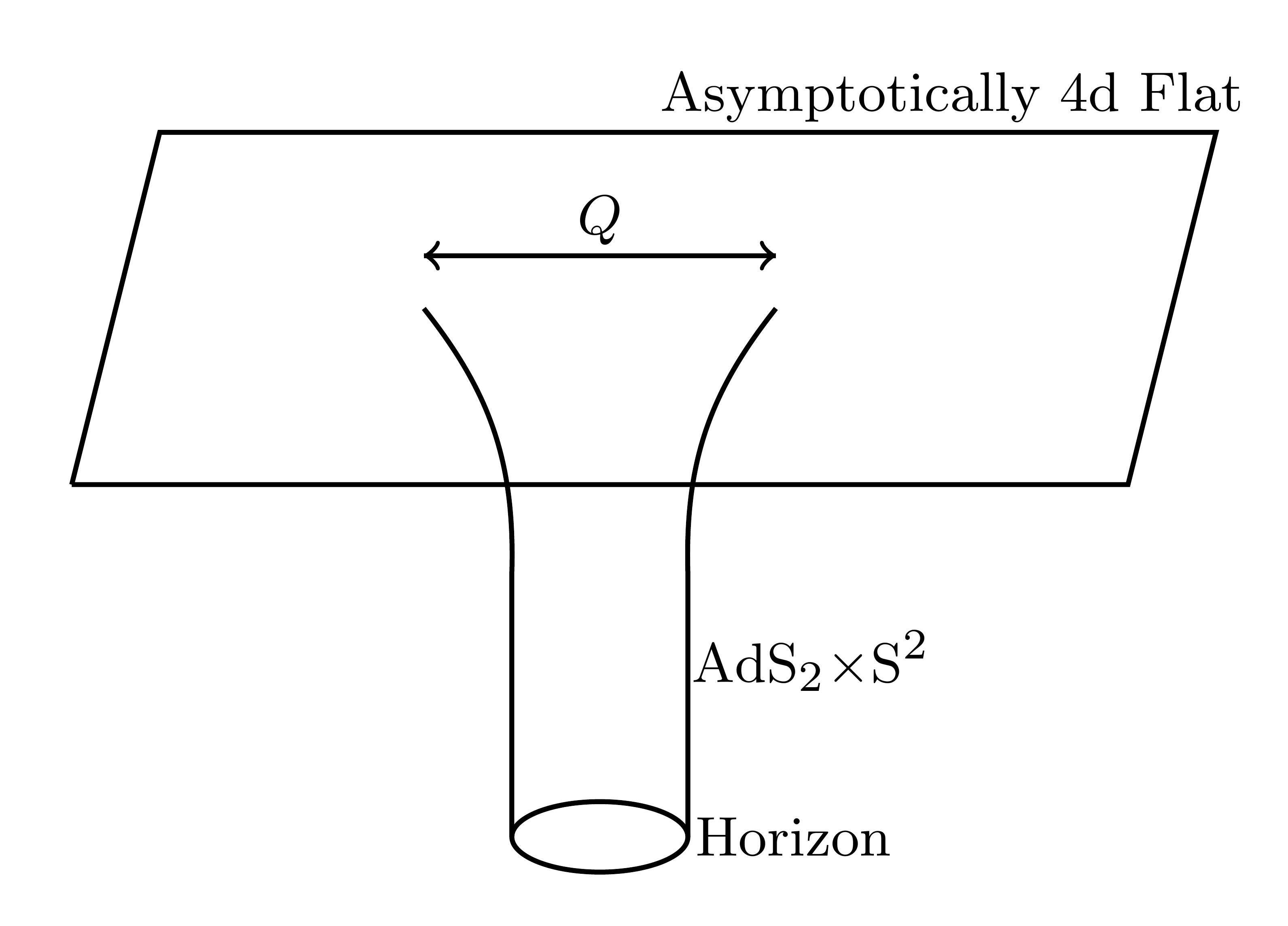}
    \caption{Spatial geometry of the Reissner-Nordstrom black hole which interpolates between flat space far away and an AdS${}_2\times$S${}^2$ throat near the horizon when the temperature is low. The size of the black hole is of order $Q$ (in appropriate units) while the length of the AdS${}_2\times$S${}^2$ throat diverges as we approach extremality. JT gravity arises as a sector of Einstein-Maxwell in 4d and is the dominant mode controlling the low energy dynamics taking place inside the throat. This is true for a large class of black holes with AdS${}_2\times$S${}^2$ throats.}
   \label{fig:my_labell}
\end{figure}

In Fig.~\ref{fig:my_labell}, we illustrate the spatial geometry of the Reissner-Nordstrom black hole which interpolates between flat space far away and an AdS${}_2\times$S${}^2$ throat near the horizon when the temperature is low. The size of the black hole is of order $Q$ (in appropriate units) while the length of the AdS${}_2\times$S${}^2$ throat diverges as we approach extremality. 

Having described the classical geometry near the horizon, we now study perturbations around it. Metric fluctuations can be parametrized by \cite{Davison:2016ngz,Nayak:2018qej,Sachdev:2019bjn}
\begin{eqnarray}\label{eqn:KKansatzRN}
ds^2 = \frac{1}{\sqrt{\chi}} g_{\mu \nu} dx^\mu dx^\nu + \chi (dy^i+A^{ij}_\mu y^j dx^\mu )( dy^i+A^{ij}_\mu y^j dx^\mu ) + \ldots ,
\end{eqnarray}
where $y=(y^1,y^2,y^3)$ with $y^iy^i=1$ parametrize the transverse $S^2$. The coordinate $x^\mu=(t,r)$ parametrizes the time and radial directions. Both $\chi$, $g_{\mu\nu}$ and $A^{ij}_\mu$ are assumed to depend only on $x^\mu$. The dots in the equation above denotes other Kaluza-Klein modes that are not turned on in the Reissner-Nordstrom solution and whose fluctuations around the background are massive, with masses $\sim Q^{-1}$ the AdS scale and hence not parametrically large for a near-extremal black hole. Therefore the correct low-energy description involves two-dimensional gravity with a large number of light $(\sim Q^{-1}$) fields.\footnote{This feature is not different from other realizations of AdS. For example in a five-dimensional description of the AdS${}_5$$\times$S${}^5$ background in type IIB string theory, there are also a large number of light fields from compactification on the sphere.} The one-forms $A^{ij}$ reduce on the 2d space to an SU$(2)$ gauge field whose charge corresponds to angular momentum $J$.

For simplicity, we work in an ensemble with fixed black hole charge and fixed vanishing angular momentum. Then the Einstein--Maxwell action \eqref{EM} dimensionally reduces to the restricted 2d dilaton-gravity dynamics (with dilaton $\chi$):\footnote{The coefficient of the Einstein term in two dimensions vanishes when $\chi=0$. This corresponds to the singularity of the 4d black hole. Since we will focus on the physics near the horizon we will not worry about that, although it is important to keep in mind.}
\begin{eqnarray}
    S[g,\chi] = \frac{1}{4 G_N^{(4)}} \int d^2 x~\sqrt{-g} \left( \chi R +  U(\chi)\right) + \ldots, \quad U(\chi) = -\frac{2Q^2}{ \sqrt{\chi}^3} + \frac{2}{\sqrt{\chi}},\nonumber
\end{eqnarray}
where we can define the 2d Newton's constant by $G_N^{(4)} = 4\pi G_N^{(2)}$. Turning on angular momentum would correct the potential by a term proportional to $J^2$. The classical solution of this theory, with appropriate boundary conditions, reproduces \eqref{eq:RN4d}. The dots in the equation above denote contribution from either matter or other Kaluza--Klein modes of the Einstein--Maxwell sector, resulting in 2d dilaton-gravity coupled to matter.

A classical solution with constant sphere area has $U(\chi=\Phi_0)=0$ and therefore $\Phi_0 = Q^2$, which corresponds to rigid AdS$_2$ $\times$ S$^2$ spacetime. This is the geometry very close to the horizon of the near-extremal black hole as presented above. JT gravity arises as we take the $S^2$ area large, and allow it to fluctuate slightly:
\begin{eqnarray}
\label{eq:dilcond}
    \chi(x) = \Phi_0 + \Phi(x) , \qquad \Phi \ll \Phi_0.
\end{eqnarray}
Under this assumption the 4d Einstein--Maxwell theory reduces near the horizon to JT gravity, after rescaling the 2d metric $g\to Q^3 g$:
\begin{equation}
    S[g,\Phi] =  \frac{\Phi_0}{4G_N^{(4)}} \int \sqrt{-g} R + \frac{1}{4G_N^{(4)}} \int \sqrt{-g}\Phi (R + 2) + \ldots.
\end{equation}
Corrections to this action are suppressed in the regime \eqref{eq:dilcond} and we end up with the JT action \eqref{JTaction} coupled to matter arising both from 4d matter as well as metric KK modes we ignored in \eqref{eqn:KKansatzRN}. (To leading order in the large $\Phi_0$ limit, the matter fields do not couple to the dilaton $\Phi$, a feature that will be important later when studying quantum effects.) Hence JT gravity arises as a sector in 4d Einstein--Maxwell theory, and it is the dominant mode controling the low-energy dynamics taking place inside the throat. 

This derivation also gives a nice interpretation of each term that appears in the lower-dimensional gravity action. We can identify the parameter $S_0$ in \eqref{JTactiontotal} with 
\begin{eqnarray}
    S_0 =  \frac{\pi\Phi_0}{G_N^{(4)}} = \frac{\pi Q^2}{G_N^{(4)}},  
\end{eqnarray}
which is the Bekenstein--Hawking area term corresponding to the 4d extremal black hole. Note that from the 2d perspective, this is $A/4G_N^{(2)}$ where $\Phi_0$ plays the role of the area $A$. We will see more evidence later that the dilaton should indeed be identified with the entropy of the black hole. We might be tempted to refer to the extremal Bekenstein-Hawking area $S_0$ as extremal degeneracy of black hole microstates, although we will see this is an incorrect interpretation both in JT gravity \cite{Stanford:2017thb} and for the higher-dimensional black hole. This point was made for the near-extremal BTZ black hole in \cite{Ghosh:2019rcj} and then generalized to the Reissner-Nordstr\"om black hole  in  \cite{Iliesiu:2020qvm, Heydeman:2020hhw}. The reason is that quantum effects from the JT mode get enhanced at low enough temperatures and, in a fixed charge ensemble, only the JT mode affects the temperature-dependence of the free energy, even at the quantum level.   

The 4d Reissner--Nordstr\"om black hole is not the only higher-dimensional black hole whose near-horizon near-extremal dynamics is governed by JT gravity, but there is in fact a large class. We will discuss this further in Sect.~\ref{sec:NEBHc}. The \textbf{universality} of JT gravity in this kinematic regime of black hole physics is hence a strong motivation for studying the JT model in more detail.

\subsection{Classical solutions}
In this section, we will solve and study the classical equations of motion of Lorentzian JT gravity, including its coupling to a matter action:
\begin{align}
\label{JTLor}
    \hspace{-0.cm}S= \frac{\Phi_0}{16\pi G_N}  \left[\int  \sqrt{-g} R + 2\oint \sqrt{-h} K\right] + S_{\rm JT}[g,\Phi]+S_m[\phi,g].
\end{align}
The first term corresponds to a constant dilaton background $\Phi_0$ and is the Lorentzian version of Eq.~\eqref{JTactiontotal} where $S_0 = \Phi_0/(4G_N)$. The second term is the JT action itself with a varying dilaton
\begin{align}
\label{JTLor2}
    S_{\rm JT}[g,\Phi] =  \frac{1}{16\pi G_N}\left[\int  \sqrt{-g} \Phi\left(  R +  2\right)+ 2\oint \sqrt{-h} \Phi(K-1)\right] .
\end{align}
Finally, $S_m[\phi,g]$ is the matter action, which we assume only couples to the metric, and not directly to the dilaton field $\Phi$. This assumption will allow for an explicit and analytic solution.\footnote{This is true when applying JT gravity to the dynamics of gravity and matter near the horizon of a near-extremal black hole in Einstein--Maxwell gravity. Another example is given by extremal dilatonic black holes in four or higher dimensions in superstring theories, where the relevant matter action coming from Ramond-Ramond fields, describing the radial dynamics, is just a free scalar field in 2d, and would indeed not have a direct coupling to the dilaton field \cite{Callan:1992rs}.}
We will not have to be explicit about the matter action in order to write down the most general solution of Einstein's equations, which is one of the benefits of working in lower dimensions.

\subsubsection{Metric solution}
\label{s:coordinates}
Now we describe general classical solutions of the matter-coupled model \eqref{JTLor}. The equation of motion one obtains by varying \eqref{JTLor} with respect to the dilaton $\Phi$ is given by
\begin{align}
\label{eqdil}
  R(x) = -2.
\end{align}
In two dimensions, knowing the value of the local Ricci scalar is sufficient to know the metric itself up to coordinate transformations, in this case some patch of the AdS$_2$ manifold. Let us make this statement more explicit.

It is convenient to describe the 2d metric in conformal gauge, which can always be reached by a coordinate transformation, where:
\begin{equation}
\label{confgauge}
    ds^2 = - e^{2\omega(u,v)}dudv,
\end{equation}
where the only metric degree of freedom is $\omega(u,v)$. The coordinates $u$ and $v$ are lightcone coordinates defined as $u=t+z$ and $v=t-z$. 
For a conformal gauge metric \eqref{confgauge}, we have $R=8e^{-2\omega}\partial_u \partial_v \omega$, and \eqref{eqdil} boils down to  Liouville's equation $4\partial_{u}\partial_{v}\omega + e^{2\omega} = 0$. The general solution to Liouville's equation is well-known:
\begin{equation}
\label{eq:solliou}
    e^{2\omega(u,v)} = \frac{4 \partial_u U(u)\partial_v V(v)}{(U(u) - V(v))^{2}},
\end{equation}   
in terms of two chiral functions $U(u)$ and $V(v)$. This solution can be restricted when imposing boundary conditions as we will implement further on in Sect.~\ref{s:bcclas}. With this solution, the metric \eqref{confgauge} is
\begin{equation}
\label{bulkgauge}
    ds^2 = -\frac{4 \partial_u U(u)\partial_v V(v)dudv}{(U(u) - V(v))^{2}} = -\frac{4 dUdV }{(U - V)^{2}},
\end{equation}
and represents the Poincar\'e patch of AdS$_2$ with lightcone coordinate $(U,V)$. The two chiral functions $U(u)$ and $V(v)$ are interpreted as the diffeomorphisms that preserve conformal gauge, and relate the lightcone coordinates $U$ and $V$ of the Poincar\'e patch to new lightcone coordinates $u$ and $v$.
This proves that all metric solutions are just different coordinate frames on the AdS$_2$ manifold. 

It is useful to summarize the geometric properties of several important classical frames of AdS$_2$. The \textbf{Poincar\'e patch} itself is described by the metric
\begin{equation}
\label{eq:Pcpatch}
    ds^2 =  \frac{-dT^2+dZ^2}{Z^2} ~=~ -R^2dT^2 + \frac{1}{R^2}dR^2 ~=~ -\frac{4 dUdV }{(U - V)^{2}}, \qquad U>V,
\end{equation}
where we defined $R=1/Z$ in the first equality. The isometry group of AdS$_2$ is PSL$(2,\mathbb{R}) \simeq $ SO$(2,1)$, which acts as M\"obius transformations on the Poincar\'e lightcone coordinates $U \to \frac{aU+b}{cU+d}, \, V\to \frac{aV+b}{cV+d}$, where $a,b,c,d \in \mathbb{R}$ and $ad-bc=1$. PSL$(2,\mathbb{R})$ is the projective subgroup of  SL$(2,\mathbb{R})$, the group of invertible $2\times 2$ matrices with determinant 1, where we identify the matrix with minus itself.

\begin{figure}
    \centering
        \includegraphics[scale=0.3]{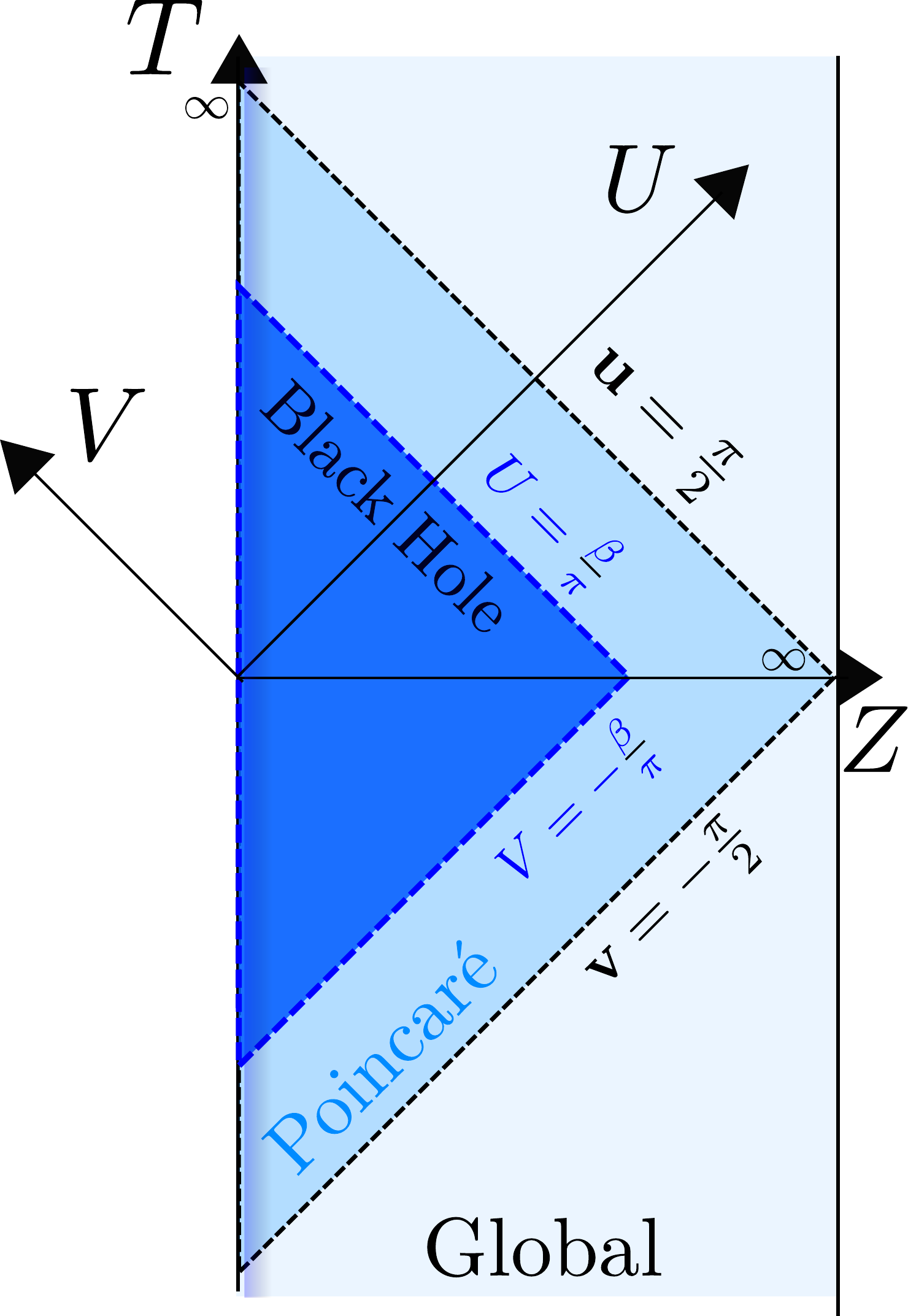} 
        \caption{Penrose diagram with the different classical patches of global AdS$_2$.}
        \label{fig:PenroseAdS}
 \end{figure}

We denote the Poincar\'e time coordinate as $T=(U+V)/2$ and the spatial coordinate as $Z=(U-V)/2$. It has a single boundary at $U=V$ (or $Z=0$). It has horizons at $U-V \to \infty$, or $U\to + \infty$ (future horizon) and at $V \to -\infty$ (past horizon). This geometry is found in the near-horizon limit of a higher-dimensional extremal black hole, the $r_h=0$ limit of Eq.~\eqref{eq:RNNH}. The horizons are at an infinite proper distance ($\int^{+\infty} \frac{dZ}{Z} \to \infty$) at the end of an infinite throat. Starting with the Poincar\'e patch, we can reach other important frames that we discuss next. We illustrate these in Fig.~\ref{fig:PenroseAdS}, and the reader can consult this figure throughout.
 
The \textbf{global frame} is found by performing the following coordinate transformation
\begin{equation}
\label{glframe}
U(\mathbf{u}) = \tan(\mathbf{u}), \quad V(\mathbf{v}) = \tan(\mathbf{v}),
\end{equation}
defining new lightcone coordinates $(\mathbf{u},\mathbf{v})$. The metric \eqref{eq:Pcpatch} is transformed into:
\begin{eqnarray}
   ds^2 = - \frac{4}{\sin^2(\mathbf{u}-\mathbf{v})}d\mathbf{u} d\mathbf{v} ,
\end{eqnarray}
and can be continued beyond the original Poincar\'e patch. We define the global spatial coordinate $\mathbf{z}=(\mathbf{u}-\mathbf{v})/2$ and time coordinate $\mathbf{t}=(\mathbf{u}+\mathbf{v})/2$. There are two boundaries where the metric blows up: this is at $\mathbf{u}=\mathbf{v}$ (or $\mathbf{z}=0$), and at $\mathbf{u}=\mathbf{v}+\pi$ (or $\mathbf{z}=\pi/2$). These coordinates cover a strip-region with two boundaries containing the Poincar\'e patch. Usually in AdS$_D$/CFT$^{D-1}$, the boundary in global coordinates is the $\mathbb{R} \times S^{D-2}$ manifold, which in this case gives a 0-sphere for the spatial manifold, or two disjoint points.

Finally, the \textbf{black hole patch} can be found by the following coordinate transformation:
\begin{equation}
\label{BHframe}
U(u) = \frac{\beta}{\pi}\tanh\left(\frac{\pi}{\beta} u\right), \quad V(v) = \frac{\beta}{\pi}\tanh\left(\frac{\pi}{\beta} v\right),
\end{equation}
with parameter $\beta$. This leads to the metric
\begin{equation}
\label{eq:bhpatch}
   ds^2 = - \frac{\pi^2}{\beta^2}\frac{4}{\sinh^2(\frac{\pi}{\beta}(u-v))}du dv .
\end{equation}
We define also the coordinates $z=(u-v)/2$ and $t=(u+v)/2$. The black hole patch \eqref{eq:bhpatch} is contained within the Poincar\'e patch \eqref{eq:Pcpatch} since the Poincar\'e coordinates only range over the restricted range $-\frac{\beta}{\pi} < U,V < \frac{\beta}{\pi}$ (with $U>V$). This geometry has horizons where $u-v \to \infty$ at finite distance ($\int^{+\infty} \frac{dz}{\sinh \frac{2\pi}{\beta} z} < \infty$). Using the radial coordinate $r =  r_h\coth \frac{2\pi z}{\beta}$ where $r_h = 2\pi/\beta$, the metric can be rewritten as
 \begin{eqnarray}
 \label{JTbh}
     ds^2 = -(r^2-r_h^2) dt^2 + \frac{dr^2}{r^2-r_h^2}.
 \end{eqnarray}
 This patch is directly found in the near-horizon regime of a higher-dimensional near-extremal black hole, as in \eqref{eq:RNNH}. 
 From the near-horizon region where $r\approx r_h$ of the Euclidean section, we can deduce the Hawking temperature $T = \beta^{-1} = r_h/2\pi$, justifying the parameter label $\beta$ as the inverse temperature. This is simultaneously the temperature of the higher-dimensional near-extremal black hole.  One further way of rewriting this patch is to use the proper distance coordinate $\rho$ to the black hole horizon, which allows us to write the same patch as
 \begin{equation}
 \label{rhoco}
  \sinh \rho = \frac{1}{\sinh \frac{\pi}{\beta}(u-v)}~~~\rightarrow~~~   ds^2 = - \frac{4\pi^2}{\beta^2}\sinh^2 \rho dt^2 + d\rho^2.
 \end{equation}
 Near the horizon where $\rho \ll 1$, this becomes the Rindler geometry.\footnote{Note that these two new radial coordinates $r$ and $\rho$ are found by doing purely spatial coordinate transformations, which hence preserve the causality structure, the thermodynamics, and the black hole interpretation of the spacetime.} Some early references on the black hole solution are \cite{Lemos:1993qn, Lemos:1993py, Lemos:1996bq}. 
 
Since the metric solution is always a patch of the AdS$_2$ manifold, our model can be interpreted in a holographic context with holographic boundary at $Z=0$. However, AdS$_2$/CFT$_1$ has been notoriously tricky to make sense of. Resolving the tensions around it, has been one of the chief successes of JT gravity as we illustrate below. For now, let us review some arguments why this version of holography is subtle. 

Firstly, for a CFT in any number of dimensions, the stress tensor is traceless: $T_\mu{}^\mu  =0$.\footnote{In an even number of spacetime dimensions, there is the famous trace anomaly that violates this equality at the quantum level, but here we are in 0+1 dimensions.} But in 0+1d there is only one index, and hence $T_{tt} =0$. Hence the energy is always zero, and we are only able to describe a theory of ground states, or extremal states. Whereas this is consistent, it is not our main interest when considering dynamical models of black hole formation and evaporation. 
Even though the metric is locally AdS$_2$, we have seen the physics in each of the patches described above is very different. Whether there is a black hole horizon or not, and the global structure of spacetime, can vary from patch to patch. So far we have not given a physical mechanism within the theory that fixes this ambiguity, something one cannot do without a (non-zero) boundary Hamiltonian $T_{tt}$. In the example in Sect.~\ref{sec:NEBH}, the black hole patch is fixed by the way the AdS$_2$ region is glued to the asymptotically flat space. We will see in the next subsection how to phrase this intrinsically in terms of JT gravity, and the price to pay is the loss of conformal invariance.

Secondly, due to a lack of dimensionful parameters, the density of states of a 1d CFT is by dimensional analysis of the form \cite{Jensen:2011su}
\begin{equation}
    \rho(E) = A \delta(E) + \frac{B}{E},
\end{equation}
where $A$ and $B$ are dimensionless.\footnote{In higher dimensions, one can have the volume of space as a further dimensionful factor, which allows for other possibilities.}
Now, either $B=0$, in which case we again have a theory of ground states as before, or $B \neq 0$ in which case the number of low-energy states diverges, and the model does not work without an additional IR cutoff.

We summarize that genuine AdS$_2$/CFT$_1$ only makes sense as a model of zero-energy states (as topological QM), but to describe dynamics, we would have to modify this framework. This is precisely what JT gravity achieves as will become apparent further on. Moreover, in examples where a genuine AdS$_2$/CFT$_1$ limit exists the bulk theory is strongly coupled and does not have a simple description \cite{Heydeman:2020hhw, Boruch:2022tno,Lin:2022rzw,Lin:2022zxd}.

\subsubsection{Dilaton solution}
\label{s:dilsol}
Let us continue with the classical solution of the JT model. Varying \eqref{JTLor} with respect to the metric, we find the equations of motion for the dilaton field:
\begin{equation}
\label{dileomcov}
\nabla_\mu\nabla_\nu \Phi - g_{\mu\nu} \nabla^2 \Phi + g_{\mu\nu} \Phi = -8 \pi G_{N} T_{\mu\nu},
\end{equation}
or when written out in conformal gauge:
       \begin{align}
       \label{dileom}
       \nonumber
    - e^{2\omega}\partial_{u}\left(e^{-2\omega}\partial_{u}\Phi\right) &= 8\pi G_N \, T_{uu} \,,\\
	-e^{2\omega}\partial_{v}\left(e^{-2\omega}\partial_{v}\Phi\right) &= 8\pi G_N \, T_{vv} \,,\\
	\nonumber
	2\partial_{u}\partial_{v}\Phi + e^{2\omega}\Phi &= 16 \pi G_N \, T_{uv},
\end{align}
where $T_{\mu\nu}$ are the stress tensor components of the matter sector in \eqref{JTLor}. Notice that in this model all backreaction of matter fields is contained in the dilaton field $\Phi$ and \emph{not} in the actual geometry itself.

These coupled equations of motion \eqref{dileom} can be solved systematically by integrating the first and second equation twice, and then inserting it in the last equation to find a consistency requirement. 

Let us first describe the vacuum solutions where $T_{uu}=T_{vv}=T_{uv}=0$. It is easy to show by directly using \eqref{dileomcov} that the quantity $\zeta^\mu =\epsilon^{\mu\nu}\partial_\nu \Phi$ is a Killing vector of the entire dilaton gravity system: $\nabla_\mu \zeta_\nu + \nabla_\nu \zeta_\mu = 0$. This is an important observation for later on: in Euclidean signature, hyperbolic surfaces ($R=-2$) at higher genus $g\geq 2$ do not have Killing vectors. So there will be no classical solutions to JT gravity on higher genus surfaces.

The most general solution of Eq.~\eqref{dileom} is in this case described by three integration constants $a,b$ and $\mu$:
\begin{equation}
\label{cldil}
    \Phi(u,v)  ~~=~~  \frac{a + b(U+V)-\mu UV}{U- V}.
\end{equation}
We note the following: the dilaton solution blows up on the holographic boundary ($U=V$). The integration constant $a$ has the dimension of length, and will be important to understand this divergent nature of the solution. The parameter $\mu$ has the dimension of energy, and will be identified with the mass of a black hole later on. The parameter $b$ will turn out to be not physically meaningful (one can set $b=0$ by using an PSL$(2,\mathbb{R})$ isometry of the metric as discussed below \eqref{eq:Pcpatch}, and we assume this has been done from here on). 

Next, we consider non-zero matter sources. For simplicity of the expressions, we restrict to a 2d conformal field theory (CFT) as our matter sector. We will make a brief comment on the more generic case later on, but we point out that for the purposes of finding good models of black hole formation and evaporation, any (conserved) matter will do. The solution with conformal classical matter, for which $T_{uv}=0$ and by energy conservation $T_{uu}(u)$ and $T_{vv}(v)$ are chiral functions as denoted, is given by \cite{Almheiri:2014cka}
\begin{align}
\label{Phisol}
 \Phi(u,v)  ~~=~~  &\frac{a}{U- V}\left( 1   - \frac{\mu}{a} UV - \frac{8\pi G_N}{a} \,( I_+ +  I_- )\right), 
 \end{align}
where $I_{\pm}$ are given in terms of the matter stress-tensor by
\begin{align}
\quad\, &I_+(u,v)  =    \int_{U}^{+\infty } d s\, {(s - U)(s - V)}\, T_{UU}(s),\\
	\nonumber
	&I_-(u,v)  =   \int_{-\infty}^{V } d s\, (s - U)(s - V) \, T_{VV}(s),
\end{align}
and $T_{UU}dU^2 = T_{uu} du^2$ and $T_{VV}dV^2 = T_{vv} dv^2$. In the rest of this subsection we point out some important features of this solution. It is useful to first work out a simple example.

\vspace{0.1cm}

\begin{figure}[t!]
\centering
        \includegraphics[width=0.3\textwidth]{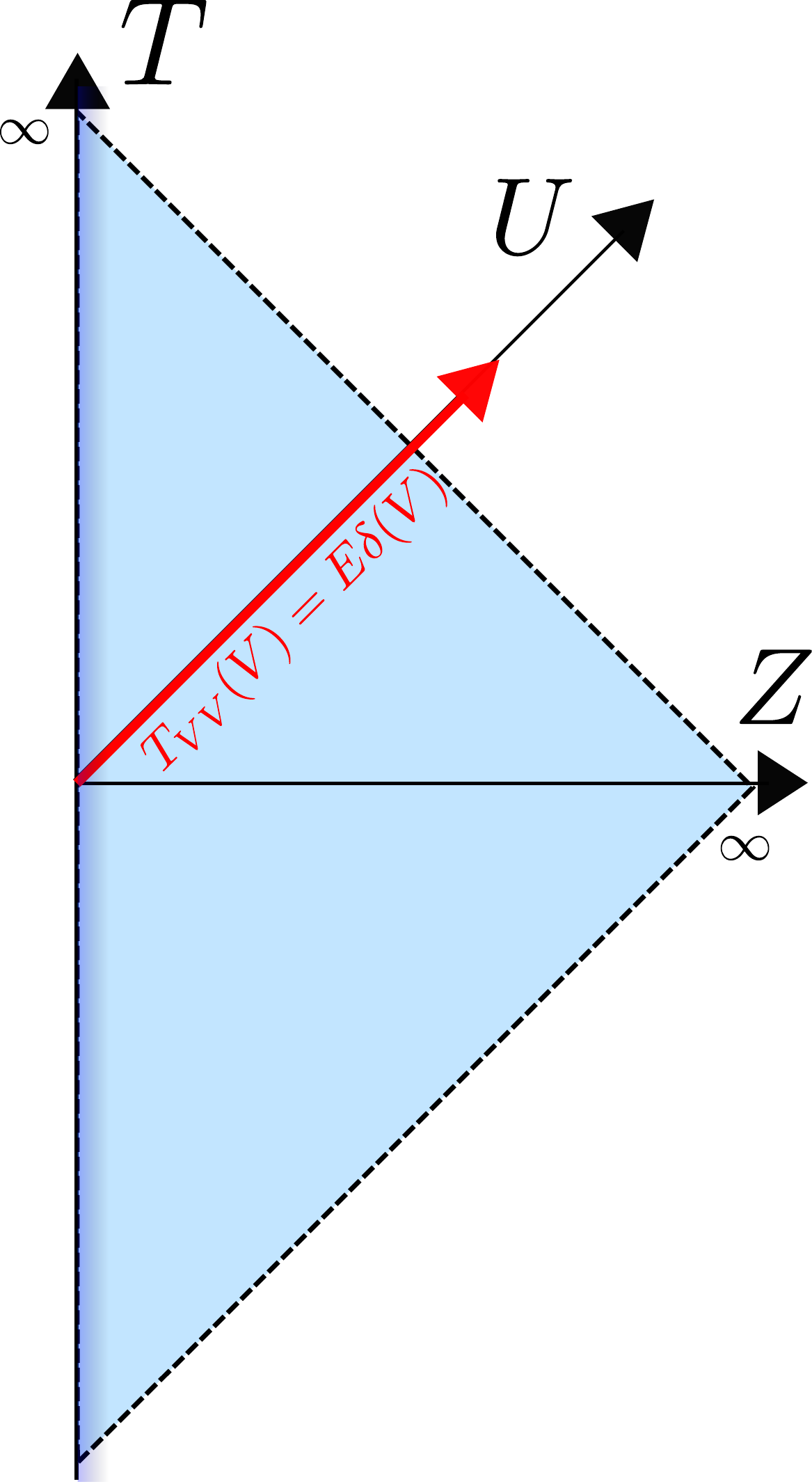}
    \caption{Penrose diagram of AdS$_2$ in the Poincar\'e patch. An energy pulse is injected at time $T=0$, indicated in red.}
    \label{Fig:pulse}
\end{figure}

\noindent \textbf{Example}:\\
Consider an infalling energy pulse of energy $E>0$ in the Poincar\'e patch, modeled as $T_{VV}(V) = E \delta( V)$ as in Fig.~\ref{Fig:pulse}. This immediately leads to $I_+= 0$ and $I_- = E UV\theta(V)$, and hence the dilaton profile
\begin{equation}
\label{dilpulse}
    \Phi = \frac{a - 8\pi G_N E UV \theta(V)}{U - V}.
\end{equation}

Comparing with \eqref{cldil}, we see that after the pulse has passed, we have created the vacuum solution where one identifies the parameter $\mu = 8\pi G_N E$. \\
\noindent We will retake this specific example at multiple instances further on as we develop the classical model. We have interpreted the parameter $\mu$ of Eq.~\eqref{cldil} in terms of the total spacetime energy $E$.

It will be useful to describe this finite $E$ solution using black hole coordinates introduced in Sect.~\ref{s:coordinates}, and it is convenient to rewrite it in several ways as:
\begin{equation}
\label{dilbh}
    \Phi = \frac{a - \mu U(u)V(v)}{U(u)-V(v)} = \sqrt{a \mu} \coth \left(\sqrt{\frac{\mu}{a}} 2 z\right) = \frac{a}{2}r,
\end{equation}
using the coordinates of Eqs.~\eqref{eq:bhpatch} and \eqref{JTbh}.
\begin{center}
---o---
\end{center}
\vspace{0.3cm}

Imagine we set $a=0$ a priori in \eqref{Phisol}, such that initially $\Phi(u,v)=0$. Then sending in some matter through $T_{uu}$ or $T_{vv}$ suddenly causes $\Phi$ at $Z=0$ to diverge. This runaway backreaction at the holographic boundary was first noticed in \cite{Maldacena:1998uz}, where it was argued to ruin Maldacena's decoupling limit \cite{Maldacena:1997re} for AdS$_2$/CFT$_1$. This is one of the bulk avatars of the difficulties with AdS$_2$/CFT$_1$. We see here that the resolution is to consider backgrounds with $a\neq 0$, where the dilaton always diverges at the boundary. The rate of divergence of the solution then contains the information on backreaction. We elaborate on this point in the next subsection. 

Moreover, in order to assess the regime of validity of the classical approximation of this section, it is useful to determine the location of possible singularities. Since the metric is always AdS$_2$, there are no curvature singularities. However, tracking the dilaton profile, there can be places where the effective Newton's constant blows up. 
This corresponds to points where $\Phi_0 + \, \eqref{Phisol} =0$. One expects the classical approximation (studied in this section) to break down far before reaching this region.\footnote{In the case considered in Sect.~\ref{sec:NEBH}, where JT gravity arises from a near-extremal black hole in 4d, the locus $\Phi_0 + \Phi=0$ corresponds to $r=0$ in coordinates \eqref{eq:RN4d}, \textit{i.e.} the black hole singularity. At this point, the expansion studied in Sect.~\ref{sec:NEBH} breaks down as well.} Demanding this locus to be far away from the holographic AdS$_2$ boundary for the solution \eqref{Phisol}, restricts us to $a > 0$, providing an independent reason to regularize the backreaction by introducing the non-zero quantity $a$.

An interesting point that emphasizes the connection between the dilaton and entanglement entropy is the following: the quantities $I_+/(U-V)$ and $I_-/(U-V)$ introduced in \eqref{Phisol} match with expressions of modular Hamiltonians of a 2d CFT in an interval between the point $(U,V)$ and a boundary at $Z=0$. This observation was explored in \cite{Callebaut:2018nlq,Callebaut:2018xfu} to show that JT gravity can be viewed as an emergent model describing the entanglement dynamics of a 2d CFT on a halfspace. 

The generic dilaton solution with arbitrary non-conformal matter is also known \cite{Joshi:2019wgi}, albeit is less elegant.

\subsection{Boundary conditions and Schwarzian dynamics}
\label{s:bcclas}
In order to proceed further, we need to implement suitable boundary conditions at the holographic boundary $u=v$. Several equivalent perspectives have been developed to both motivate and implement these efficiently. An approach rooted in hydrodynamic effective actions can be found in \cite{Jensen:2016pah}. In parallel, an insightful Euclidean geometric argument was constructed in \cite{Maldacena:2016upp}, while in \cite{Engelsoy:2016xyb}, generalizing the set-up of \cite{Almheiri:2014cka}, a real-time perspective at the level of the equations of motion was developed. These arguments have been reviewed in many works by now. Here we focus first on the approach of \cite{Engelsoy:2016xyb}, and then summarize the geometric route taken by \cite{Maldacena:2016upp}.

\subsubsection{Real-time derivation}
\label{s:lorderiv}
We impose two boundary conditions that will restrict the freedom in the coordinate transformations at the boundary as follows. 

\textbf{1. The geometry is asymptotically AdS}. We demand any bulk solution to have the leading asymptotics in Fefferman-Graham gauge:
    \begin{equation}
    \label{aspoinc}
        ds^2 = \frac{-dt^2 + dz^2}{z^2} + \text{(subleading as $z\to 0$)}.
    \end{equation}
    If we now perform an arbitrary (non-chiral) coordinate transformation in lightcone coordinates $U(u,v)$ and $V(u,v)$, then it is easy to show that in order to preserve the asymptotics \eqref{aspoinc}, we require: 
    ${}$ ($i$) $\partial_u V = 0 = \partial_v U$ at leading order as $z\to 0$, 
    ${}$ ($ii$) $U(u,v)= V(u,v)$ at leading order as $z\to 0$. 
    The first identity leads to chiral functions, and the second shows that both functions need to be the same. Translating this to new time and radial coordinates through $U\equiv F+Z$ and $V\equiv F-Z$, we obtain the near-boundary relations:
    \begin{align}
        F(t) &\equiv \frac{1}{2}\left(U(t+\epsilon) + V (t-\epsilon)\right) = U(t)+\mathcal{O}(\epsilon), \\
        Z(t) &\equiv \frac{1}{2}\left(U(t+\epsilon) - V (t-\epsilon)\right) = \epsilon F'(t)+\mathcal{O}(\epsilon^2).
    \end{align}
    where $\epsilon$ is viewed as the near-boundary cutoff $z = \frac{u-v}{2} \approx \epsilon$, regularizing the holographic boundary by moving it slightly inwards from $z=0$ to $z=\epsilon$.
    
    In Poincar\'e coordinates, this leads to a boundary trajectory determined in terms of a single function $F(t)$, as $(T=F(t), Z=\epsilon F'(t))$. This function $F(t)$ parametrizes the subset of large diffeomorphisms that preserves the asymptotics \eqref{aspoinc}, and by the usual dictionary in AdS/CFT, these generate the group of conformal transformations (i.e. arbitrary time reparametrizations in 1d) in the dual QM. However, this is not the end of the story since for all classical solutions, just like the metric, the dilaton field \eqref{Phisol} blows up as $z\to 0$ as well. 
    
\textbf{2. The dilaton field asymptotics is fixed}. We have argued that a solution with finite $\Phi$ at the boundary is not consistent with backreaction. To avoid this we choose the divergent dilaton asymptotics: 
    \begin{equation}
    \label{dilbc}
        \Phi = \frac{a}{2z}+(\text{subleading as $z\to 0$}).
    \end{equation}
    The parameter $a$ of dimension length defines the particular model we are studying. 
    
    A physical way to motivate this boundary condition, is that to define the timelike trajectory of the boundary curve, one has to impose a scalar condition. Making the boundary follow a fixed value of the dilaton field $\Phi$ is a natural coordinate-invariant way of defining this line in a dynamical situation. 
    From a higher-dimensional perspective, the dilaton field descends from the transverse area of the fixed radial hypersurface, as discussed in Sect.~\ref{sec:NEBH}. Keeping fixed the size of this surface is an invariant way of defining a radial location $r$ in a dynamical bulk geometry.
    
    Implementing this boundary condition by inserting the dilaton solution \eqref{Phisol}, we obtain the following integro-differential equation for $F(t)$:
\begin{align}
    \label{integrodiff}
        &F'(t) =  1 \hspace{-0.05cm}- \hspace{-0.05cm}\frac{8\pi G_N}{a} \left( \int_{F(t)}^{+\infty } \hspace{-0.2cm} d s\, {(s \hspace{-0.05cm}- \hspace{-0.05cm}F(t))^2} T_{UU}(s) +  \int_{-\infty}^{F(t) } \hspace{-0.2cm} d s\,(s \hspace{-0.05cm}- \hspace{-0.05cm}F(t))^2  T_{VV}(s) \right), 
\end{align}
where we have set $\mu=0$ since it can be created by the matter as discussed earlier in the example in Sect.~\ref{s:dilsol}. This procedure \textbf{breaks the conformal symmetry} of the boundary theory, and leads to a fixed and preferred boundary time frame $F(t)$ determined by the above equation.\footnote{\label{fn9} A finite-dimensional subgroup of the infinite dimensional conformal group is typically left unbroken due to the isometries of AdS$_2$. To appreciate this, consider the metric equation $g_{uv} = \frac{\partial_u F(u) \partial_v F(v)}{(F(u)-F(v))^2}$ in terms of $F$. In transfering from $g_{uv}$ to $F$, there is a redundancy since all $F$ related by PSL$(2,\mathbb{R})$ transformations as $F \to \frac{aF+b}{cF+d}$ yield the same metric, and hence have to be identified. In hyperbolic spaces with other topologies, only a subgroup of PSL$(2,\mathbb{R})$ is redundant when demanding compatibility with suitable periodicity conditions on $F$. This will become more clear in Sect.~\ref{sec:JTquantum} and in particular in Sect.~\ref{s:defects}.} This boundary condition \eqref{dilbc} resolves the tension around AdS$_2$/CFT$_1$ by deforming it into a NAdS$_2$/NCFT$_1$ of ``nearly'' AdS$_2$ (by the varying dilaton field and its asymptotics) dual to a ``nearly'' CFT$_1$ (by explicit breaking of conformal symmetry in the UV).

\noindent \textbf{Example (continued)}:\\
Let's solve equation \eqref{integrodiff} for our previous example. We have
\begin{align}
    F < 0 \,\, &\rightarrow \,\, F' = 1, \\
    F > 0 \,\, &\rightarrow \,\, F' = 1 - \frac{\mu}{a}F^2.
\end{align}
For $t<0$ the solution is trivial $F(t)=t$ and $Z(t)=\epsilon$, while for $t>0$ it is
\begin{align}
\label{eq:Fbh}
    F(t) &= \sqrt{\frac{a}{\mu}} \tanh \sqrt{\frac{\mu}{a}} t, \\
    \label{Zbh}
    Z(t) &= \epsilon F'(t)  = \frac{\epsilon}{\cosh^2 \sqrt{\frac{\mu}{a}} t}.
\end{align}

\begin{figure}[t!]
\centering
\includegraphics[width=0.3\textwidth]{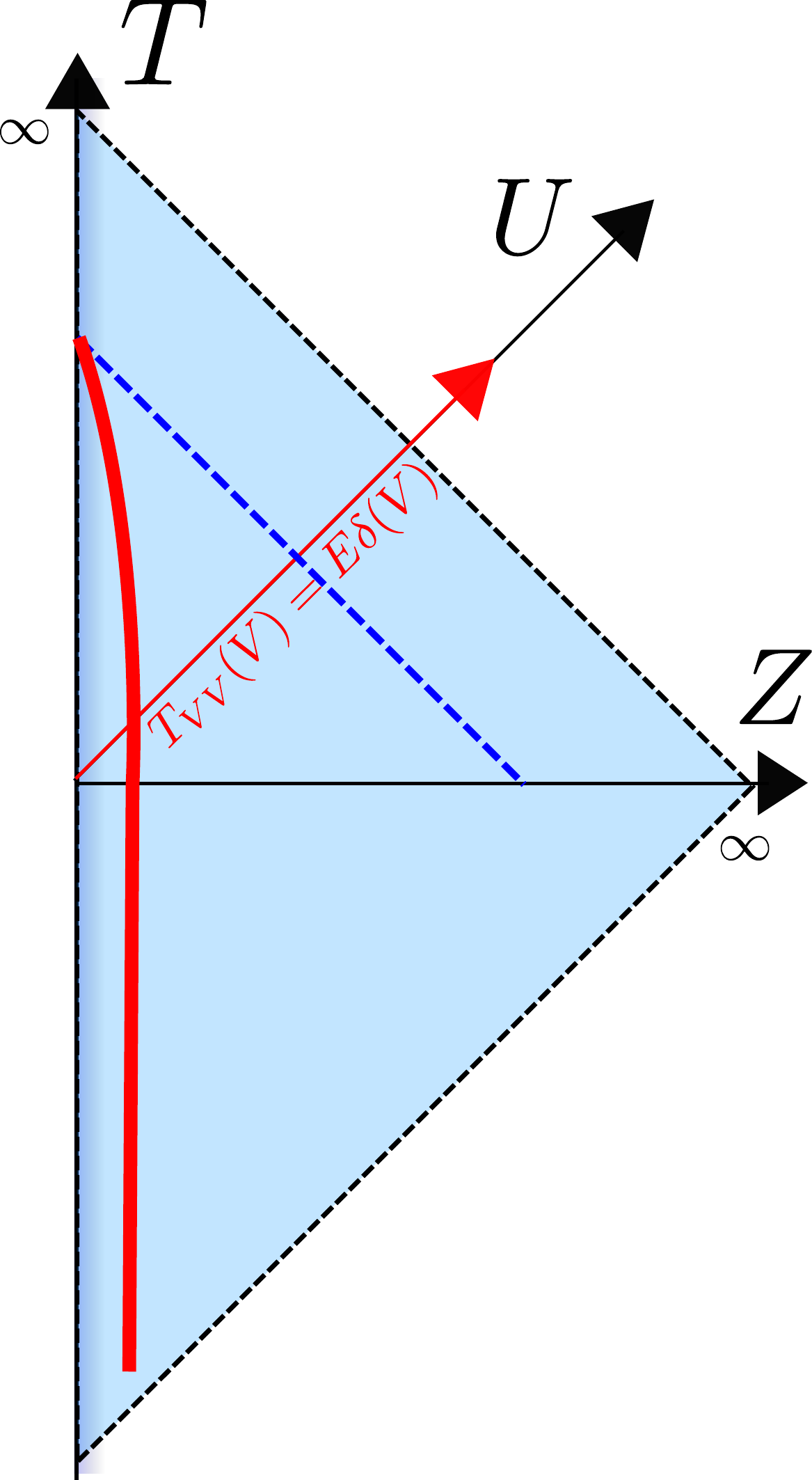}
\caption{Penrose diagram of AdS$_2$ in Poincar\'e patch before and after injecting an energy pulse. In red we show the boundary trajectory. We see this gets kicked after the insertion in such a way that a black hole horizon appears.}
\label{fig:pulse2}
\end{figure}

Notice that $Z(t) \to 0$ as $t\to \infty$ which happens at finite Poincar\'e time $F = (a/\mu)^{1/2}$. This time reparametrization is precisely the boundary limit of the black hole frame on the AdS$_2$ manifold \eqref{BHframe} we described earlier, with $\sqrt{\frac{\mu}{a}} = \frac{\pi}{\beta}$, relating the temperature $\beta^{-1}$ to the parameter $\mu$. See Fig.~\ref{fig:pulse2}.
\begin{center}
---o---
\end{center}
\vspace{0.3cm}

In order to write Eq.~\eqref{integrodiff} in a more manageable form, we first need to consider a different quantity in this model. The \textbf{holographic stress tensor} can be computed using standard techniques, supplemented here with the proper boundary coordinates \cite{Jensen:2016pah,Maldacena:2016upp,Engelsoy:2016xyb}. We will not detail the calculation itself here. In 0+1d, this quantity is the total energy contained in the spacetime at any time $t$ and, without boundary sources for massive bulk fields, can be written suggestively as:
\begin{equation}
\label{holostress}
    E(t) = - \frac{a}{16\pi G_N} \left\{F,t\right\},~~~~\left\{F,t\right\}\equiv \frac{F'''}{F'}-\frac{3}{2} \left(\frac{F''}{F'}\right)^2.
\end{equation}
The prefactor $\frac{16\pi G_N}{a}$ will appear a lot, and plays the role of the actual gravitational coupling constant in NAdS$_2$. Note that this gravitational coupling is power-counting superrenormalizable. We introduce the shorthand notation in terms of the quantity $C$ of dimension length:
\begin{equation}
\label{eq:scale}
    C \equiv \frac{a}{16\pi G_N}.
\end{equation}

Let us apply the relation \eqref{holostress} to the black hole coordinates, and study black hole thermodynamics from this perspective. For the solution \eqref{eq:Fbh} $F(t) = \sqrt{\frac{a}{\mu}} \tanh \sqrt{\frac{\mu}{a}} t$, we immediately find the stress tensor \eqref{holostress}: $E(t) = \frac{\mu}{8\pi G_N}$, agreeing with our previous analysis around Eq.~\eqref{dilpulse}. From this, we find the energy-temperature relation: 
\begin{equation}
\label{eq:firstlaw}
T = \frac{1}{\pi} \sqrt{\frac{E}{2C}}~~~\Rightarrow~~~ E(T) =2\pi^2 C  T^2.
\end{equation}
Using $\frac{\partial S}{\partial E} = \frac{1}{T}$, and identifying the zero-energy entropy with the parameter $S_0$ in the action, we can find the thermal entropy contained in the system as 
\begin{eqnarray}
\label{Sth}
    S(E) = S_0 + 2\pi \sqrt{2 C E} =S_0 + \frac{\sqrt{a\mu}}{4G_N}.
\end{eqnarray}
In the canonical ensemble, the entropy has a behavior linear with the temperature given by
\begin{eqnarray}
S(T) = S_0 + 4\pi^2 CT .
\end{eqnarray}
We can compare this entropy to the Bekenstein--Hawking (BH) entropy of a black hole $S_{\text{BH}} = \frac{A}{4G_N}$, where $A$ is the horizon area. The analogous formula in 1+1d requires the following modifications. Firstly, the area of a $d-1$ sphere is $S^{d-1} = \frac{2\pi^{d/2}}{\Gamma(d/2)}$, and hence $A \rightarrow 2$ when $d\to 1$. Because the 0-sphere consists of 2 disjoint points, we obtain $A= 1$ for a single horizon. Secondly, in our model the effective Newton's constant in dilaton gravity is spacetime-dependent: $ 1/G_{N,\text{eff}}(x) =(\Phi_0+ \Phi(x))/G_N$, which evaluates at the horizon to $(\Phi_0+ \Phi_h)/G_N$, where $\Phi_h$ is the horizon value of the dilaton field. Hence the 1+1d analog of the Bekenstein--Hawking entropy is 
\begin{equation}
\label{SBH}
    S_{\text{BH}} = \frac{\Phi_0+\Phi_h}{4 G_N}.
\end{equation}
We remind the reader that for convenience we separated the constant background piece of the dilaton $\Phi_0$ from the varying piece $\Phi$. The entropy is simply proportional to the total value of the dilaton evaluated at the horizon. 
Using now the dilaton profile for the black hole solution \eqref{dilbh}
\begin{equation}
    \Phi = \frac{a - \mu U(u)V(v)}{U(u)-V(v)} = \sqrt{a \mu} \coth \left(\sqrt{\frac{\mu}{a}} 2 z\right),
\end{equation}
we obtain $\Phi_h = \sqrt{a \mu}$, and we see that the resulting BH entropy \eqref{SBH} matches the thermal entropy \eqref{Sth}. Finally, we note that the entropy can be equally computed from the Ryu--Takayanagi (RT) formula \cite{Ryu:2006ef} as:
\begin{equation}
S = \text{Min}_z \frac{\Phi_0 + \Phi(z)}{4G_N},
\end{equation}
where the minimum occurs at the black hole horizon $z\to +\infty$. Non-trivial checks of this 2d version of the RT formula in more complicated geometries can be found in \cite{Goel:2018ubv}. 

Now let's continue our main analysis, where we come to the main point. Equipped with the formula \eqref{holostress}, taking suitable derivatives of Eq.~\eqref{integrodiff}, we can massage \eqref{integrodiff} into an ODE in terms of the energy $E(t) = - C \left\{F,t \right\}$:\footnote{The sequence applied to \eqref{integrodiff} is: differentiate, divide by $F'$, differentiate, divide by $F'$, differentiate, multiply by $F'$. Then use that $\left\{F,t\right\}'= F'\left(\left(F''/F'\right)'/F'\right)'$.}
\begin{equation}
\label{eomschw}
    \frac{d E(t)}{dt} ~~=~~ (T_{VV}(t)- T_{UU}(t))F'^2 \vert_{\partial \mathcal{M}} ~~=~~ T_{vv}(t)- T_{uu}(t) \vert_{\partial \mathcal{M}},
\end{equation}
where the right hand side is the net influx of matter from the holographic boundary in the proper coordinates $(u,v)$. Hence the differential equation that determines the frame $F(t)$ is just energy conservation, where the wiggly boundary curve $(F(t),Z(t))$ reacts to a net energy influx from the holographic boundary. In particular, injecting positive energy causes the boundary trajectory to bend towards the actual boundary as illustrated before around Eq.~\eqref{Zbh} and vice versa. With this interpretation, the right hand side of Eq.~\eqref{eomschw} can be immediately generalized to arbitrary non-conformal matter as well. This equation of motion is the main result of the classical analysis of the JT equations of motion.\footnote{This is reminiscent of hydrodynamics where the only conserved quantity is the energy \cite{Jensen:2016pah}. A related similarity is that, as we will point out in Sect.~\ref{s:defects}, the only local observable operator in the resulting quantum theory is constructed from the stress tensor.
} 

Some of the above mysterious properties of 2d dilaton gravity are naturally explained when embedding the model within higher-dimensional gravity. For near extremal charged black holes in 4d considered in Sect.~\ref{sec:NEBH}, the boundary condition on the dilaton arises from gluing the AdS$_2$ throat to the exterior region \cite{Almheiri:2016fws,Nayak:2018qej}. Another example is the observation that the dynamics of $s$-waves of 3d gravity are governed by 2d JT gravity \cite{Achucarro:1993fd}. Indeed, parametrizing a 3d spherically symmetric geometry by the metric ansatz:
\begin{equation}
    ds^2 = g_{\mu\nu}^{(2)}(x^\mu)dx^\mu dx^\nu + \Phi^2(x^\mu) d\varphi^2 , \qquad \mu,\nu= t,r, \quad \varphi \sim \varphi + 2\pi,
\end{equation}
the 3d pure gravity action reduces to JT gravity:
\begin{equation}
\frac{1}{16\pi G_N^{(3)}} \int d^3x \sqrt{-g}(R^{(3)}- \Lambda) = \frac{2\pi}{16\pi G_N^{(3)}} \int d^2x \sqrt{-g} \Phi (R^{(2)}-\Lambda),
\end{equation}
where $G_N^{(2)} = G_N^{(3)}/2\pi$, and the dilaton field $\Phi$ originated from the $g_{\varphi\varphi}$ metric component. This immediately explains the BH entropy formula \eqref{SBH} since $S_{\text{BH}} = 2\pi \Phi_h/4G_N^{(3)} = \Phi_h/4G_N^{(2)} $ where $2\pi \Phi_h$ is the horizon circumference in 3d gravity.
Moreover, our dilaton boundary condition \eqref{dilbc} can be seen to directly arise from the usual Fefferman-Graham gauge asymptotic expansion of the full 3d metric. In this sense, JT gravity resolves the tension in AdS$_2$/CFT$_1$ by being secretly a 3d gravity model in disguise.

\subsubsection{Aside: general dilaton gravity}
\label{s:aside}
It is important to know that some of this story can be generalized to arbitrary dilaton gravity models with action \eqref{gendilgrav}, and $\Phi$ in that expression denotes the total dilaton without separating into a background $\Phi_0$ and a varying piece. Starting with \eqref{gendilgrav}, one can write down the generic black hole (sourceless) solution, generalizing the JT solution \eqref{JTbh}. 
We can partially fix coordinates in the bulk by identifying $\Phi(r) =\frac{a}{2} r $, for a radial coordinate $r$ with $r\to\infty$ the boundary of our system. In this gauge, the dilaton can be thought of as a ruler providing a definition of proper radial distance.

The generic black hole solution after further gauge fixing can be written as
\begin{equation}
\label{eq:bhgen}
     ds^2 = -f(r)dt^2 + \frac{dr^2}{f(r)},~~~~f(r) =\frac{4}{a^2} \int_{\Phi_h}^{\Phi(r)} d\Phi ~U(\Phi).
\end{equation}
The location $r=r_h$ is the black hole horizon and $\Phi_h \equiv \Phi( r_h)= \frac{a}{2} r_h$. The solution near $r_h$ only has the right spacetime signature if $U(\Phi_h)\geq 0$, and the dilaton potential for $\Phi>\Phi_h$ should be such that $f(r)$ is positive everywhere outside the horizon. Expanding near $r_h$ one obtains the Hawking temperature:
\begin{eqnarray}
T = \frac{U(\Phi_h)}{2\pi a},
\end{eqnarray}
conjugate to the choice of time coordinate $t$ in \eqref{eq:bhgen}.

The location $r=0$ is the would-be singularity where $\Phi \to 0$. From the embedding of dilaton gravity models in higher dimensions, since the dilaton is proportional to the area of the transverse sphere, this is the curvature singularity of the higher-dimensional model. In 2d however, this is a strongly coupled singularity where the effective Newton's constant diverges. 
Since there is only a single curvature invariant, and since the $\Phi$ equation of motion yields
\begin{equation}
    R = - U'(\Phi),
\end{equation}
a curvature singularity in 2d can only arise when $U'(\Phi_{\rm sing.})\to \pm \infty$ for some $\Phi_{\rm sing.}$, a situation that is usually excluded by choice of dilaton potential. However, one interesting example where this does happen is in the dilaton gravity description of Liouville gravity, for which we will make some comments in Sect.~\ref{s:Liouvillegravity}.

The general $E(T)$ and $S(T)$ black hole first law equations for \eqref{eq:bhgen} are:
\begin{equation}
E(T) = \frac{1}{8\pi G_N a}\int^{U^{-1}(2\pi a T)} \hspace{-0.2cm}U(\Phi)d\Phi, ~~~~ S(T) = \frac{U^{-1}(2\pi a T)}{4G_N} = \frac{\Phi_h}{4 G_N},
\end{equation}
where $U^{-1}(.)$ is the inverse function of $U(\Phi)$. This gives a physical interpretation to the dilaton potential: $ T(S)=\frac{1}{2\pi a} U(\Phi=4G_N S) $ is the equation of state, with the entropy identified with the dilaton and the temperature with the value of the potential. The connection with the entropy is the correct interpretation of the dilaton field in 2d dilaton-gravity.

We leave it as an exercise to apply the above relations to the JT case, where in the conventions of this subsection $U(\Phi) = 2(\Phi-\Phi_0)$. We note that a linear dilaton potential is the only one where the resulting thermodynamical relations only depend on the ratio $a/G_N \sim C$. 

\subsubsection{The Schwarzian action}
We were led to a dynamical boundary curve described by \eqref{eomschw}, which in the absence of matter injections is governed by the simple equation 
\begin{equation}
\label{schweq}
    \frac{d}{dt}\left\{F,t\right\} = 0.
\end{equation}
This equation of motion can be found as the Euler-Lagrange equation of a suitable action: the \textbf{Schwarzian action}. This action describing the dynamics of JT gravity is:
\begin{equation}
\label{actSCh}
    S = -C \int dt \left\{F,t\right\}, \qquad \left\{F,t\right\} \equiv \frac{F'''}{F'} - \frac{3}{2} \left(\frac{F''}{F'}\right)^2.
\end{equation}
and contains the Schwarzian derivative as the Lagrangian. As a higher-derivative model, this looks complicated at first sight. After integrating by parts using $\int dt \, \delta \left\{F,t\right\} = -\int  dt \, \frac{\left\{F,t\right\}'}{F'} \delta F$, the equation of motion following from \eqref{actSCh} is indeed \eqref{schweq} as promised.

It is now relatively straightforward to write down the effective action whose equations of motion give the full sourced Eq.~\eqref{eomschw}. We just minimally couple a generic 2d matter Lagrangian $\mathcal{L}_m(\phi,\partial_F \phi)$, where we denoted its dependence on the fields $\phi$ and the time-derivatives $\partial_F \phi$, to the wiggly boundary curve and extrapolate the time reparametrization $t\to F(t)$ throughout the entire bulk. We can then write the total Lorentzian action in a hybrid way as\footnote{Some comments are in order here. Firstly, we can always extrapolate the boundary reparametrization into the bulk in any way we want. Here we choose a simple way. Secondly, the $Z$-integral only ranges to $Z(t)=\epsilon F'(t)$ but this is infinitesimal and does not play a role in the current argument. This term would be important for massive matter boundary sources turned on, but this will not be studied here.}
\begin{align}
S &= -C\int dt \, \left\{F,t\right\} + \int dF dZ \, \mathcal{L}_m(\phi,\partial_F \phi) \\
&= -C\int dt \, \left\{F,t\right\} + \int dt dZ \, F' \mathcal{L}_m(\phi, \frac{1}{F'}\partial_t \phi),
\end{align}
where in the second line we made the $F$-dependence of the matter part explicit. Varying with respect to $F(t)$ yields
\begin{equation}
\delta S = \int dt \left[C\frac{\left\{F,t\right\}'}{F'}\delta F - \int dZ \left(\partial_F \phi \frac{\partial \mathcal{L}_m}{\partial \partial_F \phi} -\mathcal{L}_m \right) \delta F' \right].
\end{equation}
Identifying the matter Hamiltonian density as $\mathcal{H}_m \equiv \partial_F \phi \frac{\partial \mathcal{L}_m}{\partial \partial_F \phi} -\mathcal{L}_m $,
then leads to the sourced equation of motion:
\begin{equation}
-C\left\{F,t\right\}' = F'^2\frac{d H_m}{dF}, \qquad H_m = \int dZ \, \mathcal{H}_m.
\end{equation}
The quantity on the RHS is the total change in the energy in the matter sector since by energy conservation in the matter sector $\frac{d H_m}{dF} = T_{VV}-T_{UU}\vert_{\partial \mathcal{M}}$. The explicit $F'^2$ converts the matter energy fluxes to the $t$ time coordinate. This is precisely the content of the sourced equation of motion \eqref{eomschw}. 

In order to proceed, it is convenient to summarize some of the salient mathematical properties of the Schwarzian derivative:
\begin{enumerate}
\item
By explicit calculation, we have the following representation of the Schwarzian derivative:
\begin{equation}
\label{prop1}
    \frac{1}{6} \left\{F,t \right\} = \lim_{t'\to t } \left(\frac{F'(t')F'(t)}{((F(t')-F(t))^2} - \frac{1}{(t'-t)^2}\right).
\end{equation}
\item
Composition law:
\begin{equation}
\label{prop2}
    \{F(G(t)),t\} = \{G(t),t\} + G'(t)^2\{F(G),G\}(t).
\end{equation}
This can be proven easily using the representation \eqref{prop1}. 
\item
\label{prop3}
The Schwarzian derivative itself is PSL$(2,\mathbb{R})$ invariant: if $F = \frac{a G +b}{c G + d}$, then $\left\{F,t\right\} = \left\{G,t\right\}$. This can again be easily shown by using the above identity \eqref{prop1}.
\item
\label{prop4}
The solution to the equation $\left\{F,t\right\} = 0$ is 
\begin{equation}
    F(t) = \frac{at+b}{ct+d}.
\end{equation}
To see this, note that $F(t)=t$ has vanishing Schwarzian derivative $\left\{F,t\right\} = 0$. By property \ref{prop3}, this means also $\frac{at+b}{ct+d}$ has zero Schwarzian derivative. This is a three-parameter family of solutions. The differential equation $\left\{F,t\right\} = 0$ itself is third order and hence requires three integration constants as initial conditions, $F(0)$, $F'(0)$ and $F''(0)$ which can be one-to-one mapped into the three parameters of the PSL$(2,\mathbb{R})$ transformation parametrized by $a,b,c,d$ above.
\item
A converse of property \ref{prop3}: if two functions have equal Schwarzian derivative, then they differ at most by a M\"obius transformation:
$\left\{F,t\right\} = \left\{G,t\right\}\,\, $ if $\,\,\exists \left(\begin{array}{cc} a & b \\ c & d \end{array} \right)$ such that $F = \frac{a G +b}{c G + d}$. To prove this, we first use the composition law \eqref{prop2} to rewrite the proposition as $\left\{F,G\right\} =0 $. Then by property \ref{prop4}, we immediately get $F = \frac{a G +b}{c G + d}$.
\end{enumerate}
Now we are equipped to discuss the solution of Eq.~\eqref{schweq} in more detail. We immediately have: 
\begin{equation}
    \frac{d}{dt} \left\{F,t\right\} = 0 ~~~~\Rightarrow ~~~~\left\{F,t\right\} = \text{constant}.
    \label{scheom}
\end{equation}
For any fixed value of this constant, which is interpretable as proportional to the energy \eqref{holostress}, this equation has a unique solution up to an PSL$(2,\mathbb{R})$ M\"obius ambiguity
\begin{equation}\label{eqn:SL2Rschw}
    F(t) \to \frac{aF(t)+b}{cF(t)+d}.
\end{equation}

The Schwarzian action \eqref{actSCh} has a PSL$(2,\mathbb{R})$ symmetry, under $F \to \frac{aF+b}{cF+d}$. This leads to the following conserved Noether charges:
\begin{align}
Q_{-} &= C\left[\frac{F'''}{F'{}^2}-\frac{F''{}^2}{F'{}^3}\right] , \\
Q_0 &= C\left[\frac{F'''F}{F'{}^2}-\frac{F F''{}^2}{F'{}^3}-\frac{F''}{F'}\right] , \\
Q_{+} &= C\left[\frac{F'''F^2}{F'{}^2}-\frac{F^2 F''{}^2}{F'{}^3}-\frac{2FF''}{F'}+2F'\right],
\end{align}
associated to the one-parameter subgroups generated by $b$ for $Q_-$, $a$ and $d$ for $Q_0$, and $c$ for $Q_+$ respectively.
One checks explicitly that these conserved charges form an $\mathfrak{sl}(2,\mathbb{R})$ algebra, and yield the Hamiltonian as the quadratic Casimir:
\begin{equation}
\label{casHam}
    H \equiv -C\{F,t\} = \frac{1}{2C}\left[Q_0^2 -\frac{1}{2}\left\{ Q_+,Q_{-}\right\}\right],
\end{equation}
which is also the Noether charge of Eq.~\eqref{actSCh} corresponding to time translations in the coordinate time $t \to t +c$.\footnote{The expression for the energy \eqref{holostress} is to be compared to this Noether charge, which can be used to deduce that the prefactor of the Schwarzian action is indeed $-C$ as in \eqref{actSCh}.} 

We would like to stress that even though the transformations generated by $Q$ that act on $F$ are symmetries, the transformation $t\to \frac{a t+ b}{c t + d}$ is not a symmetry unless $t\to t + b$; the boundary conformal symmetry is broken for $C\neq 0$. 

As emphasized earlier, configurations that are related by M\"obius transformations $F \to \frac{aF+b}{cF+d}$, are to be identified in the gravitational model. In order for $F(t)$ to be a good reparametrization, we also require $F'\geq 0$ anywhere on its domain. Depending on the precise situation, the set of all allowed $F(t)$ can get further restrictions as we make precise in Sect.~\ref{sec:JTquantum}. For now, let us call this space $G$, and the M\"obius subgroup (which is a redundancy) as $H = $ PSL$(2,\mathbb{R})$. Then the set of all possible $F(t)$ is described by the coset $G/H$.
This is a situation familiar from e.g. the pion effective QFT in particle physics, where we view $F$ as describing the (pseudo) Goldstone boson degrees of freedom in the symmetry breaking $G \to H$. The Schwarzian action itself \eqref{actSCh} is however only invariant under the (gauged) subgroup $H$, illustrating that the original group $G$ is broken both explicitly and spontaneously, hence the adverb \emph{pseudo}. The prefactor $C \sim a$ of the Schwarzian action hence represents the scale of explicit breaking of the 1d group of reparametrizations, i.e. the 1d conformal group. This is an interpretation we indeed encountered earlier in Sect.~\ref{s:lorderiv} when regularizing the dilaton asymptotics. 

The Schwarzian term itself is the lowest order non-trivial local term one could possibly write down as a Lagrangian that is invariant under $H$. As such, the Schwarzian model exhibits \textbf{universality}, and is fully determined by this specific pattern of symmetry breaking.

It is in this language that the Schwarzian action first emerged in A. Kitaev's work in 2015 on the Sachdev--Ye--Kitaev (SYK) model \cite{Sachdev:1992fk,kitaevTalks1,kitaevTalks2,kitaevTalks3}: the dynamics of SYK at low energies is approximated by a reparametrization $F(t)$ describing an explicitly (by UV effects) and spontaneously (by making a specific choice of $F(t)$ mod PSL$(2,\mathbb{R})$) broken symmetry.

Hence since the PSL$(2,\mathbb{R})$ symmetry is a gauge redundancy, this means only configurations with zero overall PSL$(2,\mathbb{R})$ charges are physically observable. The correct interpretation for this is that the gravitational charges have to be compensated by other sectors, either in a two-sided configuration, or by adding additional matter sectors that can carry these charges.

\subsubsection{Geometric derivation from the action}
\label{s:geomder}
There is a faster derivation of the Schwarzian description that is more geometric in nature \cite{Maldacena:2016upp} and stays at the level of the action. It will be our starting point in the next sections.
Let us go back to the Euclidean JT action:
\begin{eqnarray}
\label{eq:JTaction22}
    I_{\text{JT}}[g,\Phi] = - \frac{1}{16\pi G_N}\int  \sqrt{g} \Phi\left(  R +  2\right) - \frac{1}{8\pi G_N}\oint \sqrt{h}\Phi (K-1).
\end{eqnarray}
We first vary with respect to the dilaton field, to find $R=-2$, or a patch of the 2d hyperbolic plane $ds^2 = \frac{dT^2+dZ^2}{Z^2}$. 

\begin{figure}[t!]
\centering
\includegraphics[width=0.3\textwidth]{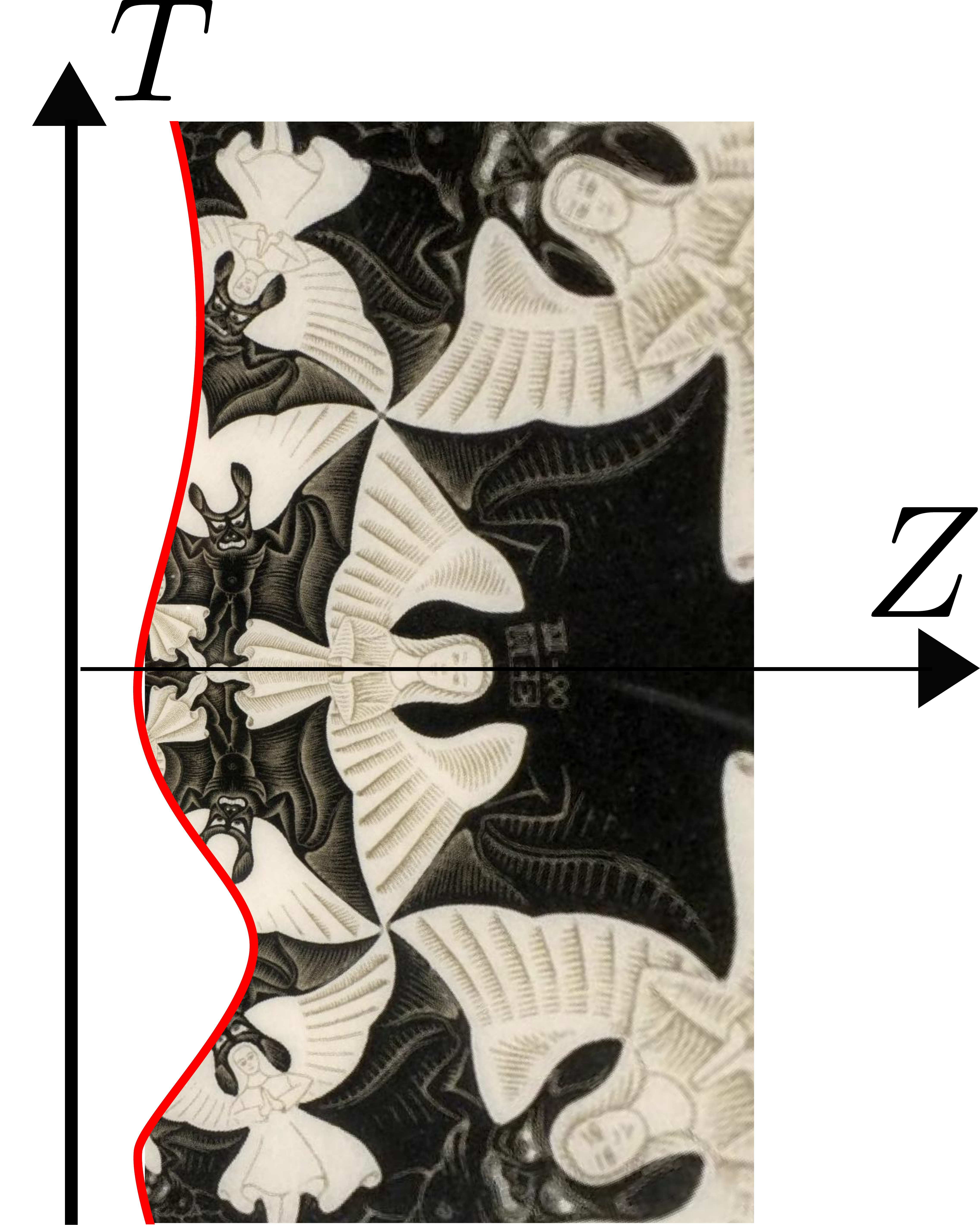} 
    \caption{Cutout of the Poincar\'e upper half plane.}
        \label{Fig:UHP}
\end{figure}

The dynamics is now fully governed by the GHY boundary term on some wiggly boundary curve. One can draw this as the hyperbolic upper half plane, where we cut out and remove a shape that is close to the actual boundary $Z=0$. This is the Euclidean counterpart of our discussion so far. We illustrate the procedure in Fig.~\ref{Fig:UHP}.

We parametrize the boundary curve as $(T=F(\tau),Z(\tau))$. Fixing the boundary metric to $\sqrt{h} =1/\epsilon$ imposes $Z(\tau)= \epsilon F'(\tau)$. We can directly evaluate the extrinsic curvature trace:
\begin{equation}
    K = \frac{F'(F'^2+Z'^2+ZZ'')-ZZ'F''}{(F'^2+Z'^2)^{3/2}} = 1 + \epsilon^2 \{F,\tau \} + \mathcal{O}(\epsilon^4),
\end{equation}
using standard techniques.\footnote{Some technical details. The non-zero Christoffel symbols are $\Gamma^{Z}_{ZZ} = -\Gamma^{Z}_{TT} = -\Gamma^{T}_{TZ} = 1/Z$. The extrinsic curvature is computed as $K = h^{tt}K_{tt} = h^{tt}M_t^\mu M_t^\nu \nabla_\mu n_\nu$ where $M^\mu_t = \partial_t x^\mu$, the pull-back metric is $h_{tt} \equiv M_t^\mu M_t^\nu g_{\mu\nu}$. The unit-normalized outwards-pointing normal co-vector is $n_\mu = \left(\frac{Z'}{Z(Z'^2+F'^2)^{1/2}},\frac{-F'}{Z(Z'^2+F'^2)^{1/2}}\right)$. The $t$-dependence of the objects $Z'$ and $F'$ can be rewritten into a $F$-dependence using the chain rule: $d(.)/dF = (.)' / F'$, making the normal co-vector a function of $(F,Z)$.}
We also fix the boundary dilaton to be $\Phi_b \equiv \Phi\vert_\partial = \frac{a}{2\epsilon}$, which appears in the GHY term in the action \eqref{eq:JTaction22}.
 Plugging this back in the Euclidean action, we immediately obtain the Schwarzian action:
\begin{eqnarray}
\label{JTaction2}
    I_{\text{JT}}[F] = - C\int d\tau \{F,\tau \}, \qquad C = \frac{a}{16\pi G_N}.
\end{eqnarray}

The added benefit is that one can directly generalize this to a Euclidean gravitational path integral argument, as we will do further on in Sect.~\ref{sec:JTquantum}. 

The Schwarzian action describes the reparametrization $F(\tau)$ as the dynamical degree of freedom. Much like in well-known studies of 3d Chern-Simons and 2d WZW boundary CFT \cite{Elitzur:1989nr} (and in the BF dimensional reduction discussed briefly in Sect.~\ref{s:BF}), the physical boundary degrees of freedom are the would-be large gauge (or diff) transformations, which have become physical and observable in the presence of a boundary.\footnote{We will encounter different boundary conditions later on for interior boundaries. We refer to \cite{Goel:2020yxl,Ferrari:2020yon} for classifications of boundary conditions.} 

Up to now, we have always described our reparametrization as referring to the Poincar\'e time coordinate $F$. This is not necessary, and in fact it is more natural to choose a different reference frame when considering the thermal system, that is the black hole frame. We define:\footnote{This is the Euclidean version of Eq.~\eqref{BHframe}, setting both $F \to -i F$  and $f \to -if$, where the prefactor $\beta/\pi$ is removed. This prefactor is part of the SL$(2,\mathbb{R})$ gauge isometry group and is hence not observable. It was however convenient to include it earlier because the time coordinate then has a clean $\beta\to +\infty$ limit back to the Poincar\'e time coordinate $F \to f$.}
\begin{equation}
\label{repther}
    F(\tau) \equiv \tan \frac{\pi}{\beta} f(\tau),
\end{equation}
in terms of a new variable $f(\tau)$, which has the properties:
 \begin{equation}
 \label{eq:peri}
f(\tau+\beta) = f(\tau) + \beta, \qquad  f'(\tau) \geq 0.
 \end{equation}
These two properties give $f(\tau)$ the interpretation of a time reparametrization of the boundary thermal circle: $f\in \text{Diff }S^1$. The metric boundary condition \eqref{aspoinc} can then be rephrased as a fixed length boundary condition, where we keep fixed the regularized boundary length to $\frac{\beta}{\epsilon}$. The rescaled (renormalized) boundary length is $\beta$, with the physical interpretation of the inverse temperature. 

Two useful graphical representations, emphasizing different perspectives, are the following.
\begin{figure}[h!]
\centering
        \includegraphics[width=0.75\textwidth]{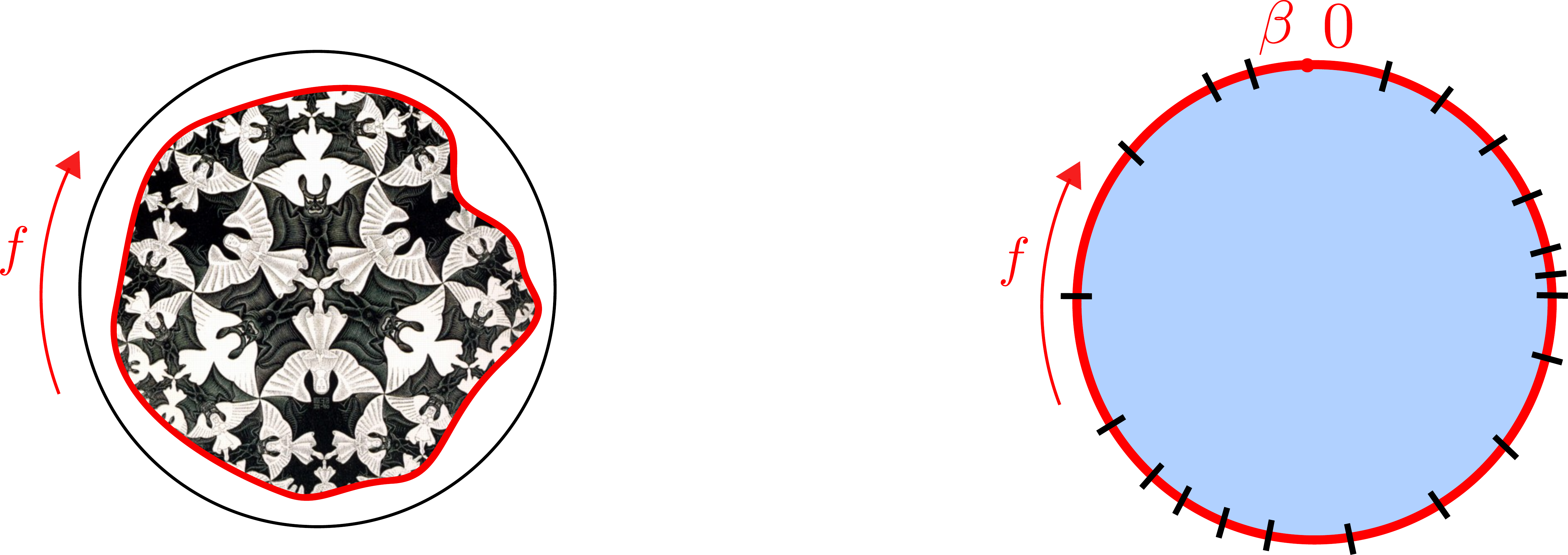}
       \caption{The left figure depicts the wiggly thermal boundary curve as a cutout of the Poincar\'e disk. The right figure shows the boundary clock ticking pattern. The number of ticks on the clock $f(\tau)$ is fixed by the periodicity constraint \eqref{eq:peri}, but their spreading along the thermal circle is not.}
\end{figure}

The classical equation of motion of Eq.~\eqref{JTaction2} for a thermal reparametrization \eqref{repther} with the above periodicity requirements, leads to the unique classical solution $f(\tau) = \tau$, up to PSL$(2,\mathbb{R})$. This is precisely the (Wick-rotated) black hole solution \eqref{BHframe}.

\subsection{Quantum matter in classical gravity}
Before starting with the quantum solution of JT gravity in the next sections, we want to present a physically interesting intermediate situation where we study quantum matter in a classical gravitational model. 

The sourced Schwarzian equation of motion \eqref{eomschw} is this model's Einstein equation $G_{\mu\nu} = 8\pi G_N T_{\mu\nu}$.
In this subsection, we promote the matter sources on the right hand side to be quantum mechanical, while retaining classicality for the gravitational sector, i.e.  $G_{\mu\nu} = 8\pi G_N \left\langle T_{\mu\nu}\right\rangle$. Such an approximation is only valid as long as the quantum effects of gravity are subdominant compared to the matter quantum effects. For a 2d matter CFT with central charge $c$, this can be realized when $c\gg 1$, and as long as any bulk black hole is not Planck-scale.

Quantum effects in a matter sector that is a 2d CFT are governed by the conformal anomaly \cite{Davies:1976ei,Christensen:1977jc}, see also the textbook \cite{Fabbri2005}. This leads to the following stress tensor components, to be inserted in \eqref{dileom}:
\begin{align}
\label{qmatstress}
\nonumber
\left\langle T_{uu} \right\rangle &~=~ -\frac{c}{12\pi}\bigl( (\partial_{u}\omega)^2 - \partial_{u}^{2}\omega\bigr) \; + \left\langle:T_{uu}(u):\right\rangle, \\
\left\langle T_{vv} \right\rangle &~=~ -\frac{c}{12\pi}\bigl( (\partial_{v}\omega)^2 - \partial_{v}^{2}\omega\bigr) \; + \left\langle:T_{vv}(v):\right\rangle, \\
\nonumber
\left\langle T_{uv}\right\rangle &~=~ -\frac{c}{12\pi}\partial_u \partial_v \omega.
\end{align}
The objects $T_{uu}$, $T_{vv}$ and  $T_{uv}$ transform as tensor components and are covariantly conserved $\nabla_\mu T^{\mu \nu} =0$, as should be for consistency of Einstein's equations. This stress tensor is decomposed as shown into a Casimir contribution, that depends solely on the metric through $\omega(u,v)$ and not on the quantum state, plus an operational part that depends on the quantum state and also on a reference state as we explain. Both transform anomalously under coordinate transformations, but their sum is a covariant tensor.

The terms on the RHS $:T_{uu}(u):$ and $:T_{vv}(v):$ are frame-dependent due to normal-ordering with respect to a specific vacuum. Let us be more explicit. Suppose we have a free boson CFT with $c=1$, with bulk field equation $\Box \phi =0$. Then $T_{uu} = \partial_u \phi \partial_u \phi$ is a composite operator in the quantum matter sector, and requires regularization and renormalization. We define the renormalized stress tensor by subtracting its expectation value in the vacuum state $\left\vert 0_u\right\rangle$ defined using positive frequency modes in the $u,v$-coordinates, as:
\begin{align}
    :T_{uu}: &~\equiv~ \partial_u \phi \partial_u \phi - \left\langle 0_u \right \vert \partial_u \phi \partial_u \phi \left\vert 0_u\right\rangle \\
    &~=~ \lim_{u \to u'}\left( \partial_u \phi(u) \partial_u \phi(u') + \frac{1}{4\pi} \frac{1}{(u-u')^2}\right),
\end{align}
and analogously for $:T_{vv}:$. This stress tensor is operationally defined and measured by local observers with measuring devices calibrated to their vacuum. Between different frames $u$ and $U$ (i.e. calibrated w.r.t. different vacua), it transforms anomalously as:
\begin{equation}
    :T_{uu}: ~=~ \left(\frac{dU}{du}\right)^2 :T_{UU}: - \frac{c}{24\pi} \left\{U,u\right\}.
\end{equation}

In our specific set-up, the influence of the quantum matter sources \eqref{qmatstress} is the following. (1) The non-zero component $\left\langle T_{uv}\right\rangle$ leads in the classical equation of motion to a simple shift in the dilaton $\Phi \to \Phi + \frac{cG_N}{3}$ \cite{Almheiri:2014cka}, which has no influence on the boundary equation of motion at $\Phi \to \infty$. (2) For the AdS$_2$ geometry with the specific formula \eqref{eq:solliou} for $\omega(u,v)$, the $T_{uu}$ and $T_{vv}$ components are still holomorphic: $T_{uu}(u)$ and $T_{vv}(v)$, leading to exactly the same sourced Schwarzian equation of motion \eqref{eomschw}.

Hence evaluating these stress tensor components \eqref{qmatstress} at the boundary, we have:
\begin{align}
\label{eq:confqn}
    \left\langle T_{uu}(t) \right\rangle\vert_{\partial \mathcal{M}} &=  \frac{c}{24\pi}\left\{F,t\right\} + \left\langle :T_{uu}(t): \right\rangle\vert_{\partial \mathcal{M}}  \\  \left\langle  T_{vv}(t) \right\rangle\vert_{\partial \mathcal{M}} &=  \frac{c}{24\pi}\left\{F,t\right\} + \left\langle :T_{vv}(t): \right\rangle\vert_{\partial \mathcal{M}}  \, ,
\end{align}
and the equation of motion \eqref{eomschw} can be written in two equivalent ways:
\begin{equation}
\label{eq:qschw}
    \frac{d E}{d t} =  \left\langle T_{vv}(t) \right\rangle - \left\langle T_{uu}(t)\right\rangle \vert_{\partial \mathcal{M}}= \left\langle:T_{vv}(t):\right\rangle - \left\langle:T_{uu}(t):\right\rangle \vert_{\partial \mathcal{M}},
\end{equation}
in terms of either the actual net influx of energy, or the net \emph{operationally measured} influx of energy, evaluated in the matter quantum state of interest.

\subsubsection{Application: Hawking-Unruh effect and information loss}\label{sec:Evaporation}
We can apply these equations directly to derive the 2d analog of Unruh and Hawking radiation, and probe information flows in the model.

Black holes in asymptotically AdS spacetimes do not typically evaporate due to standard reflecting boundary conditions at the holographic boundary. \\

\begin{figure}
\centering
\includegraphics[width=0.3\textwidth]{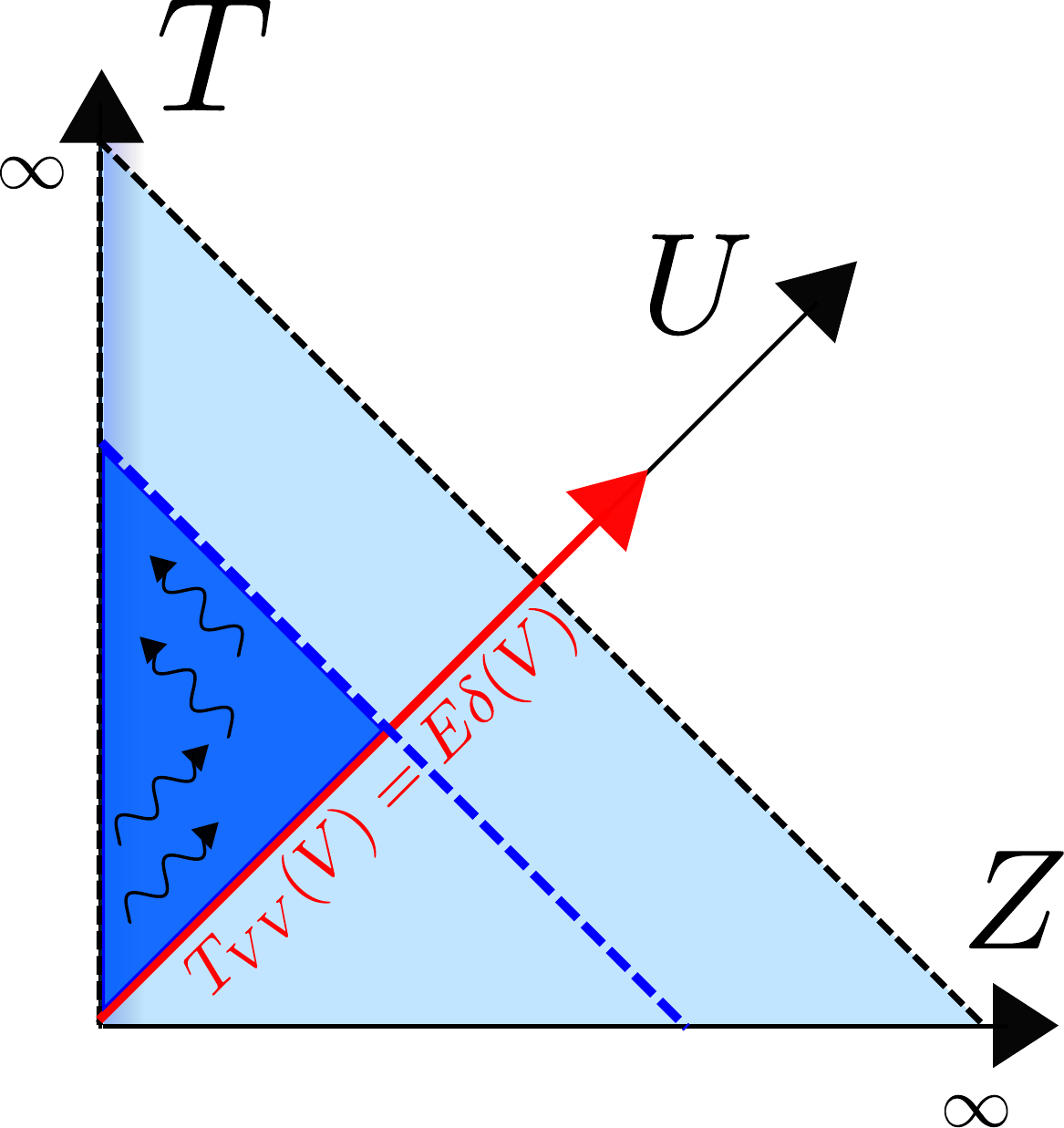}
\caption{Penrose diagram of formation and subsequent Hawking radiation of a  black hole, after injecting an energy pulse.}
\label{fig:evap1}
\end{figure}

\noindent \textbf{Example (continued)}: \\
 We impose reflecting boundary conditions for the matter sector at the holographic boundary after the initial energy injection at $t=0$, see Fig.~\ref{fig:evap1}. The matter sector state is the Poincar\'e vacuum since that is the initial vacuum state before the pulse. This leads to the boundary conditions $\left\langle T_{uu}(t) \right\rangle\vert_{\partial \mathcal{M}} = 0 = \left\langle T_{vv}(t) \right\rangle\vert_{\partial \mathcal{M}}$, and hence the net energy in the bulk spacetime does not change. 

However, the operational stress tensor components are nonzero by the conformal anomaly \eqref{eq:confqn}. Evaluating these for the frame $F(t) = \frac{\beta}{\pi}\tanh \frac{\pi}{\beta} t$, with $\beta = \pi \sqrt{\frac{8\pi G_N E}{a}}$ after the initial pulse, we have
\begin{equation}
    \left\langle :T_{uu}(t): \right\rangle\vert_{\partial \mathcal{M}}\,\, = \,\,\left\langle :T_{vv}(t): \right\rangle\vert_{\partial \mathcal{M}}\,\, = \,\, c \frac{\pi}{12 \beta^2}, \quad t>0.
\end{equation}
This is the \textbf{Unruh heat bath} with equal left- and right-moving energy flux, and no macroscopic flow of energy. This is the AdS$_2$ analog of physics in the Minkowski vacuum for flat space, or the Hartle-Hawking vacuum for the Schwarzschild metric. The energy density profile is $:T_0^0(\rho): = c \frac{\pi}{6 \beta^2} \frac{1}{\sinh^2 \rho }$ where $\rho$ is the proper distance to the horizon, as introduced around \eqref{rhoco}. The conformal anomaly technique provides a very efficient derivation of this observer-dependent physics \cite{Spradlin:1999bn}. Notice that the thermal flux immediately sets in after the black hole is created; no transient regime is present in this model and it has instant thermalization.
\begin{center}
---o---
\end{center}
\vspace{0.3cm}

To model actual evaporation, let's instead implement absorbing boundary conditions at the holographic boundary, where the observer on the boundary line removes all Hawking radiation he detects in his local frame (detector calibrated w.r.t. the vacuum associated with his own preferred time $t$ coordinate) $ \left\langle :T_{vv}(t): \right\rangle\vert_{\partial \mathcal{M}}= 0$ \cite{Engelsoy:2016xyb}. The boundary condition for the outgoing component remains the same $\left\langle T_{uu}(t) \right\rangle\vert_{\partial \mathcal{M}} = 0$.
This leads to the equation of motion:
\begin{equation}
\label{eq:dissip}
    \frac{d E}{d t} \equiv  -C \frac{d}{dt} \left\{F,t\right\} ~=~ \left\langle T_{vv}(t) \right\rangle - \left\langle T_{uu}(t) \right\rangle \vert_{\partial \mathcal{M}} ~=~ \frac{c}{24\pi}\left\{F,t\right\}, \,\, t>0,
\end{equation}
solvable by an exponentially decaying energy profile:\footnote{This profile is in agreement with a quasi-static approximation where one uses the black hole first law $T \sim \sqrt{E}$, combined with the 2d Stefan-Boltzman law $\frac{dE}{dt} \sim T^2$.}
\begin{equation}
\label{eq:disen}
    E(t) = E e^{-\frac{c}{24\pi C}t}\theta(t).
\end{equation}
From this solution, one can then find the explicit time reparametrization solution
\begin{eqnarray}
\label{eq:reparabsorb}
    F(t) = \frac{2}{\alpha A} \frac{I_0(\alpha) K_0(\alpha e^{-\frac{c}{48\pi C}t}) - K_0(\alpha)I_0(\alpha e^{-\frac{c}{48\pi C}t})}{I_1(\alpha) K_0(\alpha e^{-\frac{c}{48\pi C}t}) + K_1(\alpha)I_0(\alpha e^{-\frac{c}{48\pi C}t})},
\end{eqnarray}
in terms of modified Bessel functions $I_0$ and $K_0$, and where $\alpha = \frac{24\pi}{c} \sqrt{2C E}$.\footnote{Generalizations of the dissipating black hole when including charges and/or supercharges can be found in \cite{DeVuyst:2022bua}.}

\begin{figure}
\centering
        \includegraphics[width=0.4\textwidth]{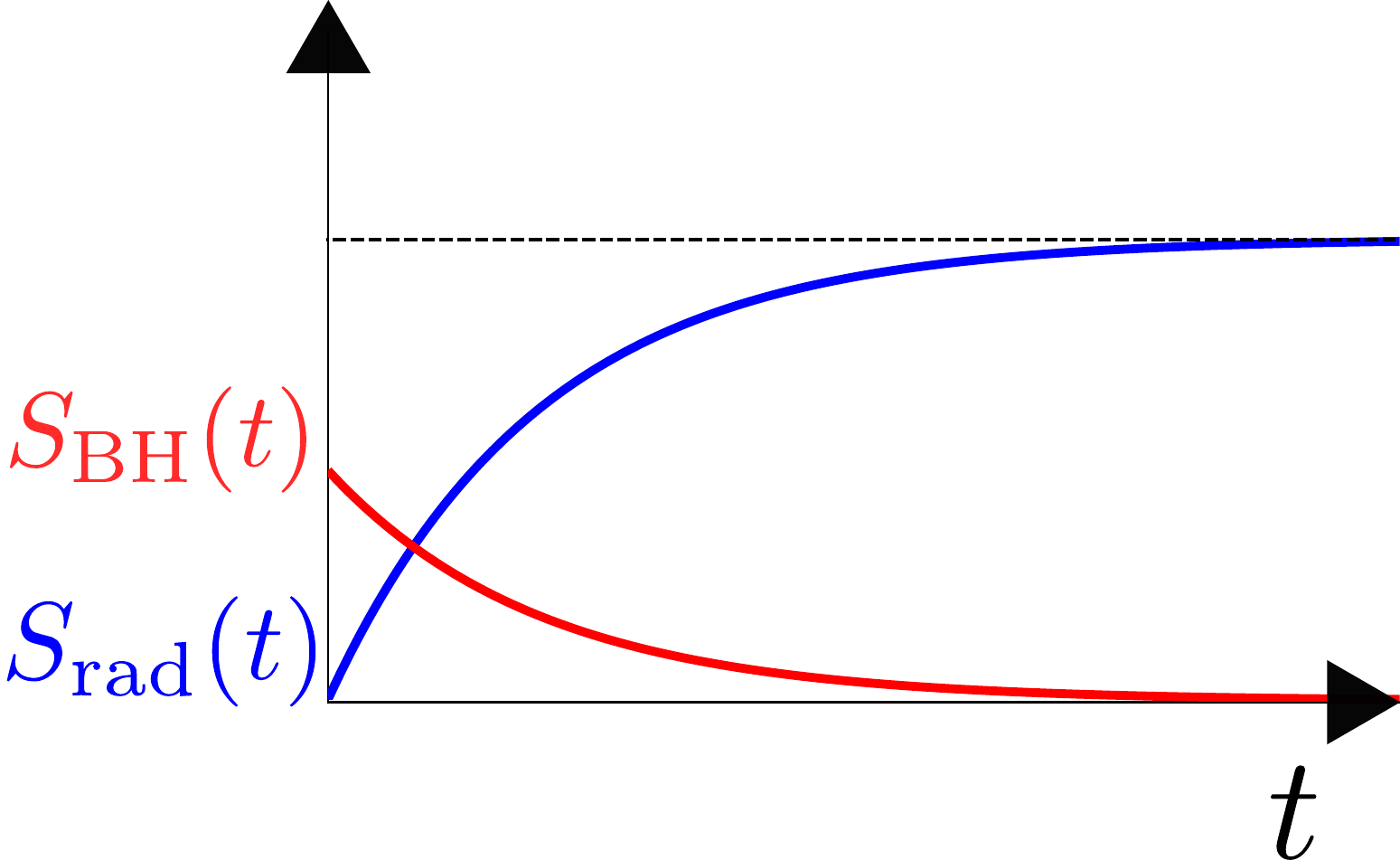}
   \caption{Coarse-grained entropy $S_{\text{BH}}(t)$ and fine-grained early-late matter entanglement entropy $S_{\text{ren}}(t)$ as a function of time $t$ during evaporation.}
        \label{fig:evap2}
 \end{figure}

From this dissipating energy profile, using \eqref{Sth}, one can likewise find an instantaneous Bekenstein--Hawking entropy as (in this argument we consider entropies above extremality, and therefore $S_0$ does not play a role)
\begin{equation}
    S_{\text{BH}}(t) = 2\pi \sqrt{2C E}e^{-\frac{c}{48\pi C}t}, 
\end{equation}
which is a measure for the course-grained entropy of the dissipating black hole during evaporation.
Using the 2d CFT entanglement entropy formula, one can also compute the fine-grained (renormalized) matter entanglement entropy between the early radiation (hitting the boundary at times $<t$) and the late radiation (arriving at the boundary at times $>t$) \cite{Mertens:2019bvy}. For a macroscopic black hole where $E \gg 1/C$, one gets:
\begin{equation}
\label{Sren}
    S_{\text{rad}}(t)  ~=~ \frac{c}{12} \ln \frac{F(t)^2}{ t^2F'(t)} ~\approx~ 4\pi \sqrt{2C E}\left( 1 - e^{-\frac{c}{48\pi C}t}\right).
\end{equation}
This profile is always increasing. Both entropies are illustrated as a function of time $t$ in Fig.~\ref{fig:evap2}. From the above, one sees that the matter entropy $S_{\text{rad}}(t)$ rises twice as fast as the black hole entropy $S_{\text{BH}}(t)$ decreases, which is no coincidence \cite{Zurek:1982zz,Fiola:1994ir}. Indeed, assuming black hole evaporation is an irreversible process of radiating into empty space, we can relate the black hole first law $dS_{\text{BH}} = d E/T$ and the 2d relativistic ideal gas law $E = ST/2$ of the thermal atmosphere, yielding the relation  $dS_{\text{BH}} = -dS/2$.

Since the matter entropy $S_{\text{rad}}(t)$ never comes down to zero, this is a clean quantitative illustration of Hawking's information loss paradox, and shows that unitarity of the full quantum system is not apparent in the semi-classical gravitational approximation.
This and related puzzles, motivates us to go beyond classical gravity, a problem to which we turn next.

\section{Quantum Jackiw--Teitelboim gravity}\label{sec:JTquantum}

The previous section has focused on aspects of JT gravity that can be understood from a classical gravity treatment. One of the attractive features of JT gravity is the fact that one can also study quantum gravitational effects exactly. We first study the quantum corrections to the black hole spectrum as derived from the gravitational path integral approach. Then we move on to the study of matter correlators including quantum gravity corrections.

\subsection{Spectrum of quantum black holes}
The gravitational path integral pioneered by Gibbons and Hawking \cite{Gibbons:1976ue} that computes the black hole partition function instructs us to integrate over smooth geometries with boundary conditions corresponding to fixed inverse temperature $\beta$, fixing the size of the thermal circle. The thermal disk or black hole partition function is given by the JT path integral:
\begin{align}
Z(\beta) &= e^{S_0} \int [Dg] [D\Phi] ~ e^{ \frac{1}{16\pi G_N}\left[\int_{\mathcal{M}} d^2 x\sqrt{g} \Phi (R+2) + 2\oint_{\partial\mathcal{M}} \sqrt{h} \Phi (K-1)\right]} \label{eqn:SchPIZJT}\\
&= e^{S_0} \int_{{\rm Diff}(S^1)/\text{SL}(2,\mathbb{R})} [Df] ~e^{C \int_0^\beta d\tau \{ \tan \frac{\pi f(\tau)}{\beta},\tau\} }. \label{eqn:SchPIZ}
\end{align} 
As explained in Sect.~\ref{s:geomder}, we reduce the bulk JT model to that of the Schwarzian theory on its boundary. Within the path integral, the procedure consists of first path-integrating over the dilaton along an imaginary contour, producing a functional delta constraint $\prod_x \delta (R(x)+2)$ reducing the integral to only hyperbolic geometries with $R(x)=-2$. The remaining degree of freedom, the Schwarzian mode, arises from the freedom to cut out an appropriate patch of the hyperbolic disk with a fixed length $\beta/\epsilon$ boundary. The cost of each configuration is given by the Schwarzian action.

Within the Euclidean gravitational path integral approach to quantum gravity, an omnipresent issue is that of the unstable conformal mode, making the Euclidean gravitational action unbounded from below, and causing the path integral to diverge. The standard way of dealing with this is to define the path integral along a specific cycle of complex metrics \cite{Gibbons:1978ac}. For JT gravity (and generic dilaton gravities), we see that the answer is to take the bulk dilaton imaginary in the Euclidean path integral. The question of unbounded Euclidean action is then transferred to the boundary Schwarzian action, which we will show explicitly does not contain negative modes.

To incorporate quantum effects, the derivation outlined in the previous paragraph is not yet complete. For Eq.~\eqref{eqn:SchPIZ} to be meaningful, it is necessary to provide a measure for the integration over $f(\tau)$, derived from the integral in Eq.~\eqref{eqn:SchPIZJT} over the metric and dilaton. This analysis was carried out in Sect.~3 of \cite{Saad:2019lba}. The proposal is to pick the measure of JT gravity to correspond to the symplectic measure in the BF formulation discussed in Sect.~\ref{s:firstorder}. This analysis leads to $[Df] = \prod_{\tau} df(\tau) / f'(\tau)$, which is invariant under reparametrizations of the Schwarzian mode.\footnote{It was shown in \cite{Moitra:2021uiv} how this measure directly arises in the second order formalism.} We will come back to the implications of this choice later.

\subsubsection{Perturbative calculation}
\label{s:diskpf}
We now compute the disk partition function in JT gravity. The Schwarzian action looks rather complicated, and a first approach is to use perturbation theory: we expand the reparametrization $f(\tau)$ in terms of its saddle $f_0(\tau)=\tau$, corresponding to a circular boundary, plus fluctuations:
\begin{equation}
    f(\tau) = \tau + \varepsilon(\tau), \qquad \varepsilon(\tau+\beta) = \varepsilon(\tau).
\end{equation}
For small $\varepsilon$ the action becomes $I_{\rm Sch} \approx \frac{2\pi^2 C}{\beta} +\mathcal{O}(\varepsilon^2)$. Setting the quantum fluctuation to zero gives the classical black hole spectrum analyzed in the previous section 
\begin{equation}
    Z(\beta) \approx e^{-I_{\rm classical}} = e^{S_0+\frac{2\pi^2 C}{\beta}}, \qquad \Rightarrow \qquad \rho(E) \approx e^{S_0+2\pi\sqrt{2CE}}.
\end{equation}
This answer reproduces the two-dimensional analog of the Bekenstein--Hawking prescription identifying the horizon value of the dilaton with the entropy of the black hole in a classical gravity approximation. 

Next we incorporate the one-loop determinant around this saddle. One of the advantages of JT gravity is the fact that this is a simple explicit computation. This is not so in higher dimensions, with the only exception being the BTZ black hole \cite{Giombi:2008vd,Maloney:2007ud}. We expand fluctuations in Fourier space:
\begin{equation}
    \varepsilon(\tau) = \frac{\beta}{2\pi}\sum_{ n\neq -1,0,1} e^{- \frac{2\pi}{\beta} i n \tau} \varepsilon_n + {\rm h.c.}
\end{equation}
The action to quadratic order is given by 
\begin{equation}
    I_{\rm Sch}= 
     \frac{2\pi^2 C}{\beta}+ \frac{4\pi^2 C}{\beta} \sum_{n>1} n^2 (n^2-1) \bar{\varepsilon}_n\varepsilon_n + \mathcal{O}(\varepsilon^3),
\end{equation}
and computing the Gaussian integral in these variables is very simple. The measure of integration around the saddle point is given by \cite{Stanford:2017thb,Moitra:2021uiv}: $[Df] = \prod_{n\geq 2} 4\pi (n^3-n) d\varepsilon_n d\varepsilon_{-n}$. We combine the quadratic action for the Schwarzian mode with the measure to obtain the partition function to a one-loop approximation 
\begin{equation}
    Z(\beta) ~~=~~ e^{S_0 + \frac{2\pi^2 C}{\beta}} \prod_{n\geq 2} \frac{\beta}{C n} 
    ~~=~~ \frac{1}{4\pi^2}\left( \frac{2\pi C}{\beta}\right)^{3/2} e^{S_0 + \frac{2\pi^2 C}{\beta}}
    .\label{eq:oneloopZSCH}
\end{equation}
We make the following observations on this result:
\begin{itemize}
    \item The measure is only defined up to an overall factor, which can be absorbed into a shift of $S_0$. This will be less trivial when considering non-perturbative contributions with spacetime wormholes in Sect.~\ref{sec:JTRMT}. \\
    \item Since there is only a single saddle, and all quadratic fluctuations are stable (positive), this means that the saddle is a global minimum and the action functional is bounded below. This is no longer true for some generalizations of the Schwarzian model to be discussed in Sect.~\ref{s:defects}. \\
    \item We have chosen zeta-function regularization, following standard practice. Other choices can be absorbed into shifts of $S_0$ or the zero-point energy (which we have set to zero). 
    The power of $3/2$ appearing in the one-loop determinant has a nice interpretation. In zeta-function regularization we have $\sum_{n\in\mathbb{Z}} 1 \to 0$, and if all Fourier modes were present, the one-loop determinant would be $\beta$-independent. Since the Schwarzian action has a PSL$(2,\mathbb{R})$ symmetry, one has to remove the $n=-1,0,1$ modes. The factor of 3 is counting the number of omitted zero-modes in the path integral. \\
    \item Potential corrections to the one-loop result appear as a Taylor expansion in $(\beta/C)$ to \eqref{eq:oneloopZSCH}. This can be seen by rescaling $\epsilon$ by $\epsilon \to \epsilon \beta/C $. Therefore this ratio $\beta/C$ can be interpreted as an effective dimensionless coupling constant.  \\
    \item A somewhat miraculous statement is that the partition function $Z(\beta)$ is in fact one-loop exact! So our expression \eqref{eq:oneloopZSCH} is the entire answer. The reason is explained by \cite{Stanford:2017thb} in terms of a localization argument. 
\end{itemize}
The result \eqref{eq:oneloopZSCH} has interesting implications. We can compute the free energy given by the logarithm of the partition function:
\begin{equation}
    -\beta F \equiv \log Z(\beta) = S_0 + \frac{2\pi^2 C}{\beta} + \frac{3}{2} \log \frac{2\pi C}{\beta} + \ldots,
\end{equation}
where the dots are temperature-independent. The first two terms on the right-hand side are classical contributions while the third is a quantum effect. The quantum effects become large as we lower the temperature $\beta \gtrsim C$. When applied to near-extremal black holes, it implies quantum gravity effects are large very close to extremality. We elaborate on this application in Sect.~\ref{sec:NEBHc}.

We can interpret \eqref{eq:oneloopZSCH} in holography, which states that a black hole can be described by a quantum system with finite entropy. The partition function of such a system with Hilbert space $\mathcal{H}_{BH}$ and Hamiltonian $H$ would be 
\begin{eqnarray}
    Z(\beta) = {\rm Tr}_{\mathcal{H}_{\rm BH}} \left[ e^{-\beta H} \right]
    = \int dE ~\rho(E)~ e^{-\beta E},~\rho(E) = \sum_n \delta(E-E_n),
\end{eqnarray}
    where $E_n$ is a discrete set of states of the quantum system describing the black hole.  
    We can use Eq.~\eqref{eq:oneloopZSCH} to infer what the density of states of this quantum system should be. An inverse Laplace transform of Eq.~\eqref{eq:oneloopZSCH} gives
\begin{equation}\label{eqn:JTexactRho}
\rho_{\rm JT}(E) = \frac{C}{2\pi^2} e^{S_0}\sinh \left( 2\pi \sqrt{2 CE} \right).
\end{equation}
Some comments on this result:
\begin{itemize}
    \item Expression \eqref{eqn:JTexactRho} is not a sum of delta functions and the spectrum is continuous. In quantum mechanics, this is usually associated to a non-compact space such that one considers the density of states per unit volume, but in our case there is no infinite spatial dimension in the boundary theory. A continuous spectrum implies the entropy in the microcanonical ensemble is infinite, such that information can be lost inside a black hole. To see signs of a discrete spectrum with finite entropy will require non-perturbative effects beyond those we have access to at the level of the disk.
    \item For large $E\gg 1/C$, the spectrum is $\rho_{\rm JT}(E) \approx e^{S_0 + 2\pi \sqrt{2 C E}}$, consistent with the classical Bekenstein--Hawking entropy of the JT black hole in \eqref{Sth}.
    \item Quantum effects are large at small energies and indeed for $E\ll 1/C$ the spectrum is strongly modified $\rho_{\rm JT}(E) \approx e^{S_0} \sqrt{2C E}$. The density of states goes to zero as $E\to 0$! This shows that it is wrong to interpret $S_0$ as a ground state degeneracy.
\end{itemize}

We reduced the gravitational description in JT gravity to a solvable boundary graviton mode. There is a derivation of the path integral at the one-loop level performed by \cite{Charles:2019tiu} that makes the relation between the boundary Schwarzian and bulk metric fluctuations more transparent. A similar analysis can also be found in \cite{Moitra:2021uiv}.

\subsection{Quantum Jackiw--Teitelboim gravity coupled to matter}
We now perform the calculation with the addition of matter (not coupled to the dilaton field $\Phi$ as studied above in Sect.~\ref{sec:JT}). For concreteness we work with a concrete example: a massive scalar field $\phi$ propagating on the JT black hole background. Its Euclidean action is
\begin{equation}
    I_{\rm matter}[\phi,g] = \frac{1}{2} \int d^2 x \sqrt{g}\left[ (\partial \phi)^2 + m^2 \phi^2\right].
\end{equation}
In the putative dual 1d nearly-conformal quantum mechanics, this field is dual to an operator $\mathcal{O}$ of scaling dimension (defined at short distances or large energies where the gravitational effects that break the conformal symmetry are small) given by $\Delta = 1/2+ \sqrt{1/4+m^2}$. We compute the gravitational path integral with Dirichlet boundary conditions for the matter field
\begin{eqnarray}
     Z(\beta; \phi_b) &=& e^{S_0}\hspace{-0.1cm} \int [Dg][D\Phi]   ~ e^{\frac{1}{16 \pi G_N}\left[\int_\mathcal{M} \Phi(R+2) + 2\oint_{\partial \mathcal{M}} \Phi (K-1)\right]} \hspace{-0.1cm}\int [D\phi]~e^{-I_{\rm matter}[\phi,g]}\nonumber\\
     &=& e^{S_0} \int [Df]~e^{-I_{\rm Sch}[f]}~ Z_{\rm matter}[\phi_b ,f],
\end{eqnarray}
where we defined the matter path integral in the hyperbolic disk described by a cut-off curve $f(\tau)$ and with boundary conditions $\phi \vert_\partial = \phi_b$ as  $Z_{\rm matter}[\phi_b,f] =\int [D\phi]~e^{-I_{\rm matter}[\phi,g]}$. This path integral depends in principle on both the Schwarzian mode and the boundary value of the scalar field. Notice the order we chose to do the path integration is important: first the dilaton, then the matter, and lastly the Schwarzian.

The undeformed theory corresponds to $\phi_b\to 0$, such that boundary sources are turned off. In the limit that the boundary curve approaches the boundary of the hyperbolic disk, i.e. its proper length goes to infinity, $Z_{\rm matter}[0,f]$ becomes independent of the Schwarzian mode to leading order. An explicit illustration of this fact is given in Appendix C of \cite{Yang:2018gdb}, for the case of a massless scalar field with $m=0$ (actually the calculation applies to any two-dimensional conformal field theory).\footnote{One can use conformal symmetry to map the cut-off hyperbolic disk with boundary curve labeled by $f(\tau)$ to a unit disk with a circular boundary. The non-trivial dependence with $f(\tau)$ arises then from the conformal anomaly involved in this transformation. An explicit calculation shows that the final answer is independent of $f(\tau)$ up to terms suppressed in the proper length of the boundary curve.}

The partition function (even in the presence of matter) is given by the Schwarzian density of states and therefore the quantum black hole spectrum is insensitive to the matter content.\footnote{This requires some fine print. If we assume $\phi_b$ is non-zero but arbitrarily small, then at low enough temperatures the contribution from the matter can be dominant in the window $1<\Delta<3/2$, see \cite{Maldacena:2016upp}.} The situation changes when we turn on a source $\phi_b \neq 0$ for the boundary dual operators $\mathcal{O}$, deforming the boundary theory. 
The source $\phi_r$ in the boundary's own time coordinate $\tau$, is defined through the relations:
\begin{equation}
\phi\vert_\partial = Z^{1-\Delta}\tilde{\phi}_r(T) =\epsilon^{1-\Delta}F'^{1-\Delta}\tilde{\phi}_r(T) = \epsilon^{1-\Delta} \phi_r(\tau),
\end{equation}
where we have first used the Poincar\'e coordinates $(T,Z)$ and its source $\tilde{\phi}_r$, and rewritten this in coordinates specified along the boundary curve defined by the Schwarzian mode. The result is the well-known generating functional for generalized free fields, but reparametrized to the fluctuating boundary line:
\begin{eqnarray}
     Z_{\rm matter}[\phi_r,f] =  e^{\frac{D}{2} \int d\tau_1 d\tau_2 \Big( \frac{f'(\tau_1) f'(\tau_2)}{\frac{\beta^2}{\pi^2}\sin^2 \frac{\pi}{\beta} [f(\tau_1)-f(\tau_2)]}\Big)^{\Delta} \phi_r(\tau_1) \phi_r (\tau_2)},\label{eq:MatterPFS}
\end{eqnarray}
where $D=\frac{(\Delta-\frac{1}{2})\Gamma(\Delta)}{\sqrt{\pi}\Gamma(\Delta-\frac{1}{2})}$. From now on we rescale the matter field to absorb the factor of $D$, $\phi_r \to \phi_r/\sqrt{D} $. Equation \eqref{eq:MatterPFS} is true up to an overall coefficient  independent of the Schwarzian mode and the source, and we absorb it in $S_0$. 

\subsection{Correlators}
\label{s:correlato}
Next, we go over the calculation of the boundary matter correlators dual to JT gravity coupled to matter fields. Some partial results were obtained in \cite{Bagrets:2016cdf, Bagrets:2017pwq} and the full solution for all correlators was derived in \cite{Mertens:2017mtv} realizing that the Schwarzian theory is a limit of Liouvile field theory, further elaborated in \cite{Mertens:2018fds, Lam:2018pvp}. These correlators were later reproduced using a variety of approaches by \cite{Blommaert:2018oro,Iliesiu:2019xuh} using the BF approach and by \cite{Kitaev:2018wpr,Yang:2018gdb, Suh:2020lco} using the particle in a magnetic field perspective on JT gravity, proposed by Kitaev in 2016.

\subsubsection{Path integral representation}
Computing correlators involves taking functional derivatives with respect to matter sources as
\begin{equation}
    \langle \mathcal{O}(\tau_1)\ldots \mathcal{O}(\tau_n)\rangle = \frac{\delta}{\delta \phi_r (\tau_1)} \ldots \frac{\delta}{\delta \phi_r (\tau_n)} Z(\beta, \phi_r) \Big\vert_{\phi_r\to 0},
\end{equation}
with generating functional
\begin{align}\label{eq:GenFuncSour}
    Z(\beta,\phi_r) =  e^{S_0} \int [Df]~e^{-I_{\rm Sch}[f]}~ e^{\frac{1}{2} \int d\tau_1 d\tau_2 \left( \frac{f'(\tau_1) f'(\tau_2)}{\frac{\beta^2}{\pi^2}\sin^2 \frac{\pi}{\beta} [f(\tau_1)-f(\tau_2)]}\right)^{\Delta} \phi_r(\tau_1) \phi_r (\tau_2)}.
\end{align}
The quantum gravity effects are encoded in the fluctuations of the Schwarzian mode.

\paragraph{Two-point function} Let us begin with the two-point function (since one-point functions vanish):
\begin{align}
     \langle \mathcal{O}(\tau_1) \mathcal{O}(\tau_2) \rangle = e^{S_0} \int [Df]e^{C \int_0^\beta \{ \tan \frac{\pi}{\beta}f,\tau\} }\left( \frac{f'(\tau_1) f'(\tau_2)}{\frac{\beta^2}{\pi^2}\sin^2 \frac{\pi}{\beta} [f(\tau_1)-f(\tau_2)]}\right)^{\Delta}.
\end{align}
The building block here is the conformal two-point function, while the coupling to gravity is accounted for by the reparametrization mode. For example, when we freeze gravity by sending $C\to\infty$, we can set $f(\tau) = \tau$ (the gravitational saddle), and we reduce to the conformal two-point function 
\begin{equation}\label{eqn:2ptsemiclass}
     \langle \mathcal{O}(\tau_1) \mathcal{O}(\tau_2) \rangle \approx Z(\beta) \left( \frac{\pi^2}{\beta^2\sin^2 \frac{\pi}{\beta}\tau_{12}} \right)^{\Delta}.
\end{equation}
We will see how this semi-classical answer emerges from the exact expression later, and how corrections are suppressed as long as $\beta,\vert \tau_{12}\vert \ll C$. \\

The previous discussion motivates introducing the following bilocal operator within the Schwarzian theory
\begin{equation}
\label{Schbil}
    G_\Delta (\tau_1,\tau_2) \equiv\left( \frac{f'(\tau_1) f'(\tau_2)}{\frac{\beta^2}{\pi^2}\sin^2 \frac{\pi}{\beta} [f(\tau_1)-f(\tau_2)]}\right)^{\Delta}.
\end{equation}
This operator has an important property: it is invariant under the SL$(2,\mathbb{R})$ transformations \eqref{eqn:SL2Rschw}, and since by \eqref{casHam} the Hamiltonian is the quadratic Casimir, it hence commutes with the Hamiltonian. It will be useful later to denote the two-point function by the following diagram
\begin{eqnarray}
\includegraphics[scale=0.14]{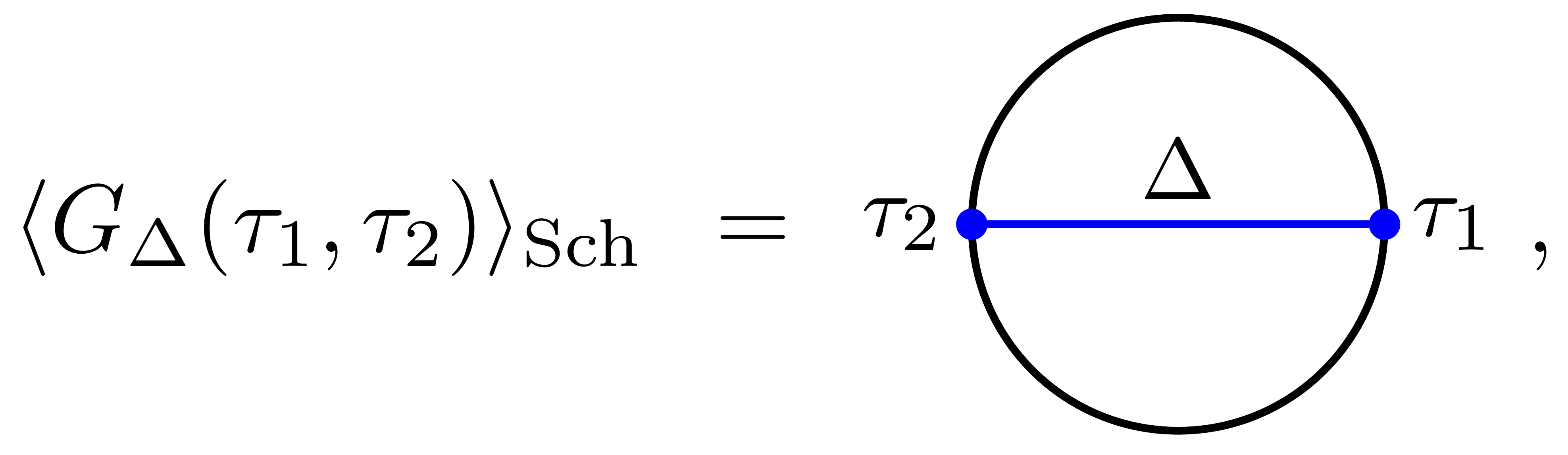}
\end{eqnarray}
where we drew the boundary thermal circle, and the points denote the insertions labeled by the times $\tau_1$ and $\tau_2$. The line going through the ``interior'' denotes the insertion of the bilocal operator $G_{\Delta}(\tau_1,\tau_2)$.

\paragraph{Four-point function}
We will also work out the four-point function in some detail. Assume for concreteness that $\tau_1<\tau_2<\tau_3<\tau_4$. The boundary four-point function of a single free field is a sum of products of two-point functions in different channels since, turning off the reparametrization mode, the action in \eqref{eq:MatterPFS} is quadratic. After applying a reparametrization, each two-point function becomes the Schwarzian bilocal. The final answer is\footnote{To simplify the discussion, we could also consider the presence of two free fields with scaling dimensions $\Delta_1$ and $\Delta_2$ dual to operators $\mathcal{O}_1$ and $\mathcal{O}_2$. Then the four point function is given by $\langle G_{\Delta_1} G_{\Delta_2}\rangle$. In this case there is only a single channel depending on the position of the two insertions and there is no need to sum over three different diagrams.}  
\begin{eqnarray}
 \langle \mathcal{O}(\tau_1)\mathcal{O}(\tau_2)\mathcal{O}(\tau_3)\mathcal{O}(\tau_4)\rangle & =& \langle G_\Delta (\tau_1,\tau_2) G_\Delta(\tau_3,\tau_4) \rangle_{\rm Sch}\nonumber\\
 &&\hspace{-3cm}+\langle G_\Delta (\tau_1,\tau_4) G_\Delta(\tau_2,\tau_3) \rangle_{\rm Sch}+\langle G_\Delta (\tau_1,\tau_3) G_\Delta(\tau_2,\tau_4) \rangle_{\rm Sch}.
\end{eqnarray}
This gives the matter four-point function as a Schwarzian correlator of a product of bilocals. We describe the answer for each term in the next section. The four-point function can be represented in a diagrammatic way as the sum of these three terms:
\begin{equation}
\includegraphics[scale=0.2]{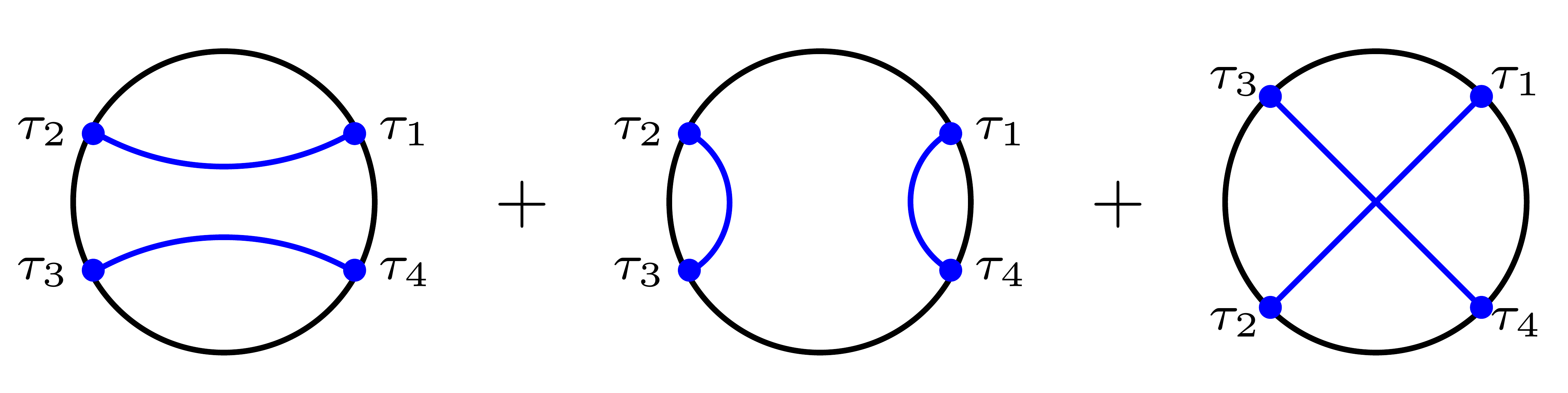}\nonumber
\end{equation}
The first two terms above are simple to compute: as mentioned above, the bilocal operator commutes with the Hamiltonian and states with the same energy propagate along the lines joining $1-4$ and $2-3$ in the first diagram, or $1-2$ and $3-4$ in the second diagram. This implies that in the energy eigenbasis the first two contributions are just products of two-point functions. The third diagram involves a new calculation due to the specific ordering of insertions and bilocals. As explained in \cite{Maldacena:2016upp}, this correlator when coupled to the Schwarzian mode will describe gravitational shockwave scattering in the bulk. 

\paragraph{Generalizations}
For free fields the generalization to higher-point functions is straightforward: they will involve higher correlators of the Schwarzian bilocal.  

Another generalization we will not pursue here is to include self-interactions between matter fields in the bulk (besides their gravitational interactions). This involves first computing the interacting matter boundary correlator in rigid AdS$_2$. Then one applies a reparametrization and integrates over the Schwarzian mode. The conformal block expansion shows that it can always be expanded in a possibly infinite number of bilocal insertions. Even in the presence of bulk interactions, the one- and two-point functions are universal. Three-point functions are universal up to a single coupling factor: the OPE coefficient, but can likewise be written in terms of a product of three bilocal operators when coupling to the gravitational Schwarzian sector.

\subsubsection{A first approach: Schwarzian perturbation theory} \label{s:pert}
As a first approach of computing Schwarzian correlation functions of bilocal operators \eqref{Schbil} explicitly, one can utilize perturbation theory as introduced before in Sect.~\ref{s:diskpf}. Schwarzian perturbation theory requires the propagator ($u = 2\pi \tau/\beta$)  \cite{Maldacena:2016hyu,Maldacena:2016upp}:
\begin{equation}
\label{propa}
\left\langle \varepsilon(\tau)\varepsilon(0)\right\rangle 
= \frac{1}{2\pi C}\left(\frac{\beta}{2\pi}\right)^{3} \left[1 - \frac{1}{2}(u-\pi)^2 + \frac{\pi^2}{6} + \frac{5}{2}\cos u + (\tau-\pi) \sin u\right],
\end{equation}
and the expansion of the bilocal operator insertion as:
\begin{equation}
\label{opexpa}
\left(\frac{F'_1F'_2}{(F_1-F_2)^{2}}\right)^{\Delta} = \frac{(1+\varepsilon_1')^{\Delta}(1+\varepsilon_2')^{\Delta}}{(\frac{\beta}{\pi} \sin \frac{\pi}{\beta}(\tau_{12} + \varepsilon_1-\varepsilon_2))^{2\Delta}}.
\end{equation}
Any propagator carries a factor $1/C$, whereas any vertex gives a factor of $C$. The resulting perturbative expansion is then a power series in $1/C$,\footnote{One convenient way of seeing this is to redefine $\sqrt{C}\epsilon \to \epsilon$ to make the propagator independent of $C$. This makes all vertices (including those coming from expanding the operator as in \eqref{opexpa}) contribute with negative integer powers of $C$.} which gets complicated very quickly since new interaction vertices appear at higher orders as well. Nonetheless, interesting computations have been done, see e.g. \cite{Maldacena:2016upp,Haehl:2017pak,Maldacena:2019cbz,Cotler:2019nbi,Qi:2019gny} for a selection of applications at lower orders in $1/C$. 

Fortunately, the situation in JT gravity is even better: exact expressions are known to which we turn next, which can be compared to the perturbative $1/C$ results as we will comment further on.

\subsubsection{Exact two-point function}\label{sec:Result2ptf}
We now state the explicit answers of Schwarzian correlation functions, and review some of their immediate gravitational properties. We postpone an overview of their derivation until Sect.~\ref{s:deriv}.

For the two-point function, the answer can be written in the following suggestive form as a double integral:
\begin{align}
    \langle \mathcal{O}(\tau_1)\mathcal{O}(\tau_2)\rangle =  e^{S_0}\int_0^\infty \hspace{-0.2cm}\prod_{i=1,2} (dk_i^2 \sinh 2 \pi k_i )~ e^{-\tau \frac{k_1^2}{2C} - (\beta-\tau)\frac{k_2^2}{2C}}\frac{\Gamma(\Delta \pm ik_1 \pm ik_2)}{8\pi^4(2C)^{2\Delta}\Gamma(2\Delta)}. \label{eqn:2ptSchwFull}
\end{align}
where we defined $ \tau =  \vert \tau_1 - \tau_2 \vert $. This is the unnormalized expression (where we did not divide by the partition function $Z(\beta)$). We introduced the useful notation that whenever $\pm$ appears inside a gamma function, it means one should take a product of that gamma function with both signs. For example $\Gamma(x \pm y ) = \Gamma(x+y) \Gamma(x-y)$ and hence the numerator in the right hand side of Eq.~\eqref{eqn:SchTwoPntME} involves a product of four gamma functions. 
As a simple check we can see that if $\Delta\to0$ then \eqref{eqn:2ptSchwFull} gives back the Schwarzian partition function, using the identities $\lim_{\Delta \to 0} \frac{\Gamma(\Delta\pm ix)}{\Gamma(2\Delta)} = 2\pi \delta(x)$ and $\Gamma(\pm 2ik) = \frac{\pi}{2 k\sinh2\pi k}$.

We can interpret this expression from the boundary holographic perspective. The quantum black hole in the bulk is supposed to be dual to a system with Hilbert space $\mathcal{H}_{\rm BH}$ defined above. Then $\mathcal{O}$ should be an operator within this theory. Quantum mechanics gives an expansion
\begin{eqnarray}
      \langle \mathcal{O}(\tau_1)\mathcal{O}(\tau_2)\rangle &=& {\rm Tr}_{\mathcal{H}_{\rm BH}} \left[ e^{-\beta H} \mathcal{O}(\tau_1) \mathcal{O}(\tau_2) \right]\\
     &&\hspace{-1cm}= \int \rho(E_1)dE_1 \rho(E_2)dE_2 e^{-\beta E_2 - \tau (E_1-E_2)} \vert \mathcal{O}_{E_1,E_2} \vert^2,\label{eqnnnn}
\end{eqnarray}
where for systems with finite entropy $\rho(E)$ is a sum of delta functions and $\mathcal{O}_{E_1,E_2}$ are the operator matrix elements between energy eigenstates. Matter does not affect the partition function to leading order, so we should identify $\rho(E) = \rho_{\rm JT}(E)$ given in Eq.~\eqref{eqn:JTexactRho}. Combining this with the expression \eqref{eqn:2ptSchwFull}, we can identify the integration variables with the intermediate energies as $E \equiv \frac{k^2}{2C}$. The remainder gives the off-diagonal matrix elements of the operator in the putative dual quantum mechanical system
\begin{eqnarray}\label{eqn:SchTwoPntME}
     \vert \mathcal{O}_{E_1,E_2}  \vert^2 =  2e^{-S_0}\frac{\Gamma\left(\Delta \pm i\sqrt{2 C E_1} \pm i\sqrt{2 C E_2}\right)}{(2C)^{2\Delta}\Gamma(2\Delta)},
\end{eqnarray}
 where the factor of $e^{-S_0}$ implies that $\mathcal{O}_{E_1,E_2} \sim \mathcal{O}(e^{-S_0/2})$ can be understood from the point of view of the eigenstate thermalization hypothesis (ETH) \cite{PhysRevA.43.2046, Srednicki:1994mfb}, as emphasized in \cite{Saad:2019pqd,Jafferis:2022uhu,Jafferis:2022wez}.\footnote{The phase of the matrix elements is arbitrary and would require knowing more complicated observables.}
 
We can extract from \eqref{eqnnnn} the two-point function in the microcanonical ensemble, essentially by stripping off the $E_2$-integral, evaluated in a state with fixed energy $E$. The answer is 
\begin{equation}
     \langle E \vert \mathcal{O}(\tau_1)\mathcal{O}(\tau_2) \vert E\rangle = \int_0^\infty dE' \rho(E') e^{-\tau (E-E')} \vert \mathcal{O}_{E,E'} \vert^2.
\end{equation}
We can verify that when the energy is of order $C$, this correlator becomes thermal with an effective temperature $\beta(E) = \sqrt{2\pi^2 C/E}$, consistent with ETH, see section 6.2 of \cite{Lam:2018pvp}.
      
At $\tau \to 0$ or $\tau\to \beta$, either the $E_1$ or the $E_2$ integral lose their exponential damping. The result is the UV divergence from the underlying matter CFT $ 1/\tau^{2\Delta}$ or $ 1/(\beta-\tau)^{2\Delta}$. This pole gets smoothed out if one embeds JT within the microscopic SYK model.
  
\begin{figure}[t!]
\begin{center}
\includegraphics[scale=0.3]{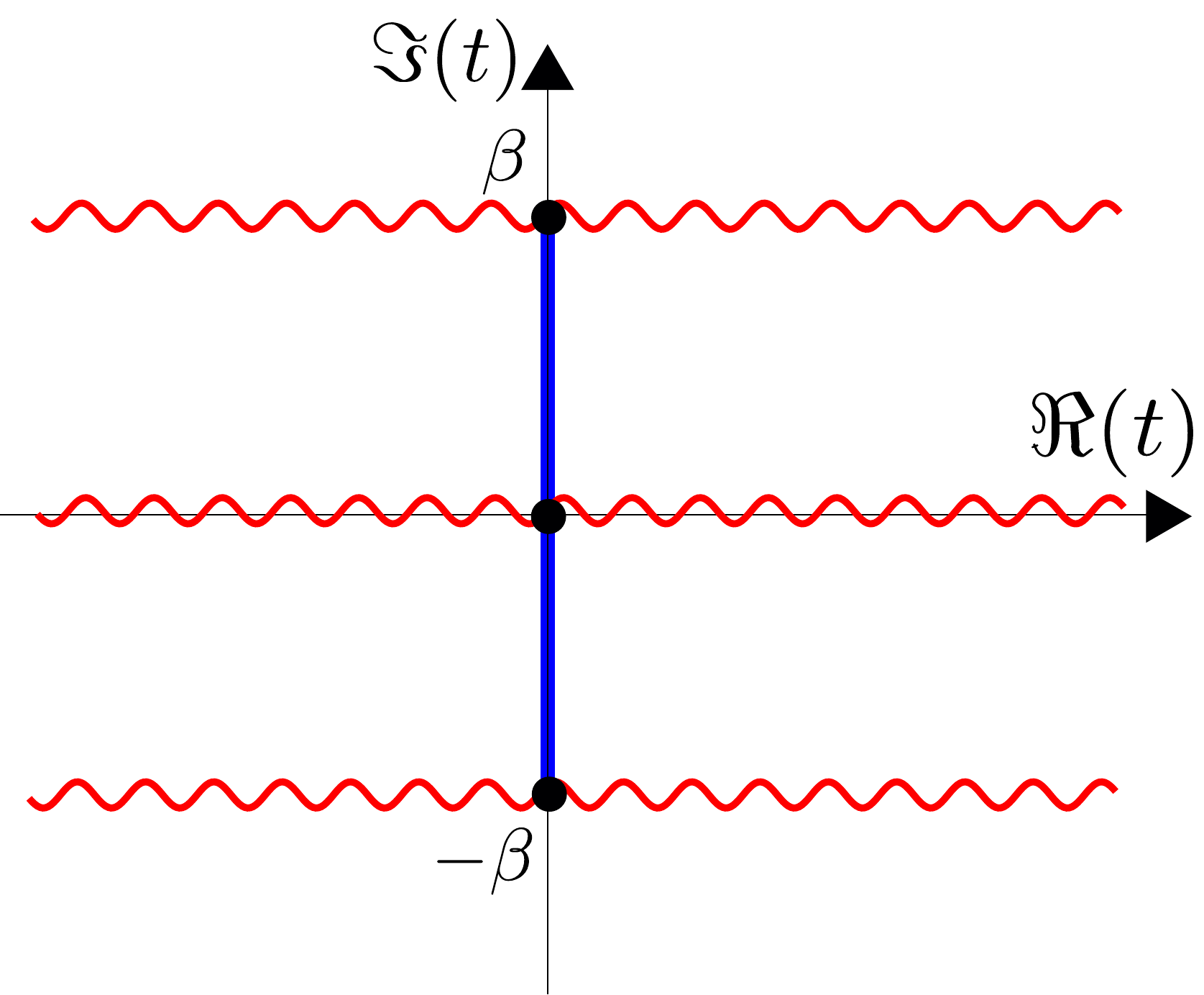} 
\end{center}
\caption{Complex time plane with Lorentzian time on the horizontal axis and Euclidean time on the vertical axis. Branch cuts of correlators are denoted by red wiggly lines.}
\label{Fig:AnalyticiTG}
\end{figure}
      
The result \eqref{eqn:2ptSchwFull} is written in Euclidean signature. It is possible to analytically continue this result to real time as follows.
The singularity as $\tau \to 0, \beta$ is actually part of a branch cut when analytically continuing to the complex plane $t$ where $\Im(t) \equiv -\tau$, with periodic cuts at $\Im(t) = n \beta$, as illustrated in Fig.~\ref{Fig:AnalyticiTG}. When continuing these expressions to real time $\Re(t)$, we have two answers depending on whether we approach the real axis from above or below.

\noindent These correspond to the two orderings of the operators $\mathcal{O}(t_i)$, and are the \textbf{real-time Wightman correlators} $G_+(t)= \langle \mathcal{O}(t_1)\mathcal{O}(t_2)\rangle$ and $G_-(t) = \langle \mathcal{O}(t_2)\mathcal{O}(t_1)\rangle$, where the $\pm$-symbol denotes the continuation from the bottom $(+)$ or top $(-)$ of Fig.~\ref{Fig:AnalyticiTG}, and $t=t_{1}-t_2$. These turn out to be complex conjugates of each other and $G_{\pm}(t)$  is given by
\begin{equation}
\label{realtime}
e^{S_0}\int_0^\infty \prod_{i=1,2} (2 k_i dk_i \sinh 2 \pi k_i )~ e^{\mp i t \frac{k_1^2}{2C} - (\beta \mp i t )\frac{k_2^2}{2C}}e^{-\epsilon \frac{k_1^2}{2C} - \epsilon \frac{k_2^2}{2C}} \frac{\Gamma(\Delta \pm ik_1 \pm ik_2)}{8\pi^4(2C)^{2\Delta}\Gamma(2\Delta)},
\end{equation}
where $\epsilon$ is a Euclidean regulator to make the integrals converge, $t=t_{1}-t_2$, and the $\pm$ symbol in the exponentials depends on which of the two correlators we consider (the Gamma-functions still involve a product over all signs). These are different since timelike separated operators generically do not commute.\footnote{We note that semi-classically both of these correlators have an infinitesimal imaginary part of opposite sign $\pm i \epsilon$. However, the full quantum expressions have finite imaginary parts and are hence different beyond infinitesimals.}

Finally, one can also separate the operators by $\beta/2$ in Euclidean time and $t$ in real time. This corresponds physically to having the operators acting on opposite halves of the thermofield double state (TFD). As visible in the above figure, the resulting real-time thermodouble correlation function is unique, reflecting the commutativity of operator insertions on opposite ends of the TFD state.
      
The semi-classical two-point function decays exponentially at large Lorentzian time. Indeed, under a Wick rotation $\tau \to i t$ of Eq.~\eqref{eqn:2ptsemiclass} we have $\langle \mathcal{O}(t) \mathcal{O}(0)\rangle \sim e^{-\Delta \frac{2\pi t}{\beta}}$. This is not so when quantum gravity is turned on. Equation \eqref{realtime} implies that at very late times $t\gg C$ the correlator is dominated by small energies $E_{1,2} \sim t^{-1}$. In this limit, $\vert \mathcal{O}_{0,0}\vert$ approaches a constant while $\rho(E) \sim \sqrt{E}$, implying that the correlators decay as 
\begin{equation}
\label{latetimeSch}
\langle \mathcal{O}(t) \mathcal{O}(0) \rangle \sim t^{-3}.
\end{equation}
This behavior was derived before knowing the full answer for the correlator in \cite{Bagrets:2017pwq}. Quantum gravity slows down the decay but does not stop it, leading to the information paradox raised in \cite{Maldacena:2001kr}. We will review later in Sect.~\ref{sec:JTRMT} how the addition of spacetime wormholes resolves this problem.

Let us finally make some comments on the large $C$ (or semi-classical) content of these expressions, and the relation to boundary graviton expansions.
\begin{itemize}
\item Quantum gravity effects break conformal invariance, and only as we take the semi-classical limit $C\to\infty$ with a \textbf{light} operator insertion $\Delta \sim \mathcal{O}(1)$, do we recover the conformal two-point function \eqref{eqn:2ptsemiclass}. This is true as long as both $\beta,\tau \ll C$. To see this, we first write $E_1 = E_2 + \omega$. Then the integral is dominated by configurations with large $E_{1,2} \sim \mathcal{O}(C)$ and small $\omega\sim \mathcal{O}(1)$. Physically, this means we have a semi-classical black hole of large energy $E_2$ that is perturbed by a small energy injection $\omega$ caused by the operator insertion. Writing $e^{-\beta E_2 - \tau(E_1-E_2)} = e^{-\beta E_2 - \tau \omega}$, in this limit the rest of the integrand becomes
      \begin{equation}
          \rho(E_1) \rho(E_2) \vert \mathcal{O}_{E_1,E_2} \vert^2 \sim e^{2 \pi \sqrt{2 C E_2}} ~\frac{e^{\pi \frac{\omega}{\sqrt{2 E_2/C}}}\Gamma\Big(\Delta \pm i \frac{\omega}{\sqrt{2 E_2/C}}\Big)}{\big( 2E_2/C\big)^{\frac{1}{2}-\Delta}\Gamma(2\Delta)},
      \end{equation}
      where we omit energy-independent prefactors. The integral over $E_2$ can be done by saddle point techniques combining the exponential in the RHS with the Boltzmann $e^{-\beta E_2}$ factor, giving the saddle identification $E_2 = 2 \pi^2 C/\beta^2$ (which is again the JT black hole first law). This produces a factor of $Z(\beta)$. The rest of the RHS is given by $\beta^{1-2\Delta}e^{\frac{\beta \omega}{2}} \Gamma(\Delta \pm i \frac{\beta \omega}{2\pi }) $. The $\omega$-integral then finally reduces to the Fourier transform of Eq.~\eqref{eqn:2ptsemiclass}.
      
      \item If the scaling dimension of the field is large in units of $C$, $\Delta \sim C$ as $C \to \infty$, the operator is called \textbf{heavy}. In this case classical gravitational backreaction is important: the saddle equation is sourced by the bilocal operator insertion since one can write the operator insertion into the action as: \begin{equation} 
      S_{\text{eff}} = \int dt \left( -C\left\{F,t\right\} - \Delta \log \frac{F_1'F_2'}{(F_1-F_2)^2} \right). \end{equation}
      Since $\Delta \sim C$, this contributes equally importantly as the Schwarzian action itself. The limit can be taken both directly in the expression \eqref{eqn:2ptSchwFull}, or using the new sourced saddle, with matching results \cite{Lam:2018pvp}. 
 \begin{figure}[t!]
  \begin{center}
\includegraphics[scale=0.15]{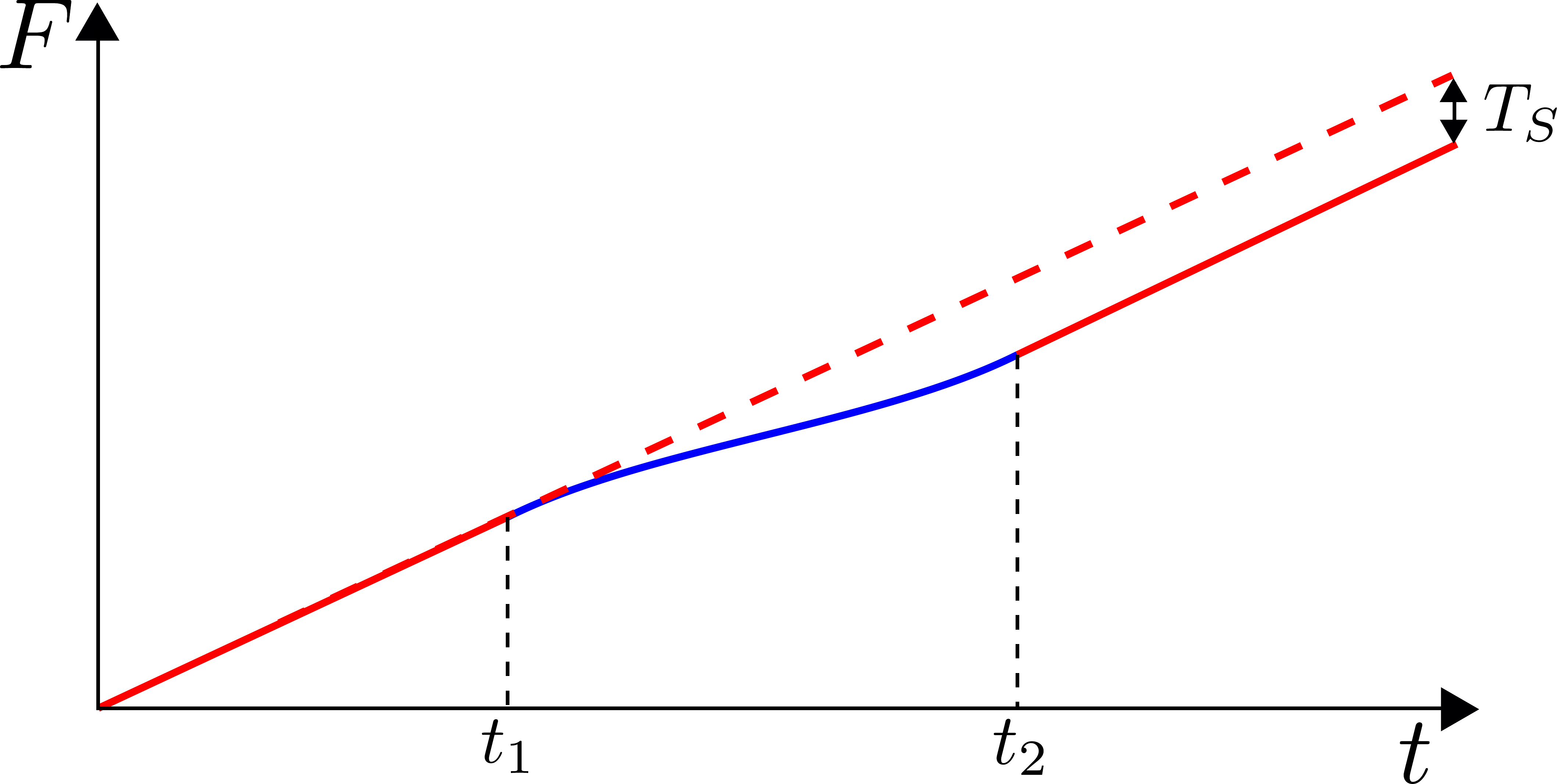} 
  \end{center}
  \caption{Classical solution for the Schwarzian mode $F(t)$ with sources inserted at time $t_1$ and reabsorbed at $t_2$. $T_S$ denotes the net time delay after this process.}
  \label{Fig:Shapiro}
 \end{figure}
An illustrative example is the case of zero temperature $\beta^{-1} \to 0$. In that case, the boundary clock starts at $F(t)=t$, transfers through a region of finite intermediate energy for $t_1<t<t_2$ between the ends of the bilocal operator, and ends up as $F(t) = t - T_S$, see Fig.~\ref{Fig:Shapiro}. The quantity $T_S > 0$ for $\Delta>0$ and is a net time delay, a Shapiro time delay, whose explicit expression depends on the time difference $t_2-t_1$ and the operator weight $\Delta$.
      
\item Beyond the large $C$ limits, the expression \eqref{eqn:2ptSchwFull} can be expanded in powers of $1/C$, and compared to the perturbative analysis. The results match \cite{Mertens:2020pfe,Griguolo:2021zsn}. Moreover, one can use this exact result to prove that the perturbative $1/C$ expansion is asymptotic for generic real values of $\Delta$, and hence cannot be captured just by the perturbative boundary graviton treatment of Sect.~\ref{s:pert}.

\item An exception occurs when $\Delta \in -\mathbb{N}/2$. In this case, the expression \eqref{eqn:2ptSchwFull} would seem to vanish due to the poles of the Gamma-function in the denominator. What actually happens, is that the numerator also degenerates to a single $k$-integral. The resulting expressions are somewhat simpler structurally, but correspond to non-unitary matter insertions \cite{Mertens:2020pfe}.\footnote{These have a natural origin in terms of the BF gauge model description (as finite-dimensional non-unitary representations of SL$(2,\mathbb{R})$), and in the Liouville CFT language (as degenerate Virasoro representations).} In this case, one can show that the $1/C$ perturbative expansion is convergent, and all physics is captured by boundary graviton Feynman diagrams.

\end{itemize}

\subsubsection{Exact four-point function}
We now describe the results for the different types of four-point functions. To separate the channels that contribute, we consider a correlator between two pairs of operators $\mathcal{O}_1$ and $\mathcal{O}_2$ with scaling dimensions $\Delta_{1,2}$, defined by
\begin{eqnarray}
    \langle \mathcal{O}_{\Delta_1}(\tau_1) \mathcal{O}_{\Delta_1}(\tau_2) \mathcal{O}_{\Delta_2}(\tau_3)\mathcal{O}_{\Delta_2}(\tau_4) \rangle.
\end{eqnarray}
We begin by considering the case with $\tau_1<\tau_2<\tau_3<\tau_4$ such that the four-point function is equal to $\langle G_{\Delta_1}(\tau_1,\tau_2)G_{\Delta_2}(\tau_3,\tau_4)\rangle_{\rm Sch}$. This corresponds to the diagram
\begin{equation}
\includegraphics[scale=0.2]{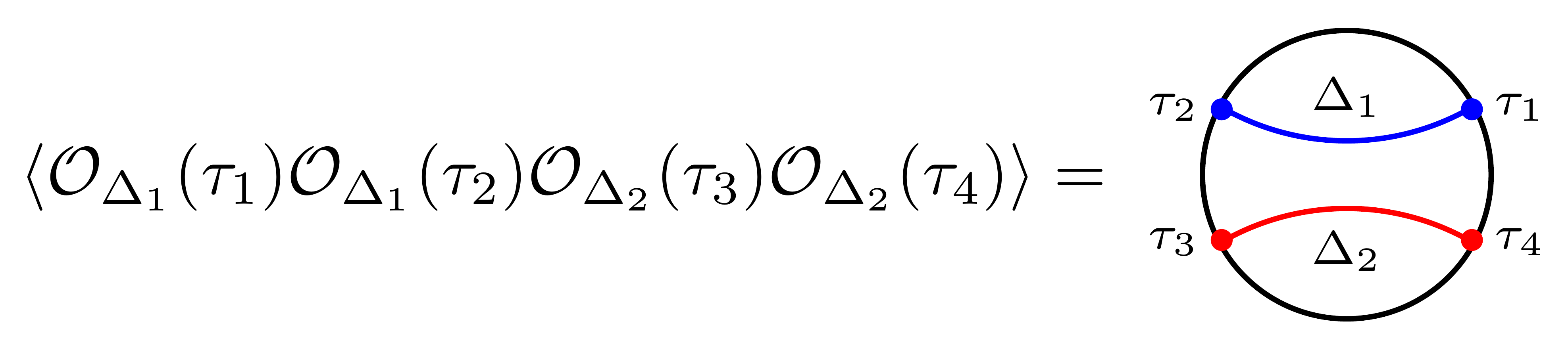}\nonumber
\end{equation}
It is convenient to rewrite the correlator in terms of a fixed energy amplitude $\mathcal{A}(k_1,\ldots,k_4)$ as 
\begin{eqnarray}
     \langle \mathcal{O}_{\Delta_1}(\tau_1) \mathcal{O}_{\Delta_1}(\tau_2) \mathcal{O}_{\Delta_2}(\tau_3)\mathcal{O}_{\Delta_2}(\tau_4) \rangle &=&\frac{e^{S_0}}{2} \int_0^\infty d\mu(k_1) d\mu(k_2)d\mu(k_3)d\mu(k_4)\nonumber\\
     &&\hspace{-3.5cm}\times e^{-\frac{k_1^2}{2C}\tau_{21}-\frac{k_2^2}{2C}\tau_{32}-\frac{k_3^2}{2C}\tau_{43}-\frac{k_4^2}{2C}(\beta-\tau_{41})}~\mathcal{A}(k_1,k_2,k_3,k_4),\label{eq:DefA4pt}
\end{eqnarray}
where $d\mu(k)\equiv  \frac{k \sinh 2\pi k}{\pi^2} dk$ and we can identify $\frac{k^2}{2C}$ with the energies propagating between each insertion: $k_1$ propagates between insertions $1-2$, $k_2$ between $2-3$, $k_3$ between $3-4$ and $k_4$ between $4-1$. Moreover it is clear that under this identification $d\mu(k)$ is proportional to the Schwarzian density of states. Then the quantity $\mathcal{A}(k_1,k_2,k_3,k_4)$ can be interpreted as a fixed energy amplitude computing the correlator. This amplitude is \cite{Mertens:2017mtv}
\begin{equation}\label{eqn:SchwExactTO4PT}
   \mathcal{A}_{\rm uncross.} = \frac{\Gamma(\Delta_1 \pm i k_1 \pm i k_2)}{(2 C)^{2\Delta_1 }\Gamma(2\Delta_1)}\frac{\Gamma(\Delta_2 \pm i k_3 \pm i k_2) }{(2 C)^{ 2 \Delta_2}\Gamma(2\Delta_2)} \frac{\pi^2 \delta(k_4-k_2)}{k_4 \sinh 2 \pi k_4}.
\end{equation}
The delta function is a manifestation of the SL$(2,\mathbb{R})$ invariance of the Schwarzian bilocal operator, and as expected the remaining factor is a product of two-point function amplitudes.

As a simple check, when $\Delta_1\to0$ \eqref{eqn:SchwExactTO4PT} the correlator reduces to the two-point function of the $\mathcal{O}_{\Delta_2}$ operator, and viceversa. We can also verify it gives the correct semiclassical limit: defining $\omega_{1,2} = (k_2^2-k_{1,3}^2)/2C$, then at large $C$ the integral over $k_2$ yields $k_2=2\pi C/\beta$ via the saddle point method, while the integrals over $\omega_{1,2}$ factorize into a product of conformal two-point functions. Finally, at very late times when quantum effects are important, the four-point function decays as $\langle \mathcal{O}(t_1) \mathcal{O}(t_2) \mathcal{O}(t_3) \mathcal{O}(t_4)\rangle \sim t_{12}^{-3} t_{34}^{-3}$ when $t_{12}, t_{34} \gg C$, similarly to the two-point function. Again this late-time power law was derived before knowing the complete answer by \cite{Bagrets:2017pwq}.

\begin{figure}
\begin{center}
\includegraphics[scale=0.17]{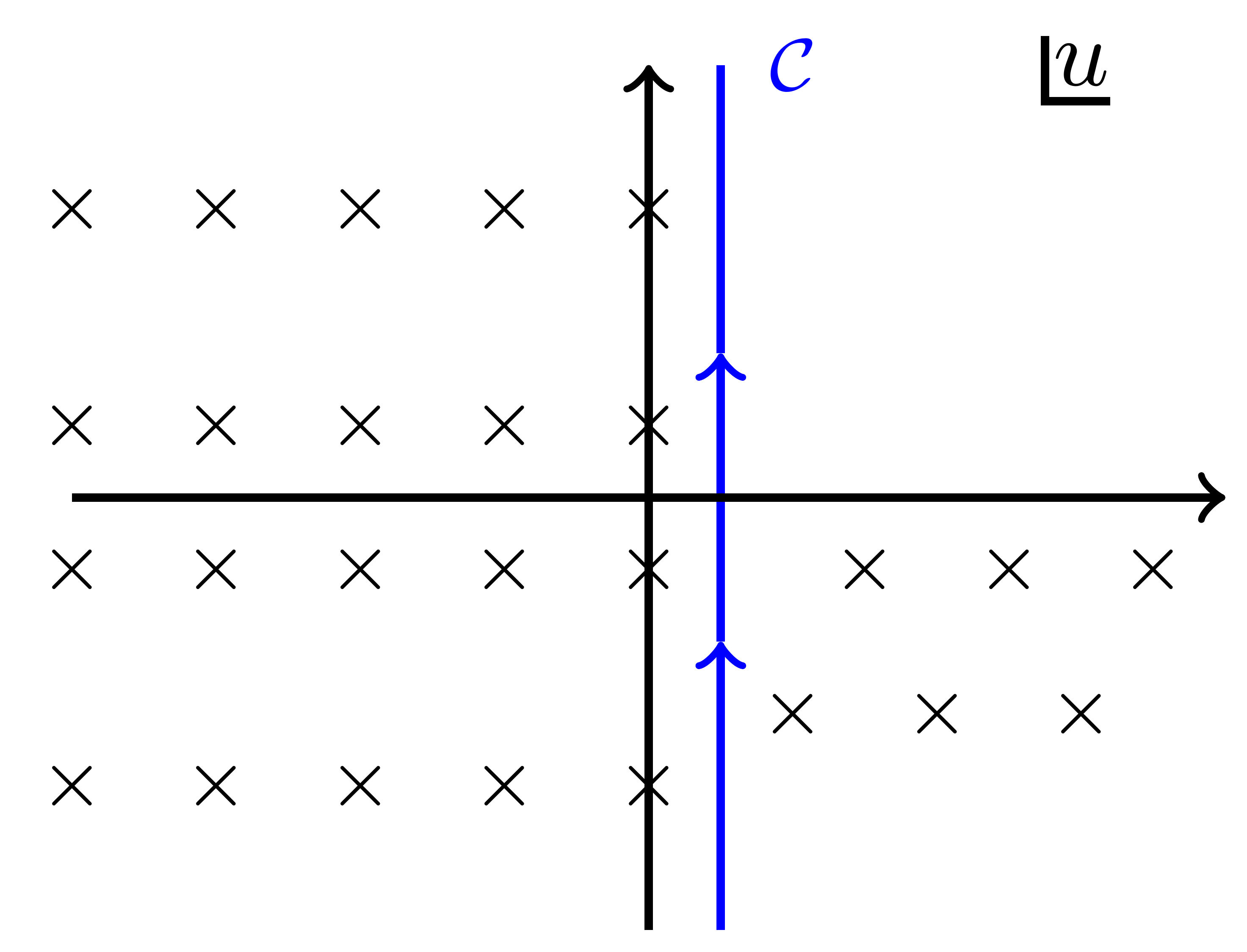}
\end{center}
\caption{Complex u-plane relevant for computing the four-point function with crossing. The contour of integration is indicated in blue while the black crosses denote the poles of the integrand.}
\label{Fig:u-integral}
\end{figure}

We now move on to the four-point function including shockwave interaction, which has a much more interesting structure. This requires a computation of $\langle G_{\Delta_1}(\tau_1,\tau_3)G_{\Delta_2}(\tau_2,\tau_4)\rangle_{\rm Sch}$ for a configuration such that $\tau_1<\tau_2<\tau_3<\tau_4$ and therefore the SL$(2,\mathbb{R})$ invariance of the bilocal does not help. This can be represented by 
\begin{equation}
\includegraphics[scale=0.17]{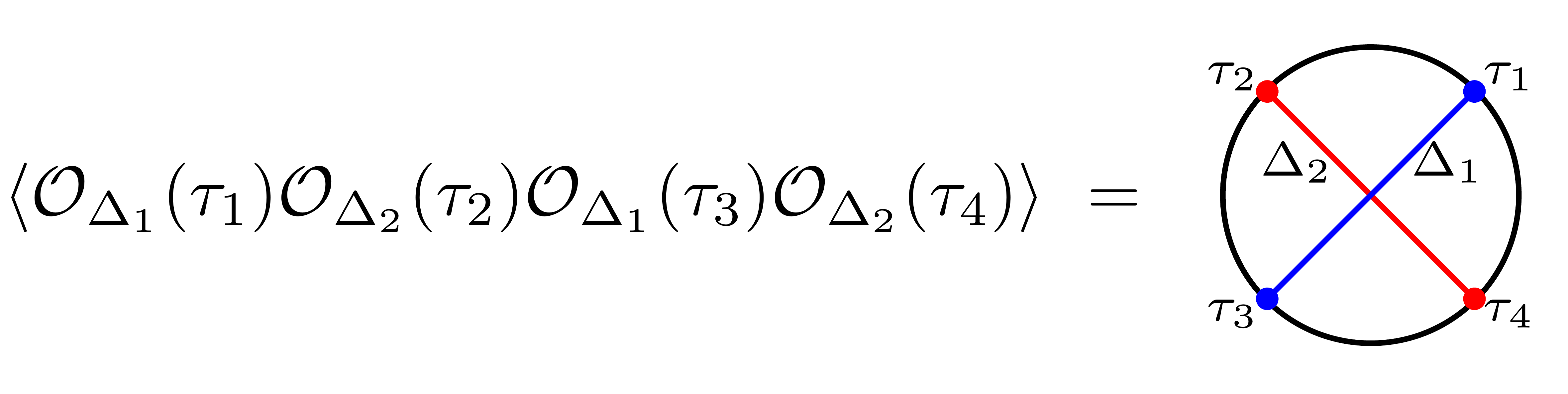}\nonumber
\end{equation}
The final result obtained in \cite{Mertens:2017mtv} is given by inserting in \eqref{eq:DefA4pt} the following amplitude:
\begin{eqnarray}
   \mathcal{A}_{\rm cross.} &=& \mbox{$ \frac{\Gamma(\Delta_1 +i k_1 \pm i k_4)\Gamma(\Delta_1 - i k_3 \pm i k_2)}{(2C)^{2\Delta_1}\Gamma(2\Delta_1)}\frac{\Gamma(\Delta_2 - i k_1 \pm i k_2)\Gamma(\Delta_2 + i k_3 \pm i k_4)}{(2C)^{2\Delta_2}\Gamma(2\Delta_2)}$}\nonumber\\
   &&\nonumber\\
   &&\hspace{-1cm}\mbox{$ \times  \int_{\mathcal{C}} \frac{du}{2\pi i}  \frac{\Gamma(u \pm ik_4)\Gamma(u +   i k_1+ik_3\pm ik_2)\Gamma(\Delta_1  -ik_1 - u)\Gamma(\Delta_2 -ik_3 - u)}{ \Gamma(u + \Delta_1 +ik_1)\Gamma(u+ \Delta_2+ik_3)}$}.\label{eq:AmpOTOC}
\end{eqnarray}
There is an extra auxiliary integral involved in the evaluation of the amplitude. The contour $\mathcal{C}$ in the complex $u$-plane is defined in the following way, illustrated in Fig.~\ref{Fig:u-integral}. The integrand has four set of poles coming from $\Gamma(u\pm i k_4)$ and $\Gamma(u+ik_1+ik_3\pm ik_2)$ that extend towards negative real $u$ and reach the ${\rm Re}(u)=0$ axis. The contour $\mathcal{C}$ is to the right of these poles. At the same time there are two set of poles coming from $\Gamma(\Delta_1 - i k_1 - u)$ and $\Gamma(\Delta_2 - i k_3 - u)$ that extend towards the right (we are assuming $\Delta_{1,2}>0$). The contour runs to the left of these poles.  

With this choice, the integral can be evaluated into a linear combination of hypergeometric functions, although it will be useful to keep it in the integral form. \\

\noindent\textbf{A check: $\Delta_1\to 0$ limit} \\

To get some practice with this amplitude we take the limit where one of the scaling dimension vanishes $\Delta_1 \to 0$, so the four-point function should reduce to a two-point function with scaling dimension $\Delta_2$. To show this, first close the contour to the right. In the $\Delta_1\to 0 $ limit only a single pole dominates given by $u=\Delta_1 - i k_1$. The residue around this pole contributes to the amplitude in the following way 
\begin{eqnarray}
\mathcal{A}_{\rm cross.}&=&\frac{\Gamma(\Delta_1 \pm i k_1 \pm k_4 )}{\Gamma(2\Delta_1)} \frac{ \Gamma(\Delta_1\pm i k_3 \pm i k_2) }{(2C)^{2\Delta_1}\Gamma(2\Delta_1)}\frac{\Gamma(\Delta_2 \pm i k_1 \pm i k_2)}{(2C)^{2\Delta_2}\Gamma(2\Delta_2)} \nonumber\\
&&\times\frac{\Gamma(\Delta_2+i k_3\pm ik_4)\Gamma(\Delta_2-\Delta_1+ik_1 - i k_3)}{\Gamma(\Delta_2 + i k_1\pm ik_2)\Gamma(\Delta_2+\Delta_1+ik_3-ik_1)}+\ldots, 
\end{eqnarray}
where the dots denote other residues that are subleading. Using again the identity $\lim_{\Delta \to 0} \frac{\Gamma(\Delta\pm ix)}{\Gamma(2\Delta)} = 2\pi \delta(x)$ and keeping track of the prefactors gives the expected answer
\begin{equation}
   \lim_{\Delta_1\to0} \mathcal{A}_{\rm cross.} = \frac{\Gamma(\Delta_2 \pm i k_2 \pm i k_4)}{(2 C)^{2\Delta_2}\Gamma(2\Delta_2)}\frac{ \pi^2\delta(k_1-k_4)}{k_4 \sinh 2\pi k_4}\frac{ \pi^2\delta(k_2-k_3)}{k_2 \sinh 2 \pi k_2}.
\end{equation}
This has the two needed delta functions to completely remove the dependence of the four-point function on $\tau_1$ and $\tau_2$, and leaves precisely the two-point amplitude.

\hspace{5cm}---o---\\

Since this correlator is of central importance in the study of quantum chaos and its relation to shockwave scattering near the black hole horizon, we will analyze some properties in the next section.

\subsubsection{Application: quantum chaos}
The exponential growth in time of double commutators is a signal of quantum chaos. This involves the computation of an out-of-time-ordered four-point function (OTOC). To compute it, we need to first analytically continue to Lorentzian signature $\tau \to \epsilon + i t$, where $t$ is Lorentzian time and $\epsilon$ is a small Euclidean time, chosen partly to produce the desired operator ordering in real time. An out-of-time-ordered correlator is  
\begin{equation}
  {\rm OTOC} \equiv  \langle \mathcal{O}_{\Delta_1} ( 0 ) \mathcal{O}_{\Delta_2} (t) \mathcal{O}_{\Delta_1}(0) \mathcal{O}_{\Delta_2}(t) \rangle , \quad t>0.
\end{equation}
From the operator placing it is clear that this involves an analytic continuation of the more interesting four-point function in Euclidean space with crossed bilocal lines. In the energy basis, the OTOC can be written as
\begin{eqnarray}
    \langle \mathcal{O}_{\Delta_1} ( t_1 ) \mathcal{O}_{\Delta_2} (t_2) \mathcal{O}_{\Delta_1}(t_3) \mathcal{O}_{\Delta_2}(t_4) \rangle  &=&\frac{e^{S_0}}{2} \int_0^\infty d\mu(k_1) d\mu(k_2)d\mu(k_3)d\mu(k_4)\nonumber\\
     &&\hspace{-4cm}\times e^{-i\frac{k_1^2}{2C}t_{12}-i\frac{k_2^2}{2C}t_{23}-i\frac{k_3^2}{2C}t_{43}-\frac{k_4^2}{2C}(\beta-it_{41})}~\mathcal{A}_{\rm cross.}(k_1,k_2,k_3,k_4),
\end{eqnarray}
where the amplitude $\mathcal{A}$ is the same as in \eqref{eq:AmpOTOC}. We also generalized the insertions to be at arbitrary four times with the understanding they are out of time ordered, meaning $t_{1,3}<t_{2,4}$.

\begin{figure}[t!]
\begin{center}
\includegraphics[scale=0.2]{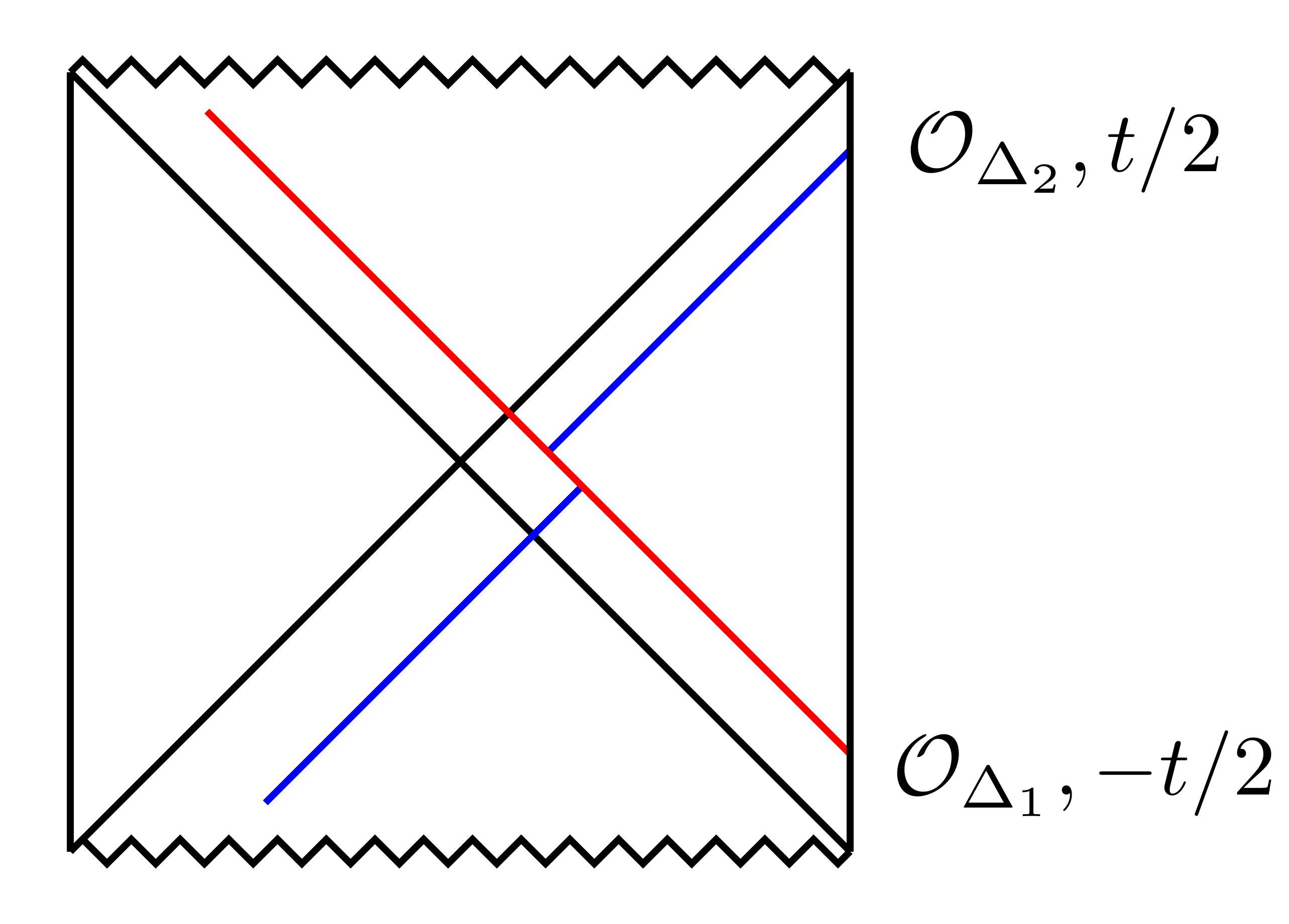}
\end{center}
\caption{Penrose diagram of an AdS black hole. A particle is inserted at very early times $-t/2$ (red) and another at very late times $+t/2$ (blue). The scattering which happens near the horizon for large time separations is captured by the out-of-time ordered correlator (OTOC). Since the scattering happens at high energies it is described by a gravitational shockwave interaction.}
\label{Fig:Shockwave}
\end{figure}

In the context of semi-classical AdS/CFT, the OTOC is computed in gravity by a four-point function in a black hole background as shown in Fig.~\ref{Fig:Shockwave}, see \cite{Shenker:2014cwa} for a detailed explanation. We have chosen $t_1=t_3=-t/2$ and $t_2=t_4=t/2$ to make it symmetric. In the semiclassical limit each insertion corresponds to a bulk particle denoted by red and blue for both types of operators.

As $t$ grows, the particles meet closer to the black hole horizon and therefore are highly boosted. The OTOC is constructed by combining the scattering S-matrix of the two particles near the horizon with their respective in- and out-wavefunctions given by bulk-boundary propagators. At high energies, this scattering is described by a shockwave where one particle (red) creates a backreaction in the geometry that shifts the position of the other particle (blue) closer to the horizon. In higher dimensions this is a genuine bulk interaction, while in 2d the particles do not really interact in the bulk: the shockwave scattering is due to the dynamics of the boundary Schwarzian mode.

The OTOC gives a signature of quantum chaos for the quantum system describing the black hole. One way to see this is to notice that the OTOC appears when computing the expectation value of $[\mathcal{O}_{\Delta_1} (t), \mathcal{O}_{\Delta_2}(0)]^2$. Systems with quantum chaos display an exponential growth in time of this quantity for times bigger than the dissipation time (order of the temperature for the black hole) and smaller than the scrambling time (logarithmic in $N$, when the double commutator saturates).\footnote{It is important for this interpretation that the four-point function involves pairwise equal insertions. The more general case is also interesting \cite{Turiaci:2019nwa} but is not captured by gravity in a simple way.}

Next we will show explicitly how all these elements of the semi-classical picture emerge from the exact OTOC expressions.

\paragraph{Semiclassical limit of OTOC}
To identify the semi-classical behavior of the OTOC, we take the large $C$ limit. In this limit all $k\sim \mathcal{O}(C)$, such that all intermediate energies are of order $C$, but energy differences are order one. As shown in \cite{Lam:2018pvp}, in this limit only the pole at $u=ik_4$ is relevant. Its residue gives a contribution to the OTOC amplitude given by
\begin{eqnarray}
\mathcal{A} &=& \frac{\Gamma(\Delta_1+ik_4 \pm i k_1)\Gamma(\Delta_1 -i k_3 \pm i k_2)}{(2C)^{2\Delta_1}\Gamma(2\Delta_1)} \frac{\Gamma(\Delta_2 - i k_1 \pm i k_2)\Gamma(\Delta_2 + ik_4 \pm i k_3)}{(2C)^{2\Delta_2}\Gamma(2\Delta_2)}  \nonumber\\
&& \quad \times  \Gamma(-2ik_4) \Gamma(-ik_4 \pm i k_2+ik_1+ik_3 )+\ldots
\end{eqnarray}
Following the procedure applied for the two-point function we parametrize the energies as 
\begin{equation}
    \frac{k_1^2}{2C}=E+\omega_1 , \quad \frac{k_2^2}{2C}=E-\omega_3-\omega_4, \quad \frac{k_3^2}{2C}=E-\omega_4, \quad \frac{k_4^2}{2C}=E,
\end{equation}
where $E$ is taken to be of order $C$ and $\omega\sim \mathcal{O}(1)$. The first observation is that the $E$-dependence simplifies and gives ${\rm OTOC} = \int dE ~e^{2\pi \sqrt{2CE} - \beta E} \times \mathcal{O}(1)$, where the order one factor is a function of the $\omega$'s and $E/C$. Therefore, just like for the two-point function, we can perform the $E$-integral at large $C$ by saddle point techniques, relating $E$ to the temperature $E=2\pi^2 C/\beta^2$. We can now write the order one term as
\begin{eqnarray}
\frac{{\rm OTOC}}{Z(\beta)} &=& \int \prod_{i=1}^4\frac{d\omega_i}{2\pi}~ e^{\sum_{i=1}^4i\omega_i t_i}e^{\frac{\beta}{4}(\omega_1-\omega_3-2\omega_4)}\Big(\frac{2\pi}{\beta}\Big)^{2\Delta_1+2\Delta_2-3}\nonumber\\
&&\times  \frac{\Gamma(\Delta_1-i\frac{\beta\omega_1}{2\pi})\Gamma(\Delta_1-i\frac{\beta\omega_3}{2\pi})\Gamma(\Delta_2+i\frac{\beta\omega_2}{2\pi})\Gamma(\Delta_2+i\frac{\beta\omega_4}{2\pi})}{\Gamma(2\Delta_1)\Gamma(2\Delta_2)}\nonumber\\
&&\times\Big( \frac{\beta}{4\pi C}\Big)^{-i\frac{\beta(\omega_1+\omega_3)}{2\pi}}\Gamma\Big(i\frac{\beta(\omega_1+\omega_3)}{2\pi}\Big) ~2\pi \delta\Big(\sum_{i=1}^4\omega_i\Big).\label{eq:OTOCsemic2}
\end{eqnarray}
This expression has a nice holographic interpretation. The OTOC involves a scattering of two particles and $\omega$ measures their energy according to an asymptotic observer. The factors in the first and second line of Eq.~\eqref{eq:OTOCsemic2} are bulk-to-boundary propagators and the particles interact through an S-matrix given by the final line of Eq.~\eqref{eq:OTOCsemic2}. 

The physical interpretation becomes more transparent when written in a basis of wavefunctions with fixed Kruskal energy. In this basis the S-matrix should be the Dray--'t~Hooft shockwave S-matrix $\mathcal{S} = e^{i\frac{\beta}{4\pi C}p_- q_+}$, where $p_-$ and $q_+$ are the Kruskal momenta of the two particles being scattered. Applying the change of basis between the Dray--'t~Hooft S-matrix from Kruskal to Schwarschild energies, which can be found in \cite{Shenker:2014cwa}, gives the expression:
\begin{eqnarray}
\mathcal{S}(\omega_1,\omega_2,\omega_3,\omega_4) &=&\int_0^\infty \frac{dq_+}{q_+} \int_0^\infty \frac{dp_-}{p_-} q_+^{\frac{i\beta (\omega_1+\omega_3)}{2\pi}} p_-^{-\frac{i\beta (\omega_2+\omega_4)}{2\pi}} e^{i\frac{\beta}{4\pi C} p_- q_+}  \nonumber\\
&&\hspace{-2.3cm} =\Big( \frac{\beta}{4\pi C}\Big)^{-i\frac{\beta(\omega_1+\omega_3)}{2\pi}}e^{-\frac{\beta}{4}(\omega_1+\omega_3)}\Gamma\Big(\frac{i\beta(\omega_1+\omega_3)}{2\pi}\Big) ~2\pi \delta\Big(\sum_{i=1}^4\omega_i\Big).
\end{eqnarray}
The remainder of the calculation is to match the other factors in \eqref{eq:OTOCsemic2} with bulk-boundary propagators. The details of the calculation are in \cite{Lam:2018pvp}.

After rewriting Eq.~\eqref{eq:OTOCsemic2} in the Kruskal basis, we can do the integrals and obtain a closed expression for the semiclassical OTOC. This was done in \cite{Maldacena:2016upp} without the full OTOC answer and by working directly in the semi-classical limit (which can be matched to Eq.~\eqref{eq:OTOCsemic2}). The answer is:
\begin{eqnarray}
\frac{{\rm OTOC}}{Z(\beta)} =\frac{U(2\Delta_1,1+2\Delta_1 -2\Delta_2, 1/z)}{z^{2\Delta_1}(\frac{\beta}{\pi}\sin  \frac{\pi it_{13}}{\beta})^{2\Delta_1}(\frac{\beta}{\pi} \sin  \frac{\pi it_{24}}{\beta})^{2\Delta_2}},
\label{OTOCf}
\end{eqnarray}
where $U(a,b,x)$ is the Tricomi confluent hypergeometric function and we normalize by the partition function $Z(\beta)$. We have also defined
\begin{equation}
    z = \frac{i\beta}{16 \pi C} \frac{e^{\pi(t_2+t_4-t_1-t_3)/\beta}}{\sinh \frac{\pi t_{13}}{\beta} \sinh \frac{\pi t_{24}}{\beta}}.
\end{equation}
This answer also matches with specific Virasoro vacuum blocks computed in \cite{Chen:2016cms}, for reasons explained in \cite{Lam:2018pvp}. We see from the expression for $z$ that the correlator is trivial at large $C$ unless we scale times such that $e^t \sim C/\beta$. This can also be seen by noting that the effective Dray-'t Hooft scattering phase is actually $\mathcal{S}_{\text{eff}} = e^{i\frac{\beta e^t}{4\pi C}p_- q_+ }$, which goes to $1$ unless $e^t \sim C/\beta$.

\paragraph{Behavior of OTOC}
We assume $C$ is large and look at how the OTOC depends on time. It is convenient to choose the operators evenly spread along the thermal circle:
\begin{eqnarray}
t_1 = -i\frac{\beta}{2} ,~t_2 = t-i\frac{\beta}{4},~t_3 = 0,~t_4 =t+i\frac{\beta}{4},~~\Rightarrow~~z=\frac{\beta}{16\pi C}e^{\frac{2\pi}{\beta}t}.
\end{eqnarray}
This regularizes divergences that appear at coincident points by Euclidean damping factors. The following quantity will play an important role 
\begin{eqnarray}
 t_{\rm sc} = \frac{\beta}{2\pi} \log \frac{2\pi C}{\beta},
    \end{eqnarray}
    and is called the \textbf{scrambling time} of the black hole. 
    
\begin{figure}[t!]
   \begin{center}
      \includegraphics[scale=0.5]{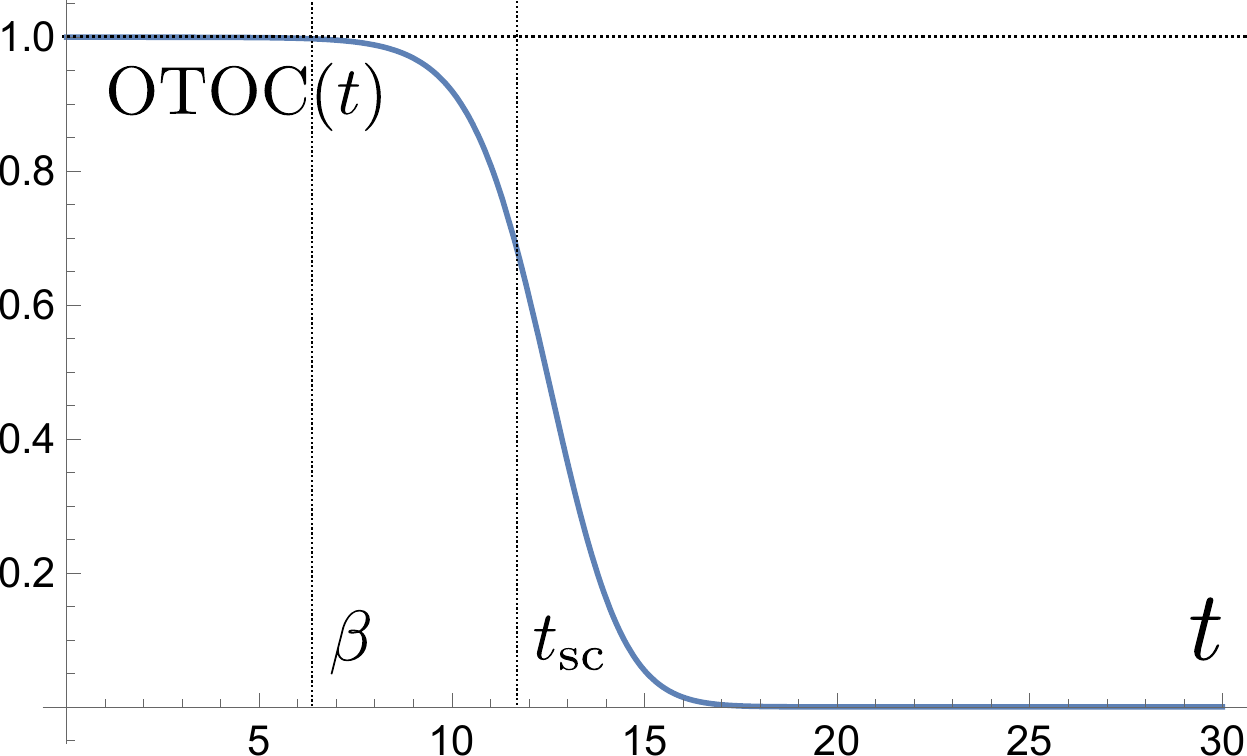} 
      \end{center}
      \caption{Plot of the OTOC as a function of time. We chose the parameters as $\Delta_1=\Delta_2=1$, $\beta =2\pi$ and $8C = 10^6$. The vertical lines separate different physical behavior. The first corresponds to time shorter than the dissipation time $\sim \beta$. Then the chaos regime kicks in at times $\beta \ll t \ll t_{\rm sc}$. Finally after the scrambling time $t_{\rm sc}$, the OTOC decays exponentially. For very late times $t\gg C$ (not shown) quantum gravity effects become important.}
      \label{Fig:OTOC}
   \end{figure}

With these choices, we show a plot of the normalized OTOC$(t)$ \eqref{OTOCf}, where we additionally divide out the product of two-point functions, such that it starts at 1 for $t\to -\infty$.
      We marked the thermal time $\beta$ and scrambling time $t_{\rm sc}$ for this configuration. This is illustrated in Fig.~\ref{Fig:OTOC}.
The following qualitative features can be distinguished.
\begin{itemize}
    \item We can first consider early times with small $z$, or equivalently $t\ll t_{\rm sc}$. Using the expansion $z^{-a}U(a,b,1/z) \approx 1 - a(1+a-b) z$ for small $z$, we get
    \begin{equation}
        \frac{{\rm OTOC}(t)}{Z(\beta)} \sim (\pi/\beta)^{2\Delta_1+2\Delta_2}\left(1 - \frac{\Delta_1 \Delta_2 \beta}{4\pi C}~e^{\frac{2\pi}{\beta}t}+\ldots\right).
    \end{equation}
    This is valid for times smaller than the thermal scale, where the OTOC becomes the product of two-point functions (the ``$1$'' in the above expansion), and is also valid at times $t \gtrsim \beta$ where it displays exponential deviations from this: this is chaotic \textbf{Lyapunov behavior} with maximal chaos exponent $\frac{2\pi}{\beta}$. This interpretation arises from the fact that 
    \begin{eqnarray}
    \langle [\mathcal{O}_{1}(t),\mathcal{O}_{2}(0)]\rangle \sim e^{\lambda t}, \qquad \lambda=\frac{2\pi}{\beta },
    \end{eqnarray}
    indicating the chaotic spread of an initial perturbation with Lyapunov exponent $\lambda$. The value for black holes saturates the chaos bound of \cite{Maldacena:2015waa}, and black holes are therefore maximally chaotic.
    \item The correlator transitions to a different behavior at the scrambling time $t\gg t_{\rm sc}$ defined such that $z$ is order one and the OTOC is small. In this regime $U(a,b,1/z)\approx \frac{\Gamma(1-b)}{\Gamma(1+a-b)}+z^{b-1}\frac{\Gamma(b-1)}{\Gamma(a)}$ when $z\gg1$ and therefore, assuming for concreteness that $\Delta_1>\Delta_2$, the OTOC decays exponentially as 
    \begin{eqnarray}
    {\rm OTOC}(t) \sim  e^{-2 \Delta_2 \frac{2\pi}{\beta}t}.
    \end{eqnarray}
    This is \textbf{quasinormal mode decay}. The reason is that the kinematical effect from the bulk-boundary propagation at these times dominates over the effect of the shockwave scattering.
    \item The scrambling time is logarithmic in $C$. A final transition happens when $t\gg C$ (not shown on the figure above). At such late times quantum gravity effects are important, invalidating \eqref{OTOCf}, and we need to go back to the exact expression. The low energy behavior of Eq.~\eqref{eq:AmpOTOC} and the density of states near the edge determines a power-law decay $
    {\rm OTOC}(t) \sim 1/t^{6}. $
    This behavior has a quantum gravity origin and cannot be understood classically.
\end{itemize}

\subsection{Diagrammatic rules for exact correlators}
More complicated correlators can be computed by following a simple set of diagrammatic rules, developed in \cite{Mertens:2017mtv} and \cite{Lam:2018pvp}. These can be used to compute any bilocal Schwarzian correlator, or equivalently the boundary correlators with quantum gravity effects. 

The first step is to draw the boundary thermal circle with the corresponding insertions, joined by bilocal lines in the configuration of interest. 
The second step is to assign a positive real number $k$ to each region bounded by the boundary or bilocal lines in the diagram. In the example of the OTOC four-point function, there are four $k_1$, $k_2$, $k_3$ and $k_4$. Each of these variables is integrated with the Schwarzian density of states measure $d\mu(k)=\frac{1}{\pi^2} k dk \sinh 2\pi k$. 

For each boundary line bounding a region with parameter $k$, we assign the exponential factor: 
\begin{equation}
\label{frules}
\includegraphics[scale=0.2]{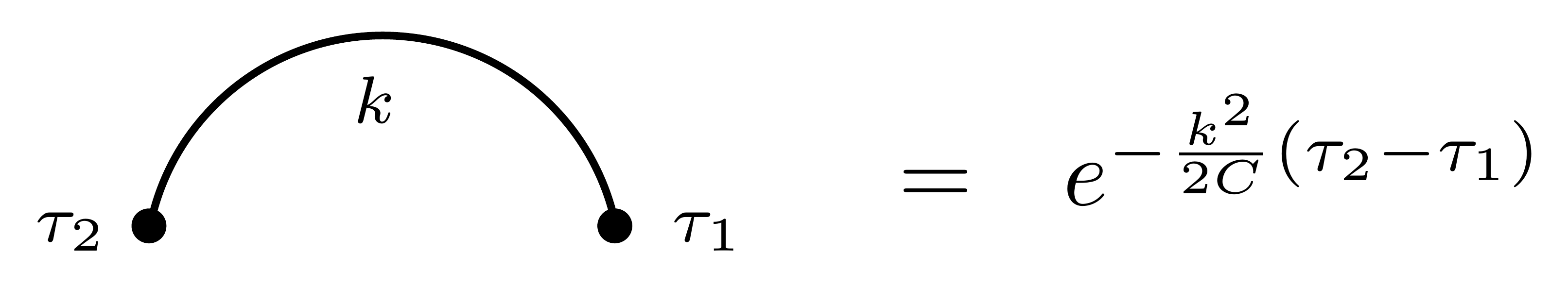}
\vspace{0cm}
\end{equation}
This factor represents the usual Hamiltonian time evolution of the intermediate energy eigenstates. Each point where a bilocal line with conformal weight $\Delta$ hits the boundary bounding two regions with $k_1$ and $k_2$, we insert a vertex factor
\begin{equation}
\label{frules2}
\includegraphics[scale=0.25]{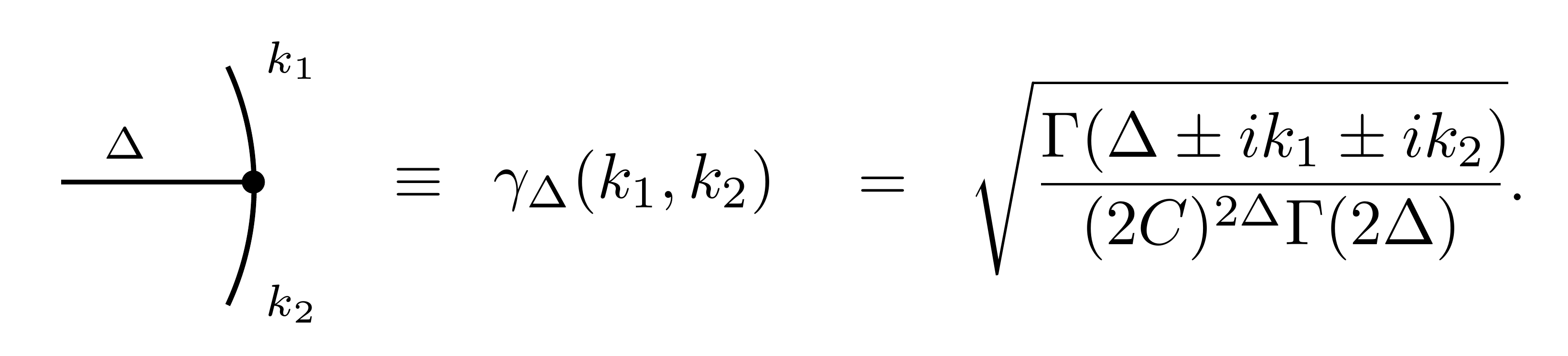}
\end{equation}
This vertex factor represents the matrix element of an endpoint of the bilocal operator between the corresponding two energy eigenstates. 

Finally, for each crossing of bilocal lines in the interior of the diagram, we insert the following factor which depends on the four energies $k$ surrounding it, and the scaling dimensions of the two crossing lines: 
\begin{eqnarray}
\label{crossing}
\includegraphics[scale=0.17]{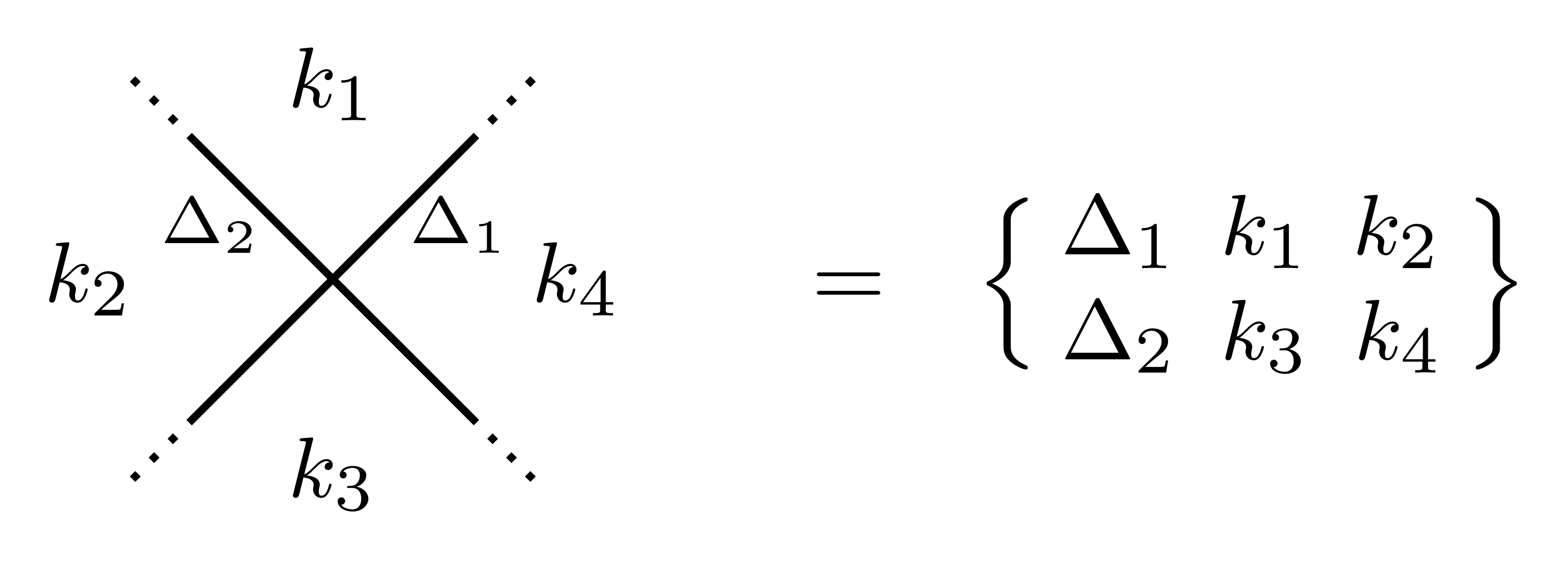}
\end{eqnarray}
The quantity on the right hand side depends on six variables. It is the $6j$-symbol of SL$(2,\mathbb{R})$, and its explicit expression is
\begin{eqnarray}
~~\sixj{\Delta_1}{\Delta_2}{k_1}{k_3}{k_2}{k_4} &=& \mbox{$\sqrt{\frac{\Gamma(\Delta_1+ik_1 \pm ik_4)\Gamma(\Delta_2-ik_1\pm ik_2)\Gamma(\Delta_1-i k_3 \pm ik_2) \Gamma(\Delta_2+ik_3\pm ik_4)}{\Gamma(\Delta_1-ik_1 \pm ik_4)\Gamma(\Delta_2+ik_1\pm ik_2)\Gamma(\Delta_1+i k_3 \pm ik_2) \Gamma(\Delta_2-ik_3\pm ik_4)}}$}\nonumber \\[2mm]
\! & \!\! & \!\hspace{-1cm}\mbox{$\times \int\limits_{-i\infty}^{i\infty}\!\! \frac{du}{2\pi i} \,
\frac{\Gamma(u \pm ik_4)\Gamma(u +   i k_1+ik_3\pm ik_2)\Gamma(\Delta_1  -ik_1 - u)\Gamma(\Delta_2 -ik_3 - u)}{ \Gamma(u + \Delta_1 +ik_1)\Gamma(u+ \Delta_2+ik_3)}$}.\nonumber \\[-4mm]
\label{appfinal}
\end{eqnarray}

Multiplying the vertex factors with the 6j-symbol involved in the crossing of lines for the case of the four-point function gives 
\begin{eqnarray}
\mathcal{A}_{\rm cross.} =   \, \gamma_{\Delta_1}(k_1,k_4) \gamma_{\Delta_2}(k_4,k_3) \gamma_{\Delta_1}(k_3,k_2) \gamma_{\Delta_2}(k_2,k_1)\sixj{\Delta_1}{\Delta_3}{k_1}{k_3}{k_2}{k_4}.
\end{eqnarray}
which is precisely the amplitude quoted in \eqref{eq:AmpOTOC}. The final step consists in multiplying the whole correlator by $e^{S_0}/2$. Note that we are computing correlators without normalizing by the partition function $Z(\beta)$. 

Some further examples of these rules, including higher-point OTOCs, can be found in \cite{Lam:2018pvp}.

\subsection{Pure states and end-of-the-world (EOW) branes} \label{Sec:EOWbranes}

Most of this section studies JT gravity in the hyperbolic disk (which corresponds to the Euclidean thermal path integral of the boundary theory). Half of the hyperbolic disk can be used to prepare a state: the thermofield double state. This state is dual to the Einstein--Rosen bridge obtained from the maximally extended black hole solution: 
\begin{eqnarray}
\includegraphics[scale=0.2]{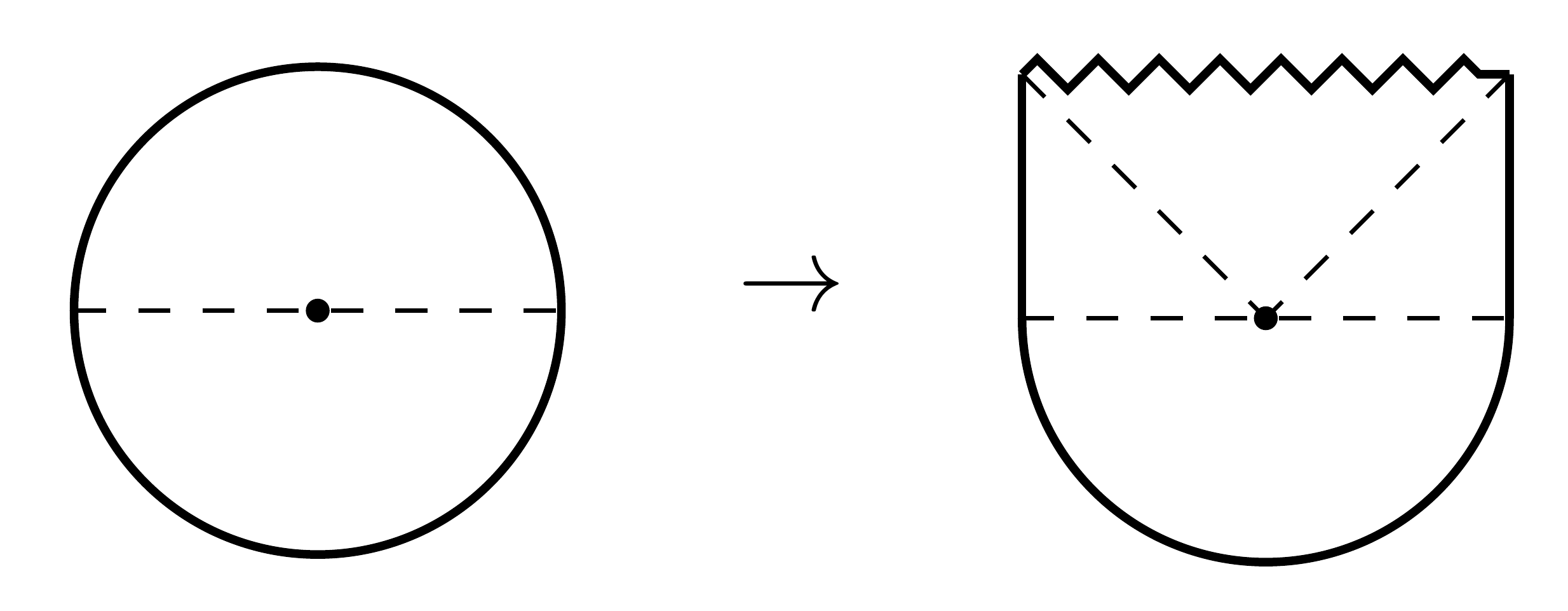}
\end{eqnarray}
This state is represented by $\vert {\rm TFD}\rangle = \sum_n e^{-\beta E_n/2} \vert E_n \rangle_L \otimes \vert E_n \rangle_R$. After integrating out the left boundary, this gives a thermal density matrix for the right system. 

To instead produce a pure state on the right, one can project the left state onto a chosen state $\vert B \rangle_L$ such that we get $\vert \Psi \rangle_R = {}_L \langle B \vert {\rm TFD}\rangle$. The construction of such states was done explicitly in the SYK model. The projection on specific states can be modeled by a heavy operator with scaling dimension of order $C$. The diagram proposed in \cite{Kourkoulou:2017zaj} is 
\begin{eqnarray}
  \includegraphics[scale=0.2]{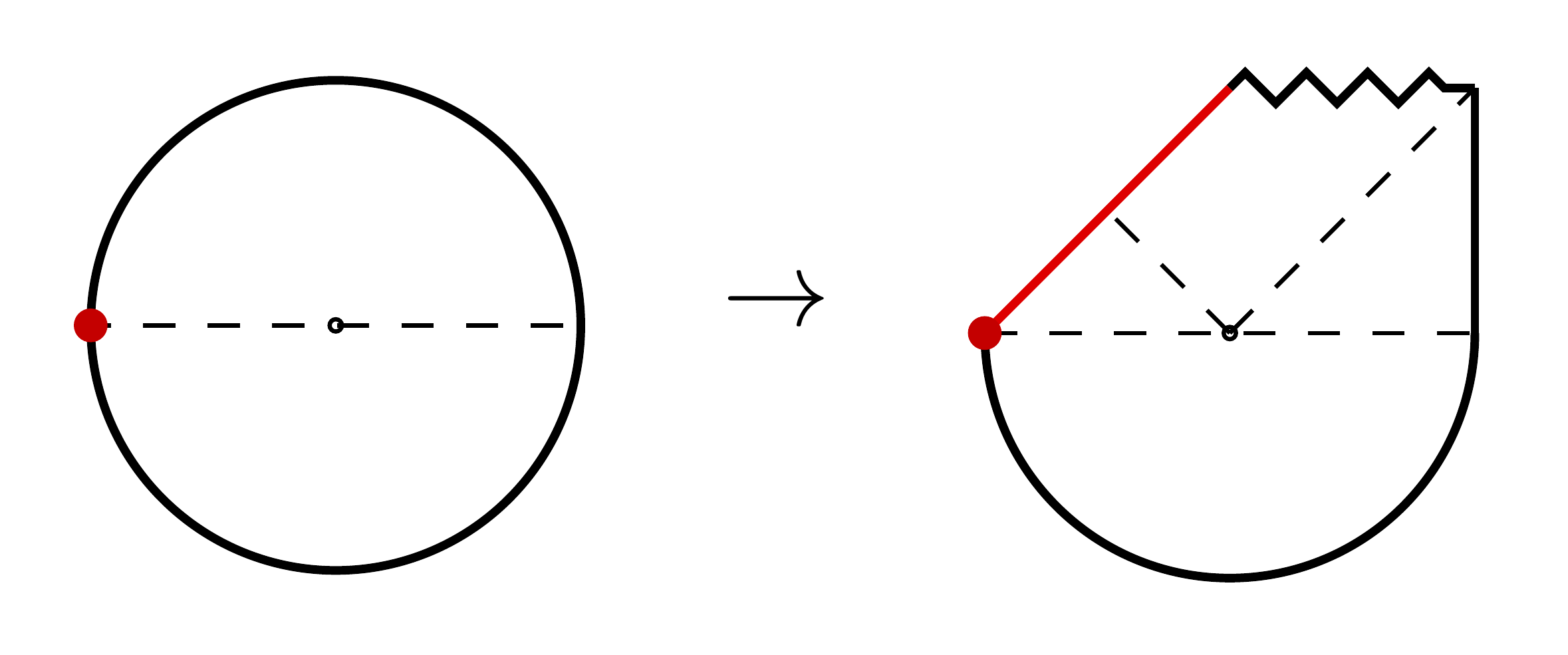}
\end{eqnarray}
On the left we show the Euclidean preparation with the projection operator inserted in red. On the right is the diagram after continuing to Lorentzian signature, and shows how the projection creates a high-energy shockwave in the bulk that acts as a so-called EOW brane behind the black hole horizon. We could also consider the Penrose diagram fully in Lorentzian time: the EOW brane emerges from the past singularity, reaches the left boundary and then falls to the future singularity. 

More general projections that act non-locally near the left boundary can also be considered. These are dual to states described geometrically by 
\begin{eqnarray}
  \includegraphics[scale=0.2]{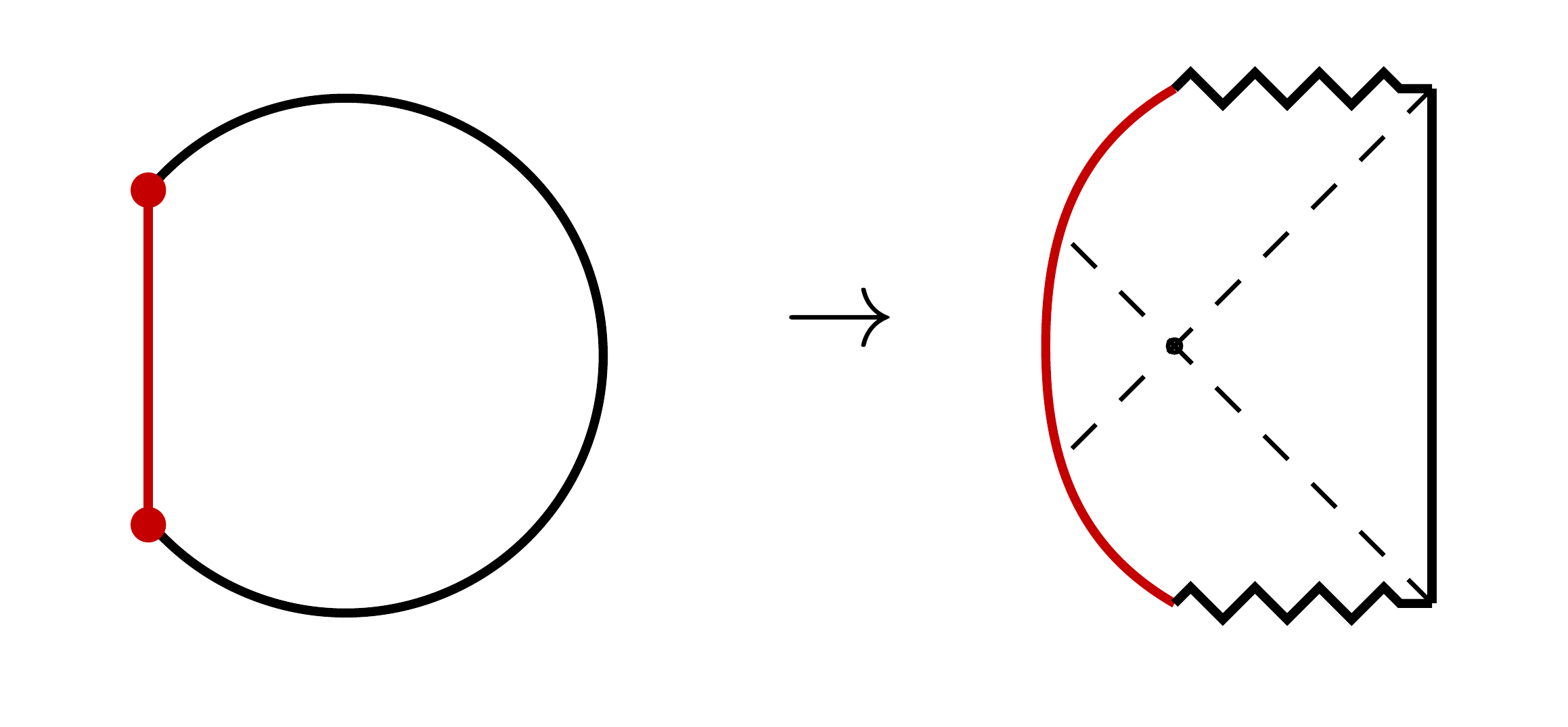}
\end{eqnarray}
The EOW brane moves behind the horizon, but does not reach the left boundary. In Euclidean signature, this can be obtained by a boundary curve made of two pieces: a holographic boundary of renormalized length $\beta$ and a geodesic (corresponding to the EOW brane) of renormalized length $\ell$. The Euclidean action describing this set-up is \begin{equation}
I=I_{\rm JT}+ m \int_{\text{\footnotesize brane}} ds,
\end{equation}
where the second term is integrated over the EOW brane worldline and $m$ is the tension. This term effectively imposes the boundary conditions $\partial_n \Phi = m$ and $K=0$ along the EOW worldline, with $\partial_n$ the normal derivative.  
The path integral in Euclidean signature over the disk with a partly holographic boundary of renormalized length $\beta$ and an EOW brane with geodesic length $\ell$ is then given by \cite{Harlow:2018tqv, Yang:2018gdb}:
\begin{eqnarray}
\label{gluel}
 \includegraphics[scale=0.2]{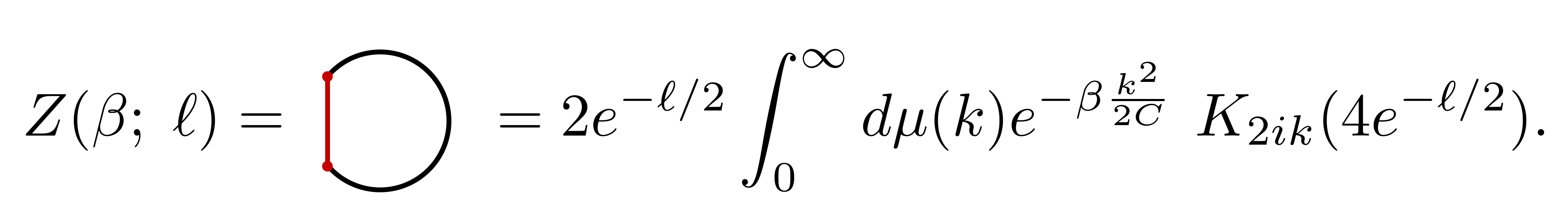}
\end{eqnarray}
As a check of this formula, one can verify that gluing two such disks along the geodesic, adding a matter propagator $e^{-\Delta \ell}$ and integrating over the geodesic length $\int d\ell Z(\tau;\ell)Z(\beta-\tau;\ell)e^{-\Delta \ell}$ reproduces the two-point function \eqref{eqn:2ptSchwFull}. Equation \eqref{gluel} can also be interpreted as the wavefunction for creating an ER bridge of length $\ell$ in a TFD state of inverse temperature $\beta$.

There have been several applications in JT gravity involving EOW branes. The first was already mentioned \cite{Kourkoulou:2017zaj}. One can also consider intermediate situations of partially entangled states which can be modeled by (not EOW) branes gluing to patches of AdS$_2$ \cite{Goel:2018ubv}. More recently, they were used in \cite{Penington:2019kki} to model pure state black hole evaporation in a simple model of a system with a Page curve. Embedding of dynamical EOW branes in a random matrix framework appeared in e.g. \cite{Gao:2021uro,Blommaert:2021etf}.

\subsection{Other operator insertions in JT gravity}
\label{s:defects}
The bilocal operator studied so far is not the only operator one can insert into the gravitational path integral. Here we go through other possibilities. \\

In fact, the simplest \emph{local} operator insertion that is PSL$(2,\mathbb{R})$ invariant is the \textbf{local energy operator} $E(t) = - C \left\{F,t \right\}$. Its correlators are relatively straightforward to determine and are piecewise constant up to contact terms. Within the diagrammatic formalism, they represent the energy $k^2/2C$ in the different sectors of the diagram, depending on where we insert them in the Schwarzian path integral. To derive the expressions, one can write down a Ward identity relating a correlator with such an insertion to one without, see \cite{Stanford:2017thb,Mertens:2017mtv,Mertens:2019bvy} for details from various perspectives on this result.

Mixing up these operator insertions with the bilocals, reinforces the physical interpretation of the bilocal operator itself in holography: in the sector between the endpoints of the bilocal, an energy has been injected into the gravitational bulk, only to be removed again at the second endpoint of the bilocal operator. A single endpoint of these operators can hence be compared to the classical energy pulses we studied in Sect.~\ref{sec:JT}. \\

A structurally more interesting class of operators is the so-called \textbf{defects} in the JT bulk to which we turn next \cite{Mertens:2019tcm}. We have previously shown that the thermal disk partition function is described by writing $F(\tau) = \tan \frac{\pi}{\beta} f(\tau)$, where $f(\tau) \in \frac{G}{H} \equiv \frac{\text{Diff } S^1}{\text{PSL}(2,\mathbb{R})}$. This integration space is special since it is symplectic, but it is also precisely the coadjoint orbit of the identity element of the Virasoro group \cite{Witten:1987ty,Balog:1997zz}. This is no coincidence. In fact, there is a host of Schwarzian models that can be found by considering all of the different Virasoro orbits, with different choices of the preserved subgroup $H$. All of them can be interpreted gravitationally. Let us give an overview. In all cases, we write the Schwarzian action as:
\begin{equation}
\label{eqintroaction}
S_H[f]=-C \int_0^\beta d\tau \{ F \circ_H f(\tau), \tau\}.
\end{equation}
The different models are at a crude level distinguished by a monodromy matrix $M \in \text{PSL}(2,\mathbb{R})$, defined by the periodicity relation $F(\tau+\beta) = M \cdot F(\tau) = \frac{aF(\tau)+b}{cF(\tau)+d}$, and we label the orbits by the conjugacy class of this matrix $M$. The stabilizer $H$ is then defined as the subgroup of all $h \in \text{PSL}(2,\mathbb{R})$ satisfying $[h,M]=0$. We can distinguish the following different models:\footnote{Some fineprint is due. Firstly, there is no agreed-upon naming of the different orbits, so care must be taken when comparing different references. Secondly, the difference between orbits and conjugacy classes is mainly that we need to keep track of the winding number of the elliptic conjugacy class. This means the parameter $\theta \in \mathbb{R}$ instead of $\theta \in [0,1)$. Finally, we only consider orbits that have a constant representative. The reason is that the other orbits, by definition, have no solution to the equation $\frac{d}{d\tau} \left\{F,\tau\right\}=0$, but this is precisely the saddle equation of all of these models. Such orbits would hence be highly quantum, and we are not aware of any use of these orbits (in any context).}

\begin{itemize}
\item \textbf{Elliptic} $H=U(1)_\theta$. Elliptic orbits are parametrized by
\begin{equation}\label{intro:ell1}
F\circ_\theta f = \tan \frac{\pi}{\beta} \theta f, \qquad M = \left(\begin{array}{cc}
\cos(\pi \theta) & \sin(\pi\theta) \\
-\sin(\pi\theta) & \cos(\pi \theta) \\
\end{array}\right) \, \in \text{PSL}(2,\mathbb{R}).
\end{equation}
In the bulk, one can choose conformal gauge and continue this time reparametrization into the bulk as in \eqref{bulkgauge} by setting $U(u)=F(u)$ and $V(v)=F(v)$. This choice is arbitrary, but coordinate-invariant results we obtain from it have meaning. This leads to the metric:
\begin{equation}
\label{saddlem}
ds^2 = 4 \left(\frac{\pi \theta}{\beta}\right)^2\frac{d\tau^2+dz^2}{\sinh^2 \frac{2\pi}{\beta} \theta z}.
\end{equation}
This geometry can be readily shown to have a conical singularity of periodicity $2\pi \theta$.

Within the path integral, one can show that the result is again one-loop exact:
\begin{align}
\label{dosell}
\includegraphics[scale=0.22]{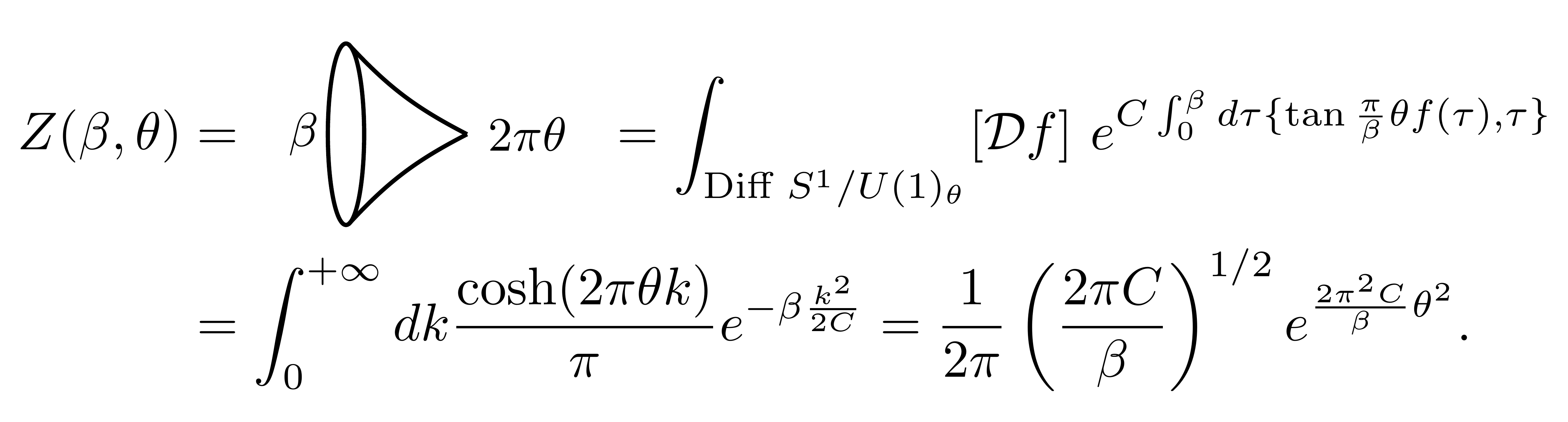}
\end{align}
The calculation only has a single zero-mode in this case, since the only isometry of the metric are rotations around the defect.

The action functional for these models precisely matches with the energy functional in Virasoro coadjoint orbit theory. In particular, bounds of this functional in the different orbits have been analyzed \cite{Balog:1997zz}, which can then be used to show that for $\theta>1$ negative modes exist and the path integral evaluation is formal. Nonetheless, the answer can be derived by other means (e.g., from limits of Virasoro characters \cite{Mertens:2019tcm}, or by complexifying the unstable fluctuations), see \cite{Alekseev:2020jja} for some further speculation on an interpretation in terms of a change of integration cycle.\footnote{We present some details on the boundedness of the Schwarzian action. We first write the Schwarzian action as $-C\int_0^\beta d\tau \left\{F,\tau\right\} = -C\int_0^\beta d\tau \left\{f,\tau\right\} - \frac{2\pi^2 C}{\beta^2} \int_0^\beta d\tau f'^2$. One can show the following inequalities for the Schwarzian action of $f \in \text{Diff }S^1$:
\begin{equation}
-C\frac{2\pi^2}{\beta^2} \int_0^\beta d\tau (1-f'^2) \leq -C\int_0^\beta d\tau \left\{f,\tau\right\} = \frac{C}{2} \int_0^\beta  d\tau \left(\frac{f''}{f'}\right)^2.
\end{equation}
The rightmost equality shows that the Euclidean action is unbounded from above by making $f'$ have strong fluctuations. The inequality on the left can then be used directly to derive the following lower bound on the action:
\begin{equation}
-C\int_0^\beta d\tau \left\{F,\tau\right\} \geq -\frac{2\pi^2\theta^2C}{\beta} + \frac{2\pi^2C}{\beta^2}(1-\theta^2)\int_0^\beta d\tau (f'-1)^2
\end{equation}
When $\theta^2<1$, the second term on the RHS is positive for all $f$, and the action is bounded below by $-\frac{2\pi^2\theta^2C}{\beta}$. On the other hand, for $\theta^2>1$, one can readily find examples that lower the action without bound. 
The result is that for $\theta > 1$, the path integral has negative modes and the one-loop computation is formal.}

The following observation made in  \cite{Mertens:2019tcm} will be important in Sect.~\ref{sec:MTW}. The presence of a defect at the point $x_c$ of the hyperbolic disk means the scalar curvature has a delta-function singularity $R(x)=-2 + 4\pi(1-\theta)\delta^2(x-x_c)$. This is equivalent to the insertion in the JT gravity path integral of an operator $e^{\frac{1}{4 G_N}(1-\theta)\Phi(x_c)}$ at the location of the defect. \\

\item \textbf{Exceptional Elliptic} $H=\text{PSL}^n(2,\mathbb{R})$. When $\theta \in \mathbb{N}$, the stabilizer is enhanced:
\begin{equation}
F\circ_n f = \tan \frac{\pi}{\beta} n f,~~~ M = \left(\begin{array}{cc}
1 & 0 \\
0 & 1 \\
\end{array}\right) \, \in \text{PSL}(2,\mathbb{R}).
\end{equation}
The conical singularity is $2\pi n$, and this represents a replicated version of the original disk. There are three zero-modes (just like for the $n=1$ case where there is no conical deficit), and we obtain the one-loop exact amplitude:
\begin{align}
\label{dospsl}
Z(\beta,n) &\equiv\int_{\text{Diff } S^1/\text{PSL}^n(2,\mathbb{R})} [\mathcal{D}f]~e^{C \int_0^\beta d\tau \{ \tan{ \frac{\pi}{\beta}n f(\tau)}, \tau\} }\nonumber\\
&= \int_{0}^{+\infty}dk \frac{k\sinh(2\pi n k)}{2\pi^2} e^{-\beta \frac{k^2}{2C}} =  \frac{n}{4\pi^2} \left( \frac{2 \pi C}{\beta} \right)^{3/2} e^{\frac{2 \pi^2 C}{\beta} n^2}.
\end{align}

\item \textbf{Hyperbolic} $H=U(1)_\lambda$. The hyperbolic orbit can be viewed as an analytic continuation of the elliptic case where in essence $\theta \to i\lambda$:
\begin{equation}
F \circ_\lambda f = \tanh \frac{\pi}{\beta}\lambda f,~~~ M = \left(\begin{array}{cc}
\cosh(\pi \lambda) & \sinh(\pi\lambda) \\
\sinh(\pi\lambda) & \cosh(\pi \lambda) \\
\end{array}\right) \, \in \text{PSL}(2,\mathbb{R}).
\end{equation}
The geometry is of the form:
\begin{equation}
\label{globalgeom}
ds^2 = 4 \left(\frac{\pi \lambda}{\beta}\right)^2\frac{d\tau^2 + dz^2}{\sin^2\frac{2\pi}{\beta} \lambda z},
\end{equation}
and describes instead a macroscopic tube, with a minimal length circle (the neck of a wormhole) at $z=\frac{\beta}{4\lambda}$. This neck is a geodesic with a circumference $b \equiv 2 \pi \lambda$.

Once again, the answer is one-loop exact:
\begin{align}
\label{doshyp}
\includegraphics[scale=0.23]{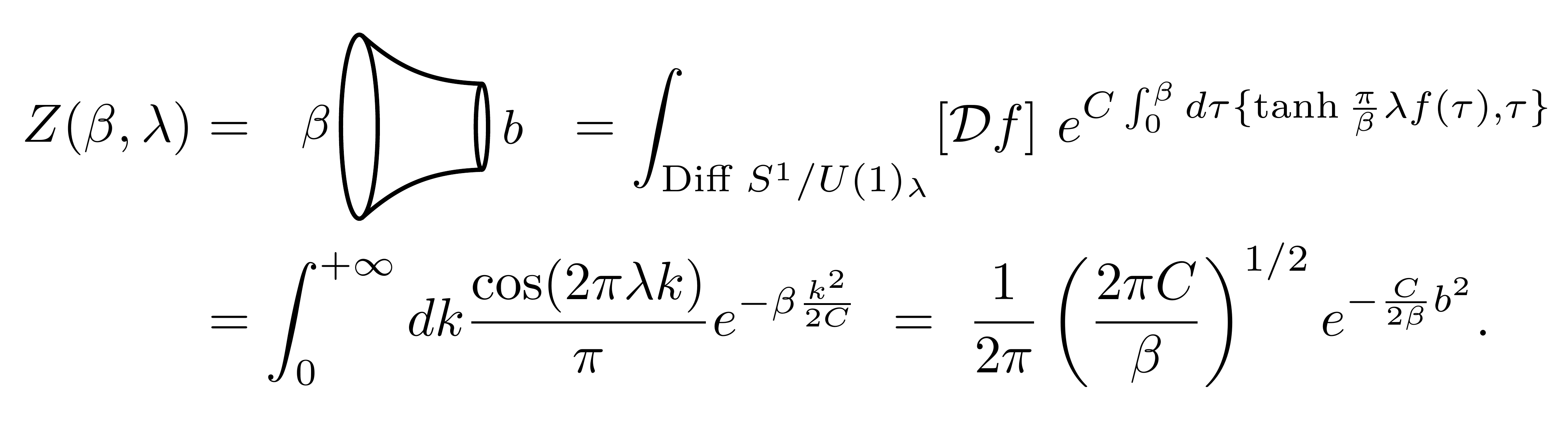}
\end{align}

\item \textbf{Parabolic} $H=U(1)_0$. Finally, we have the parabolic orbit:
\begin{equation}
\label{intro:par1}
F\circ_0 f =  f,~~~ M = \left(\begin{array}{cc}
1 & 1 \\
0 & 1 \\
\end{array}\right) \, \in \text{PSL}(2,\mathbb{R}),
\end{equation}
which is found as a limiting case of both elliptic and hyperbolic defects. It corresponds geometrically to an infinite cusp-singularity, identifiable as thermal AdS$_2$ in Poincar\'e coordinates, with one-loop exact amplitude:
\begin{equation}
\label{eq:orbpara}
\includegraphics[scale=0.2]{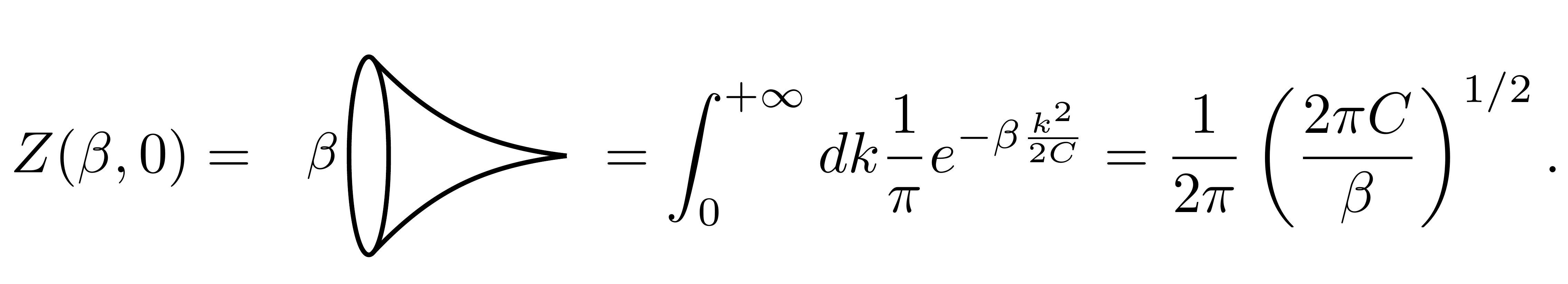}
\end{equation}
\end{itemize} 
One can mix in these defect insertions with bilocal operator insertions by replacing the usual JT spectral density $d\mu(k_i)$ in the sector of the diagram of interest by those found in \eqref{dosell}, \eqref{dospsl}, \eqref{doshyp}, or \eqref{eq:orbpara}. Some example expressions were written down in \cite{Mertens:2019tcm}.

 A useful diagrammatic representation is to include a cross in the relevant sector of the diagram, labeled by the orbit parameter. An example for the OTOC with a hyperbolic defect $\lambda$ in one of the four sectors is the diagram:
\begin{equation}
\includegraphics[scale=0.13]{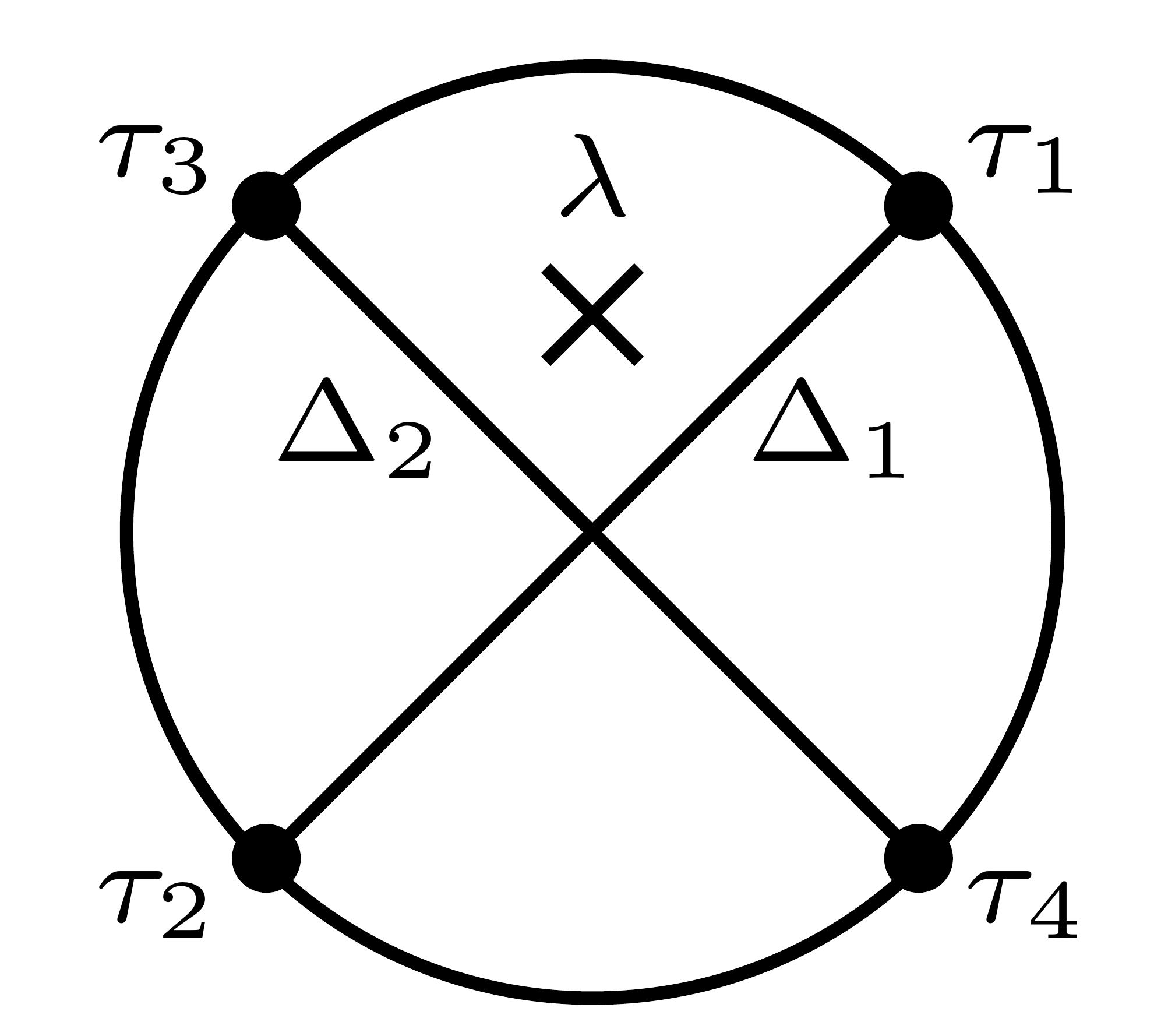}
\end{equation}

\subsection{Outline of derivations in the literature}
\label{s:deriv}
In this subsection we briefly outline and summarize the main approaches available to derive the JT amplitudes discussed in this section, and refer to the original papers for more details. Each of the studied approaches has benefits and downsides, which we describe, and it is worthwhile to have some understanding of all of them.

\subsubsection{Free particle approach}
Using the redefinition $F'=e^\varphi$, which is always possible since $F' \geq 0$ for a suitable time reparametrization, one immediately sees that the thermal Schwarzian action can be rewritten into a free boson action:
\begin{equation}
\label{eq:freesch}
S = -C \int_0^\beta dt \left\{F,t\right\} = \frac{C}{2} \int_0^\beta dt (\partial_\tau \varphi)^2.
\end{equation}
Implementing the (formal) periodicity requirement $F(\tau+\beta) = F(\tau)+\infty$ leads to the constraint $\int_0^\beta dt e^\varphi = +\infty$, which can be implemented after regularization using a Lagrange multiplier, and which leads to an exponential potential added to the otherwise free boson system \eqref{eq:freesch}.

Bilocal operator insertions become more complicated non-local operators since
\begin{equation}
\left(\frac{F_1'F_2'}{(F_1-F_2)^2} \right)^\Delta = \frac{e^{\Delta \varphi_1} e^{\Delta \varphi_2}}{\left(\int_{\tau_1}^{\tau_2} d\tau e^{\varphi}\right)^{2\Delta}}.
\end{equation}
Correlators can then be computed by using the identity $A^{-p} = \frac{1}{\Gamma(p)}\int_0^\infty d\alpha \alpha^{p-1}e^{-\alpha A}$, which leads to an exponential potential with piecewise constant prefactor \cite{Bagrets:2016cdf, Bagrets:2017pwq}, with the prefactor changing each time a bilocal endpoint is encountered.

As a closely related approach, one can instead use the black hole time coordinate $f'=e^\varphi$ and rewrite the Schwarzian action again as a particle in exponential potential(s). This is however less useful when including operator insertions, and the full expression for the correlators has not been obtained from this approach, except in some special cases.

\begin{table}[ht]
\centering
\begin{tabular}{p{0.45\linewidth} | p{0.45\linewidth} }
\emph{Strong points} & \emph{Weak points} \\
\hline
${\color{blue}\rightarrow}$ Fast and intuitive, since it relates to the 1d free boson. & ${\color{blue}\rightarrow}$ Subtle in terms of boundary conditions and handling of PSL$(2,\mathbb{R})$ gauge group. 

${\color{blue}\rightarrow}$ More difficult for more complicated correlators, not useful to derive final answers.
\end{tabular}
\end{table}
\vspace{-0.5cm}

\subsubsection{Limit of Liouville theory}
\label{s:Liou}
\begin{figure}[t!]
   \begin{center}
\includegraphics[width=0.9\textwidth]{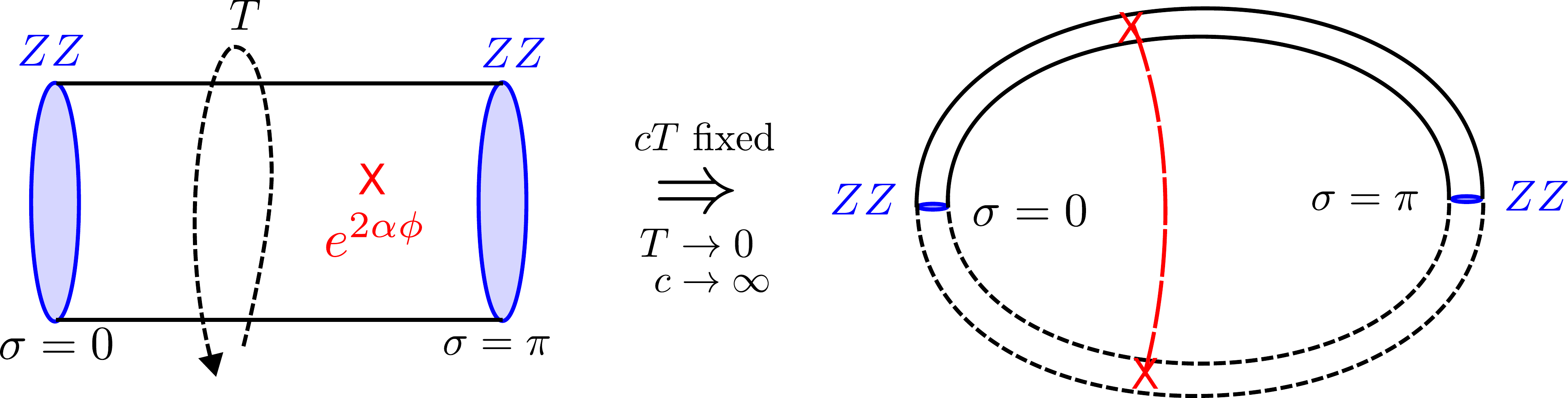}
      \end{center}
      \caption{2d Liouville CFT between vacuum branes and its double-scaled Schwarzian limit. Operator insertions can be added and are depicted in red.}
      \label{fig:Liou}
\end{figure}

Schwarzian correlation functions can be found by taking a suitable double-scaling limit of Virasoro CFT \cite{Mertens:2017mtv}. Let us sketch how that works. Virasoro CFT at large central charge can be dynamically realized by the 2d Liouville CFT, described by the classical action:
\begin{equation}
\label{liou}
S_{\text{L}} = \frac{1}{4\pi b^2} \int_0^T d\tau \int_0^\pi d\sigma\, \left[ (\partial \phi)^2 +  4 \mu e^{2 \phi} \right],
\end{equation}
on a cylindrical surface with $\sigma \in (0,\pi)$ and Euclidean time direction $\tau$ of period $T$. The central charge is $c=1+6(b+b^{-1})^2$. In 2d Liouville CFT, one can consider branes for the worldsheet to end on. In particular, generalizing Cardy's construction of boundary states to irrational CFT, considering a pair of identity or ZZ-branes \cite{Zamolodchikov:2001ah} at the ends of the cylinder worldsheet at $\sigma=0,\pi$ gives an amplitude that is the Virasoro vacuum character:
\begin{equation}
\label{eq:vac}
\chi_0 (q) = \frac{q^{\frac{1-c}{24}} (1-q)}{\eta(iT/2\pi)}, \quad q=e^{-T}.
\end{equation}
A convenient trick in dealing with identity branes is to use the mirror doubling trick with periodic boundary conditions around the doubled space. The procedure is now to study this amplitude in the limit where the cylinder becomes long and narrow, in conjunction with the classical large $c$ limit ($b\to0$). This means that in this limit, the cylinder becomes a narrow (doubled) circular tube of length $2\pi$ that degenerates. This is precisely the thermal circle where the Schwarzian model lives! The entire set-up is given in Fig. \ref{fig:Liou}.
\vspace{0.15cm}

One can use this procedure at the level of the Liouville path integral directly to find both the Schwarzian action and path integral measure \cite{Mertens:2017mtv,Mertens:2018fds} using an older field redefinition of Gervais and Neveu \cite{Gervais:1981gs,Gervais:1982nw,Gervais:1982yf,Gervais:1983am}. In the process, one encounters the 2d ancestor of the Schwarzian model as the so-called Alekseev-Shatashvili geometric action for coadjoint orbits of the Virasoro group \cite{Alekseev:1988ce,Alekseev:1990mp}.\footnote{This structural relation with Virasoro CFT is useful also in its relation with 3d gravity, see e.g. \cite{Cotler:2018zff}.}

We can furthermore consider Liouville primary vertex operator insertions $e^{2\alpha\phi(u,v)}$ on the cylinder worldsheet in the above double-scaling limit. This operator pairs up with its mirror image to form a single bilocal line \eqref{Schbil}, as drawn above. This calculation leads precisely to the JT disk boundary correlators as shown in detail in \cite{Mertens:2017mtv,Mertens:2018fds,Lam:2018pvp}.

Diagrams with crossing bilocal lines are found by incorporating the Schwarzian double-scaling limit of the 2d braiding $R$ matrix of Virasoro conformal blocks. Intuitively, if we have two uncrossed bilocal lines of the type drawn above, we can generate the crossed diagram if we swap either of the endpoints of both bilocals. This process acts on a chiral half of the Liouville CFT calculation and can be achieved by using the braiding $R$ matrix of the relevant modular tensor category.

Furthermore, defect insertions can be found by further enriching the story by adding Verlinde loops wrapped along the $\tau$-direction of the cylinder \cite{Mertens:2019tcm}. This is a generalization of studying different brane types (beyond ZZ) at either end of the worldsheet.

Finally, local 2d CFT stress tensor insertions reduce to insertions of the energy operator $E(t)$ in the Schwarzian limit. \\

Let us show the procedure in more detail for the partition function. We can evaluate \eqref{eq:vac} in the dual ``closed'' channel as
\begin{equation}
\chi_0 \left(q\right) = \int_0^\infty \, dP \, S_0{}^P~\chi_P(\tilde{q}), \qquad \chi_P(\tilde{q}) =  \frac{ \tilde{q}^{P^2}}{\eta(i2\pi/T)}, \qquad \tilde{q} = e^{-4\pi^2/T},
\end{equation}
with the Virasoro modular S-matrix element:
\begin{equation}
S_0{}^P \sim \sinh \bigl( 2 \pi b P\bigr) \sinh\left( \frac{2 \pi P}{b} \right).
\end{equation}
Parametrizing $P = bk$, we take the \textbf{Schwarzian double-scaling limit} \cite{Mertens:2017mtv}, where we let the central charge go to infinity $c\to \infty$ (or $b\to 0$) and simultaneously let $T \to 0$ keeping the combination $T c = \frac{24\pi^2 }{\beta}$ fixed where $\beta$ will be the inverse temperature of the resulting 1d model. We get:
\begin{equation}
\chi_0 \left(q\right) \to 2\pi b^2 \int dk k \sinh 2\pi k e^{-\beta k^2},
\end{equation}
which is the JT disk partition function, up to an immaterial prefactor that can be absorbed into $S_0$ again. Notice that descendants scale out as is suitable for a classical limit, and that the momentum $k$ comes from the ``closed'' channel and hence flows along the boundary circle.

\begin{table}[ht]
\centering
\begin{tabular}{p{0.45\linewidth} | p{0.45\linewidth} }
\emph{Strong points} & \emph{Weak points} \\
\hline
${\color{blue}\rightarrow}$ Explains presence of 2d CFT structure in JT and Schwarzian.

${\color{blue}\rightarrow}$ One can use and exploit the past knowledge of 2d Liouville results. & ${\color{blue}\rightarrow}$ Obscures the bulk JT disk picture, since the auxiliary Liouville CFT lives on a different space than JT.
\end{tabular}
\end{table}

\subsubsection{Boundary particle approach}
A different approach uses a direct rewriting of the GHY boundary term of the gravity path integral using the 2d Gauss-Bonnet theorem into a worldline path integral as \cite{Kitaev:2018wpr,Yang:2018gdb, Suh:2020lco}:
\begin{equation}
\sim \int [D X] e^{-\int_0^{\beta} d\tau \frac{1}{2}\frac{T'^2+Z'^2}{Z^2} + q \frac{T'}{Z}}, \qquad q = \frac{a}{2\epsilon} \to +\infty,
\end{equation}
which is a non-relativistic particle in 2d hyperbolic space $(T,Z)$ in an imaginary infinite magnetic field. One can reach the same conclusion in a Hamiltonian formulation of the model. \\

This non-relativistic system has been solved by A. Comtet and P. J. Houston \cite{Comtet:1984mm,Comtet:1986ki}, with the spectral measure leading to the JT disk spectral density in the $q\to+\infty$ limit:
\begin{equation}
\rho(k) = \frac{1}{4\pi^2}\frac{k \sinh 2 \pi k}{\cosh 2\pi q + \cosh 2\pi k} \quad \underset{q \to + \infty}{\to} \quad  \sim k \sinh 2\pi k,
\end{equation}
with an infinite but $k$-independent prefactor in the Schwarzian limit.

Using the known propagator of the particle in the magnetic field in the same $q\to +\infty$ limit, one can derive the following suggestive expression for JT boundary correlators for the simplest case of ordered operators \cite{Yang:2018gdb}:
\begin{align}
 &\left\langle \mathcal{O}(\tau_1) \hdots \mathcal{O}(\tau_n) \right\rangle  \\
 &=\int \frac{\prod_{i=1}^n\frac{d^2x_i}{Z_i^2}}{V(\text{SL}(2,\mathbb{R}))} G(\tau_{12};x_1,x_2) \hdots G(\tau_{n1};x_n,x_1) \left\langle \mathcal{O}(x_1) \hdots \mathcal{O}(x_n) \right\rangle_{\text{QFT}},
 \nonumber
\end{align}
where the ingredients are the bulk propagators of the particle in magnetic field model $G(\tau_{ij};x_i,x_j) \equiv \left\langle x_i \vert e^{-\tau_{ij} H} \vert x_j\right\rangle$ in proper time $\tau_{ij}$ (between points in the hyperbolic disk $x_i$ and $x_j$) and the ``undressed'' bulk matter correlator $\left\langle \mathcal{O}(x_1) \hdots \mathcal{O}(x_n) \right\rangle_{\text{QFT}}$ that can be determined in terms of Witten diagrams.

\begin{table}[ht]
\centering
\begin{tabular}{p{0.45\linewidth} | p{0.45\linewidth} }
\emph{Strong points} & \emph{Weak points} \\
\hline
${\color{blue}\rightarrow}$ Direct approach.

${\color{blue}\rightarrow}$ Makes the JT bulk explicitly visible, and allows for bulk interactions to be treated universally since we start with the generic undressed $\left\langle \mathcal{O}(x_1) \hdots \mathcal{O}(x_n) \right\rangle_{\text{CFT}}$. & ${\color{blue}\rightarrow}$ Structural and symmetry properties less visible a priori in more complicated correlators.
\end{tabular}
\end{table}

\vspace{-0.5cm}

\subsubsection{Two-dimensional gauge theory}
\label{s:BF}
We have previously mentioned the BF formulation of JT gravity on a closed manifold in Sect.~\ref{s:firstorder}.
For a group $G$ BF model on a manifold $\mathcal{M}$ with a boundary $\partial \mathcal{M}$, we can include the boundary term and condition \cite{Mertens:2018fds}:
\begin{equation}
I_\text{BF} = -\int_\mathcal{M} \text{Tr}(B F) + \frac{1}{2} \oint_{\partial \mathcal{M}} d\tau \text{Tr}(B A_\tau), \qquad B = A_\tau \vert_{\partial \mathcal{M}}.
\label{jtac}
\end{equation}
With this boundary term, we have a well-defined variational principle. Structurally, one can motivate this also by realizing one gets precisely this term when dimensionally reducing 3d Chern-Simons theory to the 2d BF model.

Path-integrating over $B$ along an imaginary contour makes $F$ pure gauge on the disk, and $A_\mu = g^{-1}\partial_\mu g$. The dynamics reduces to a pure boundary dynamics given by the action of a non-relativistic particle on a group manifold:
\begin{equation}
\label{pog}
I_\text{BF} = \frac{1}{2} \oint_{\partial \mathcal{M}} d\tau  \text{Tr}(g^{-1} \partial_\tau g)^2.
\end{equation}
Notice the similarity of this argument with that for JT gravity of Sect.~\ref{s:geomder}.

Interesting operator insertions in this model are \textbf{boundary-anchored Wilson lines}. Since they intersect the boundary twice, these correspond to bilocal operator insertions in the boundary model defined by the action \eqref{pog}. Correlation functions of any number of these bilocal operator insertions can be determined using techniques very similar to 2d Yang--Mills theory \cite{Migdal:1975zg,Witten:1991we,Cordes:1994fc}. The result of the disk partition function, and that with a single Wilson line insertion,
in this model are given by \cite{Mertens:2018fds,Blommaert:2018oro,Iliesiu:2019xuh}:
\begin{align}
\label{BFpf}
Z(\beta) &= \sum_R (\text{dim R})^2 e^{-\beta  C_R}, \\
\label{BFtwop}
\left\langle G^{R}_{MM}(\tau,0) \right\rangle &= \hspace{-0.4cm} \sum_{R_1,R_2,m_1,m_2} \hspace{-0.4cm} \text{dim R}_1 \text{dim R}_2 \, e^{-C_{R_1} \tau}e^{-C_{R_2} (\beta-\tau)}  \threej{R_1}{m_1}{R_2}{m_2}{R}{M}^2,
\end{align}
where the $R_i$ labels run through all unitary irreducible representations of the group $G$ of interest, with $\text{dim R}_i$ and $C_{R_i}$ its dimension and quadratic Casimir respectively. The $m$-label runs through the different states within each representation. Finally, the $3j$-symbols ($\threej{R_1}{m_1}{R_2}{m_2}{R}{M}$ in the above example), and $6j$-symbols can appear for Wilson line endpoints and bulk crossings respectively  \cite{Mertens:2018fds,Blommaert:2018oro,Iliesiu:2019xuh}. One can directly see the analogy with the JT correlation functions by setting $\sum_R \to \int_0^{+\infty}dk$ and interpreting:\footnote{One might worry about the $(\text{dim R})^2$ in the disk amplitude \eqref{BFpf}. One way to understand how to reproduce the JT answer, is that gravity has asymptotic Brown-Henneaux boundary conditions, which mathematically lead to a coset model instead of a pure group model. For a coset model, the indices running in loops are somewhat constrained, removing one (\text{dim R}) factor and leading directly to an expression similar as in JT gravity.}
\begin{equation}
\text{dim R} \to \frac{1}{\pi^2}k\sinh 2\pi k, \quad
3j \to \sqrt{\frac{\Gamma(\Delta \pm ik_1 \pm i k_2)}{(2C)^{2\Delta}\Gamma(2\Delta)}}, \,\,\,\,\,
6j \to \sixj{\Delta_1}{\Delta_2}{k_1}{k_3}{k_2}{k_4}.
\end{equation}
These observations can be made more explicit in terms of a BF model closely related to the non-compact SL$(2,\mathbb{R})$ group.\footnote{Around \eqref{gluel} we noted one can get the two-point function by performing an integral over the boundary-to-boundary geodesic length $\ell$. That integral has a direct interpretation in terms of a group integral of three representation matrices.} There are two proposals: one based on the positive semigroup SL$^+(2,\mathbb{R})$ \cite{Blommaert:2018oro,Blommaert:2018iqz}, and one based on an analytic continuation of the universal cover of SL$(2,\mathbb{R})$ \cite{Iliesiu:2019xuh}. \\

Defect insertions can also be studied and correspond to insertions of $\text{Tr}_R e^{2\pi B}$, which descend from ``vertical'' Wilson loops in irrep $R$ in the Chern-Simons ancestor of the BF model.
Finally, insertions of the quadratic Casimir itself correspond directly to energy operator insertions in the JT model. \\

It should be noted that one could also be interested in the case of compact group $G$ in its own right, and add this sector to the gravitational JT sector. We present some details later on in Sect.~\ref{s:JTgauge}. This describes more general models of 2d gravity that include conserved charges, of relevance to higher-dimensional black hole physics.

Both the approach of this section and that of Sect.~\ref{s:Liou} above have a structurally natural embedding of the different JT operator insertions, and from both perspectives, these operator insertions exhaust the interesting possibilities.

\begin{table}[ht]
\centering
\begin{tabular}{p{0.45\linewidth} | p{0.45\linewidth} }
\emph{Strong points} & \emph{Weak points} \\
\hline
${\color{blue}\rightarrow}$ Intuitive picture of natural operator insertions.

${\color{blue}\rightarrow}$ Group theoretical structure explicit. 

${\color{blue}\rightarrow}$ Provides efficient generalizations to models that include other conserved quantum numbers. 

& ${\color{blue}\rightarrow}$ Subtleties in the group theoretic structure make this approach a priori difficult. 

${\color{blue}\rightarrow}$ Higher topology moduli space is Teichm\"uller space, instead of the moduli space of Riemann surfaces.
\end{tabular}
\end{table}
\vspace{-0.5cm}

\section{Spacetime wormholes and random matrices}\label{sec:JTRMT}
In this section, we review how to compute wormhole contributions to gravitational amplitudes, and how they further correct the previous answers towards results that are more in line with a finite entropy boundary system in holography. This section heavily draws on material from the seminal work by \cite{Saad:2019lba}.

\subsection{Motivation: information loss and late time decay}
\label{s:infoloss}
First, let us take stock of where we are up to this point and how our JT amplitudes are integrated into a holographic framework of a microscopic UV-complete system. 

Consider a discrete microscopic holographic model, i.e. a 0+1d quantum mechanical system with a bulk dual. Its boundary two-point correlator can be expanded in the energy eigenbasis as:
\begin{equation}
\label{2pfdiscre}
\text{Tr} \left[e^{-\beta H} \mathcal{O}(t)\mathcal{O}(0)\right] = \sum_{n,m}e^{-\beta E_n} \vert \left\langle n\vert\mathcal{O}\vert m\right\rangle \vert^2 e^{-i t (E_m-E_n)},
\end{equation}
where this truly is a discrete sum over energy eigenstates labeled by $n$ and $m$. Due to the energy phase factors, this expression oscillates erratically as a function of time $t$. At late times, the terms with $n=m$ dominate the correlator, and we find the late-time mean non-zero value:\footnote{Physically, we can motivate this dominance of diagonal contributions by introducing a late-time averaging procedure to smoothen erratic fluctuations. E.g. $\frac{1}{\Delta}\int_{T-\Delta/2}^{T+\Delta/2} e^{-it (E_m-E_n)}dt = \frac{2 \sin \frac{1}{2}\Delta (E_m-E_n)}{\Delta (E_m-E_n)}e^{-iT (E_m-E_n)}$, which is suppressed as long as we average over multiple periods of the oscillating signal: $\Delta \gg (E_m-E_n)^{-1}$. Diagonal contributions ($E_m=E_n$) are not suppressed by this procedure.}
\begin{equation}
\text{Tr} \left[ e^{-\beta H}\mathcal{O}(t)\mathcal{O}(0)\right] \approx \sum_{n}e^{-\beta E_n} \vert \left\langle n\vert\mathcal{O}\vert n\right\rangle\vert^2.
\end{equation}

This late-time behavior of oscillations around a non-zero mean, is to be contrasted with our results in JT gravity so far. Indeed, semi-classically we have the late-time quasinormal mode exponential decay (as in Eq.~\eqref{eqn:2ptsemiclass}):
\begin{equation}
\frac{\left\langle \mathcal{O}(t)\mathcal{O}(0) \right\rangle_\beta}{Z(\beta)} = \Big(\sinh \frac{\pi}{\beta}t\Big)^{-2\Delta} \sim e^{-\frac{2\pi}{\beta} \Delta t}.
\end{equation}
Quantum gravitational corrections discussed up to this point, would modify this into a power-law decay instead \eqref{latetimeSch}:
\begin{equation}
\frac{\left\langle \mathcal{O}(t)\mathcal{O}(0) \right\rangle_\beta}{Z(\beta)} \sim t^{-3}.
\end{equation}
Both of these are in conflict with the behavior of a discrete finite entropy boundary system \eqref{2pfdiscre}.
This mismatch is one of the manifestations of information loss \cite{Maldacena:2001kr}, related to a continuum of energies in the energy spectrum due to the infinitely large near-horizon black hole region, or to the infinite redshift of near-horizon matter according to an asymptotic observer. 

In order to address this, it is useful to first simplify a bit the observable to its core essence. For sufficiently simple operators $\mathcal{O}$ that have zero one-point-function,\footnote{This can be achieved either by subtracting the one-point functions explicitly, or by the presence of e.g. a $\mathbb{Z}_2$-symmetry mapping $\mathcal{O} \to - \mathcal{O}$.} by ETH one expects the squared energy matrix elements $\vert \left\langle n \vert \mathcal{O} \vert m \right\rangle \vert^2$ to be smooth functions of the energy \cite{PhysRevA.43.2046, Srednicki:1994mfb}. Hence the main qualitative features we want to study here solely come from the erratically oscillating phase factors in \eqref{2pfdiscre}. So, as a zeroth-order approximation, we remove these energy matrix elements and study instead the \textbf{spectral form factor}:
\begin{equation}
\label{eq:SPFf}
Z(\beta + it) Z(\beta-it) = \sum_{n,m}e^{-(\beta+it) E_n} e^{-(\beta-it)E_m},
\end{equation}
which is the modulus squared of an analytically continued partition function. This quantity starts off at $t=0$ as $\vert Z(\beta) \vert^2$ and oscillates erratically around a late-time mean $Z(2\beta)$, which is suppressed by $e^{S}$.\footnote{Poincar\'e recurrences are expected at much later times $e^{e^{S}}$, and will not play a role in this review.} These high-frequency oscillations have a period on the scale of the energy difference between adjacent levels. This is for a system whose energy levels are sufficiently random, such as a chaotic system, with the hallmark example being random matrix ensembles. For integrable systems on the other hand, conspiracies between energy levels can create regular oscillations and recurrences. 

\begin{figure}[t!]
\begin{center}
\includegraphics[scale=0.15]{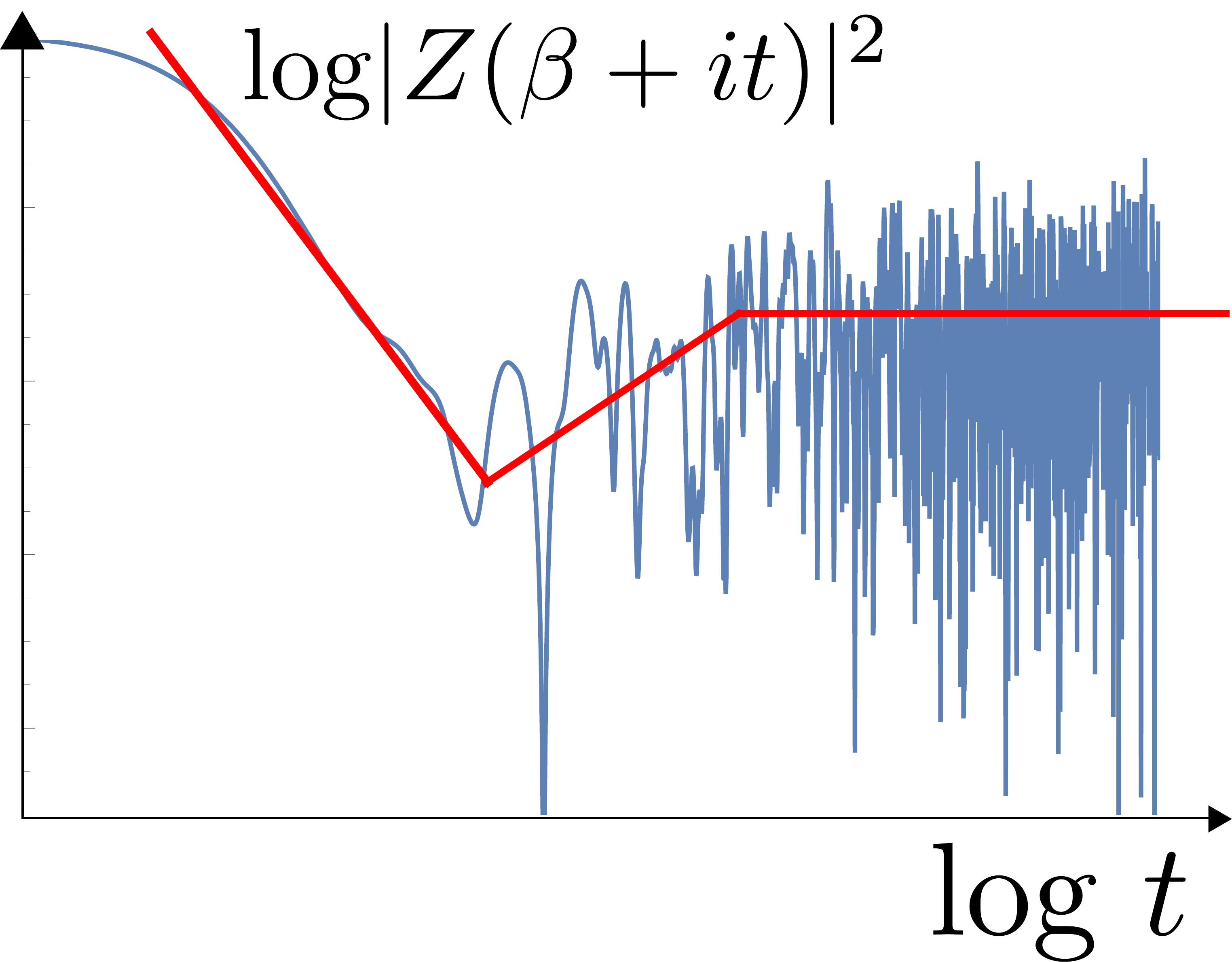} 
\end{center}
\caption{Spectral form factor for a matrix drawn from the GUE, with the averaged result in red.}
\label{FIG:SSSGUE}
\end{figure}

 The spectral form factor \eqref{eq:SPFf} for a chaotic system has a very characteristic shape as a function of time \cite{Cotler:2016fpe}. We sketch an example where we draw the spectrum from the Gaussian Unitary Ensemble (GUE) in Fig. \ref{FIG:SSSGUE}. We can see an initial downward slope, followed by an erratic rising ramp and erratic oscillations around a final plateau. The proposal of \cite{Cotler:2016fpe} is that the black hole spectrum is chaotic and the level spacing statistics is captured by random matrix theory.

In order to see any indications of this underlying discreteness and chaotic behavior of a microscopic holographic system in its bulk gravitational description, we will have to include quantum gravity effects beyond those we have studied up to this point: gravitational configurations whose topology is different from the disk, \textit{i.e.} spacetime wormholes. We first turn to describing how one computes such amplitudes gravitationally in JT gravity. We will come back to the spectral form factor and signs of discreteness for JT gravity in Sect.~\ref{s:sff}. Some aspects of this connection between wormholes and chaos generalize to black holes in any dimension through the double-cone wormhole of \cite{Saad:2018bqo}.

\subsection{Multiboundary higher genus amplitudes}
\label{s:sss}
In this section we review the computation of the gravitational path integral on geometries with spacetime wormholes, and sum over topologies. We consider first the case of pure JT gravity studied by P. Saad, S. Shenker and D. Stanford (SSS) in \cite{Saad:2019lba,Saad:2018bqo}.

As discussed at length in sections \ref{sec:JT} and \ref{sec:JTquantum}, computing the partition function using the gravitational path integral involves considering the situation with a Euclidean geometry with a boundary curve (in 2d) such that the proper length $L$ and the boundary dilaton $\Phi_\partial$ both diverge at a constant rate $L,\Phi_\partial \to \infty$ with $L/\Phi_\partial = \beta/ C$. The parameter $C$ is fixed once and for all and sets the units in which the inverse temperature $\beta$ is measured, see Sect.~\ref{sec:JT} for more details. 

It will be convenient for the remainder of this section to generalize this set-up to the case of $n$ boundaries, each described by their own inverse temperatures $\beta_1,\ldots, \beta_n$. We will denote the gravity path integral with $n$ boundaries by $Z_{\rm grav}(\beta_1,\ldots, \beta_n)$. The aim of this section is to analyze whether 
\begin{eqnarray}\label{eq:JTwhat}
Z_{\rm grav}(\beta_1,\ldots,\beta_n) \overset{?}{=}~ {\rm Tr}_{\mathcal{H}_{\rm BH}}\left( e^{-\beta_1 H} \right)\ldots {\rm Tr}_{\mathcal{H}_{\rm BH}}\left( e^{-\beta_n H} \right)
\end{eqnarray}
is true, where $H$ is a Hamiltonian acting on the putative black hole Hilbert space $\mathcal{H}_{\rm BH}$. The purpose of this subsection is to compute the left hand side of this expression, while the relation on the right hand side will be analyzed in the next subsection. 

To start the analysis, we repeat here the Euclidean action of pure JT gravity including the topological term:
\begin{eqnarray}
I &=& -\frac{S_0}{4\pi} \left[ \int_{\mathcal{M}} \sqrt{g}R + 2 \oint_{\partial\mathcal{M}} \sqrt{h}K \right]\nonumber\\
&&\hspace{1cm} - \frac{1}{16 \pi G_N} \left[ \int_{\mathcal{M}} \sqrt{g}\Phi(R+2) + 2 \oint_{\partial\mathcal{M}}\sqrt{h} \Phi (K-1) \right]\label{eqn:JTactions4},
\end{eqnarray}
where $g$ is the metric on the two-dimensional space $\mathcal{M}$ and $h$ the induced metric on its boundary $\partial \mathcal{M}$. We focus on a connected surface here. Each time we add a handle to the Euclidean geometry, we  increase the genus $g$ of the surface. These handles are often called \textbf{spacetime wormholes} in this context.\footnote{This nomenclature is to distinguish them from purely spatial wormholes that live at a fixed timeslice, such as the Einstein--Rosen bridge of an eternal black hole. Spacetime wormholes on the other hand extend in the full (Euclidean) spacetime. See e.g. \cite{Kundu:2021nwp} for a recent review.} Since the first line of the action $I$ is proportional to the Euler characteristic $\chi = 2- 2g -n$, with $g$ the number of handles and $n$ the number of boundaries, this allows us to expand the full path integral as a genus expansion
\begin{eqnarray}
Z_{\rm grav, conn}(\beta_1,\ldots,\beta_n)= \sum_{g=0}^\infty e^{S_0(2-2g-n)} Z_{g,n}(\beta_1,\ldots,\beta_n).
\end{eqnarray}
Notice the suppressing prefactor at higher genus $g$ for fixed $n$. In this expression $Z_{g,n}$ denotes the contribution from geometries with fixed topology. For convenience we focus solely on connected contributions, which are the building blocks for the full answers. \\

Before directly computing $Z_{g,n}$, it will be useful first to consider the case of a surface of genus $g$ with $n$ \emph{geodesic} boundaries of lengths $b_i,i=1\hdots n$, and no holographic (or Schwarzian) boundaries. A geodesic boundary by definition has zero extrinsic curvature trace: $K=0$. The JT path integral on such a surface becomes
\begin{equation}
\int \frac{[D g_{\mu\nu}][D \Phi]}{V(\text{Diff})}e^{-I_{\text{JT}}} = \int \frac{[D g_{\mu\nu}]}{V(\text{Diff})} \delta(R+2),
\end{equation}
where no boundary actions appear. Here we mod out by bulk diffeomorphisms. The RHS is suggestive: the JT path integral on such a manifold is a volume integral (since the integrand is ``1'') over all hyperbolic $(R=-2)$ metrics mod diffeomorphisms on the given surface. This is the volume of the moduli space of Riemann surfaces. We denote this moduli space as $\mathcal{M}_{g,n}(\vec{b})$. The real dimension of this moduli space is $\text{dim }(\mathcal{M}_{g,n}(\vec{b})) = 6g-6+2n$.\footnote{This is explained above equation \eqref{Vform} but another way to derive this is as follows. We start with the two-sphere $S^2$. Increasing the genus by one can be done by cutting out two disks from a surface and gluing them together. This requires the specification of four real moduli for the locations of both disks, one real moduli for their relative size, and one real moduli for their relative angle. This gives $6g$ real moduli overall. The $-6$ comes from the $SL(2,\mathbb{C})$ global conformal transformations on the starting $S^2$, producing the same surface. Finally, a boundary has three real moduli: two for its location, and one for its length. Here we keep the length fixed, so we are left with two real moduli per geodesic boundary. Note that one can get a puncture if we let $b_i \to 0$.} 

The measure in the integral over $\mathcal{M}_{g,n}(\vec{b})$ requires a calculation of the one-loop determinant arising after integrating out the dilaton and restricting to hyperbolic metrics. Insightful discussions can be found in Sect.~3.2 of \cite{Saad:2019lba} in a perturbative string language, and in Sect.~3 of \cite{Stanford:2019vob}, using the relation between analytic and combinatoric torsion. The upshot is that the correct measure over $\mathcal{M}_{g,n}(\vec{b})$ that appears in JT gravity is the one induced from the Weil--Petersson symplectic form as follows. We can decompose any hyperbolic surface with genus $g$ into a set of $2g-2+n$ pair-of-pants with $3g-3+n$ connecting tubes with lengths and twists $(\widetilde{b}_i,\tau_i)$ for $i=1,\ldots,3g-3+n$. The Weil--Petersson form in these length-twist or Fenchel-Nielsen coordinates, is   
\begin{eqnarray}
\label{Vform}
\omega_{g,n} = \sum_{i=1}^{3g-3+n} d\widetilde{b}_i \wedge d\tau_i.
\end{eqnarray}
The volume form extracted from this Weil--Petersson form is $\frac{\omega_{g,n}^{3g-3+n}}{(3g-3+n)!}$. Even though this measure is very simple, it is not easy to determine the region of integration over the variables $(\widetilde{b}_i,\tau_i)$ when modding by the group of so-called large diffeomorphisms, i.e. the mapping class group. These are by definition the set of diffeomorphisms that are not continuously connected to the identity. We have to do this to make sure geometries are not overcounted. We will come back to this later. 

Now we are ready to present the computation of $Z_{g,n}$.

\subsubsection{The disk $Z_{0,1}(\beta)$} 

There are two cases we need to work out separately. The first is the disk with $g=0$ and $n=1$ (and hence $\chi=1$), which we already computed in Sect.~\ref{s:diskpf}: 
\begin{eqnarray}
Z_{0,1}(\beta) = Z_{\rm Sch, disk}(\beta)= \frac{C^{3/2}}{(2\pi)^{1/2} \beta^{3/2}} e^{ \frac{2\pi^2 C}{\beta}}.\label{eq:Z01}
\end{eqnarray}
This gives a contribution of order $e^{S_0}$ to the partition function, since $\chi=1$. We can use this to extract the leading order density of states: 
\begin{eqnarray}
\rho_0 (E) = \frac{C}{2\pi^2}\sinh \left( 2\pi \sqrt{2 C E} \right).
\end{eqnarray}
where we have defined the $g=0$ density $\rho_0(E)$ with the overall factor $e^{S_0}$ removed (compare with \eqref{eqn:JTexactRho}). When summing over topologies this is only the first term in an $e^{-2S_0}$ expansion. 

\subsubsection{The cylinder $Z_{0,2}(\beta_1,\beta_2)$} 

The second case we need to consider separately is the genus zero two-boundary amplitude. The quantity $Z_{\rm grav}(\beta_1,\beta_2)$ gets two contributions. The first comes from two disconnected disks $e^{2S_0}Z_{0,1}(\beta_1)Z_{0,1}(\beta_2)$. More interesting is the contribution from the connected geometry without handles, a cylinder (or annulus) between the two boundaries, $Z_{0,2}(\beta_1,\beta_2)$. Its Euler characteristic is $\chi = 0$, and for large $S_0$ it is at first sight strongly suppressed compared to the disconnected two disks. This conclusion, however, is too quick and can be invalidated in certain parametric regimes as we will explain later on in Sect.~\ref{s:sff}. This geometry is so important that it got a new name: the \textbf{double trumpet}. \\

To compute this quantity we first integrate out the dilaton, imposing that the metric is hyperbolic. We can write down the two-parameter family of hyperbolic metrics on the double trumpet as:
\begin{equation}
    ds^2 = dr^2 + \cosh^2 r [ b dx +\tau \delta(r) dr]^2,~~x\sim x+1.
\end{equation}
where $b>0$ and $0<\tau<b$. For this metric, $r$ is the radial distance and the boundaries are at large negative $r$ (left boundary) or large positive $r$ (right boundary). The precise radial cutoff is picked to match the boundary conditions. 

In the center at $r=0$ there is a geodesic with length $b$. Finally the parameter $\tau$ denotes a possible twist when crossing the middle geodesic. Given $(b,\tau)$, the geometry can be depicted as:
\begin{eqnarray}
\centering
\includegraphics[scale=0.13]{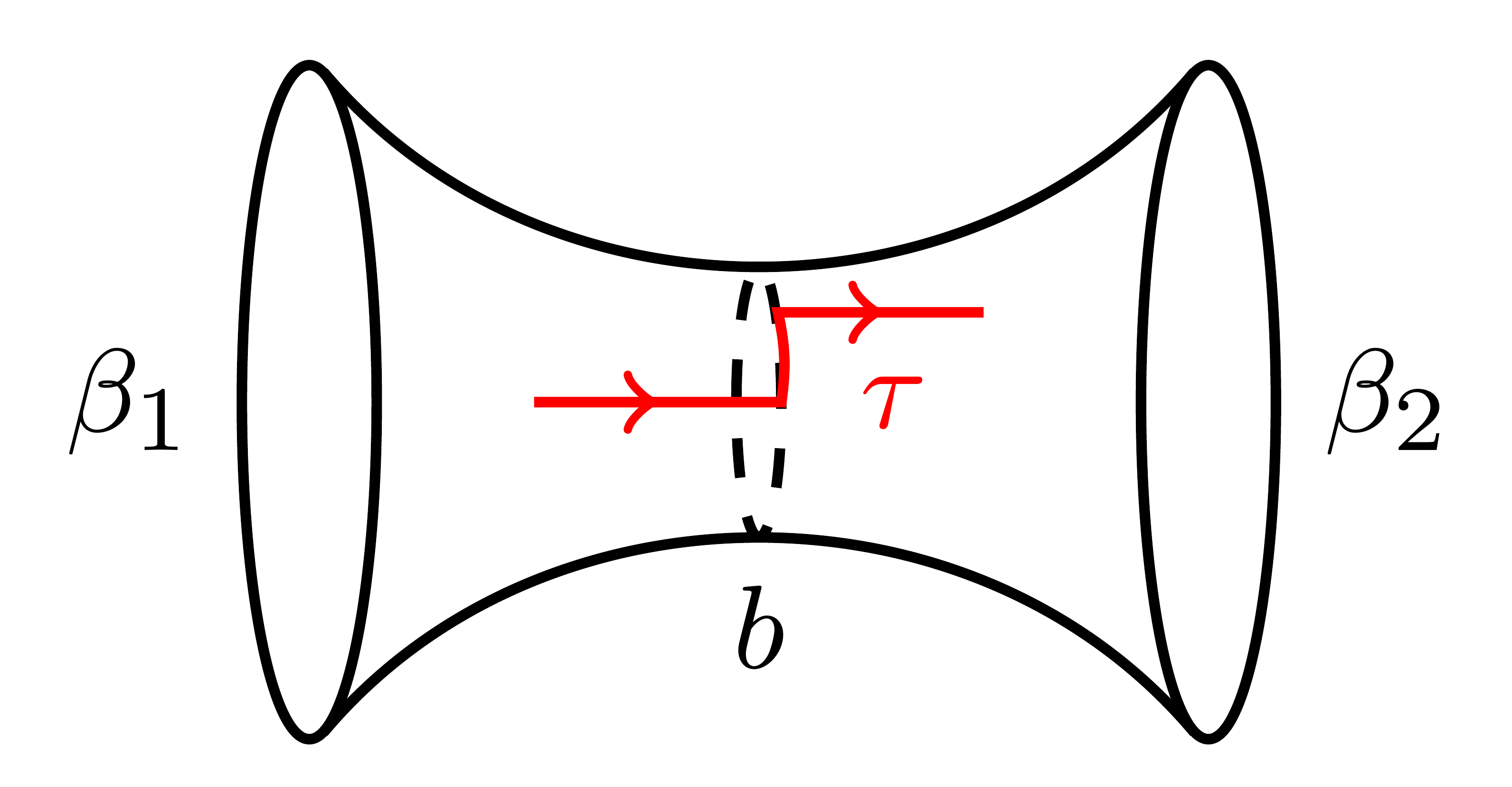}
\end{eqnarray}

We first compute the path integral on this cylinder with fixed $(b,\tau)$, and afterward integrate over these parameters. This can be done by separating the cylinder into two \textbf{single trumpets}: one for $r<0$ and another for $r>0$. Each trumpet is a hyperbolic cylinder bounding a holographic boundary and a geodesic. This procedure can be depicted as
\begin{eqnarray}
\includegraphics[scale=0.2]{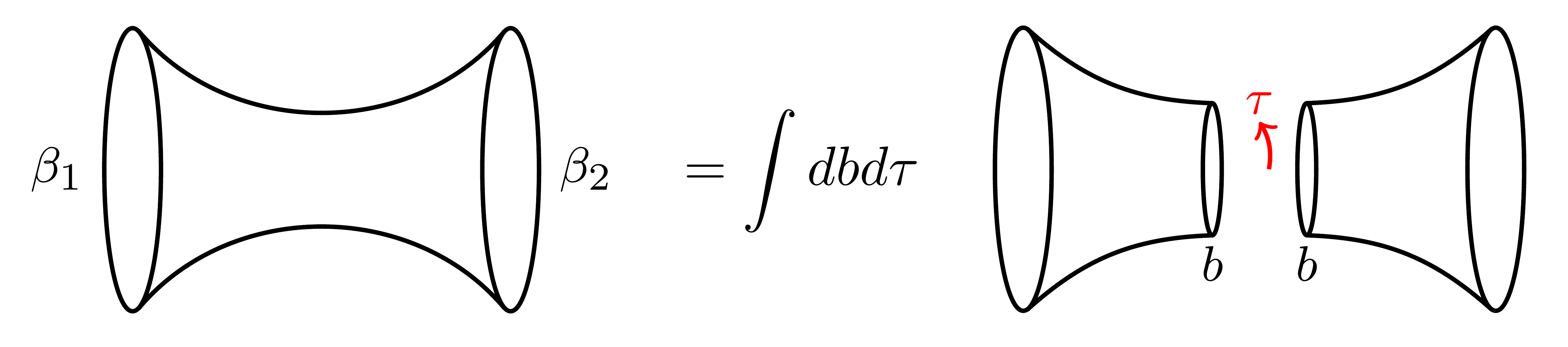}
\end{eqnarray}
On each single trumpet, the path integral is given by the contribution of the Schwarzian mode on the holographic boundary.\footnote{On the inner geodesic boundary there are no dynamical degrees of freedom since the extrinsic curvature is being fixed to zero.} The boundary mode of the single trumpet is not the same as the disk, it corresponds to an integral over the hyperbolic orbit ${\rm Diff}(S^1)/U(1)$. This orbit is labeled by a continuous parameter $b$. We quote the result \eqref{doshyp} from Sect.~\ref{sec:JTquantum}:
\begin{eqnarray}
\label{hyporbit}
\includegraphics[scale=0.22]{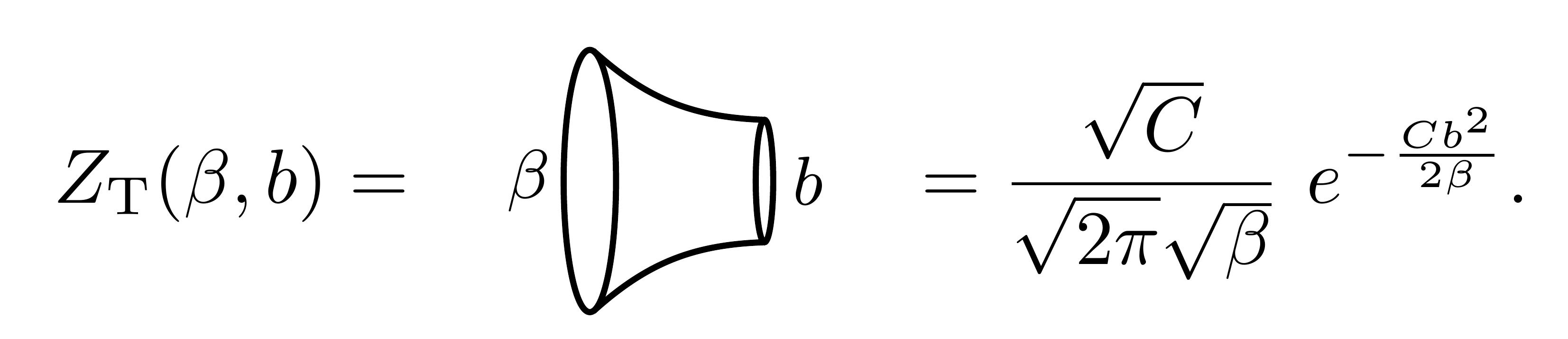}
\end{eqnarray}
The final step is to glue the two trumpets into the complete cylinder, which requires knowing the correct measure on the moduli space of hyperbolic surfaces as derived earlier. Locality guarantees that the same measure and volume form \eqref{Vform} should be used here when gluing along the pair of geodesics: the volume form is thus $dbd\tau$. Since the integrand is independent of the twist, we can integrate over it producing a factor of $\int_0^b d\tau = b$, giving a total measure of $bdb$.\footnote{If we utilize Teichm\"uller space instead of $\mathcal{M}_{g,n}(\vec{b})$, the integral over twists would give a divergent but $b$-independent factor. Teichm\"uller space is the universal cover of $\mathcal{M}_{g,n}(\vec{b})$, where we ``undo" the modding by the mapping class group of large diffeomorphisms. Another way of representing Teichm\"uller space is by adding a Moore-Seiberg graph to a given Riemann surface. Using the local gluing procedure in the BF gauge theoretical TQFT framework, leads precisely to the Teichm\"uller space gluing integrals \cite{Blommaert:2018iqz}.} The final answer is:
\begin{eqnarray}
    Z_{0,2}(\beta_1,\beta_2) =
    \int_0^\infty bdb ~Z_{\rm T}(\beta_1,b) Z_{\rm T}(\beta_2,b) = \frac{1}{2\pi} \frac{\sqrt{\beta_1 \beta_2}}{\beta_1 + \beta_2}.\label{eq:Z02}
\end{eqnarray}
We note that this is a completely off-shell calculation: there is no saddle point value determining $b$ and stabilizing the wormhole, as can be seen from \eqref{hyporbit} where the exponential factor pushes $b$ to zero.\footnote{Although one could think of $b=0$ as a saddle on the boundary of the moduli space \cite{Witten:2021nzp}.} A general argument for this lack of classical solution can be found in \cite{Maldacena:1998uz} and \cite{Qi:2019gny}.
This is not necessarily a problem, but it does complicate things when trying to generalize to higher-dimensional physics, where we do not have an exact quantum solution. Related to this, we make two further comments:
\begin{itemize}
\item Fixing $b$ does allow for a saddle to exist. This saddle is hence the solution of a constrained extremization problem, and is called a \textbf{constrained instanton}. With this constraint, wormhole saddle solutions have been found explicitly in higher dimensions  \cite{Cotler:2020lxj,Cotler:2022rud}.
\item Analytically continuing $\beta \to \beta +it$, we can find a saddle at $\beta=0$, a genuine saddle if we also transfer to a microcanonical version of Eq.~\eqref{eq:Z02} \cite{Saad:2018bqo}. This geometry has been constructed explicitly and looks like a \textbf{double cone} of two cones joined at the tip, and where Lorentzian time is periodically identified with period $t$. Stability of this saddle has been investigated in \cite{Mahajan:2021maz}. This wormhole solution can actually be written for black holes in any number of dimensions.
\end{itemize}

\subsubsection{The general case $Z_{g,n}(\beta_1,\ldots, \beta_n)$} \label{s:genWP}
We can now describe a formula valid for all other $Z_{g,n}$ as long as $(g,n) \neq (0,1),(0,2)$. To derive it, first notice that in hyperbolic geometry we can always find geodesics homologous to each holographic boundary, such that these $n$ geodesics bound an inner hyperbolic surface in the bulk. For example, for three boundaries at genus one: 
\begin{equation}
\nonumber
\includegraphics[scale=0.17]{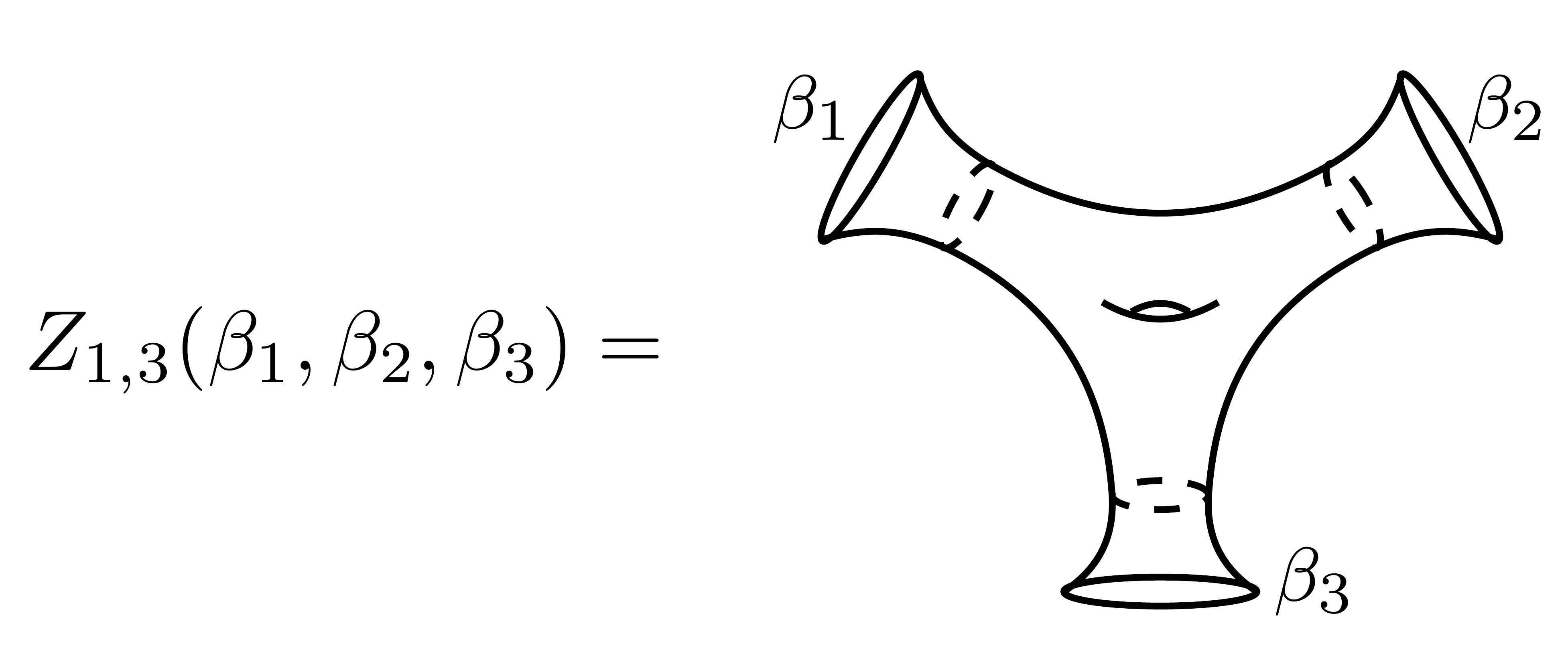}    
\end{equation}
Each of the three boundaries has an associated geodesic, drawn as a dashed line, bounding a single trumpet. The interior remainder is a hyperbolic surface with $g$ handles (in the example just one) bounded by geodesics. The hyperbolic surface now is described by two types of moduli: the one associated to the inner surface $\mathcal{M}_{g,n}(\vec{b})$, and $(b_1,\tau_1,\ldots, b_n,\tau_n)$ that determine how the ``outer'' geodesics are glued to the boundary trumpets. Using this decomposition, after integrating over the boundary twists $\tau_1, \ldots \tau_n$, we find the following expression \cite{Saad:2019lba}:  
\begin{equation}
    Z_{g,n}(\beta_1,\ldots,\beta_n) =\int_0^\infty \left[\prod_{i=1}^n b_idb_i~Z_{\rm T}(\beta_i,b_i) \right] V_{g,n}(b_1,\ldots ,b_n).\label{eq:SSSglue}
\end{equation}
The integrals on the right hand side are over the lengths of the geodesics homologous to each boundary and include the single trumpet contributions. The last factor $V_{g,n}(b_1,\ldots,b_n)$ is the volume of the interior moduli space $V_{g,n}(b_1,\ldots,b_n)={\rm Vol}(\mathcal{M}_{g,n}(\vec{b}))$, or the Weil--Petersson volumes. \\

The last ingredient we need is to provide a way to compute efficiently these Weil--Petersson volumes, so that the general expression for $Z_{g,n}$ can become useful. There are two approaches to this problem, both developed by M. Mirzakhani in \cite{mirzakhani2007simple, Mirzakhani:2006eta}. The first consists in using the following identity:
\begin{eqnarray}\label{eqn:WPinter}
    V_{g,n}(b_1,\ldots,b_n) = \int_{\overline{\mathcal{M}}_{g,n}} e^{\omega_{g,n} + \frac{1}{2} \sum_i b_i^2 \psi_i},
\end{eqnarray}
where $\overline{\mathcal{M}}_{g,n}$ is the Deligne-Mumford compactification of the moduli space of hyperbolic surfaces of genus $g$ and $n$ punctures (which corresponds to a geodesic boundary with $b\to0$) and $\omega_{g,n}$ is the Weil--Petersson form on the moduli space of genus $g$ and $n$ punctures. $\psi_i$ is defined as follows: let $\mathcal{L}_i$ be the cotangent space to the $i$th puncture. As we move along the moduli space, $\mathcal{L}_i$ varies as the fiber of a complex line bundle over moduli space. The psi-class is the first Chern class of this line bundle $\psi_i = c_1(\mathcal{L}_i)$. A simple recursion formula to compute this type of integrals over psi-classes and Weil--Petersson form (or more generally kappa-classes) was conjectured by \cite{Witten:1990hr} and proven by \cite{Kontsevich:1992ti}. This point of view was exploited in the context of JT gravity mainly by \cite{Okuyama:2019xbv,Okuyama:2020ncd}.

Instead, we will follow the approach originally taken by SSS which has a more geometric origin. First of all, even though the volume form is very simple, taking into account the mapping class group to determine the region of integration in the $(\vec{b},\vec{\tau})$ space is a very complicated problem. In \cite{mirzakhani2007simple} this was circumvented by using the following trick.\footnote{A useful presentation of the argument aimed at physicists can be found in Appendix D of \cite{Stanford:2019vob}.} First we pick a (geodesic) boundary, labeled as $1$ below, and start shooting geodesics perpendicular to the boundary, along the entire circumference of the boundary. There are three options: (A) the curve intersects itself at some point or returns to the same geodesic boundary, and we truncate it as soon as either happens; (B) the geodesic reaches a different boundary geodesic; (C) the geodesic continues forever in the interior without self-intersection and without reaching any boundary. Now we can write the length of the geodesic $b_1$ as a sum over the length of segments where geodesics of type A, B and C start. It was proven by Birman and Series that the set C has measure zero so we can ignore its contribution to the length. When the hyperbolic surface is decorated by a geodesic of either type A or B, we can single out a pair of pants (i.e. a three-holed sphere surface) that in case A bounds the geodesic $1$ and two internal ones; or in case B the geodesic $1$, another boundary geodesic and an internal one. This decomposes the surface of genus $g$ and $n$ boundaries into either one or two surfaces with either lower genus or a smaller number of boundaries. Using the fact that there is only a single hyperbolic surface of genus zero with three given geodesic lengths: $V_{0,3}(b_1,b_2,b_3)=1$, and a short calculation in hyperbolic geometry reviewed in \cite{Stanford:2019vob}, one arrives at Mirzakhani's recursion relation:
\begin{eqnarray}
b_1 V_{g,n}(b_1,B) &=& \frac{1}{2}\int_0^\infty b' db' b'' db'' D(b_1,b',b'')  V_{g-1}(b',b'',B),\nonumber\\
 &&+ \frac{1}{2}\int_0^\infty b' db' b'' db'' D(b_1,b',b'') \sum_{\rm stable} V_{h_1} (b',B_1) V_{h_2} (b'',B_2),\nonumber\\
&&+\sum_{k=2}^{n} \int_0^\infty b' db' (b_1-T(b_1,b',b_k)) V_g(b',B/b_k),\label{eq:MirzakhaniRecursion}
\end{eqnarray}
where $B=\{b_2,\ldots,b_n\}$, and we defined $T(b_1,b_2,b_3)=\log \frac{\cosh \frac{b_3}{2} + \cosh \frac{b_1+b_2}{2}}{\cosh \frac{b_3}{2} + \cosh \frac{b_1-b_2}{2}}$ and $D(b_1,b_2,b_3) = b_1 - T(b_1,b_2,b_3) - T(b_1,b_3,b_2)$. These functions arise from computing within each pair of pants the length along $b_1$ of segments of types A or type B in hyperbolic geometry. The three lines in \eqref{eq:MirzakhaniRecursion} correspond graphically to the decomposition of surfaces:
\begin{equation}\label{eqn:diagramtrr}
\nonumber
\includegraphics[scale=0.2]{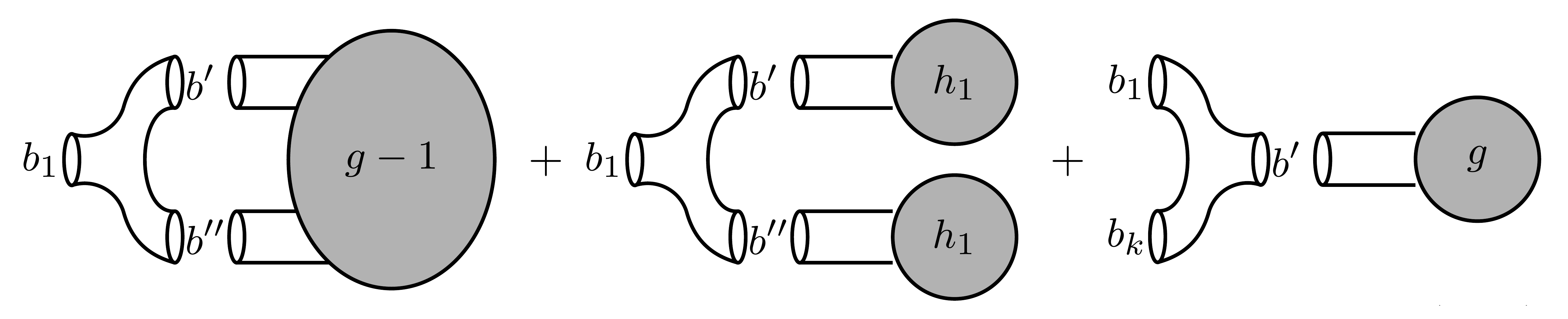}
\end{equation}
where the grey blobs denote the remainder hyperbolic Riemann surface in this decomposition. The first line in \eqref{eq:MirzakhaniRecursion} corresponds to case A, cutting the pair of pants involves only one boundary and does not divide the surface, but lowers the genus to $g-1$. The second line in \eqref{eq:MirzakhaniRecursion} corresponds also to case A, when removing the pair of pants divides the surface in two with genus $h_1$ and $h_2$ and boundaries $(b',B_1)$ and $(b'',B_2)$, such that $h_1+h_2=g$ and $B_1 \union B_2 = B$.\footnote{The sum is restricted to only stable surfaces. This means surfaces on which a hyperbolic metric exists; this excludes for example genus zero with less than three boundaries or genus one with no boundary.} Finally, the third line in \eqref{eq:MirzakhaniRecursion} corresponds to case B where the pair of pants to be removed involves two boundaries $b_1$ and $b_k$. Then $B/b_k$ denotes the set $B$ with the element $b_k$ removed.

In spite of not having an explicit closed expression for the Weil--Petersson volumes,\footnote{An exception is the case $g=0$. An explicit formula for $V_{0,n}(b_1,\ldots, b_n)$ was recently derived in \cite{Mertens:2020hbs}, equation (7.54).} one can immediately derive the following important property: $V_{g,n}(b_1,\ldots,b_n)$ is a symmetric multivariate polynomial of degree $3g-3+n$ in the $b_i^2$. This is most easily seen from \eqref{eqn:WPinter}. \\

Using the recursion relation above, one can compute any Weil--Petersson volume and combine it with the exterior trumpets to compute any desired $Z_{g,n}$. This completes the exact gravitational solution of JT gravity in a topological (or genus) expansion, i.e. by summing over all spacetime wormhole configurations. \\

As a final comment, going back to the intersection numbers approach of Eq.~\eqref{eqn:WPinter} and inserting that directly in \eqref{eq:SSSglue}, one can obtain a formula for $Z_{g,n}$ as an integral over moduli space with punctures:
\begin{equation}
    Z_{g,n}(\beta_1,\ldots,\beta_n) = \frac{\sqrt{\beta_1\ldots \beta_n}}{(2\pi C)^{n/2}} \int_{\overline{\mathcal{M}}_{g,n}}  \frac{e^{\omega_{g,n}}}{\prod_{i=1}^n(1-\frac{\beta_i}{C} \psi_i)}.
\end{equation}
This expression can be understood as an application of localization to JT gravity, as explained in \cite{Eberhardt:2022wlc}. This approach generalizes nicely to chiral gravity in three dimensions.

To illustrate this rather abstract treatment, we will work out explicitly the example of the one-wormhole correction to the disk partition function $Z_{1,1}(\beta)$.

\textbf{Example: $Z_{1,1}(\beta)$:}\\
The only option to decompose the surface with one boundary and one handle is: 
\begin{eqnarray}
\nonumber
\includegraphics[scale=0.175]{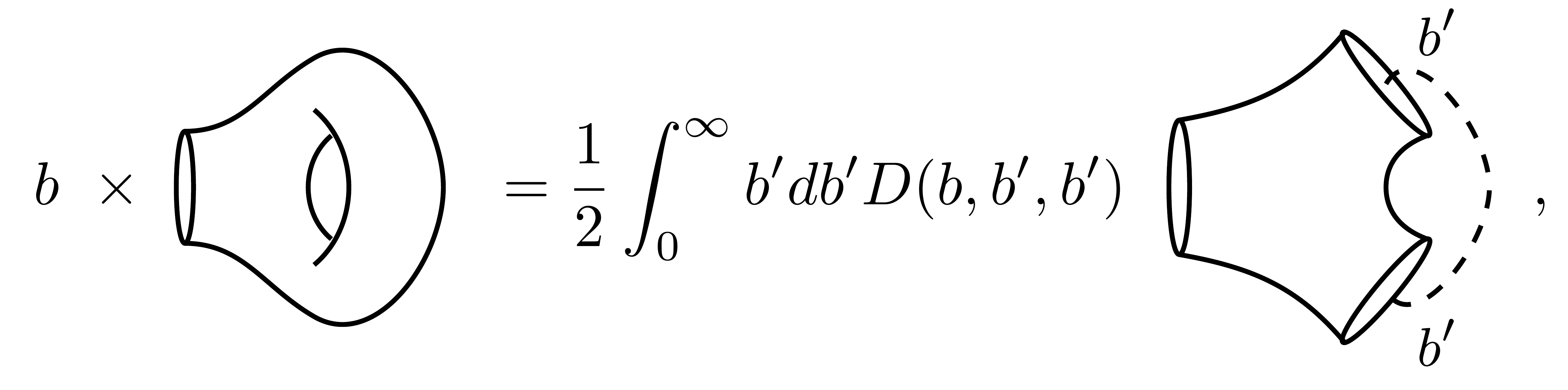}
\end{eqnarray}
where in the right hand side we have a single pair of pants with the two geodesics of length $b'$ identified (indicated by the dashed line). The equation for the volume becomes
\begin{eqnarray}\label{eq:vol11}
 V_{1,1}(b) = \frac{1}{2b} \int_0^\infty b'db' D(b,b',b')=\frac{b^2 + 4\pi^2}{48}.
\end{eqnarray}
This integral is actually divergent. The standard way of dealing with this is to compute instead $\partial_b (b V_{1,1}(b))$, which involves $\partial_b D(b,b',b')$ for which the integral converges. After integrating the result over $b$, the integration constant can be fixed by demanding $V_{1,1}(b)$ has a smooth $b\to0$ limit. 

This case does not quite work the same as in the generic case of Eq.~\eqref{eq:MirzakhaniRecursion}. Firstly, there is no volume factor appearing on the right hand side of the recursion and also one less integral since when gluing two ends of a single pair of pants those lengths are identified. Note that if we define $V_{0,2}(b_1,b_2) =\frac{1}{b_1} \delta(b_1-b_2)$ then this could be written as a special case of Eq.~\eqref{eq:MirzakhaniRecursion}. This case is special for a second reason: the torus with one hole has a $\mathbb{Z}_2$ symmetry and the volume given above is taking this symmetry into account. 

Using \eqref{eq:vol11} we can plug it in \eqref{eq:SSSglue} and obtain the one-wormhole correction to the partition function 
\begin{align}
\includegraphics[scale=0.29]{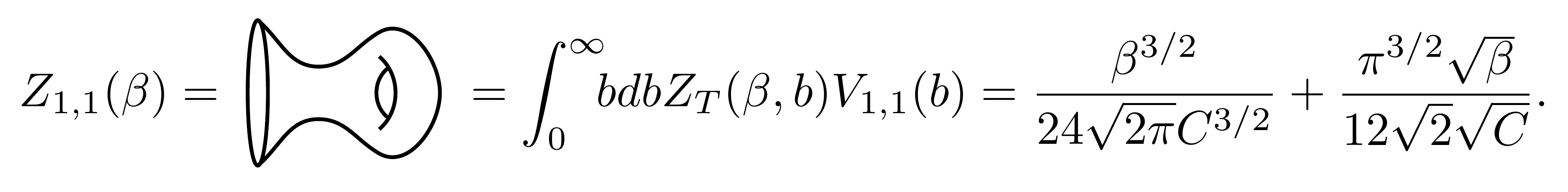}
\end{align}
\begin{center}
---o---
\end{center}
\vspace{0.5cm}

The above example already displays some interesting and generic features. Firstly, the amplitude with one wormhole is a polynomial in $\sqrt{\beta}$. This is in contrast to the disk which has an essential singularity at $\sqrt{\beta}=0$. The lack of an exponential term in $Z_{1,1}(\beta)$ is related to the fact that there is no classical solution of a disk with one wormhole. Much like the double trumpet, all of these configurations have no saddle point in the JT path integral. The reason was already pointed out around \eqref{dileomcov}: the equation of motion for the metric leads to a Killing vector which does not exist on higher genus hyperbolic surfaces. Secondly, this correction is of order one and therefore it is suppressed by $e^{-S_0}$ compared to the disk with no wormhole. Nevertheless this is too naive. Adding the disk and one-wormhole contributions, we find schematically for small temperatures $Z \sim e^{S_0}\beta^{-3/2}\# + e^{-S_0} \beta^{3/2}\#$, and therefore the one-wormhole contribution becomes dominant when $\beta\sim \mathcal{O}(e^{2S_0/3})$. However, at such low temperatures we cannot even trust the perturbative expansion in $e^{-S_0}$ to begin with.\footnote{In general, the large $\beta$ behavior is dominated by the large $b$ behavior of the volumes. When one of the boundaries, say the first, has a very large length $b_1\gg 1$ one can show $V_{g,n}(\vec{b}) \sim \frac{b_1^{6g-6+2n}}{(24)^g g!2^{3g-3+n} (3g-3+n)!}$ (see Appendix~A of \cite{Maxfield:2020ale}) and therefore in the case of one boundary $Z_{g,1}(\beta) \sim \frac{C^{3/2}}{\sqrt{2\pi}\beta^{3/2}}\frac{(\beta/C)^{3g}}{(24)^g g!}$ at large $\beta$. We see that at higher genus, the divergence at low temperatures becomes stronger. This can be interpreted as a perturbative $e^{-S_0}$ correction to the zero-point energy.}

\subsection{JT and random matrices}
\label{s:JTRMT}
In this subsection, we turn towards the connection between JT gravity and random matrix models in the double-scaling limit. This connection can be seen in the intersection number approach to computing Weil--Petersson volumes \eqref{eqn:WPinter}, since those can be obtained in a double-scaled matrix model through the Witten--Kontsevich theorem \cite{Witten:1990hr, Kontsevich:1992ti}. A discussion along these lines can be found in \cite{Dijkgraaf:2018vnm}. Instead we will follow SSS \cite{Saad:2019lba}, and use Mirzakhani's recursion relation and compare it with the loop equations of the matrix model. 

\subsubsection{SSS duality}
\label{s:SSS}
We come back to the original question of whether a relation such as \eqref{eq:JTwhat} is true in pure JT gravity. From the calculations done in the previous section, it is immediately clear that such a relation cannot be correct since, (1) there is no indication (in perturbation theory at least) that the density of states is discrete in any way which is in tension with (finite-volume) holography, and (2) the connected two-boundary path integral due to the connected wormhole amplitude does not factorize:
\begin{eqnarray}
Z_{\rm grav}(\beta_1,\beta_2) \neq Z_{\rm grav}(\beta_1) Z_{\rm grav}(\beta_2).
\end{eqnarray}
This is in tension with \eqref{eq:JTwhat}, which would predict a factorized answer for a given $H$ and $\mathcal{H}_{\rm BH}$. This is the \textbf{factorization puzzle} for holography raised first by \cite{Witten:1999xp}. Both problems can be resolved by relaxing \eqref{eq:JTwhat} in the following way. If pure JT gravity is not dual to a single quantum system with a discrete spectrum, can it be dual to an ensemble average of quantum systems? The results of SSS give a positive answer to this question. This resolves the tension (1) since the continuous spectrum can arise after averaging, and also resolves tension (2) since the non-factorization is due to the statistics of the ensemble.

We now describe the holographic dual of JT gravity. Take the black hole Hilbert space $\mathcal{H}_{\rm BH}$ to be of dimension $L$, with $L$ a large integer which will be later related to $e^{S_0}$. A quantum theory is described by an $L\times L$ Hermitian Hamiltonian matrix $H$, acting on $\mathcal{H}_{\rm BH}$. The JT gravitational path integral is then equal to an ensemble average over theories, in this case an integral over all $L\times L \sim e^{S_0} \times e^{S_0}$ Hermitian matrices:
\begin{equation}\label{eqn:SSSDUA}
    Z_{\rm grav} (\beta_1,\ldots,\beta_n) = \int dH\, P(H)~{\rm Tr}\left( e^{-\beta_1 H}\right) \ldots {\rm Tr}\left( e^{-\beta_n H} \right),
\end{equation}
where $P(H) \propto e^{-L {\rm Tr}V(H)}$ is the probability distribution over theory space (in this case over the Hamiltonian matrix itself), and it is normalized such that $\int dH P(H) =1$. The quantity $V(H)$ is the potential, which is taken to be a sum (possibly infinite) of positive powers of $H$.
Hence the claim is that \textbf{JT gravity is a matrix integral} \cite{Saad:2019lba}!

Matrix models and techniques to solve them have a long history that we will not do justice here. We refer the reader to \cite{Mehta}, and to \cite{Eynard:2015aea} for a pedagogical review of more recent techniques. Shorter accounts can also be found in reviews on Liouville gravity and non-critical string models \cite{Ginsparg:1993is,DiFrancesco:1993cyw,Nakayama:2004vk}, a topic that also has a deep link with matrix models, which turns out not to be a coincidence as we will point out later on in Sect.~\ref{s:Liouvillegravity}.

Let us continue with our main story. On the left hand side of Eq.~\eqref{eqn:SSSDUA}, the JT gravity path integral can be expanded as a sum over topologies, so the first question is: what is the meaning of this expansion on the matrix model side? 

Matrix integrals have natural insertions given by the resolvent $R(E)$, the ``partition sum'' $Z(\beta)$ and the spectral density $\rho(E)$:
\begin{equation}
R(E) \equiv {\rm Tr}\,\frac{1}{E-H}, \qquad Z(\beta) \equiv {\rm Tr}\left( e^{-\beta H}\right), \qquad \rho(E) \equiv \sum_{m=1}^L \delta(E-E_m),
\end{equation}
where we denoted the eigenvalues of $H$ as $E_m$. There are simple relations between these different quantities as:
\begin{gather}
\label{rel1}
\int_0^\infty e^{\beta E} Z(\beta) = R(E), \qquad\qquad Z(\beta) = \int dE \rho(E) e^{-\beta E}, \\
\label{rel2}
R(E+i\epsilon) - R(E-i\epsilon) = -2\pi i \rho(E),
\end{gather}
so it suffices to study one of them. We then define correlation functions on the matrix integral side with $n$ of these insertions:
\begin{eqnarray}
R_{\rm MM}(E_1,\ldots, E_n) &=& \left\langle R(E_1)\ldots R(E_n) \right\rangle,\\
Z_{\rm MM}(\beta_1,\ldots, \beta_n) &=& \left\langle Z(\beta_1) \ldots Z(\beta_n)\right\rangle,\\
\rho_{\rm MM}(E_1,\ldots, E_n) &=& \left\langle\rho(E_1) \ldots \rho(E_n)\right\rangle,
\end{eqnarray}
where $\langle X \rangle$ denotes the ensemble average $\langle X \rangle = \int dH P(H) X $. Correlation functions of resolvents in the matrix integral are known to have a $1/L$ expansion in the large $L$ limit, of the form 
\begin{equation}
    R_{\rm MM}(E_1,\ldots, E_n)_{\rm conn.} \simeq \sum_{g=0}^\infty \frac{R_{g,n}(E_1,\ldots, E_n)}{L^{2g+n-2}},
\end{equation}
for resolvents and similarly for partition functions. The left hand side of this equation denotes the connected piece only of the $n$-resolvent correlator. The series on the right hand side is only asymptotic and there are non-perturbative corrections, but we can take it as a definition of the coefficients $R_{g,n}$. This expansion arises from a perturbation theory in 't Hooft double line diagrams \cite{tHooft:1973alw,Brezin:1977sv}. In this context, $n$ is the number of insertions while $g$ is the genus of the surface needed to embed the particular double-line diagram into. This means the large $L$ expansion of the matrix model should be identified with the large $e^{S_0}$ expansion of JT gravity, and the coefficients should match term by term. 

To make the identification \eqref{eqn:SSSDUA} precise, hence requires a precise relation between $L$ and $S_0$, and the probability distribution $P(H)$ through the matrix potential $V(H)$. We address these two points in the next subsections and complete the derivation of Eq.~\eqref{eqn:SSSDUA}.

\subsubsection{Double-scaling limit of matrix integrals}

Consider the spectral density of the matrix model: $\rho(E) = \sum_{i=1}^L \delta(E-E_i)$. Even if each realization for $H$ has a discrete spectrum for finite $L$, after averaging over $H$ and taking the large $L$ limit, the resulting density of states will be continuous. 

In the simplest models, the leading density of states at large $L$ has support on a single interval $E \in [a_-,a_+]$ with $a_-$ and $a_+>a_-$ real numbers.\footnote{More general models can have several disjoint cuts.} In these one-cut cases, there is a simple relation between the function $V(H)$ and $\rho_0(E)$, see \cite{Eynard:2015aea} for details. The prototypical example where this applies is a Gaussian model with leading density of states
\begin{equation}
\label{eq:Gauss}
    V(H)= \frac{8}{a^2} \left(H-\frac{a}{2}\right)^2 ~~\rightarrow~~\ \rho_0(E) \equiv \langle \rho(E)\rangle_{L\to\infty} =  \frac{8 L}{\pi a^{3/2}}\sqrt{\frac{E(a-E)}{a}},
\end{equation}
where the brackets mean we average over $H$ with the measure determined by $V(H)$. This is the famous \textbf{Wigner semicircle} distribution. We chose the potential such that $a_-=0$ and $a_+=a>0$ and the overall normalization of the density of states is such that its integral gives $L$. \\

To apply this to the JT gravity case, we should find a potential $V(H)$ such that its density of states to leading order at large $L$ matches with JT gravity on the disk. This is not possible, since JT gravity has an infinite support over $E\in [0,+\infty)$. The resolution of this puzzle is that JT gravity is dual not to an ordinary matrix integral, but to a \textbf{double-scaled matrix integral}.

Let us see first how the double scaling limit works in the case of the Gaussian matrix model \eqref{eq:Gauss}. We wish to obtain a distribution that has support on the entire positive real axis. So we should take the limit of $a\to \infty$. This runs into an issue: since we have a finite number of eigenvalues $L$ to be distributed on the whole real line, the spectral density $\rho_0(E)$ goes to zero. To resolve this, we need to send the number of eigenvalues to infinity $L\to\infty$ at the same time, such that we keep the following quantity $8L/\pi a^{3/2} \equiv e^{S_0}$ fixed. This gives the double-scaled density of states $\rho_0(E) \approx e^{S_0} \sqrt{E}$, which is smooth and has support on the positive energy axis, although it is not normalized: the number of eigenvalues is infinite. Much like the original Gaussian matrix model \eqref{eq:Gauss}, the resulting double-scaled matrix model is exactly solvable and is called the \textbf{Airy model}. Schematically, the operation we realize on the leading density of states is:
 \begin{eqnarray}
 \nonumber
\includegraphics[scale=0.27]{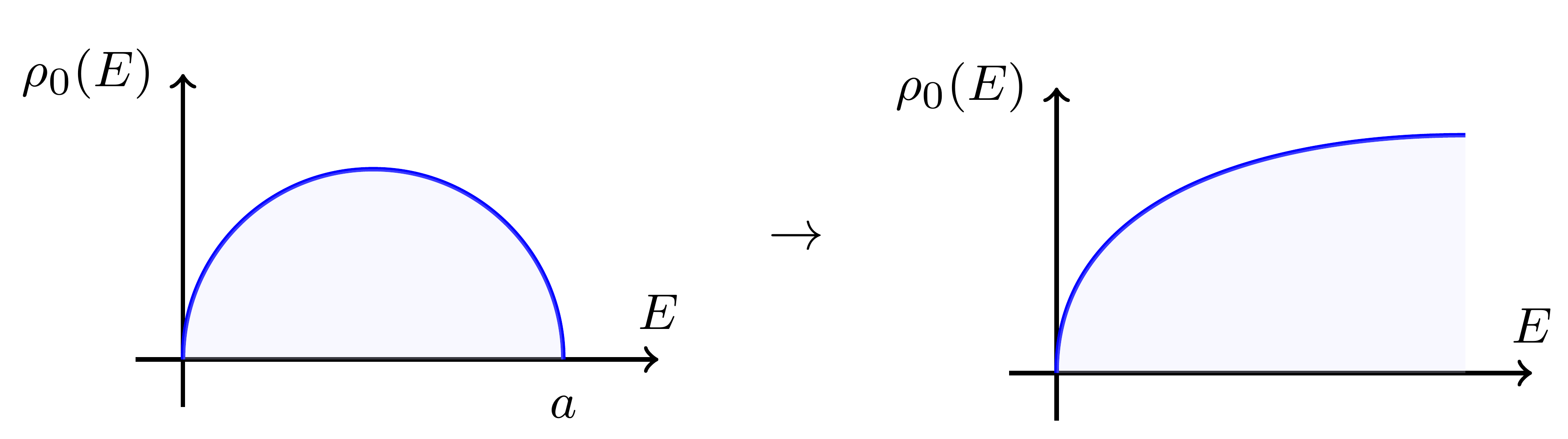}
\end{eqnarray}
 This operation can be translated as due to a particular fine-tuning of the matrix potential.
 
 This double-scaling limit affects the topological expansion of the matrix integral in a simple way. Instead of weighing different topologies by $L$, now the expansion is in the new parameter $e^{S_0}$ which is large but fixed as $L,a\to\infty$. In terms of resolvents, for example, we can write:
 \begin{equation}
    R_{\rm MM}(E_1,\ldots, E_n)_{\rm conn.} \simeq \sum_{g=0}^\infty \frac{R_{g,n}(E_1,\ldots, E_n)}{(e^{S_0})^{2g+n-2}}.\label{eq:DSLtopexp}
\end{equation}
It requires a separate calculation to justify that once $L$ is scaled in the right way as $a\to\infty$, with $E$ fixed, all the terms in the topological expansion have a well-defined limit  \eqref{eq:DSLtopexp}. The quantity $e^{S_0}$ controls the scale of the eigenvalue \emph{density}, and $e^{-S_0}$ hence controls the average eigenvalue spacing. \\

We now have the tools to describe the holographic dual of JT gravity. Since standard matrix integrals in the large $L$ limit only have finite support, we begin by ``regularizing'' the disk density of states, and constructing a matrix potential $V(H)$, such that to leading order in $1/L$ e.g.
 \begin{equation}
     \rho_0(E) \equiv \langle \rho(E) \rangle_{L\to\infty} = e^{S_0} \frac{C}{2\pi^2}\sinh\left( 2\pi \sqrt{2C E\frac{a-E}{a}}\right),
 \end{equation}
 where $e^{S_0}$ is a function of $L$ and $a$ that can be determined by fixing normalization. We do not need to know this relation explicitly, since we are next going to take $a\to\infty$ and $L\to \infty$, such that $e^{S_0}$ is fixed. Note that there are many ways to ``regularize'' the JT disk density of states.
Schematically, this gives us: 
 \begin{eqnarray}
 \nonumber
\includegraphics[scale=0.27]{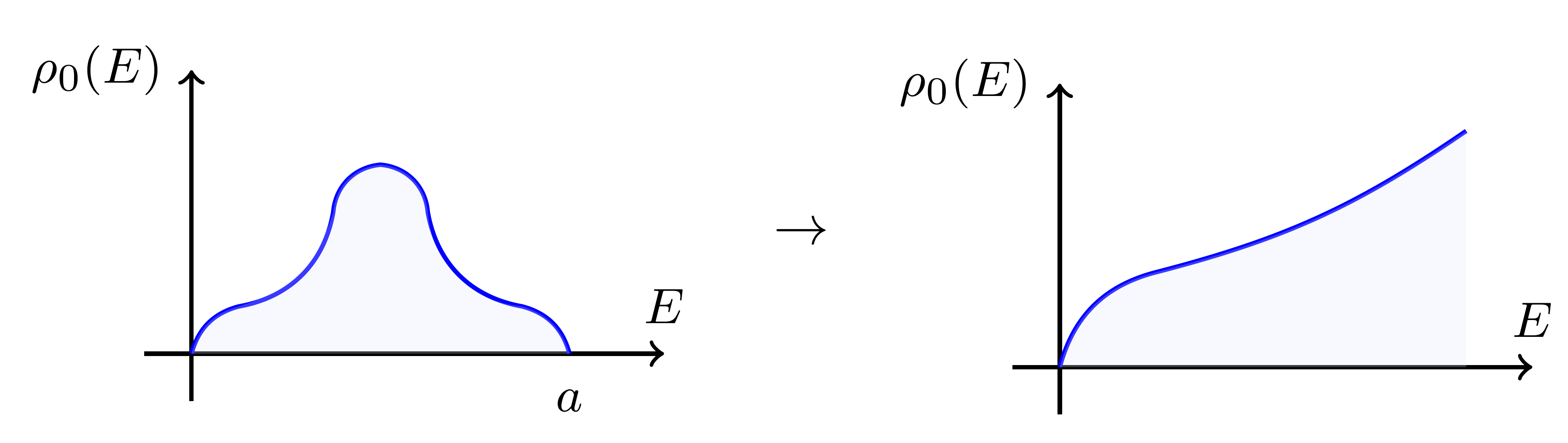}
\end{eqnarray}
The topological expansion for $Z_{\rm MM}(\beta_1,\ldots,\beta_n)$ is now in terms of $e^{S_0}$ instead of $L$. It is this double-scaled parameter $S_0$ that we identify with the prefactor of the topological term in the JT gravity action \eqref{eqn:JTactions4}, that weighs different topologies in gravity by their Euler characteristic $\chi$. 

The non-trivial statement we will describe next is that, once the matrix potential is tuned to give the disk JT gravity density of states in the large $e^{S_0}$ limit, all other subleading terms in the topological expansion will automatically match between the matrix model and the gravity calculation we described in Sect.~\ref{s:sss}.

\subsubsection{Derivation of the duality}
A very direct way to derive the equivalence between JT gravity and a double-scaled matrix integral is through the so-called loop equations or topological recursion relations of the matrix model, which provide a recursive relation that determines all $R_{g,n}$ or equivalently $Z_{g,n}$, in the $1/L$ or $e^{-S_0}$ expansion. The loop equations themselves were studied a long time ago \cite{Migdal:1975zg}, but they were put in its most useful form by \cite{Eynard:2004mh}. The strategy will be to relate this recursion relation to the geometrical one used to compute Weil--Petersson volumes \eqref{eq:MirzakhaniRecursion}.

To state the topological recursion relation in the matrix model, it is convenient to introduce the following quantities related to the resolvent correlators:
\begin{eqnarray}
W_{g,n}(z_1,\ldots,z_n) \equiv (-2)^n z_1 \ldots z_n R_{g,n}(-z_1^2,\ldots,-z_n^2)_{\rm conn.},E=-z^2.
\end{eqnarray}
Just as in the gravitational genus expansion, the matrix model topological expansion requires previous knowledge of two special cases $W_{0,1}$ and $W_{0,2}$. These are related to the leading $\langle \rho(E) \rangle_{L\to\infty}$, which is determined by the potential, and to the density-density correlator $\left\langle \rho(E)\rho(E')\right\rangle_{L \to \infty}$, which is universal (independent of the matrix potential). They are explicitly given by
\begin{equation}
\label{eq:seeds}
W_{0,1}(z) 
=2zy(z) ,~~~~~W_{0,2}(z_1,z_2)=\frac{1}{(z_1-z_2)^2}.
\end{equation}
The function $y(z)=-i\pi \rho_0(-z^2)$ determines the \textbf{spectral curve}, given by the locus $(-E(z)=z^2,y(z))\subset \mathbb{C}^2$. Once these two quantities are determined, all other corrections to the resolvent correlators are determined by:
\begin{eqnarray}
W_{g,n}(z_1,Z)&=& \underset{z\to 0}{{\rm Res}} \Big\{ \frac{1}{z_1^2-z^2}\frac{1}{4y(z)} \Big[ W_{g-1,n+1}(z,-z,Z)  \nonumber\\
&&~+\sum_{\rm stable} W_{h_1,1+\vert Z_1\vert}(z,Z_1)W_{h_2,1+\vert Z_2\vert }(-z,Z_2)\Big]\Big\},\label{eq:RRW}
\end{eqnarray}
where $Z=\{ z_2,\ldots,z_n \}$. The sum in the second line is over sets $Z_{1,2}$ such that $Z_1 \cup Z_2 = Z$ and $g=h_1 + h_2$, and the sum is again over stable curves. This form of the loop equations was derived in \cite{Eynard:2004mh}, and is written for the case of a spectrum with support on the positive real line $[0,+\infty)$. For more general spectral support $[a_-,a_+]$, the expression is the same, but with the residue taken at both endpoints. 

We now compare this with JT gravity, for which the spectral curve is chosen as
\begin{equation}
y(z) = \frac{C \sin(2\pi \sqrt{2C} z)}{2\pi}.
\end{equation}
First of all, $W_{0,1}$ and $W_{0,2}$ are related to \eqref{eq:Z01} and \eqref{eq:Z02} respectively by a Laplace transform that turns the resolvents into thermal partition functions. More generally, $W_{g,n}$ is related to $Z_{g,n}$ by the $n$-fold Laplace transform:
\begin{equation}
    W_{g,n}(z_1,\ldots,z_n) =  \left(\prod_{i=1}^n 2 z_i \int_0^\infty  d\beta_i e^{-\beta_i z_i^2}\right)Z_{g,n}(\beta_1,\ldots,\beta_n).\label{eq:WintZ}
\end{equation}
Expressing $Z_{g,n\neq (0,1),(0,2)}$ in terms of the Weil--Petersson volumes in JT gravity \eqref{eq:SSSglue}, one finds:
\begin{equation}\label{eq:defW}
W_{g,n}(z_1,\ldots, z_n) =(2C)^{n/2}\left(\prod_{i=1}^n \int_0^\infty  b_idb_i e^{-\sqrt{2C}b_i z_i}\right) V_{g,n}(b_1,\ldots,b_n).
\end{equation}
Since the volumes are polynomials in $b_i^2$, these integrals are convergent for $z\neq 0$. Now comes the non-trivial part of the proof. \cite{Eynard:2007fi} started from Mirzakhani's recursion relation \eqref{eq:MirzakhaniRecursion} and applied to it the Laplace transform \eqref{eq:defW} to derive a recursion relation for $W_{g,n}$. After a lengthy calculation, they obtained precisely the recursion \eqref{eq:RRW} with  the spectral curve $y(z) = \frac{C \sin(2\pi \sqrt{2C} z)}{2\pi}$ used in JT gravity. This result completes the derivation that JT gravity is equivalent, order by order in the topological expansion, to a double-scaled matrix integral. \\

\textbf{Example: $Z_{1,1}(\beta)$ revisited:} \\
We repeat now the example worked out in Sect.~\ref{s:genWP} but from the matrix model side. To do this we use the matrix model loop equation in the form \eqref{eq:RRW} to obtain the genus one contribution to the resolvent
\begin{eqnarray}
W_{1,1}(z_1) = \underset{z\to 0}{{\rm Res}}\left[ \frac{1}{z_1^2-z^2} \frac{1}{4y(z)} W_{0,2}(z,-z)\right] = \frac{4 \pi ^2 C z_1^2+3}{48 \sqrt{2} C^{3/2} z_1^4}.
\end{eqnarray}
The matrix model prediction for the genus one correction to $Z(\beta)$ is then given by solving for $Z_{1,1}(\beta)$ in the relation \eqref{eq:WintZ}. The answer is
\begin{equation}
Z_{1,1}(\beta)  = \frac{\beta ^{3/2}}{24 \sqrt{2 \pi } C^{3/2}}+\frac{\pi ^{3/2} \sqrt{\beta }}{12 \sqrt{2} \sqrt{C}}.
\end{equation}
This is precisely the same as the gravitational answer found in subsection \ref{s:genWP}.
\begin{center}
---o---
\end{center}

\subsection{Non-perturbative effects in topological expansion}
\label{s:nonpt}
By the above matching of recursion relations, the SSS duality is proven order-by-order in the topological expansion in powers of $e^{-S_0}$. Since in gravity $S_0 \sim 1/G_N$, these are non-perturbative effects in $G_N$. From the matrix integral side, since the size of the matrix is $L\sim e^{S_0}$, these are perturbative effects in the (double-scaled) 't Hooft expansion. 

We know this cannot be the whole story: from either perspective, the genus expansion is divergent since each term grows for large genus as $(2g)!$. This indicates the need for doubly non-perturbative effects of order $e^{-\# e^{S_0}}$, as originally suggested by \cite{Shenker:1990uf} in the context of the non-critical string. In that context, this issue has a beautiful resolution: these doubly non-perturbative effects can be accounted for by adding D-branes to string theory. 

In JT gravity, the interpretation of these doubly non-perturbative corrections is more mysterious. When seen as a limit of the non-critical string (as we review in Sect.~\ref{s:Liouvillegravity}), the spacetime of JT gravity is identified with the string worldsheet. Therefore, D-branes create holes in the JT gravity spacetime. However, as usual in D-brane physics \cite{Polchinski:1994fq}, we need to sum over configurations where the worldsheet (= spacetime here) ends on a single D-brane an arbitrary number of times, so a universe with an arbitrary number of boundaries. This obscures an intuitive bulk picture, but we can still write down the equations. A D-brane insertion in the matrix integral (in the non-critical string of FZZT type) is a determinant insertion:
\begin{eqnarray}
\text{det}(E-H) = e^{\text{Tr Log } (E-H) }.
\end{eqnarray}
Here we wrote it as the exponential of a new object.
The insertion $\text{Tr Log } (E-H) = \int \frac{d\beta}{\beta} e^{\beta E} Z(\beta)$ in the matrix integral can be viewed as an unmarked fixed energy boundary.\footnote{It is unmarked due to the dividing by $\beta$ as familiar in worldsheet string theory. The resolvent $R(E)$ is marked since it does not have this $1/\beta$ present as in \eqref{rel1}.} The exponentiation of this insertion then shows that we consider an arbitrary number of these unmarked boundaries, accounting for their indistinguishability by $1/n!$ in the Taylor expansion. Computations can be done for such insertions, and we refer to \cite{Saad:2019lba}.

In this subsection, we will review how the doubly non-perturbative effects work from the matrix integral perspective, and leave their interpretation in bulk physics open (see \cite{Saad:2019lba} for some discussion). For concreteness, we will analyze these corrections in the context of computing the black hole density of states and spectral form factor.

\subsubsection{Density of states}

Consider the exact computation of $\left\langle \rho(E) \right\rangle$ in the random matrix framework for large $e^{S_0}$. 
As previously explained, perturbative corrections are suppressed by powers of $e^{-S_0}$ and non-perturbative corrections are naively even further suppressed as $e^{-\# e^{S_0}}$. However, it is known that counter-intuitively the non-perturbative corrections can actually be more important than the perturbative ones, essentially due to the factor $\#$ in the exponential being imaginary. 

\begin{figure}[t!]
\begin{center}
\includegraphics[scale=0.19]{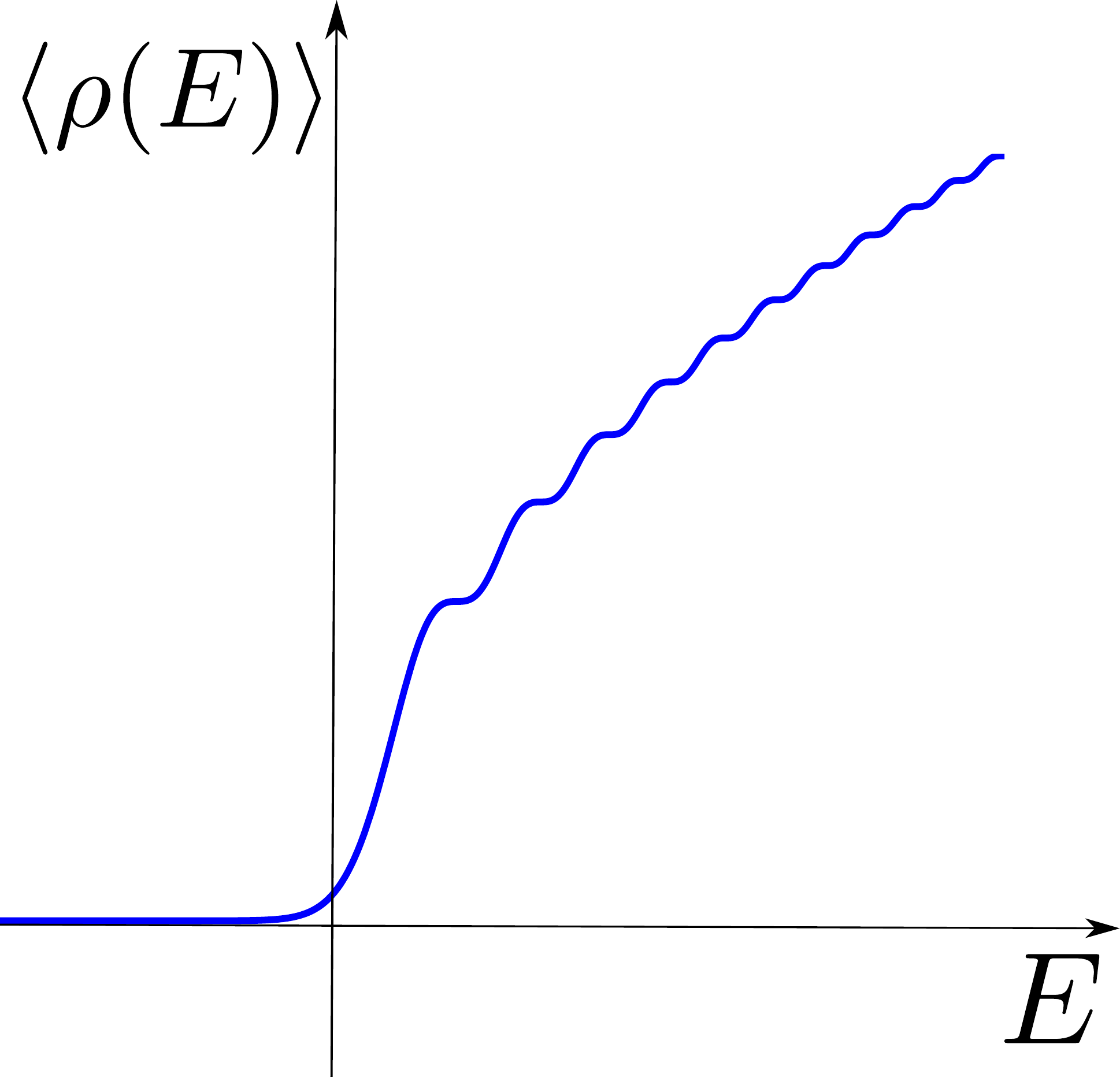} 
\end{center}
\caption{Sketch of full spectral density of JT gravity including non-perturbative effects.}
\label{FigNPRHO}
\end{figure}

In the large $S_0$ regime, the full spectral density of the JT matrix model (for $E \lesssim 1/C$) is sketched in Fig.~\ref{FigNPRHO}. There are several ways to find it. One approach is that of \cite{Johnson:2019eik}, which uses numerical techniques. This requires some random matrix machinery from the '90s that we will not explain here. Another approach, only valid at large $S_0$, is to approximate the JT answer using only contributions from topological disks and cylinders.

This ``disk and cylinder'' approximation was developed first for determinant insertions in \cite{Maldacena:2004sn}, but can be translated to density insertions.\footnote{In \cite{Saad:2019lba}, the authors went from single determinant (or brane) insertions to resolvents and then evaluated their discontinuity \eqref{rel2} to get the density in the large $S_0$ regime. In \cite{Blommaert:2019wfy}, density operator insertions were directly determined from brane pair insertions. The results are the same.}
This approach does not work at small $E$, but one can match it there onto the exactly solvable Airy model mentioned earlier. \\

Let us discuss the qualitative features of the spectral density $\left\langle \rho(E) \right\rangle$.

\paragraph{Corrections to forbidden region $E<0$}
We begin by analyzing the classically forbidden region $E<0$. It is easy to show that any perturbative corrections will never give a non-zero answer in this region: $\rho_{g,1}(E<0)=0$ for all $g$. However, there is a non-perturbative correction to the density of states, computable using the ``disk and cylinder'' method, in the classically forbidden region $E<0$:
\begin{equation}\label{eq:ZZeffects}
\langle \rho(E<0)\rangle =
\frac{e^{-2e^{S_0} \int_0^{\vert E\vert } dx~ y(\sqrt{x})}}{8\pi \vert E\vert }  ,
\end{equation}
which is a very small exponential tail going all the way to $E\to -\infty$.\footnote{For $E\approx 0$, this expression is no longer useable and one has to resort to other means, as mentioned above.} The function in the exponential $V_{\rm eff}(E)\equiv 2e^{S_0} \int_0^{\vert E\vert } dx~ y(\sqrt{x})$ is the effective potential felt by an eigenvalue placed at that energy $E<0$. The one-eigenvalue instanton corresponds to the locations where this effective potential is stationary \cite{Shenker:1990uf}.\footnote{See \cite{Alexandrov:2003nn} for a string theory perspective as a ZZ brane contribution.}

If the effective potential would be positive in the forbidden region, this would have a simple interpretation, giving a doubly exponentially suppressed in $e^{-e^{S_0}}$ probability to find a black hole state with an energy below the classically allowed region. Nevertheless this is not so simple for JT gravity, since the effective potential is $V_{\rm eff}(E) = \frac{e^{S_0}}{4\pi^3}[\sin (2\pi \sqrt{2 C \vert E \vert}) - 2 \pi \sqrt{2C \vert E\vert} \cos ( 2\pi \sqrt{2 C \vert E \vert})]$ for $E<0$. This means that the matrix model dual to JT gravity is non-perturbatively ill-defined. An option proposed in \cite{Saad:2019lba} is to choose a steepest-descent contour for the matrix eigenvalues that runs along the real axis only for $E>-1/8C$ and at $E=-1/8C$ goes to the complex plane in a convergent direction. A different option is to change its random matrix completion to remove the forbidden region even non-perturbatively. This approach has been developed in a series of papers, nicely reviewed in \cite{Johnson:2022wsr}.

\paragraph{Corrections to allowed region $E>0$}
The non-perturbative corrections in the allowed region $E>0$, computed using the ``disk and cylinder'' method, give instead
\begin{equation}\label{eq:FZZTeffects}
\langle \rho(E>0)\rangle =
e^{S_0}\tilde{\rho}_0 (E) - \frac{1}{4\pi E} \cos \Big( 2\pi e^{S_0} \int_0^E dE' \tilde{\rho}_0(E')\Big),
\end{equation}
where the first term is the disk answer with the $e^{S_0}$ written explicitly.\footnote{From the non-critical string perspective, \eqref{eq:FZZTeffects} can be thought of as coming from FZZT-brane corrections to the the partition function.} This correction is not small, but is instead extremely rapidly oscillating $e^{i \# e^{S_0}}$ (For example, it is larger than the leading one-wormhole correction to the disk $Z_{1,1}$ which is of order $e^{-S_0}$).These rapid oscillations with frequency $\sim e^{S_0}$, of the order of the average separation between matrix eigenvalues, are an indication of the underlying discreteness of the system.

The rapidly oscillating corrections are subleading after performing an average over an energy window, for example when computing the thermal partition function. In this sense, they can be ignored compared with the perturbative corrections in the topological expansion. Nevertheless, as emphasized in \cite{Saad:2019lba}, these oscillations can become very important in some observables, such as the very late time behavior of the spectral form factor, which we analyze next.

\subsubsection{Late time decay of spectral form factor}
\label{s:sff}
As our second example, let us look at the late-time behavior of the spectral form factor in JT gravity. Within the random matrix ensemble, the spectral form factor is the average  $\left\langle Z(\beta+it)Z(\beta-it)\right\rangle$.
First we double Laplace transform to write
\begin{equation}
\label{doubleL}
    \left\langle Z(\beta+it)Z(\beta-it)\right\rangle = \int dE dE'  \left\langle \rho(E)\rho(E')\right\rangle e^{-\beta(E+E')}e^{-it(E-E')}.
\end{equation}
The pair density correlator $\left\langle \rho(E)\rho(E')\right\rangle$ in random matrix theory is of monumental importance. At large $e^{S_0}$, for small energy separation $\vert E- E'\vert \ll 1$, and sufficiently far away from spectral edges such that the spectral density $\rho_0(E)$ is not wildly varying, it is given by the expression:
\begin{equation}
\label{pair1}
\left\langle \rho(E)\rho(E')\right\rangle \approx
\rho_0(E)\rho_0(E') - \frac{\sin^2(\pi \rho_0(E)(E-E'))}{\pi^2(E-E')^2} + \rho_0(E) \delta(E-E').
\end{equation}
This equation holds for \emph{any} random matrix potential $V(H)$ in the unitary universality class, which is of relevance to oriented bosonic JT gravity.\footnote{See \cite{Stanford:2019vob} for a thorough treatment of the relation between the other ensembles and gravity models.} This is a direct manifestation of \textbf{random matrix universality}, governing universal level statistics at small energy separations.

The RHS contains the famous sine kernel $\frac{\sin^2(\pi \rho_0(E)(E-E'))}{\pi^2(E-E')^2}$, implementing level repulsion in a chaotic system. Indeed, as $E\to E'$, the first two terms in \eqref{pair1} cancel out as $\sim (E-E')^2$, indicating energy levels do not wish to be on top of each other.

It is useful to first mention the scaling in $e^{S_0}$ of the different terms, hidden within $\rho_0(E) \sim e^{S_0}$. The first, factorized, term scales as $e^{2S_0}$ and seems to vastly dominate the other terms in the regime of interest $e^{S_0} \gg 1$: the sine kernel oscillates rapidly but does not scale in amplitude with $e^{S_0}$, and the contact term scales as $e^{S_0}$. However, this is no longer true if the energy separation is extremely tiny: $E-E' \sim e^{-S_0}$. In that case, the sine kernel can compete against the factorized contribution. Since small energy separation means late times, this is precisely what happens at very late times as we will now explain.

It will be convenient to explicitly extract the average value of the sine kernel by rewriting $\left\langle \rho(E)\rho(E')\right\rangle$ \eqref{pair1} as:
\begin{equation}
\rho_0(E) \rho_0(E') - \frac{1}{2\pi^2(E-E')^2} + \frac{\cos(2\pi \rho_0(E)(E-E'))}{2\pi^2 (E-E')^2} + \rho_0(E) \delta(E-E').
\label{pairdensity}
\end{equation}
We go through all four terms of Eq.~\eqref{pairdensity} and look at their late-time behavior and gravitational interpretation.

\begin{itemize}
    \item The first term $\rho_0(E) \rho_0(E')$ is just the factorized contribution we studied before, and arises gravitationally from two disconnected disks $\sim e^{2S_0}$. It leads to the expression:
\begin{equation}
Z(\beta+it) = \frac{e^{S_0}}{4\pi^2}\left(\frac{2\pi C}{\beta+it}\right)^{3/2}e^{\frac{2\pi^2C}{\beta+it}} ~~\to~~  \langle{\vert Z(\beta+it) \vert}^2\rangle \to \frac{e^{2S_0}}{2\pi} \frac{C^3}{t^3}.
\end{equation}
The spectral form factor decays with a power-law expression as studied extensively for the more complicated boundary two-point functions in Sect.~\ref{sec:JTquantum}. This is the so-called \textbf{slope}. \\
\item
Inserting the second term $- \frac{1}{2\pi^2(E-E')^2}$ of Eq.~\eqref{pairdensity} into \eqref{doubleL}, and using the Fourier transform identity 
\begin{equation}
-\int_{-\infty}^{+\infty} dx \frac{1}{2\pi^2 x^2} e^{i t x }= \frac{1}{2\pi} \vert t \vert,
\end{equation}
this leads to a linear growth in time $\langle{\vert Z(\beta+it) \vert}^2\rangle \supset \frac{1}{4\pi \beta}t$. The term $- \frac{1}{2\pi^2(E-E')^2}$ has no $e^{S_0}$ scaling, and should gravitationally then be sought for in the connected piece of $Z_{0,2}$, where $\chi=0$. We already computed this in \eqref{eq:Z02}. Setting $\beta_1=\beta+it$ and $\beta_2=\beta-it$, we get at late times:
\begin{equation}
Z_{0,2}(\beta_1,\beta_2) = \frac{1}{2\pi} \frac{\sqrt{\beta_1 \beta_2}}{\beta_1 + \beta_2} = \frac{1}{2\pi} \frac{\sqrt{\beta^2+t^2}}{2\beta} ~~\to ~~ \frac{1}{4\pi \beta} t,
\end{equation}
indeed gravitationally accounting for the linear growth in late time $t$ with precisely the correct prefactor to match the random matrix result. This is the so-called \textbf{ramp}. Hence we can summarize that the ramp is gravitationally explained by the double trumpet amplitude.\footnote{An alternative proposal was made in \cite{Germani:2022rac}, but we disagree with their approach.} In a microcanonical ensemble version of the spectral form factor, there is a solution of this form, the double cone we mentioned above. This contribution has a linear growth in time regardless of spacetime dimension, hinting at a random matrix universality of the black hole spectrum in all dimensions.

One can wonder about higher genus corrections $Z_{g,2}$ at late times. One can readily show starting with \eqref{eq:seeds} and \eqref{eq:RRW} that only $W_{0,2}$ is diverging at small energy \emph{differences}, and for which hence the enhancement can compensate the suppression by $e^{-2g S_0 }$ at late times. \\


The decreasing slope $(\sim \frac{e^{2S_0}}{t^3})$ and rising ramp  ($\sim t$) intersect at a late time $t_d = (2\beta C^3)^{1/4} e^{S_0/2}$. This is called the \textbf{dip time} $t_d$ and its precise location depends on details of the dynamics of the model. Note that its scaling with the entropy $t_d \sim e^{S_0/2}$ is more robust since it only depends on the topologies responsible for the slope and ramp. \\

\item 
The third term of Eq.~\eqref{pairdensity} is wildly oscillating: $
    \frac{\cos(2\pi \rho_0(E)(E-E'))}{2\pi^2 (E-E')^2}$ due to $\rho_0(E) \sim e^{S_0} \gg 1$, 
and non-perturbative in $e^{-S_0}$ (due to exp$(i\# e^{S_0})$). As mentioned before, this is actually doubly non-perturbative in $G_N \sim 1/S_0$. Its Fourier transform yields a linear downward piece, starting at the time $t\sim C e^{S_0}$, and perfectly canceling the slope linear growth. This location is smoothed out due to the remaining Laplace transform in \eqref{doubleL}. This causes the spectral form factor at times $\sim C e^{S_0}$ to flatten out and reach the \textbf{plateau}. \\

\item
Finally, the height of the plateau is determined by the contact term $\rho_0(E)\delta(E-E')$ in \eqref{pairdensity}, which when inserted into \eqref{doubleL} leads to the final non-zero value $Z(2\beta)$ as required for the late-time average as discussed in Sect.~\ref{s:infoloss}.
\end{itemize}

\begin{figure}[t!]
\begin{center}
\includegraphics[width=0.75\textwidth]{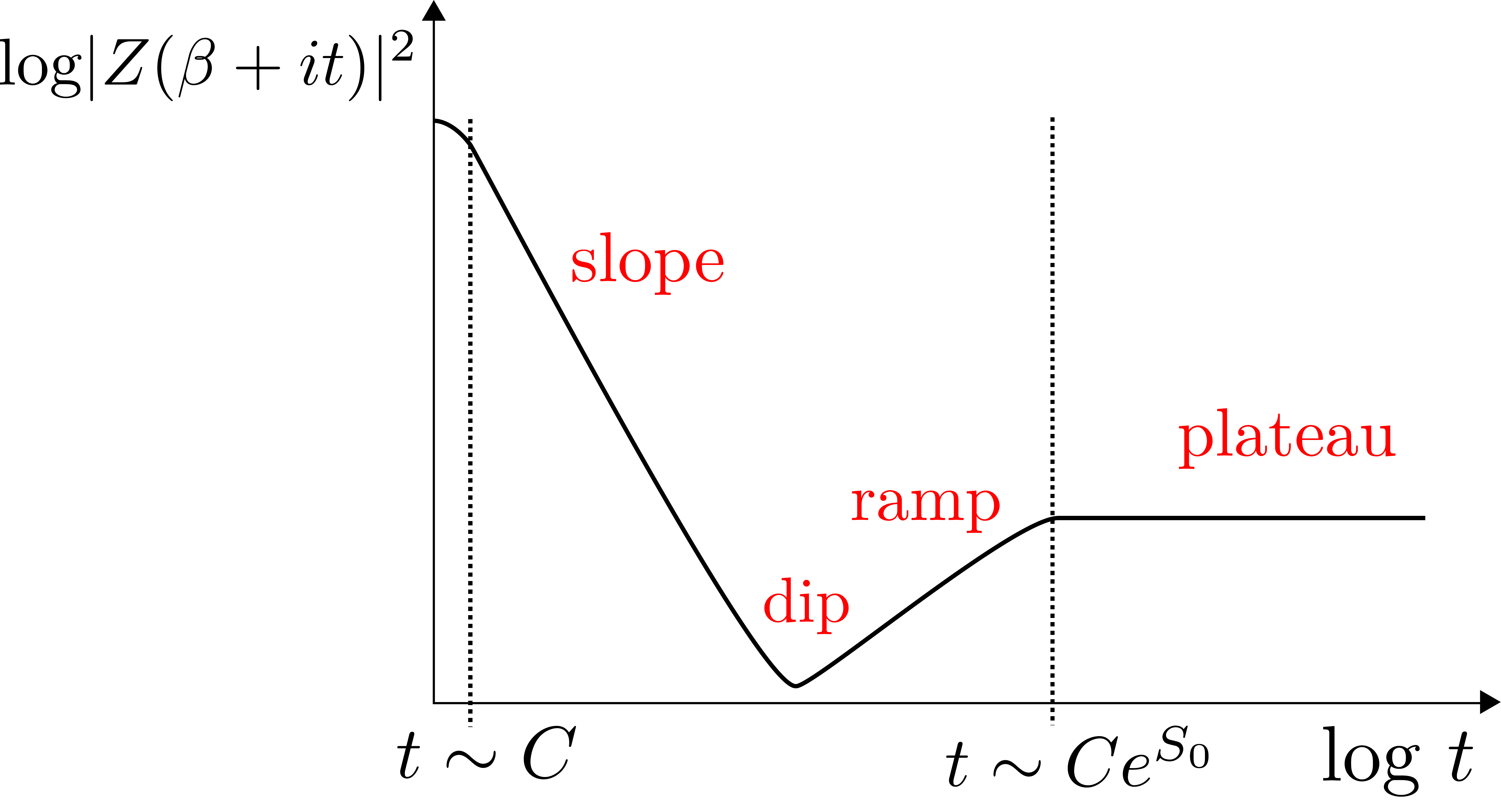}
\end{center}
 \caption{Averaged behavior of the spectral form factor from random matrix theory.}
\label{Fig:sloperamp}
\end{figure}

The full late-time behavior of the spectral form factor is conveniently summarized in a log-log plot, and is qualitatively of the shape shown in Fig. \ref{Fig:sloperamp}. 
These random matrix results resolve the tension between the perpetual late-time decay of the semi-classical correlators and unitarity, as mentioned before in Sect.~\ref{s:infoloss}.

We understand the origin of the slope and ramp from gravity: it is an exchange in dominance from the disconnected pair of disks to the leading wormhole contribution. The gravity origin of the plateau is more mysterious, although there is recent progress indicating it might be reproduced by resumming a subset of the terms that appear in the topological expansion \cite{Blommaert:2022lbh, Saad:2022kfe}.

This is the shape predicted for a discrete system for which some sort of late-time averaging has been done to remove the erratic oscillations (such as the red curve drawn in the figure in Sect.~\ref{s:infoloss}). It is still an open question at the time of writing to understand the gravitational origin of the erratic oscillations, and fully make contact with an unaveraged discrete system. We will present some directions on this in Sect.~\ref{s:discrete} later on.

\subsection{Generalization} 
We finish this section with a brief outline of some key generalizations on the duality between two-dimensional gravitational models and matrix integrals.

\subsubsection{Other ensembles}
The SSS duality has on one side an integral over Hermitian matrices, interpreted as a disordered Hamiltonian of a holographic theory, and on the other JT gravity on orientable manifolds with arbitrary topology. This involves the unitary ensemble, for which the canonical example is the Gaussian Unitary Ensemble (GUE). There are a total of ten types of random matrix ensembles characterized by \cite{Altland:1997zz}. The simplest generalizations are the GOE and GSE ensembles which are dual to JT gravity on unorientable manifolds. For all the ten ensembles, it is possible to consider gravitational duals as modifications of JT gravity, either by adding topological theories in the bulk, or by working with $\mathcal{N}=1$ supersymmetric JT gravity. The details of all dualities can be found in \cite{Stanford:2019vob}. 

As an example consider $\mathcal{N}=1$ supersymmetric JT gravity. Its disk density of states is known \cite{Stanford:2017thb, Mertens:2017mtv}
\begin{eqnarray}
\rho_{\mathcal{N}=1,{\rm disk}}(E) =e^{S_0}  \frac{2C}{\pi} \frac{\cosh\left( 2\pi \sqrt{2 C E}\right)}{\sqrt{ C E}},\label{eq:DOSN1}
\end{eqnarray}
and in the topological expansion one also has to sum over spin structures. Whereas bosonic JT gravity has the universal square root spectral edge $\rho \sim \sqrt{E}$, \eqref{eq:DOSN1} has instead the universal $\rho \sim 1/\sqrt{E}$ small $E$ behavior. The dual is a Hermitian matrix integral over the $L\times L$ supercharge matrix $Q$ such that the Hamiltonian is $H=Q^2 \geq 0$.\footnote{Notice that the Hilbert space does not separate into bosonic and fermionic subsectors: the $(-1)^F$ symmetry is anomalous.} The eigenvalues of $Q$ have support over the whole real axis in the double-scaling limit and have no edge. The $\rho \sim 1/\sqrt{E}$ behavior is just coming from transforming the $Q$ eigenvalue density to the $H$ spectral density.

There is a topological theory in the two-dimensional bulk that weighs odd and even spin structures differently, and we can add this theory to $\mathcal{N}=1$ JT gravity. This has the same disk density of states as \eqref{eq:DOSN1} but a different topological expansion. The Hilbert space of the dual theory separates in two equal sized blocks (fermionic and bosonic, the $(-1)^F$ is a symmetry) and the supercharge acts as $Q={\footnotesize \Big(\begin{matrix}
0 & M^\dagger \\
M & 0
\end{matrix}\Big)}$, with $M$ an arbitrary complex matrix. The ensemble average is done over $M$ and the Hamiltonian is $H=Q^2$. The universal edge of this ensemble is $\rho \sim 1/\sqrt{E}$, consistent with \eqref{eq:DOSN1}. This realizes another ensemble in the Altland-Zirnbauer classification.

It is possible to define theories of pure JT gravity with $\mathcal{N}=2$ or $\mathcal{N}=4$ supersymmetry. These theories of gravity also appear when describing near-extremal black holes in string theory, in either asymptotically flat space or higher-dimensional AdS. The matrix ensemble dual to pure $\mathcal{N}=2$ JT gravity was derived in \cite{Turiaci:2023jfa} and surprisingly only involves the Altland-Zirnbauer ensembles as building blocks.

\subsubsection{2d dilaton gravity with general potential}\label{sec:MTW}
We can consider other theories of pure 2d dilaton gravity on orientable manifolds. At the two-derivative level, we reviewed in Sect.~\ref{sec:DilGravMod} that such theories of gravity are parametrized by a single function, the dilaton potential $U(\Phi)$. On the other side of the duality, matrix integrals that we consider are parametrized as well by a single function: the spectral density (or equivalently the matrix potential $V(H)$, before double scaling). 

It is natural to expect a duality, similar to the JT case, between pure 2d dilaton gravity and a matrix integral by finding a relation between the dilaton potential and the matrix potential. This duality was proven in \cite{Maxfield:2020ale} and independently in \cite{Witten:2020wvy}, and a precise relation was found between these two functions.

The idea is the following. Starting from JT gravity, each time a defect with opening angle $2\pi \theta$ is added to the geometry at point $x$ in the two-dimensional spacetime, it is equivalent to inserting an operator $e^{-2\pi(1-\theta)\Phi(x)}$ \cite{Mertens:2019tcm}. Integrating over the position of the defect is equivalent to inserting the integrated operator $\int d^2x \sqrt{g} e^{-2\pi(1-\theta)\Phi(x)}$. Finally, we can insert an arbitrary number of these defects and sum over their number, introducing a defect `fugacity' $\lambda$. The effect of this operator insertion in the JT path integral is very simple: the summation over the number of defects exponentiates the operator and corresponds to a deformation of the JT action that can be encoded in a shift of the dilaton potential from $U_{\rm JT}(\Phi)=2\Phi$ to 
\begin{equation}
    U(\Phi) = 2 \Phi + W(\Phi), \quad W(\Phi)= \int_0^1 ~d\theta \lambda(\theta )e^{-2\pi(1-\theta)\Phi},
\end{equation}
where we allowed for an arbitrary distribution of deficit angles and weights (For example a discrete set of defect flavors would correspond to $\lambda(\theta) = \sum_i \lambda_i \delta(\theta-\theta_i)$). We can now compute the gravitational path integral for theories with such dilaton potentials by summing over defects in the JT path integral. This requires a generalization of the Weil--Petersson volumes to include the possibility of defects, geodesic boundaries and handles. These volumes are well-understood for the case that all $\theta <1/2$ (sharp defects) and that allowed \cite{Maxfield:2020ale} to prove the equivalence with a random matrix integral. The density of states in this case, even at the disk level, requires a non-trivial resummation over defects and the answer is 
\begin{equation}
    \rho_0(E; W) =\frac{e^{S_0}}{2\pi} \int \frac{dy}{2\pi i} e^{2\pi y} \tanh^{-1} \left( \frac{\sqrt{E-E_0}}{\sqrt{y^2-2 W(y) -E_0}}\right),
\end{equation}
where now the support is $E\in (E_0,+\infty)$ and $E_0$ is the largest solution of the equation $\int dy e^{2\pi y}(y-\sqrt{y^2-E_0-2W(y)})=0$. The integration over the auxiliary variable $y$ is on a contour that runs along the imaginary axis such that all singularities of the integrand are placed to the left of it. 

The expression above for $\rho_0(E;W)$ and the duality of 2d dilaton gravity with matrix models can only be proven when the defect distribution $\lambda(\theta)$ has support for $\theta \in (0,1/2)$. It was proposed in \cite{Turiaci:2020fjj} by comparing this result with a limit of the minimal string that the expression above is valid for all $\theta \in (0,1)$, and justified in \cite{Eberhardt:2023rzz}. This matches the answer originally found in \cite{Maxfield:2020ale,Witten:2020wvy} for sharp defects, but the dependence with the deficit angle  can be highly non-analytic. 

Since the correction to the dilaton potential decays exponentially at large $\Phi$, and since $\Phi = r$ classically as discussed in Sect.~\ref{s:aside}, all these theories are asymptotically AdS. It would be interesting to extend this approach to other spaces. 

This argument has been extended to $\mathcal{N}=1$ supersymmetric dilaton gravity theories in \cite{Rosso:2021orf}. 

\subsubsection{JT gravity coupled to gauge fields}
\label{s:JTgauge}
Having considered a general class of 2d dilaton gravity models, the next step is to generalize the 2d black hole/random matrix duality to the case where matter is present. A simple realization of this is to couple JT gravity to 2d gauge fields \cite{Davison:2016ngz, Mertens:2019tcm}  (another option previously discussed is to add also fermions using supersymmetry). These models of gravity coupled to gauge fields are also dual to double-scaled matrix models \cite{Iliesiu:2019lfc, Kapec:2019ecr} in a way we now describe. 

Let us begin with the gravity side. For simplicity, we consider JT gravity coupled to a group $G$ BF theory (similar considerations can be made for 2d Yang--Mills). We work in the grand canonical ensemble. 
The total Euclidean action is \eqref{jtac}:
\begin{equation}
    I = I_{\rm JT} - \int_{\mathcal{M}} {\rm Tr}~BF + \frac{K}{2} \oint_{\partial \mathcal{M}} d\tau {\rm Tr} \left[(A_\tau+ \mu)^2\right],
\end{equation}
supplemented with the chemical potential $\mu$ (an element of the Lie algebra) and $K$ is a dimensionful coupling constant (sometimes called the compressibility \cite{Davison:2016ngz}). We impose the following mixed boundary conditions\footnote{In higher dimensions, working in the grand canonical ensemble implies keeping fixed the holonomy of the gauge potential at the boundary. The 2d case is special in this regard and requires a boundary condition mixing $B$ and $A$ in this fashion.}
\begin{eqnarray}
B= K(A_\tau + \mu) \vert_{\partial \mathcal{M}}.
\end{eqnarray}
The exact solution of 2d Yang--Mills theory is known explicitly \cite{Witten:1991we,Witten:1992xu}, from which the BF amplitudes can be determined almost immediately: some amplitudes were written down in Eqs.~\eqref{BFpf} and \eqref{BFtwop} in Sect.~\ref{s:BF}. We first need to obtain the disk partition function. Integrating out the $B$ field (along an imaginary direction) results in an integral over flat connections $A=g d g^{-1}$. We mod out by bulk (small) gauge transformations, but the resulting action depends on the boundary value $g(\tau)$ giving the particle-on-a-group action:\footnote{The field transformation from $A$ to $g$ has a redundancy under right multiplication $g(\tau)\to  g(\tau) U $ with $U\in G$. This results in an integration space ${\rm Loop}(G)/G$. This is not true for left multiplication which acts as a true global symmetry. This redundancy is the gauge theory version of that discussed in footnote \ref{fn9} earlier.}
\begin{eqnarray}
I = I_{\rm JT} + \frac{K}{2} \int_0^\beta d\tau ~{\rm Tr}\left[\left(g \partial_\tau g^{-1}+\mu \right)^2 \right],
\end{eqnarray}
 where $g(\tau+\beta) = g(\tau)$. This theory is one-loop exact \cite{picken1989propagator}, and the partition function is given by $$Z_{\rm Disk, BF}= \frac{1}{{\rm vol}(G)}\sum_R {\rm dim}(R) \chi_R(e^{\beta \mu}) e^{-\beta \frac{C_2(R)}{2K}},$$ where $R$ are the irreducible representations of $G$, $C_2(R)$ the quadratic Casimir, and $\chi_R(V)$ the character in representation $R$ of group element $V\in G$. Combining this result with the JT gravity sector we obtain the total disk partition function:
\begin{eqnarray}
Z_{\rm Disk, JT+BF}(\beta,\mu) = \sum_{R} \int_{\frac{C_2(R)}{2K}}^\infty dE~\rho_R(E)\chi_R(e^{\beta \mu})~e^{-\beta E},
\end{eqnarray}
with a density of states per representation given by \cite{Mertens:2019tcm}
\begin{equation}\label{eq:DOSJTBF}
\rho_R(E) = \frac{{\rm dim}(R)}{{\rm Vol}(G)} \frac{e^{S_0}C}{2\pi^2}\sinh\left(2\pi\sqrt{2C\Big(E- \frac{C_2(R)}{2K}\Big)}\right).
\end{equation}
This is precisely the Schwarzian spectrum with a representation-dependent gap proportional to $C_2(R)$. 

The calculation can be extended to higher genus surfaces in the following way. Compute first the JT and BF path integral on single trumpets, where we additionally fix the gauge holonomy of $A$ along the geodesic boundaries. Next, we need to perform the path integral over the interior surface with geodesic boundaries with fixed length and holonomies. Similarly to JT, there are no boundary terms on these gluing cycles and the result is the volume of the moduli space of flat $G$ connections. These are easy to compute locally (since there is no issue analogous to the mapping class group) by cut-and-glue TQFT techniques \cite{Witten:1991we,Witten:1992xu}. The final step is to glue on the single trumpets by integrating over geodesic lengths and holonomies.

Now we move on to the holographic dual which is a random matrix model. In holography, when a group $G$ gauge field is present in the bulk, a global symmetry $G$ should commute with the boundary Hamiltonian. This implies that the boundary Hilbert space $\mathcal{H}_{\rm BH}= \oplus_{R} \mathcal{H}_{\rm BH}^R$ breaks up into separate subspaces in different representations $R$ of $G$. We can define separate Hamiltonians $H_R$ acting on each subspace labeled by the representation $R$. The  dual of JT gravity coupled to a group $G$ gauge field, derived in \cite{Iliesiu:2019lfc, Kapec:2019ecr}, corresponds to an ensemble average over Hamiltonians, where each $H_R$ is taken from a matrix ensemble described by a disk density of states $\rho_R(E)$ given by \eqref{eq:DOSJTBF}. The Hamiltonians acting on different representation subsectors are statistically independent. It is a non-trivial calculation to verify that this matrix integral reproduces the higher genus bulk calculations described in the previous paragraph.

This story can be easily generalized to a more general 2d dilaton gravity theory coupled to gauge fields, by applying the defect expansion of Sect.~\ref{sec:MTW} to JT gravity with gauge fields. As far as we know, this has not been studied in the literature, but it is straightforward.

\subsubsection{JT gravity coupled to matter}
One further extension is to describe matter boundary correlation functions of the type studied in Sect.~\ref{s:correlato} including wormholes. Increasingly detailed investigations of these were performed in \cite{Saad:2018bqo,Blommaert:2019hjr,Saad:2019pqd}, see also \cite{Blommaert:2020seb,Iliesiu:2021ari}. A concrete random matrix dual of JT gravity with matter was proposed in \cite{Jafferis:2022uhu,Jafferis:2022wez}.

We focus here on the boundary two-point function. The algorithm is the following. We first cut open the disk along the bilocal line:
\vspace{0.3cm}
\begin{center}
\includegraphics[width=0.55\textwidth]{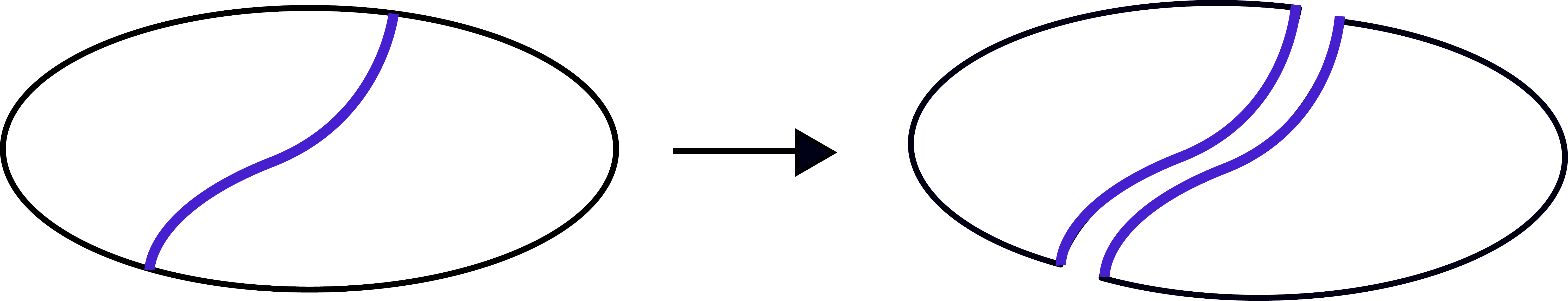}
\end{center}
\vspace{0.3cm}
In a second step, we can add all wormhole corrections in all possible ways to the resulting cut geometry:
\vspace{0.3cm}
\begin{center}
\includegraphics[width=0.95\textwidth]{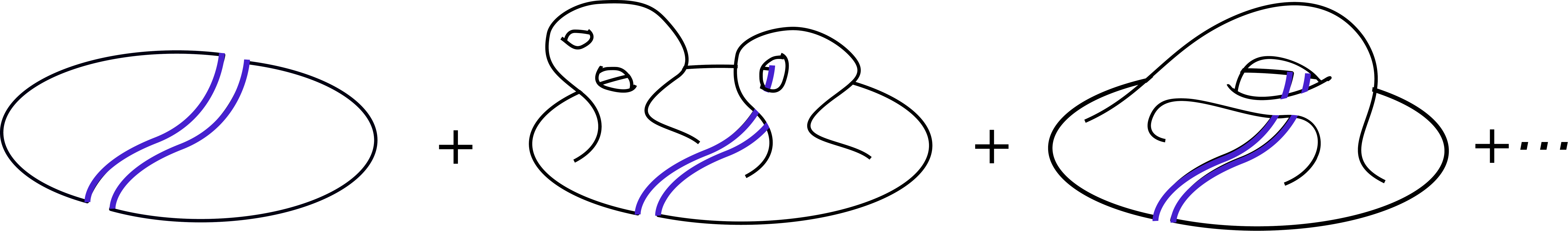}
\end{center}
\vspace{0.3cm}
It was shown in \cite{Saad:2019pqd} that this procedure is identical to first starting with a given higher genus hyperbolic surface, and then summing over all topologically distinct bilocal lines one can draw on the surface. The result is quite elegant, and upon summing over genus combines into
\begin{align}
      \langle \mathcal{O}(\tau_1)\mathcal{O}(\tau_2)\rangle = \int dE_1 dE_2 \, \langle \rho(E_1,E_2)\rangle\, e^{-\beta E_2 - \tau (E_1-E_2)} \vert \mathcal{O}_{E_1,E_2} \vert^2,
\end{align}
where, comparing to \eqref{eqnnnn}, we only replace the product of densities $\rho_0(E_1)\rho_0(E_2)$ at the disk level, by the pair density correlator $\langle \rho(E_1,E_2)\rangle$ of the JT random matrix ensemble described in Sect.~\ref{s:JTRMT}. This matches with inserting simple operators $\mathcal{O}(\tau)$ with given matrix elements $\langle E_1 \vert \mathcal{O} \vert E_2 \rangle$ between $H$-eigenvalues $E_{1,2}$, directly into the JT random matrix integral. The decomposition of the pair density correlator \eqref{pairdensity} in random matrix theory, mirrors the gravitational amplitudes. In particular, the first two diagrams above correspond to the disconnected contribution to $\langle \rho(E_1,E_2)\rangle$. The last diagram contains the connected contributions, which at lowest genus leads to a linearly increasing ramp for the boundary two-point function at late times \cite{Blommaert:2019hjr,Saad:2019pqd}. Investigations beyond the two-point function can be found in \cite{Blommaert:2020seb,Stanford:2021bhl}. \\

There are still some open questions surrounding these matter correlation functions. One open problem is the role of ``self-intersecting'' bilocal lines, whose intersection is topologically supported by wrapping around the wormhole. These were not included in the above prescription. At late times, we expect these are subdominant, but it is an open question how to think of them from the random matrix ensemble, and even whether they should be included in the first place. Another problem is that when considering dynamical matter in the bulk, one can have matter loops encircling wormhole necks. It is easy to show that the resulting negative Casimir energy causes a divergence due to the small wormhole UV region $b\to 0$, essentially the tachyon divergence familiar from string theory\footnote{Details of this calculation can be found in section 2.2 of \cite{Jafferis:2022wez}.}. The recent work by \cite{Jafferis:2022uhu,Jafferis:2022wez} studies a way to regularize this divergence by deforming the model directly at the matrix level.

A final comment is that the matrix integral language does not a priori know about the dynamical ingredients in correlation functions of these local boundary operators (the vertex factors \eqref{frules2} and the crossing kernel \eqref{crossing}), and these have to be supplied externally. In this sense, the matrix integral is merely a machine that produces the correct combinatorics of higher genus contributions in a non-perturbative framework.

\section{Applications and future directions}\label{sec:Generalization}
In this last section, we provide several applications and generalizations of the results we reviewed up to this point on the exact solution of JT gravity. This section is not aimed to be entirely self-contained, we instead refer to the literature for more details.

\subsection{The entropy of Hawking radiation}\label{sec:Islands}

Since the discovery of Hawking radiation \cite{Hawking:1975vcx}, it was realized that semi-classical black hole evaporation is inconsistent with quantum mechanical unitarity \cite{Hawking:1976ra}. Holography however implies that black holes behave as a quantum system, and it is important to resolve this tension and determine the fate of spacetime inside and around the horizon. This tension manifests in two ways when naively computing the gravitational path integral semi-classically. 

The first is that semi-classical gravity does not produce a discrete black hole spectrum, manifested e.g. in the late-time decay behavior described in Sect.~\ref{s:infoloss}. This is inconsistent with the microscopic interpretation of the Bekenstein--Hawking entropy. We already discussed in Sect.~\ref{s:sff} how it can be improved via the inclusion of spacetime wormholes. Obtaining an entirely discrete spectrum and seeing the erratic wiggles discussed in Sect.~\ref{s:infoloss} is not yet well-understood. We come back to this in Sect.~\ref{s:discrete} further on.

The second is that the radiation entropy of an evaporating black hole grows forever, see $S_{\text{rad}}(t)$ plotted in the figure below Eq.~\eqref{Sren} for the JT computation. If the evaporation is unitary and the initial state is pure, the black hole microstates can only purify as much as their own entropy. This was recently resolved as well by the inclusion of spacetime wormholes that appear when computing the entropy using the replica trick, sometimes called \textbf{replica wormholes} \cite{Penington:2019kki,Almheiri:2019qdq}. 

The result of this discussion is the following proposal for a formula that computes the radiation entanglement entropy in a gravitating system  \cite{Penington:2019npb,Almheiri:2019psf,Penington:2019kki,Almheiri:2019qdq}:
\begin{eqnarray}
    S = {\rm min}_X \left\{ {\rm ext}_X \left[ \frac{{\rm Area}(X)}{4G_N} + S_{\rm semicl}(\Sigma_X)\right]\right\}.\label{eqn:island}
\end{eqnarray}
The meaning of the right hand side is the following: first extremize the generalized entropy $S_{\rm gen}(X) = \frac{{\rm Area}(X)}{4G_N} + S_{\rm semicl}(\Sigma_X)$ over a codimension-two surface $X$ such that it defines a codimension-one surface $\Sigma_X$, whose boundary is $X$, on the region whose entropy we are computing. After finding all extrema $X$ for $S_{\rm gen}$, the one with the minimum value is selected. Finally, the quantity $S_{\rm semicl}(\Sigma_X)$ is the radiation entanglement entropy computed in the semiclassical state. The regions in $\Sigma_X$ disconnected from the semi-classical radiation region are called \textbf{entanglement islands}.

The above island formula resolves the issue with the entropy of Hawking radiation: when the semiclassical radiation entropy $S_{\rm semicl}(\Sigma_X)$ grows larger than the Bekenstein--Hawking entropy $\frac{{\rm Area}(X)}{4G_N}$, the entropy is computed by choosing $X$ to be a surface slightly in the interior of the black hole. In the end, this means that one essentially follows the \emph{minimum} of both curves plotted earlier below Eq.~\eqref{Sren} in Sect.~\ref{sec:Evaporation}. We note that the procedure of swapping between extrema as a function of time is a non-perturbative (in $G_N$) tunneling process in quantum gravity.

In the coming three subsections, we first perform this computation in JT gravity, followed by reviews on the role of replica wormholes in understanding these developments, and in particular deriving formula \eqref{eqn:island}.

\subsubsection{Entanglement islands in JT gravity}

Having simple solvable black hole systems as JT gravity was fundamental for uncovering the entropy formula \eqref{eqn:island} to begin with. We described a setup in JT gravity of evaporating black holes in Sect.~\ref{sec:Evaporation} that we retake here. We supplement that set-up by adding an asymptotically flat region, acting as a bath collecting the radiation. On this flat region, we use the metric $ds^2 = -du dv$ where $u=t+z$ and $v=t-z$ in terms of the proper coordinates defined using the near-boundary asymptotics in Sect.~\ref{s:bcclas}. Use of this time coordinate $t$ allows a continuous gluing to the bulk region.
Here we will sketch the computations that lead to a non-trivial entanglement island and resolve the tension with quantum unitarity. We follow a simplified combination of \cite{Almheiri:2019psf,Goto:2020wnk}.

\begin{figure}
\centering
\includegraphics[width=0.3\textwidth]{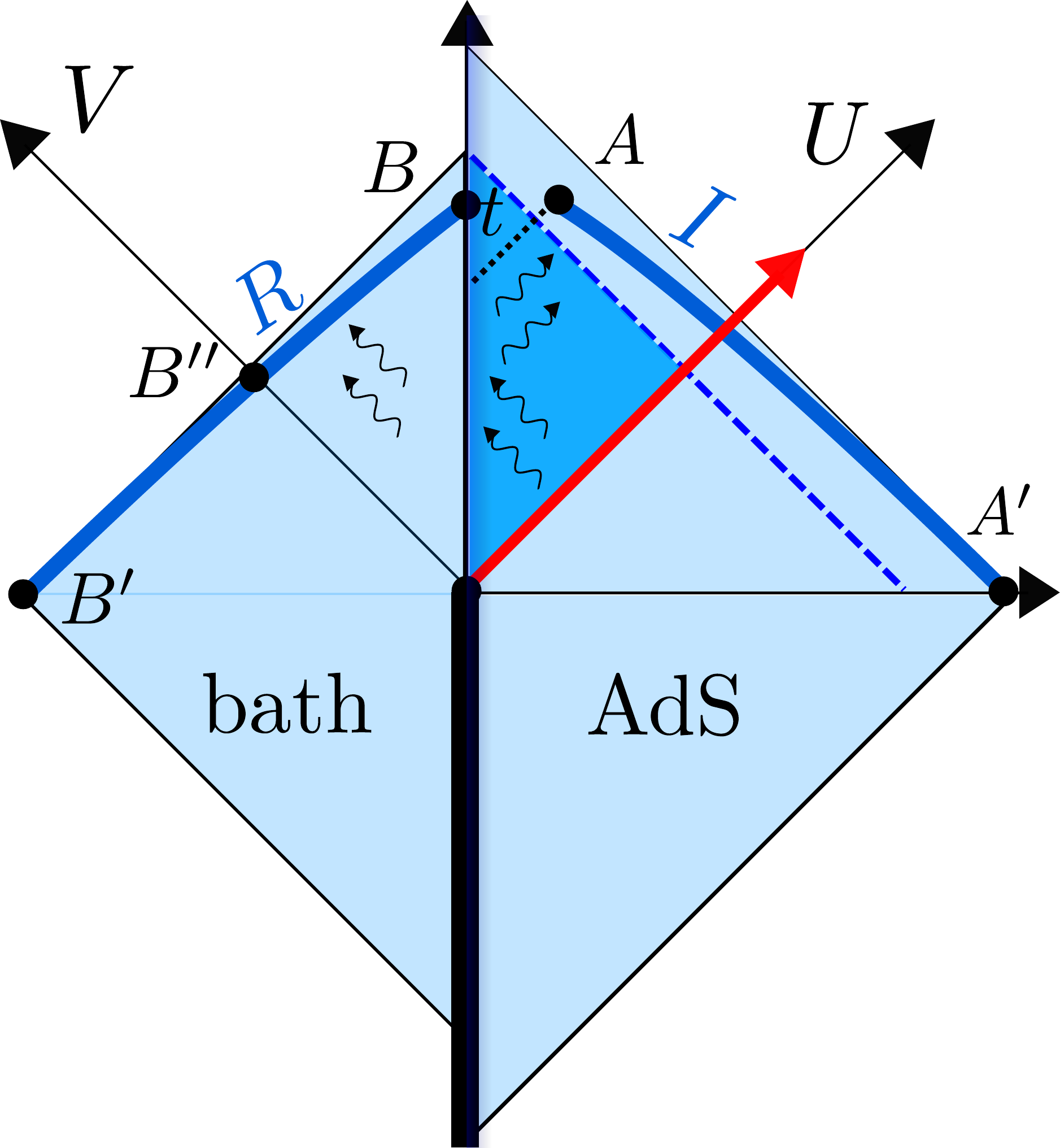}
\caption{Evaporating AdS$_2$ black hole coupled to a flat space bath to the left.}
\label{fig:Bathads}
\end{figure}

We consider the (naive, semi-classical) radiation region $R$ consisting of the flat space bath spacelike interval $R=[B',B]$, where we will imagine taking $B$ on the boundary (at measured time $t$) for simplicity, although our equations are more general. The bath is decoupled for $t<0$, and the interface is made transparent for $t\geq 0$. Next to this, we allow for a non-trivial spacelike island $I=[A,A']$ ranging from the Poincar\'e horizon $A'$ to some point $A$ that could be behind the black hole horizon, as illustrated in Fig.~\ref{fig:Bathads}. 

Our goal is to find the location $A=X$ keeping fixed all other quantities by following the above prescription \eqref{eqn:island}. In this enlarged set-up, we can continue with the expressions obtained in Sect.~\ref{sec:Evaporation}.
We can read the stress tensor vevs around \eqref{eq:dissip} as those corresponding to the vacuum state defined by the coordinate $(U=F(u),v)$ after the pulse, where $F(u)$ is the explicit solution \eqref{eq:reparabsorb}.\footnote{To compare notation with \cite{Almheiri:2019psf,Goto:2020wnk}, our $(u,v)$ are their $(y^+,y^-)$. Our $(U,V)$ are their $(x^+,x^-)$.}

First let us consider the case of no island ($A \to A'$). Then we only need to consider the entanglement entropy of the radiation region $R$. The emitted Hawking radiation is only captured there from $U>0$, and we can consider the entanglement entropy of the interval $[B'',B]$. There is an additive reference (and uninteresting) entanglement entropy in the state where the bath and system would remain decoupled forever. If one were to remove this, we would obtain the same result as before \eqref{Sren}. 

If a non-trivial island endpoint $A$ can be found, then the radiation region consists of both the naive one $R$ and the island region $I$. As the black hole evaporates, the union $R \union I$ increases in size, and the entanglement entropy decreases with time, yielding a result consistent with unitarity.

Let us work this out more explicitly. The computation is more straightforward when trying to solve the closely related problem of finding a quantum extremal surface at point $A$, where we extremize the same quantity as before \eqref{eqn:island}, except that we take the QFT entanglement entropy in the complementary interval $[A,B]$ instead, and extremize for $A$. It is straightforward to show that in the case of no ingoing flux $:T_{vv}:=0$, the dilaton solution \eqref{Phisol} can be written explicitly in terms of the solution $F(t)$ of Eq.~\eqref{eq:qschw} as \cite{Goto:2020wnk}:
\begin{align}
\label{eq:dilbh}
\Phi(U_A,v_A) = \frac{a}{2}\left(\frac{F''(v_A)}{F'(v_A)} + \frac{2F'(v_A)}{U_A-V_A(v_a)}\right).
\end{align}
For the metric $ds^2 = -\Omega^{-2}dz^+dz^-$, and for the vacuum state defined in the $(z^+,z^-)$ coordinates, the curved spacetime 2d CFT entanglement entropy for an interval between points $(z_A^+,z_A^-)$ and $(z_B^+,z_B^-)$ is given by \cite{Fiola:1994ir}
\begin{equation}
\label{eq:SCFT}
S_{\text{CFT}} = \frac{c}{6} \log \frac{\vert(z_A^+-z_B^+)(z_A^--z_B^-)\vert}{\epsilon_{A}\epsilon_{B}\Omega(z_A)\Omega(z_B)},
\end{equation}
where the UV cutoffs $\epsilon_{A,B}$ are measured in the inertial coordinates at the endpoints. In our case, the coordinates $(U,v)$ are those with respect to which we choose the vacuum. The somewhat peculiar feature here is that one endpoint ($A$) is in the AdS bulk, whereas the other endpoint ($B$) is in the glued flat region. Writing the bulk AdS metric as
\begin{equation}
ds^2 = \frac{-4dUdV}{(U-V)^2} = \frac{-4F'(v) dUdv}{(U-V)^2}\quad \Rightarrow \quad \Omega(z_A) = \frac{1}{2}(U_A-V_A) \frac{1}{\sqrt{F'(v_A)}},
\end{equation}
we can read off $\Omega(z_A)$ in the bulk point $A$ as written.
Analogously, in the flat region, we write the metric as:
\begin{equation}
ds^2 = -du dv = - \frac{1}{F'(u)}dUdv, \quad \Rightarrow \quad \Omega(z_B) = \sqrt{F'(u_B)}.
\end{equation}
With this information, we evaluate the QFT entanglement entropy \eqref{eq:SCFT} as
\begin{equation}
S_{\text{CFT}} = S_{\text{semicl}} = \frac{c}{6} \log \frac{2(U_A-U_B)(v_B-v_A)\sqrt{F'(v_A)}}{\epsilon_{A} \epsilon_B (U_A-V_A)\sqrt{F'(u_B)}}.
\end{equation}
Adding this to the gravity contribution \eqref{eq:dilbh}, we write the entropy functional $S_{\text{gen}}(X)$ as:
\begin{equation}
S_0 + \frac{a/2}{4G_N}\left(\frac{F''(v_A)}{F'(v_A)} + \frac{2F'(v_A)}{U_A-V_A(v_a)}\right) +  \frac{c}{6} \log \frac{2(U_A-U_B)(v_B-v_A)\sqrt{F'(v_A)}}{\epsilon_{A} \epsilon_B (U_A-V_A)\sqrt{F'(u_B)}}.
\end{equation}
The UV cutoffs $\epsilon_{A,B}$ yield additive terms that can be reabsorbed into a renormalization of $G_N$. Now, for fixed $B$-coordinates, this expression is to be extremized for the island endpoints $U_A$ and $v_A$. Upon doing this, the following important conclusions are found:
\begin{itemize}
\item At early times $t<t_{\text{Page}}$, the minimal $S_{\text{gen}}(X)$ is found to be the no-island situation where $A \to A'$, located on the initial Poincar\'e horizon. This leads to zero contribution from the BH entropy in the entropy formula, and the rising Hawking semi-classical radiation entropy \eqref{Sren} is retrieved:
\begin{equation}
S \approx S_{\text{rad}}(t).
\end{equation}
\item
At late times $t> t_{\text{Page}}$, a non-trivial island endpoint $A$ minimizes the entropy functional. Plugging the resulting location back into the entropy formula, one finds 
\begin{equation}
S \approx S_{\text{BH}}(t),
\end{equation}
and the black hole entropy decreases in time, following roughly the quasi-statically decreasing Bekenstein--Hawking entropy. The resulting Page curve is then just the minimum of these two entropies, as illustrated in Fig.~\ref{fig:evap3}.
\begin{figure}
\centering
\includegraphics[width=0.4\textwidth]{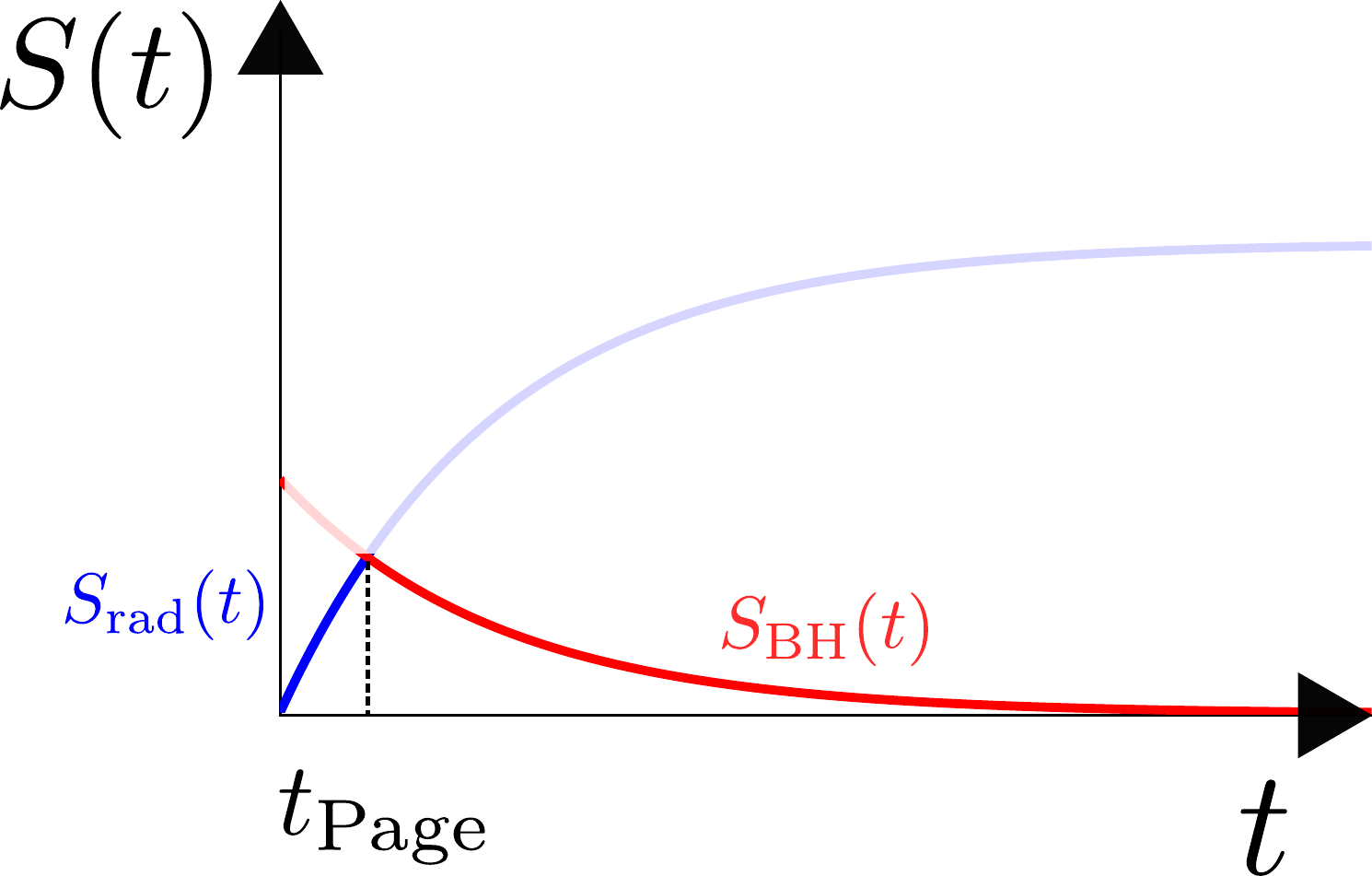}
\caption{Page curve of evaporating black hole using the island prescription.}
\label{fig:evap3}
\end{figure}
\item The Page time itself gets its dimensions from $C$, and is inversely proportional to the (assumed large) matter central charge $c$, as $t_{\text{Page}} \sim C/c$.
\item The location $U_A > F(t\to +\infty)$, putting the island endpoint behind the black hole horizon.\footnote{For the eternal black hole, the island endpoint is located outside of the black hole horizon \cite{Almheiri:2019yqk}.} 
\item For the other lightcone component, one can write \cite{Penington:2019npb,Almheiri:2019psf}:
\begin{equation}
\label{eq:trail}
v_A \sim t - \frac{1}{2\pi T(t)} \log \Big( 16\frac{S(t)-S_0}{c} \Big),
\end{equation}
where $T(t) = \frac{1}{\pi}\sqrt{\frac{E}{2C}} e^{- \frac{c}{48\pi C}t} \theta(t)$ is the quasi-static temperature determined from the energy profile \eqref{eq:disen} and the energy-temperature relation \eqref{eq:firstlaw}. The island $v_A$-coordinate \eqref{eq:trail} trails behind a scrambling time behind the time $t$ when we measure the entropy at the boundary. Both this and the previous feature were illustrated in the Penrose diagram above.
\end{itemize}

The apparent conflict with unitarity is hence resolved by taking into account entanglement islands at times greater than the Page time.

\subsubsection{Replica wormholes}
Next we review ways to motivate and derive the island formula \eqref{eqn:island}.

A first route is to consider the case where the matter CFT sector is on its own strongly coupled and holographic. This is a version of double holography studied in \cite{Almheiri:2019hni}. The quantum extremal surface or island formula \eqref{eqn:island} reduces to the classical Ryu--Takayanagi formula in this higher-dimensional space. The 2d JT model is then described by a dynamical metric on the ``Planck brane'' (or Randall--Sundrum brane \cite{Randall:1999ee,Karch:2000ct}).

There is a more generic derivation of the island rule \eqref{eqn:island} based on Euclidean replica trick techniques. This approach was investigated by two groups \cite{Almheiri:2019qdq,Penington:2019kki}. Here we provide a short summary of the first one, referred to as the ``east coast model'', the other one follows in the next subsection. 

In QFT, the $n$'th R\'enyi entropy $S_n \equiv \text{Tr}(\rho_R^n)$ of an interval 
is computed by considering $n$ replicas of the original manifold, glued together cyclically across branch cuts, with branch points at all endpoints of the interval of interest. We depict the $n=2$ case for a single endpoint, where the color-coding denotes how one moves between replicas as one crosses a branch cut:
\begin{equation}
\centering
\nonumber
\includegraphics[width=0.3\textwidth]{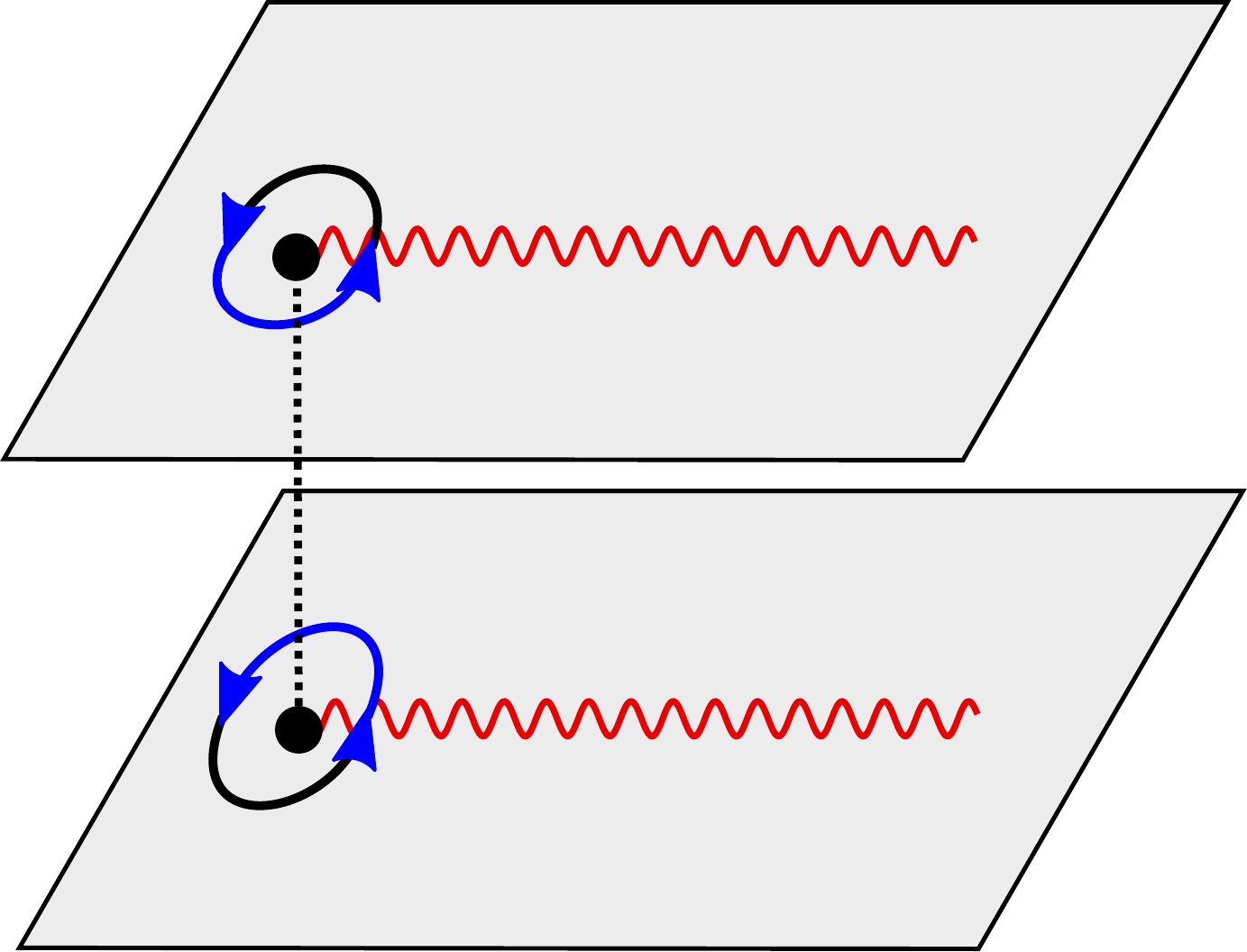}
\end{equation}
However, if there is a gravitational region in the space, we are instructed by holography (or quantum gravity more generally) to fill in the gravitational interior in all possible ways, and then sum over these possibilities with the Euclidean path integral. One of these possibilities is to connect all of the different replicas together into a single connected geometry, forming a \textbf{replica wormhole}. In this context, this is a saddle solution of the replicated equations of motion. For example, for the specific case of the two-sided eternal black hole in AdS, for which the radiation region contains an interval in both the left- and right flat bath region, we sketch the 2-replica configuration below: 
\begin{equation}
\nonumber
\includegraphics[width=0.7\textwidth]{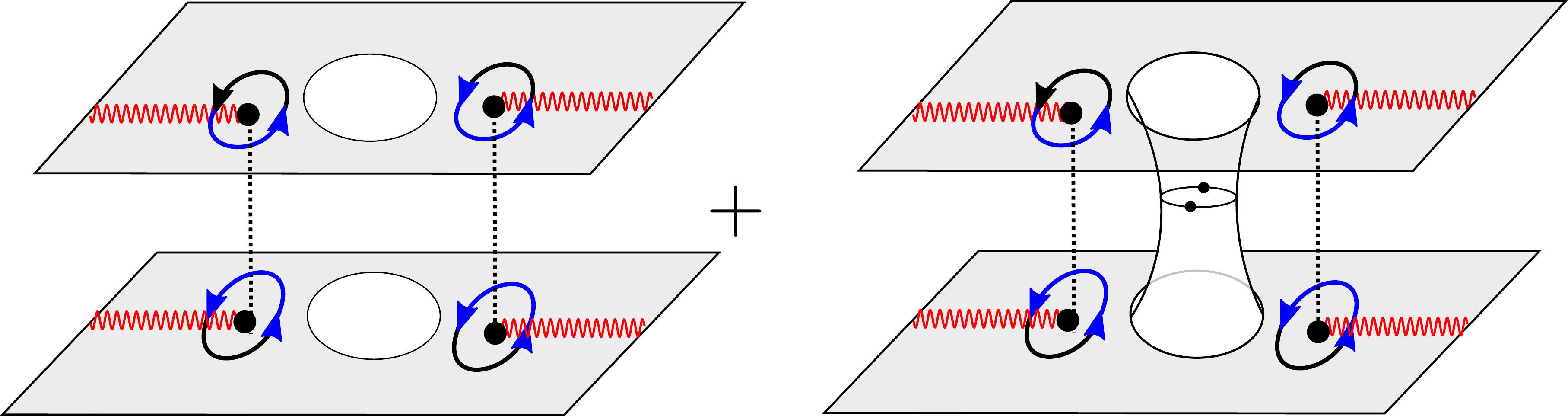}
\end{equation}
The grey region is the non-gravitating region (the flat bath) in Euclidean signature. Locations on opposite sides of the eternal black hole (the two intervals) are in Euclidean signature separated by a rotation by 180 degrees, as drawn. The white region is a gravitating region for which we are instructed to sum over geometries, for which the two saddle topologies are drawn.
For replica number $n$, the connected saddle is topologically suppressed as $\sim e^{S_0(2-n)}$, but can get enhanced at late times. 

For a $\mathbb{Z}_n$-symmetric fully connected replica saddle configuration, one can represent the replica wormhole in two ways. As a single smooth surface as above; or by modding by the $\mathbb{Z}_n$ symmetry as a surface of disk topology that has two twist points present (drawn in the figure for $n=2$), around which one has a $2\pi/n$ conical singularity:
\begin{equation}
\centering
\nonumber
\includegraphics[width=0.3\textwidth]{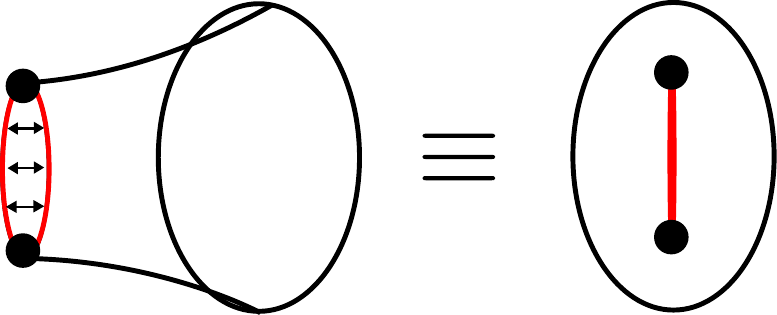}
\end{equation}

With these ideas, the following conclusions can be drawn:
\begin{itemize}
\item The island formula \eqref{eqn:island} follows directly from a generalization of the argument of \cite{Lewkowycz:2013nqa} that was used to prove the Ryu--Takanayagi formula. 
\item The twist operator locations in the replica geometry are also part of the extremization procedure, and these locations are precisely the island endpoints in the gravitational bulk.
\item The disconnected replica geometry dominates at early times (before the Page time), whereas the fully connected replica geometry dominates at late times (after the Page time).
\end{itemize}
This approach was first used for the information paradox for the eternal black hole in \cite{Almheiri:2019qdq}, for which we drew the 2-replica geometries above, and generalized to the evaporating case in \cite{Goto:2020wnk}.

\subsubsection{Replica wormholes and EOW branes}
We next review a very simple model proposed in \cite{Penington:2019kki}, commonly referred to as the ``west coast model'', to study the role of wormholes in replica trick computations. It consists of a black hole in a pure state using the end-of-the-world (EOW) branes introduced in Sect.~\ref{Sec:EOWbranes}. To produce an information-like paradox, we can introduce a very large number $k$ of internal states of the EOW brane, entangled with an auxiliary system $R$, schematically as: 
\begin{eqnarray}
\nonumber
 \includegraphics[scale=0.13]{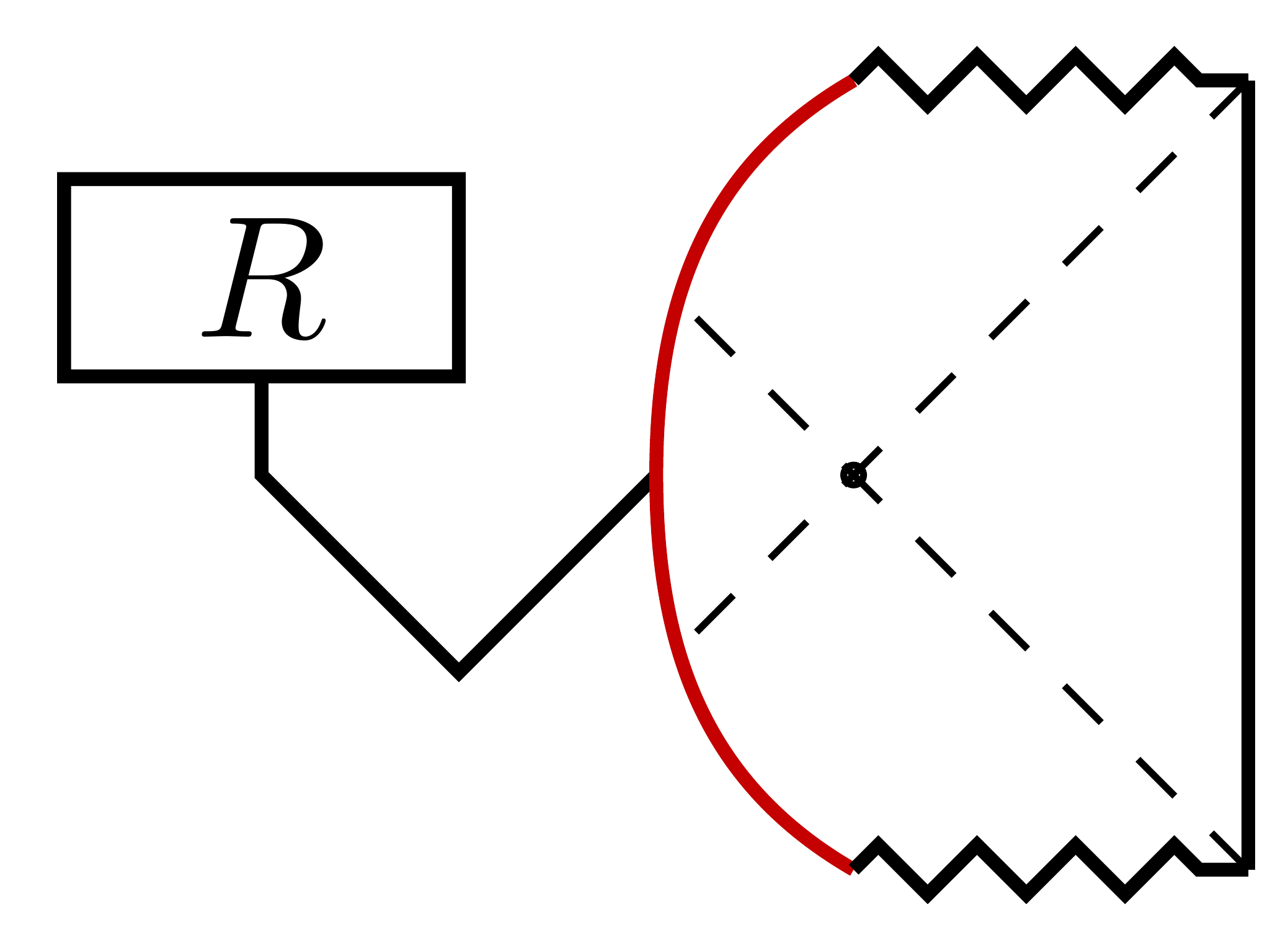}
\end{eqnarray}
The motivation of \cite{Penington:2019kki} to consider this model was to interpret the EOW states as the interior radiation modes and the auxiliary system R as the early Hawking radiation. The state of the whole system and the radiation density matrix are therefore
\begin{eqnarray}
\label{eq:rhoR}
\vert \Psi \rangle = \frac{1}{\sqrt{k}} \sum_{i=1}^k \vert \psi_i \rangle_{\rm B} \vert i \rangle_{\rm R},~~\rightarrow~~\rho_{\rm R} = \frac{1}{k} \sum_{i,j=1}^k \vert j \rangle \langle i \vert_{\rm R} \langle \psi_i \vert \psi_j\rangle_{\rm B},
\end{eqnarray}
where the state of the black hole with an EOW brane in state $i$ is $\vert \psi_i \rangle_{\rm B}$. It is easy to see that when wormholes are not included, and if we pick an orthogonal basis of EOW states, gravity will give:
\begin{eqnarray}
\includegraphics[scale=0.13]{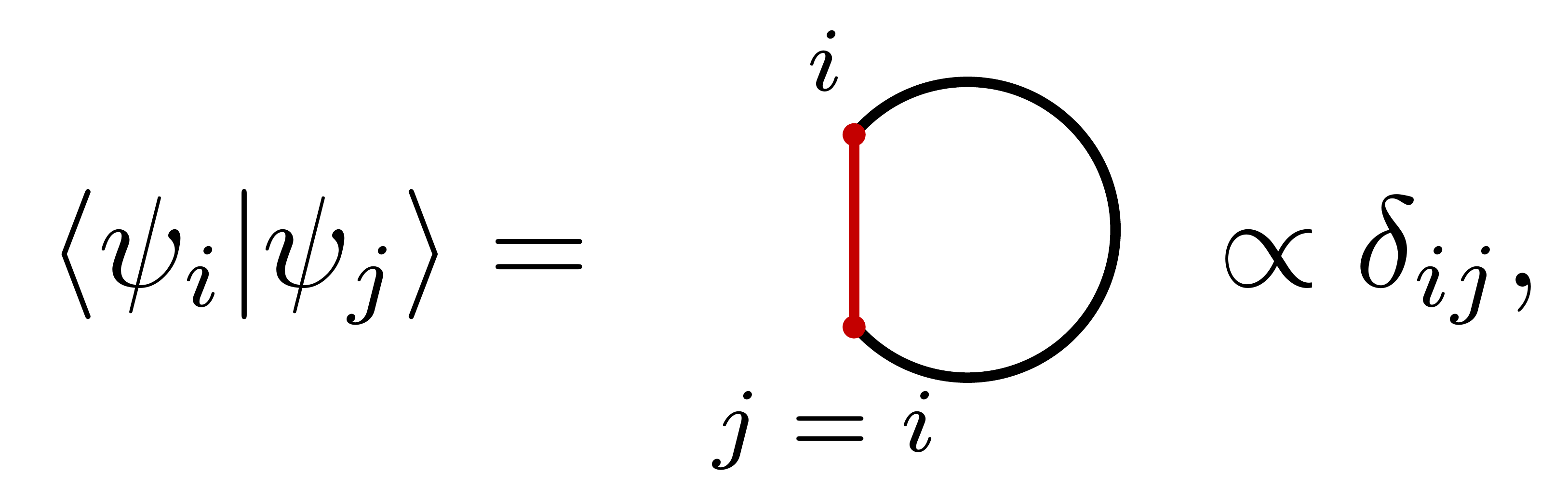}
\end{eqnarray}
implying $\rho_{\rm R} = k^{-1} \sum_{i=1}^k \vert i \rangle \langle i \vert$ is maximally mixed with entropy $\log(k)$. This is the information paradox: as we increase $k$ the entropy of the radiation can grow and become bigger than the black hole entropy $S_{\rm BH}$. If the black hole is described by a quantum system with $e^{S_{\rm BH}}$ degrees of freedom, this is an inconsistency: for system B to purify R, the dimension of B should be at least as large as that of R.  

Instead of considering entropies, it is instructive to study the purity of a state defined as ${\rm Tr}\left( \rho_{\rm R}^2 \right)$. This quantity is one for a pure state and strictly smaller than one for a mixed state. Ignoring wormholes, the fact that the entropy of Hawking radiation grows indefinitely is related to the fact that the purity can decay without bound in time. Using the radiation density matrix \eqref{eq:rhoR} gives
\begin{equation}
{\rm Tr}\left( \rho_{\rm R}^2 \right) = \frac{1}{k^2} \sum_{i,j=1}^k \vert \langle \psi_i \vert \psi_j \rangle \vert^2.
\end{equation}
If we take $\langle \psi_i \vert \psi_j \rangle \propto \delta_{ij}$ then the purity is ${\rm Tr} \left( \rho_{\rm R}^2 \right) = 1/k$ and decays indefinitely as $k$ grows. Computing this quantity in gravity involves a computation of $ \vert \langle \psi_i \vert \psi_j \rangle \vert^2$, which in the presence of wormholes gets the following contributions to leading order:  
\begin{eqnarray}
\includegraphics[scale=0.23]{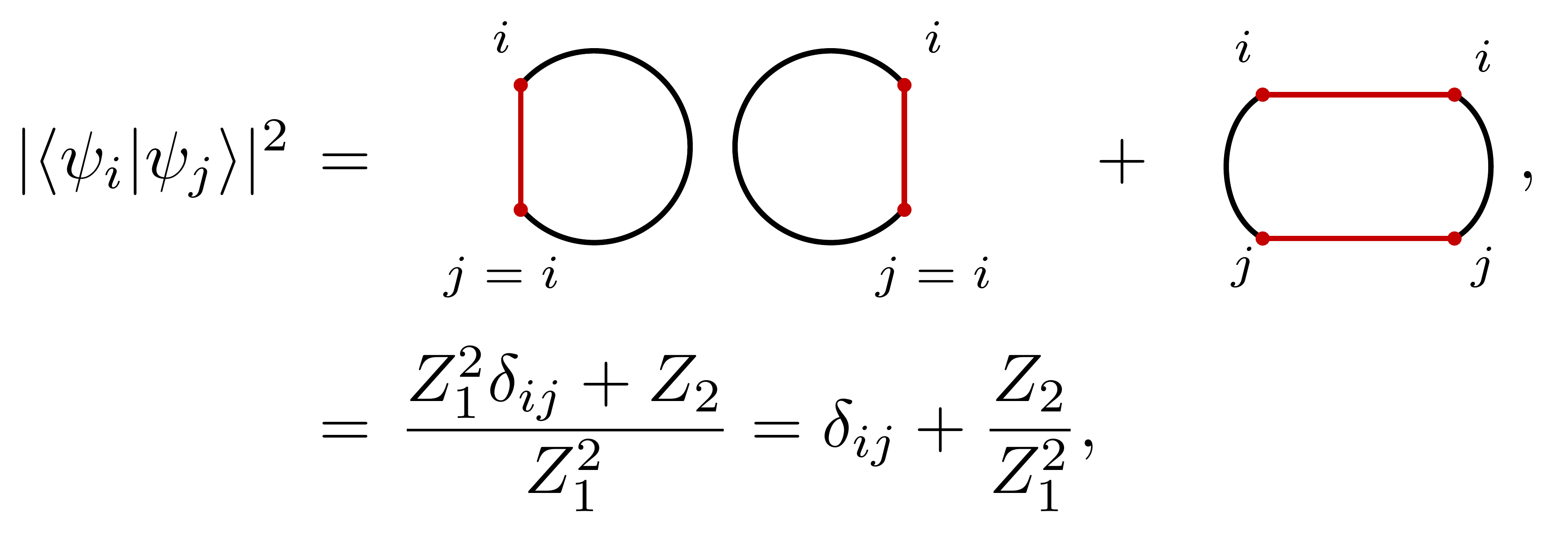} 
\end{eqnarray}
where the path integral on a disk with one holographic boundary and one EOW brane is denoted by $Z_1$ (only non-zero when $i=j$) and discussed in Sect.~\ref{Sec:EOWbranes}, while the path integral on a surface with two holographic boundaries and two EOW branes is denoted by $Z_2$ and is independent of the EOW brane flavor indices $i,j$. The denominator $Z_1^2$ is only a choice of normalization of the states $\vert \psi_i \rangle$. There are two important features we need for this discussion. The first is that both $Z_1 = \tilde{Z}_1 e^{S_0}$ with $\tilde{Z}_1$ of order one, $Z_2 = \tilde{Z}_2 e^{S_0}$. The second is that while the disconnected diagram imposes $j=i$ and therefore is given by $Z_1^2 \delta_{ij}$, the second diagram is non-zero for any $i$ and $j$, giving an off-diagonal component to the inner product. We can use this result to compute the purity
\begin{eqnarray}
{\rm Tr}\left( \rho_{\rm R}^2\right) &=& \frac{k Z_1^2 + k^2 Z_2}{(k Z_1)^2} = \frac{1}{k} + \frac{Z_2}{Z_1^2},\\
&=& \frac{1}{k} + \frac{1}{e^{S_0}}\frac{\tilde{Z}_2}{\tilde{Z}_1^2},~
\end{eqnarray}
with $\tilde{Z}_2/\tilde{Z}_1^2$ an order one number computed in \cite{Penington:2019kki}. We now see how the wormhole term resolves this version of the information paradox: while the first term $1/k$ decays without bound as $k\to \infty$ whenever the number of states is of order $k \gtrsim e^{S_0}$ the purity will saturate to ${\rm Tr} \left( \rho_{\rm R}^2 \right) \sim e^{-S_0}$ thanks to the wormhole contribution.

The previous argument can be applied to the computation of Renyi entropies involving ${\rm Tr} \left( \rho_{\rm R}^n \right)$ and used to extract the entanglement entropy of the radiation $S({\rm R}) = - {\rm Tr} \left( \rho_{\rm R} \log \rho_{\rm R}\right)$. The result is:
\begin{eqnarray}
\nonumber
\includegraphics[scale=0.22]{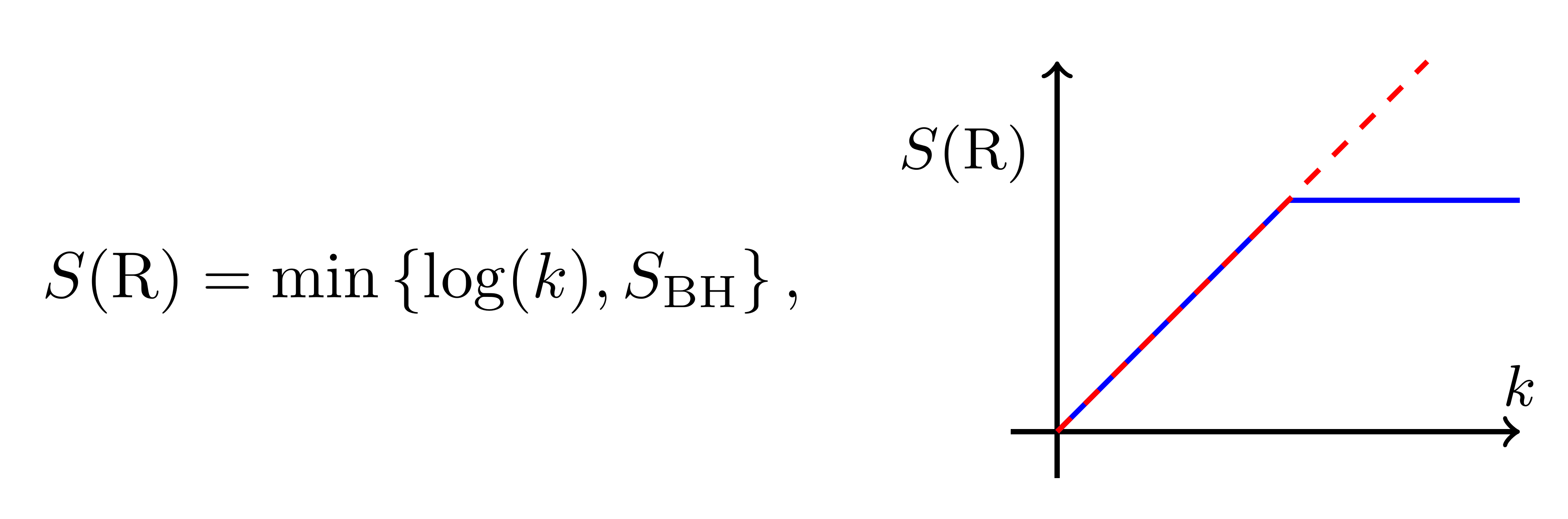}
\end{eqnarray}
which is precisely what one obtains from the ``island rule'' defined in Eq.~\eqref{eqn:island}, when applying that discussion to an eternal black hole (see e.g. \cite{Almheiri:2019yqk}) and interpreting $k$ loosely as time. In the right we show the analog of the Page curve for this model. This answer is accurate except when we are close to the transition where $k\sim e^{S_{\rm BH}}$. That same work \cite{Penington:2019kki} also found the answer that precisely interpolates between the early and late behaviors in this simple toy model. This is a striking result: without gravity, if we want to compute the entropy of a quantum system R it is enough to know $\rho_{\rm R}$. If the auxiliary system R is entangled with a system with gravity, such as the pure JT black hole, then the entanglement wedge of R includes the region behind the black hole horizon with an area term contributing $S_{\rm BH}$.

Finally, another interesting feature of this calculation is related to the interplay between wormholes, the lack of factorization and the need for ensemble average. In terms of the matter density matrix, the derivation above does not seem to make sense since $\langle \psi_i \vert \psi_j \rangle = \delta_{ij}$ while $\vert \langle \psi_i \vert \psi_j \rangle \vert^2 = \delta_{ij}+Z_2/Z_1^2$. The interpretation is that gravity is computing ensemble averages over these matrix elements, and wormholes capture the variance around the leading order delta function.

\subsection{Factorization, discreteness, and ensemble averaging in gravity}
\label{s:discrete}
The non-perturbative features of random matrix theory contain high frequency wiggles of temporal size $\sim e^{-S_0}$ that are an indication of the underlying discreteness of the matrix integral. 
However, JT gravity's matrix model description is double-scaled and moreover evaluated in a $e^{-S_0}$ expansion, where the actual discreteness of the underlying system is washed out. But if JT gravity is truly a low-energy approximation of some microscopic model (such as SYK), then it should be possible to describe the discreteness of that UV model in the gravitational low-energy language. There have been several works attempting to reintroduce discreteness in JT gravity 
by introducing additional brane-like objects in the gravitational bulk, see e.g. \cite{Blommaert:2019wfy,Johnson:2022wsr,Blommaert:2021fob} for various approaches. 

One of the important features of this link with a discrete spectrum is the so-called \textbf{factorization problem} mentioned in Sect.~\ref{s:SSS} before. The paradox goes as follows. Suppose we have an $n$-boundary amplitude
$\left\langle Z(\beta_1) \hdots Z(\beta_n) \right\rangle$. Then the boundary theories are disconnected and the amplitude should factorize. However, holography instructs us to sum over all possible geometric interiors that end on these $n$ boundaries, disconnected \emph{and} connected.  How does gravity know in this set-up that it ultimately should factorize microscopically, since many individual geometries do not? One possibility is that there are actually additional saddles in the bulk path integral that do not appear in an averaged theory, but would appear for any specific microscopic version of it. 
A simple analog of the mechanism is to consider the spectral form factor (at $\beta =0$) of a discrete system:
\begin{equation}\left(\sum_i e^{i E_i t}\right) \left(\sum_i e^{-i E_j t}\right) = \sum_{i=j}1 + \sum_{i \neq j} e^{i (E_i-E_j) t}.
\end{equation}
After averaging (e.g. in time), the off-diagonal terms on the RHS cancel out due to the heavy fluctuations; only the diagonal terms survive. But this diagonal contribution on its own does not factorize as the LHS does. Hence, the non-factorization is an artifact of the averaging procedure, but is not there in the truly microscopic model. For a simplified version of the SYK model, this mechanism was investigated in \cite{Saad:2021rcu,Mukhametzhanov:2021nea,Saad:2021uzi}, where the new configurations were dubbed ``half-wormholes''. A geometric interpretation of these is still lacking at the time of writing.

This discussion is related to the deep problem of how one should really think about quantized gravity when including wormholes. There have been several distinct indications \cite{Harlow:2018tqv,Penington:2019kki,Almheiri:2019qdq,Giddings:2020yes,Stanford:2020wkf,Engelhardt:2020qpv,Johnson:2021rsh,Saad:2021uzi} that the gravitational path integral is not equivalent to an ordinary quantum mechanical system (where we interpret the boundary circles as representing Euclidean time). 

As a further example of this tension, we noted at several points that the higher genus sectors in the gravitational path integral do not contain classical saddle points in dilaton gravity. However standard canonical quantization starts by solving the classical equations of motion, which means that, in canonical quantization, it is seemingly impossible to see effects from any surfaces besides disconnected disks! 

The new interpretation that we seem to be led to, is that gravity should instead be viewed as an ensemble average that happens at the very end of the computation. This is precisely the same situation as in disorder averages in spin glass systems in condensed matter physics. For example, for the free energy of the system, one should compute the quenched free energy $\left\langle \log Z \right\rangle$, and not the annealed free energy $\log \left\langle Z \right \rangle$, where the brackets denote the gravitational path integral. The situation is exactly the same in gravity \cite{Engelhardt:2020qpv,Johnson:2021rsh,Alishahiha:2020jko}.

\subsection{Near-extremal black holes}\label{sec:NEBHc}
In Sect.~\ref{sec:NEBH}, we gave a brief overview of how JT gravity describes physics close to the horizon of a near-extremal black hole, at the classical gravity level. This is an old observation which was recently developed further thanks to an improved understanding of JT gravity. A partial collection of references that explore this connection is \cite{Nayak:2018qej, Moitra:2018jqs, Hadar:2018izi, Castro:2018ffi, Larsen:2018cts, Kolekar:2018sba, Moitra:2019bub, Sachdev:2019bjn, Hong:2019tsx, Castro:2019crn, Charles:2019tiu,Larsen:2020lhg,Narayan:2020pyj,Castro:2021wzn, Castro:2021fhc,Castro:2021csm}.

In \cite{Ghosh:2019rcj} and \cite{Iliesiu:2020qvm}, it was explained how the link between JT gravity and near-extremal black holes can be made also at the quantum level. This means that the leading low-temperature behavior of quantum gravity corrections from large fluctuations of a certain mode in the metric of the higher-dimensional black hole can be approximated by the quantum effects in JT gravity we analyzed in Sect.~\ref{sec:JTquantum}. This leads to interesting conclusions such as the fact that there are no classical extremal black holes in gravity and the Bekenstein--Hawking area law for extremal black holes is meaningless, except when the black hole is embedded in a supersymmetric theory and preserves enough supersymmetry \cite{Heydeman:2020hhw,Boruch:2022tno}. A more detailed review on these results can be found in the upcoming review \cite{ROPreviewNEBH}.

\subsection{Supersymmetric JT}
JT gravity has since its start been generalized to include supersymmetry. Next to providing interesting solvable models of supergravity, whose solution can be done in parallel with the bosonic case, they are also of direct relevance to match with higher-dimensional supersymmetric black hole physics, as briefly mentioned in Sect.~\ref{sec:NEBH}.

Here we summarize what has been computed in supersymmetric generalization of JT gravity at the time of writing this review.

The simplest case with $\mathcal{N}=1$ has one supercharge and describes on the boundary side the symmetry breaking pattern associated to ${\rm OSp}(2\vert1)\supset {\rm SL}(2,\mathbb{R}) $. The super-Schwarzian action has been derived using a fluctuating boundary curve analysis in \cite{Forste:2017kwy}. At the level of the pure $\mathcal{N}=1$ JT gravity partition function, it has been computed for all genus and matched with random matrix models in \cite{Stanford:2019vob}. In the presence of matter, at the disk level, the time-ordered correlators were computed in \cite{Mertens:2017mtv,Fan:2021wsb} while the OTOCs are unknown. Defects have been classified in \cite{Fan:2021wsb}.

The next case has two supercharges, $\mathcal{N}=2$ JT gravity describing the symmetry breaking pattern of SU$(1,1\vert 1)\supset {\rm SL}(2,\mathbb{R})\times U(1)_R$. The super-Schwarzian analysis appears in \cite{Forste:2017apw}. The disk partition function was computed in \cite{Stanford:2017thb}, and the density of states in \cite{Mertens:2017mtv}. This theory can have fractional charge and also be supplemented by a theta angle, see \cite{Boruch:2022tno} and \cite{Heydeman:2022lse}. The time-ordered correlators were computed recently in \cite{Lin:2022zxd, Lin:2022rzw}. The OTOCs are unknown. The contribution from higher topologies and its matrix dual were recently proposed in \cite{Turiaci:2023jfa}.

The case with four supercharges, $\mathcal{N}=4$ JT gravity, describing the breaking pattern of PSU$(1,1\vert2)\supset {\rm SL}(2,\mathbb{R})\times SU(2)_R$, was defined and the disk partition function computed in \cite{Heydeman:2020hhw}. Matter correlators have not been computed and the sum over topologies is not known, although some preliminary calculations are in \cite{Turiaci:2023jfa}. 

\subsection{Two-dimensional cosmology}\label{sec:JTdS}

JT gravity with positive cosmological constant coupled to matter was considered in \cite{Maldacena:2019cbz, Cotler:2019nbi}. The action takes the form \eqref{gendilgrav} with a dilaton potential $U(\Phi) = - 2 \Phi$, supplemented by a $\Phi_0$ term. This theory has asymptotically two-dimensional de Sitter solutions and it therefore models lower-dimensional cosmology. 

There are several set-ups one can analyze in JT gravity with positive cosmological constant. One option is to follow the Hartle-Hawking no-boundary proposal \cite{Hartle:1983ai}. In a frame where the metric is rigid dS$_2$, there is a conformal symmetry that is broken by the (asymptotics of the) dilaton which is slowly increasing with time (similar pattern of symmetry breaking as in inflation). The value of the dilaton acts as a clock, and we fix its value on the future boundary $\Phi_b= \Phi \vert_\partial$. The wavefunction of the universe depends on the proper length of the spatial universe which we call $\ell$ (analogous to $\beta/\epsilon$ in the AdS case), and it is computed by the geometry
\begin{eqnarray}
\includegraphics[scale=0.21]{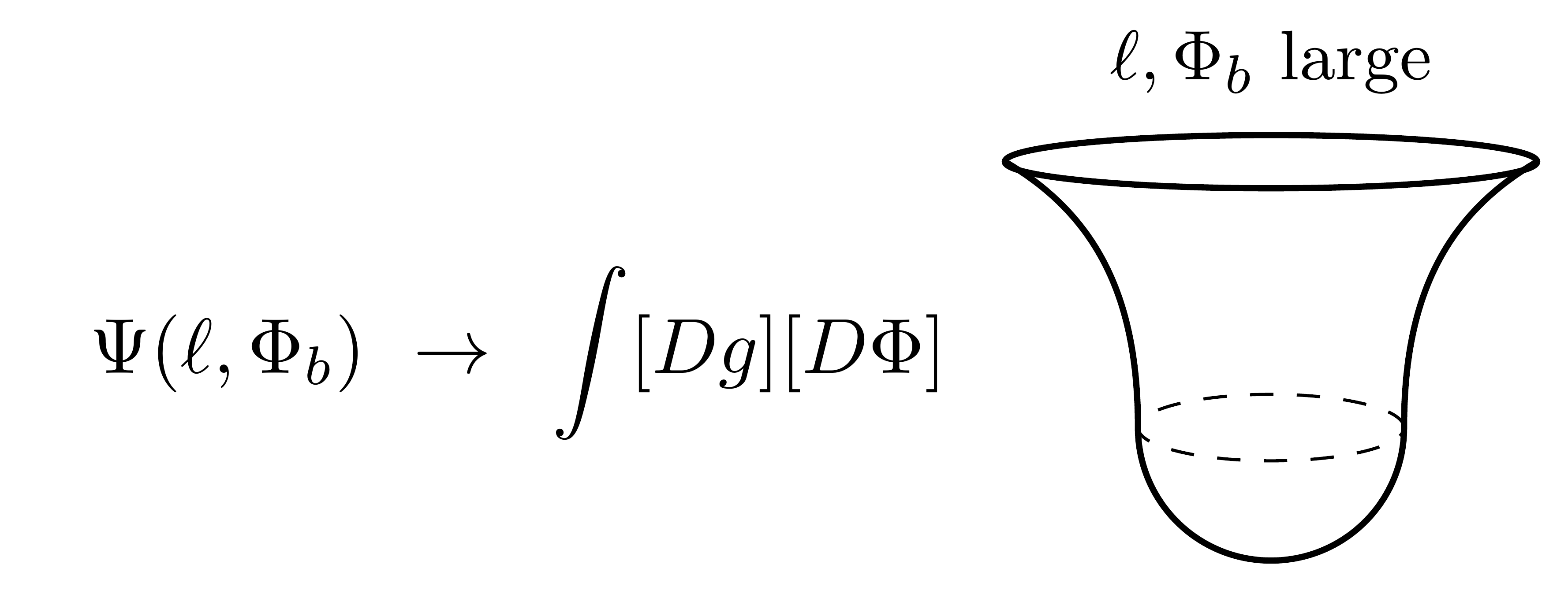}
\end{eqnarray}
Considering $\Phi_b$ and $\ell$ large is an invariant way of specifying we want to compute the wavefunction at late times. The result can be computed by reducing JT gravity to a space-like Schwarzian mode living in the future boundary. The result is $\Psi (\ell,\Phi_b) \propto \frac{\Phi_b^{3/2}}{\ell^{3/2}} e^{- i \Phi_b \ell + S_0+ i \frac{2\pi^2 \Phi_b}{\ell} }$. In this context the parameter $S_0$ can be interpreted as half of the dS entropy. An interpretation of this formula can be found in \cite{Maldacena:2019cbz, Cotler:2019nbi}. When matter is present, it is possible to compute quantum gravity corrections to late time matter correlators using the Schwarzian action \cite{Maldacena:2019cbz}. Since dS correlators involve an integral over the wavefunction (instead of derivatives as in AdS) \cite{Maldacena:2002vr}, this is considerably harder than the AdS case and the full answer for these correlators is not known.

For the case of pure JT gravity, the sum over non-trivial topologies when computing this wavefunction of the universe can be reproduced by a matrix model similar to SSS \cite{Maldacena:2019cbz, Cotler:2019nbi}. The ultimate role of these cosmological wormholes is not understood.  Finally, when matter is present there is a paradox due to the fact that the matter entropy can grow larger than the de Sitter entropy. A simplified version of a resolution of this paradox uses wormholes \cite{Chen:2020tes} not too different from the ones discussed in Sect.~\ref{sec:Islands}.

JT gravity with positive cosmological constant describes the physics near the horizon of a near-Nariai black hole in higher-dimensional dS. This limit is very different than the near-extremal near-horizon counterpart: it does not require charge and it is a solution of pure gravity. Four-dimensional black holes in dS have a maximal mass and when this is approached, a dS$_2$ $\times$ S$^2$ throat emerges close to the black hole and cosmological horizons.

Having discussed AdS and dS it is important to point out that some works also studied JT gravity in flat spacetime, see \cite{Dubovsky:2017cnj, Godet:2021cdl}, and even proposed a matrix model dual \cite{Kar:2022vqy}.

\subsection{Traversable wormholes}
The main ingredient of wormhole traversability in a holographic system is a specific coupling between two boundary systems that allows an initial state, localized to one boundary system, after a complicated evolution become a simple final state localized in the second boundary system. 
The holographic dual of such a process, is geometrically smooth travel through a traversable wormhole. We need to distinguish between two classes of such wormholes, the first containing black hole horizons and the second containing no horizons.

\paragraph{Traversable ER wormholes}
The thermofield double state in holography has a spatial wormhole, which is the AdS analog of the Einstein--Rosen (ER) bridge. This wormhole is famously not traversable by any bulk probe, but it is very close.

It was argued in \cite{Gao:2016bin} that one can make this TFD wormhole traversable by introducing a double trace boundary interaction, leading to a schematic path integral insertion $e^{ig O_L(0) O_R(0)}$ at a reference time $t=0$ for simple boundary operators $O_{L,R}$ localized to either the left ($L$) or right ($R$) boundary, and with suitable sign of the coupling $g$. This results in a negative bulk energy insertion violating the null energy condition and making the wormhole traversable. This procedure was explored in detail in the NAdS$_2$ context in \cite{Maldacena:2017axo}. The physics of the resulting traversability can be conveniently captured by thinking about the wiggly boundary curve experiencing a kick whenever bulk matter is injected. In particular, after the interaction happens, the left- and right boundaries are pushed towards each other due to the negative energy injection, making a bulk signal able to traverse the wormhole and arrive at the other boundary.

\begin{figure}
\centering
\includegraphics[width=0.55\textwidth]{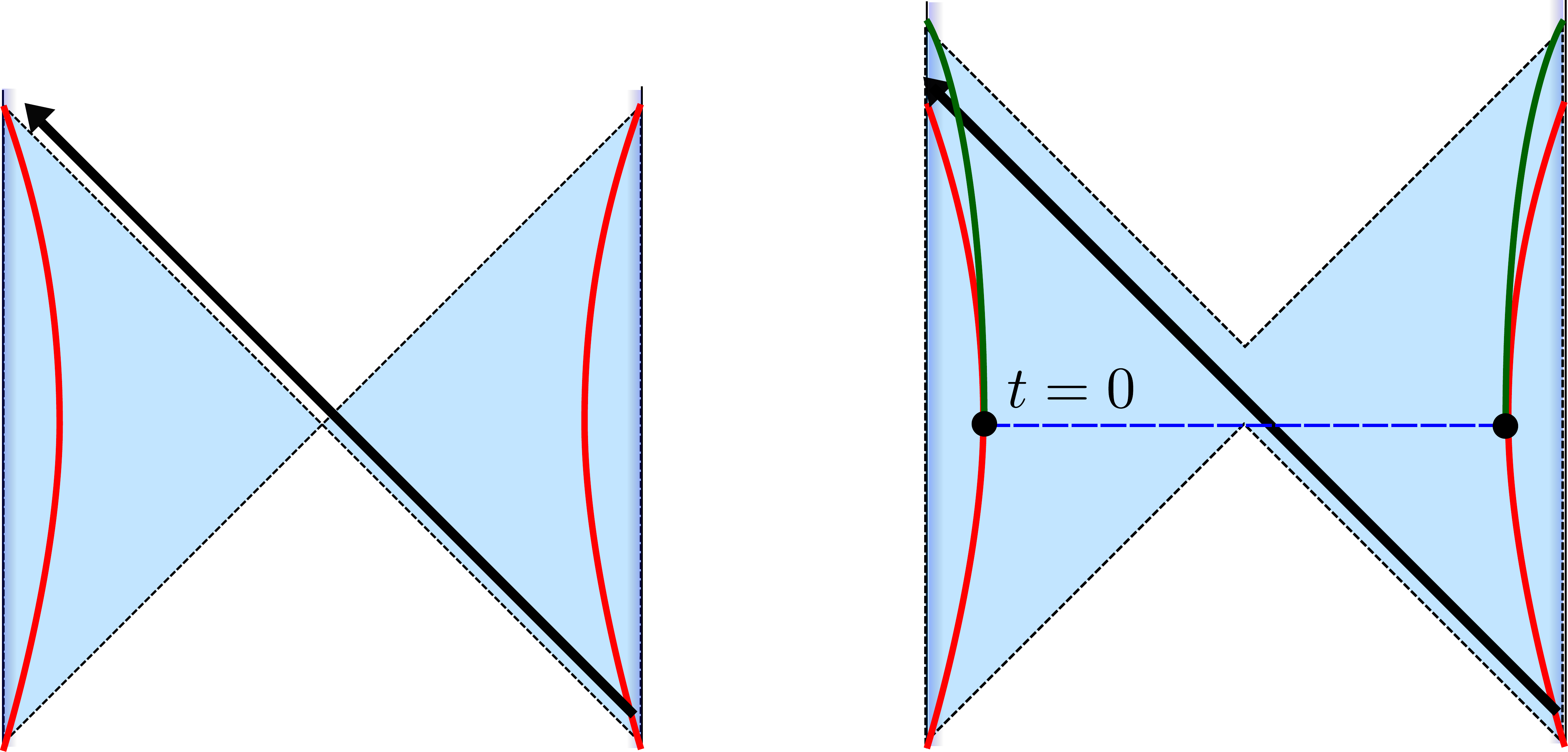}
\caption{Left: two-sided black hole as a non-traversable wormhole, or Einstein--Rosen bridge. Right: an interaction between the two sides can open up the wormhole.}
\label{fig:tw1}
\end{figure}

In Fig.~\ref{fig:tw1}, the left panel illustrates the non-traversability of any signal through the Einstein--Rosen bridge. The right panel of the same figure illustrates that after negative energy injection through the double trace interaction (blue), the boundary is kicked inwards (green) and the wormhole becomes traversable.

The bulk particle, in turn, also backreacts (not shown above) on the boundary curve, kicking the boundary curve towards the asymptotic boundary, and hence limits the amount of information that can be sent through the wormhole. 

\paragraph{Eternal traversable wormholes}
There is a second class of traversable wormholes that are eternal and horizonless, first described in the JT framework in \cite{Maldacena:2018lmt}. These can be described in NAdS$_2$ by understanding how to interpret both global boundaries as NAdS boundaries. 
The vacuum solution for the dilaton field $\Phi$ in global coordinates is $\Phi = a \cot 2 \mathbf{z}$,\footnote{We set $\mu = -a$ (and $b=0$) in Eq.~\eqref{cldil} and use the global frame \eqref{glframe}. Note that because $\mu <0$, the global frame has a negative energy $-a/8\pi G_N$ compared to the Poincar\'e patch. This is very similar to the situation in 3d gravity.} where the right boundary is at $\mathbf{z}=0$ and leads to $\Phi \to + \infty$ as before. The left boundary on the other hand is at $\mathbf{z}=\pi$ and hence $\Phi \to - \infty$. This boundary hence does not satisfy the boundary conditions of Sect.~\ref{s:lorderiv}. 

\begin{figure}
\centering
\includegraphics[width=0.25\textwidth]{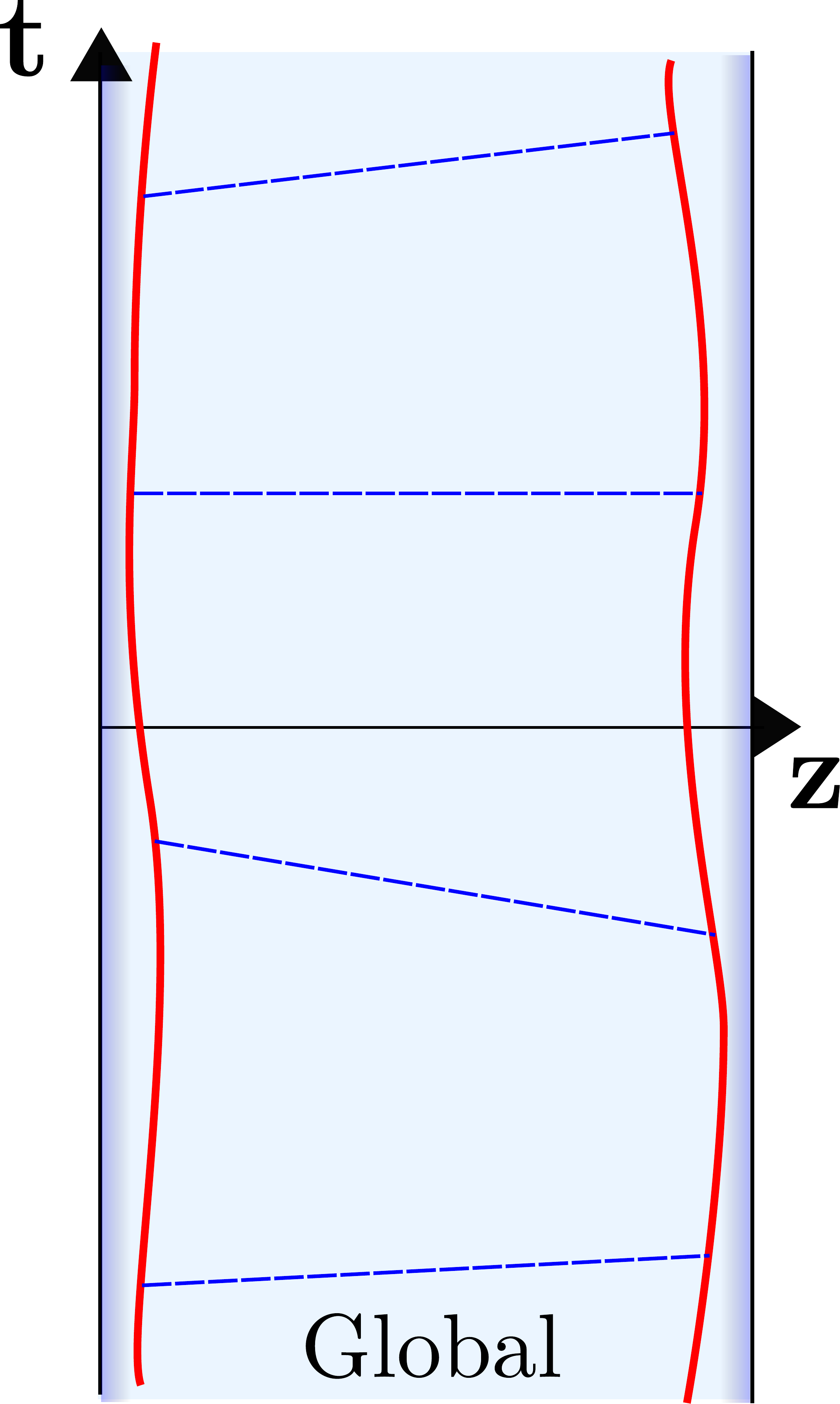} \caption{Global AdS$_2$ as an eternal traversable wormhole.}
\label{fig:tw2}
\end{figure}

In order to make sense of the two-boundary system in NAdS$_2$/NCFT$_1$, we need to have $\Phi\to +\infty$ at both ends. To implement this, one has to violate the null energy condition of the matter sector, since by integrating the first equation of motion in \eqref{dileom} along the $u$-direction from boundary to boundary, we get $
    -(\Phi_b^L+\Phi_b^R) = 8\pi G_N \int_{-\infty}^{+\infty} T_{\mathbf{x}^+\mathbf{x}^+}d\mathbf{x}^+$, where $\mathbf{x}^+$ is the affine coordinate along the null line $d\mathbf{x}^+ = e^{2\omega}d\mathbf{u}$. The LHS is negative, but the RHS is positive if the null energy condition is satisfied. 
Violating the null energy condition can again be implemented by a left-right coupling of similar type as before: $S_{\text{int}} = g \int dt \sum_{i=1}^N O_i^L(t)O_i^R(t)$. At leading order in large $N$ and small $g$ with fixed $Ng$, the effect of this coupling can be captured in a double Schwarzian language in terms of left and right global time reparametrizations $f_L$ and $f_R$ of the respective boundaries. The global frame is shown in Fig.~\ref{fig:tw2}, with bulk negative energy interactions (blue) and two wiggly curves at the holographic boundaries.

The effective action describing this coupled system is
\begin{align}
    I = -C\int dt \Biggr[\left\{\tan \frac{f_L}{2},t\right\} +\left\{\tan \frac{f_R}{2},t\right\} - \frac{gN}{2^{2\Delta}C} \left(\frac{f'_L(t)f'_R(t)}{\cos^2 \frac{f_L(t)-f_R(t)}{2}}\right)^{\hspace{-0.05cm}\Delta}\Biggr],
\end{align}
where there is only a single ``diagonal'' M\"obius redundancy.
For the analysis of this system, and an alternative derivation from the low-energy limit of two coupled SYK models, we refer to the original work by \cite{Maldacena:2018lmt}.

Physically, the resulting wormhole is horizonless. From the higher-dimensional perspective, it has been argued that one could obtain it by gluing together the throats of two near-extremal black holes by integrating out sufficiently energetic field modes. For further investigations into the presence of these kinds of solutions in our universe, we refer to \cite{Maldacena:2018gjk}.

\subsection{Finite cut-off and $T\bar{T}$}
The $T\bar{T}$ deformation is a solvable irrelevant deformation of two-dimensional conformal field theories \cite{Smirnov:2016lqw}. It was proposed in \cite{McGough:2016lol} that the AdS dual interpretation is to move the holographic boundary inside the bulk with Dirichlet boundary conditions. Even though this has passed many tests, it is still not fully understood. 

In \cite{Iliesiu:2020zld, Stanford:2020qhm} the problem of computing the partition function of JT gravity in AdS with a finite-length boundary was studied, using different methods. Up to now we considered boundary conditions where both the dilaton $\Phi_b$ and the boundary proper length $L$ diverge, with a fixed ratio $L/\Phi_b$ given in terms of the boundary temperature. This corresponds to sending the boundary to infinity in AdS$_2$. The finite cutoff version of this calculation is to work with a fixed dilaton and proper length not parametrically large. The results found by the two groups are different, and it is still an open question to resolve it. The result in \cite{Iliesiu:2020zld} is given by 
\begin{eqnarray}
Z(L,\Phi_b) = \frac{L \Phi_b^2}{L^2 + 4\pi^2} e^{- L \Phi_b} K_2 \Big( - \sqrt{ \Phi_b^2 (L^2 + 4\pi^2) } \Big),
\end{eqnarray}
and was matched to a quantum-mechanical version of the $T\bar{T}$ deformation introduced in \cite{Gross:2019ach} applied to the Schwarzian theory. 

A problem with this theory is that it has a non-unitary spectrum, an issue anticipated in \cite{McGough:2016lol}. The result in \cite{Stanford:2020qhm} is more complicated, but has a more transparent geometric interpretation.

Further references include \cite{Griguolo:2021wgy} which applies resurgence techniques to the problem, and \cite{Ebert:2022ehb} which analyzes the connection with 3d gravity.


\subsection{Volume of black hole interior and complexity}
A further success of JT gravity is the explicit evaluation of the volume of the black hole interior, including quantum gravity corrections. 
The extremal volume of a codimension-one slice, reduces to the geodesic length in one spatial dimension. In the hyperbolic two-plane $ ds^2=\frac{dZ^2+dT^2}{Z^2}$, geodesics are circle segments. For two bulk points $P=(T_1,Z_1)$ and $Q=(T_2,Z_2)$, the length $\ell$ along the circle geodesic connecting these two points is readily computed
\begin{equation}
\ell(P,Q)=\int_P^Q ds 
= 2 \, \text{Arcsinh}\, \sqrt{\frac{(Z_1-Z_2)^2+(T_1-T_2)^2}{4Z_1Z_2}}.
\end{equation}
In the case where both points are on the wiggly boundary curve, and hence $Z = \epsilon T' = \epsilon F'$, we get
\begin{equation}
\label{geodist}
\ell = \ln \frac{(F_1-F_2)^2}{F_1' F_2'} \, - \, \ln \epsilon^2.
\end{equation}

If we finally plug in the black hole solution $F(\tau) = \tan \frac{\pi}{\beta}\tau$, move one endpoint to the other side of the TFD state by letting $\tau_1 \to \tau_1 + \beta/2$, and Wick-rotate as $t\equiv i\tau_1-i\tau_2$, we obtain the renormalized real-time result:
\begin{equation}
\ell_{\text{ren}}(t) = \ln \cosh^2 \frac{\pi}{\beta}t  \quad \underset{t\to \infty}{\to} \quad \frac{2\pi}{\beta} t,
\end{equation}
which is linear in the time separation $t$ for large times \cite{Brown:2018bms,Goto:2018iay,Akhavan:2018wla}. This geodesic length is growing indefinitely due to the unbounded growth of the black hole interior as time evolves \cite{Hartman:2013qma}.

The beauty of the expression \eqref{geodist} is that we can go beyond classical gravity, and plug it into the Schwarzian path integral directly. The resulting path integral evaluation can be done exactly by relating it to the boundary two-point function using the trick: $\frac{\partial}{\partial \Delta} G_\Delta (\tau_1,\tau_2)\vert_{\Delta=0} = -\ln \frac{(F_1-F_2)^2}{F_1'F_2'}$.
The resulting expression is written in \cite{Yang:2018gdb} and somewhat surprisingly shows a continued linear growth. This means the quantum gravity fluctuations described in Sect.~\ref{sec:JTquantum} do not cause a change in very late time behavior.\footnote{The same late time growth can be shown to hold in the $\mathcal{N}=1$ supersymmetric version of JT gravity as well \cite{Fan:2021wsb}.}

Going beyond the Schwarzian, one can include wormhole corrections to this computation. As previously, the calculation can be done by directly relating it to the boundary two-point function \cite{Iliesiu:2021ari}. The result is a saturation of the growth of the black hole interior size at the timescale where non-perturbative random matrix effects kick in $t\sim C e^{S_0}$. Within the complexity = volume conjecture \cite{Susskind:2014rva}, it was argued by L. Susskind and collaborators \cite{Brown:2017jil} that this late time rise and saturation is expected for the growth of computational complexity of the dual boundary quantum system.

\subsection{Bulk correlators and observables}

Within any theory of quantum gravity, defining bulk observables is a delicate business. Namely, when coordinate transformations (or diffeomorphisms) are treated as a gauge symmetry, only diff-invariant operators are physically observable. Defining such diff-invariant operators requires \textbf{gravitationally dressing} a bulk matter operator. There are uncountably many ways of doing this. Here we focus on one particular choice that leads to explicit results in JT gravity \cite{Blommaert:2019hjr,Mertens:2019bvy}. Using the wiggly boundary curve for any off-shell $F(t)$, we use a radar procedure to define a bulk point as $(U=F(u),V=F(v))$ in terms of two proper times $u$ and $v$ on the boundary clock. This then defines an observable matter operator $\mathcal{O}(U=F(u),V=F(v))$ that has implicit dependence on the gravitational degrees of freedom through $F(t)$. 

For instance, for a massless scalar field in the bulk, the CFT bulk two-point function with Dirichlet boundary conditions at the AdS boundary equals:
\begin{equation}
    \left\langle \phi(u_1,v_1)\phi(u_2,v_2)\right\rangle_{\text{CFT}} = - \frac{1}{4\pi} \ln \frac{(F(u_1)-F(u_2))(F(v_1)-F(v_2))}{(F(u_1)-F(v_2))(F(v_1)-F(u_2))}.
\end{equation}
This object is then viewed in a second step as a gravitational operator to be inserted in the Schwarzian path integral. To actually evaluate it, we can make use of the following trick:
\begin{equation}
\label{trick2}
       \ln \frac{(F(u_1)-F(u_2))(F(v_1)-F(v_2))}{(F(u_1)-F(v_2))(F(v_1)-F(u_2))} = \int_{v_1}^{u_1} d t_1\int_{v_2}^{u_2}dt_2\,\frac{ F'(t_1)F'(t_2)}{(F(t_1)-F(t_2))^2},
\end{equation}
relating it to a double integral of the known boundary two-point function. 

The technical trick \eqref{trick2} is actually the HKLL bulk reconstruction formula \cite{Hamilton:2005ju,Hamilton:2006az} applied to a massless scalar in AdS$_2$. This prescription is hence elevated here into generic off-shell configurations $F(t)$.

One can Fourier transform these expressions to get the spectral occupation of the bulk matter modes, leading to quantum gravitational corrections to the Unruh heat bath \cite{Mertens:2019bvy,Blommaert:2020yeo}. These dressed bulk observables seem to correspond to measurements done by static (fiducial) observers in the quantum bulk.

The near-horizon region is very far from the holographic boundary and corresponds to an IR region. This means that correlators of these bulk observables experience strong quantum gravitational fluctuations in the near-horizon Rindler region. Hence, insisting on using such observable operators leads to tension with assuming quantum field theory in curved spacetime close to the black hole horizon, one of the main assumptions in arguing for information loss during black hole evaporation. This comment is independent from other relations to the information paradox discussed in earlier sections, which relied on wormhole configurations.

\subsection{Liouville gravity and minimal string}
\label{s:Liouvillegravity}
When considering solvable models of 2d gravity, next to JT gravity, there is an older model that attracted considerable attention in the 1980s and 1990s: that is Liouville gravity. An important recent result is that these models are not independent, but JT gravity can be embedded in Liouville gravity. This was first noticed for the spectrum in \cite{Saad:2019lba}, and subsequently developed in detail in \cite{Mertens:2020hbs,Mertens:2020pfe,Turiaci:2020fjj,Fan:2021bwt,Suzuki:2021zbe,Hirano:2021rzg,Artemev:2022hvu}.

Liouville gravity, or the non-critical string, is a model of two-dimensional quantum gravity, where a 2d CFT matter sector $S_M[\chi]$ is coupled to gravity as
\begin{equation}
    Z=\sum_{\rm topologies} \int \frac{[\mathcal{D}g] [\mathcal{D}\chi]}{{\rm Vol}({\rm Diff})} e^{ - S_M[\chi,g] - \mu_0 \int_\Sigma d^2x \sqrt{g}}.
\end{equation}
The conformal factor $e^{2\omega}$ of the metric gets dynamics at the quantum level governed by the Liouville CFT $S_L$ \eqref{liou} \cite{Polyakov:1981rd,Distler:1988jt, David:1988hj}.
This gauge fixing leads to additional $bc$ ghosts, and the total model is hence 2d Liouville + matter + ghost CFT, with the conformal anomaly constraint $c_M + c_L + c_{\rm gh} =0$, where $c_L = 1+6Q^2 = 1 +6(b+b^{-1})^2$. 

Within this string theoretic framework, the fixed length amplitude on the disk worldsheet, in suitable units, equals
\begin{align}
    \label{Lioudos}
    Z(\beta) \sim \int_{1}^{\infty} dE ~e^{-\beta E}\rho_0(E),~~~~\rho_0(E) = \sinh\Big(\frac{1}{b^2} \text{arccosh} E\Big).
\end{align}
We now interpret this in the same way as in JT gravity as a canonical partition function of a quantum black hole with temperature $\beta^{-1}$, but with a modified density of states.
At low energies, $\rho_0(E)$ limits to the JT density of states. To see this, we identify new kinematic variables as $ E = 1+ 2\pi^2b^4 E_{\rm JT}$ and $\beta =\frac{\beta_{\rm JT}}{2 \pi^2 b^4}$. Zooming in on the region close to the spectral cutoff by $b\to 0$, keeping $E_{\rm JT}$ and $\beta_{\rm JT}$ fixed, we reduce to the JT density of states $\rho_0 (E) \approx \sinh 2 \pi \sqrt{E_{\rm JT}}$.
At high energies, the Liouville gravity density of states \eqref{Lioudos} deviates from JT and rises with a power-law as $\sim E^{1/b^2}$.

This observation of embedding JT in the IR regime of Liouville gravity is no coincidence, and substantial more evidence can be gathered for it:
\begin{itemize}
    \item Bulk and boundary vertex operators can be inserted on the disk, and their fixed length amplitudes computed. The results match in each case with the JT expressions in the IR kinematic regime \cite{Mertens:2020hbs}.
    \item Describing $S_M$ with a timelike Liouville field $\chi$, one can perform a field redefinition from the Liouville gravity variables $(\phi,\chi)$ into the dilaton gravity variables $(\omega,\Phi)$ by using $\phi = b^{-1} \omega - b\pi  \Phi$, and $\chi = b^{-1} \omega + b\pi  \Phi$. This transformation leads to a dilaton gravity model of the type \eqref{gendilgrav} with dilaton potential
    \begin{equation}
    \label{sinhdilpot}
        U(\Phi) \sim \sinh (2\pi b^2 \Phi),
    \end{equation}
    which in the deep IR region, where $\Phi = r$ is small, becomes the JT dilaton potential. Close to the boundary $r\approx +\infty$, the dilaton potential diverges leading to a curvature singularity at the holographic boundary.
    \item The amplitudes of this model are governed by an algebraic structure related to the $U_q(\mathfrak{sl}(2,\mathbb{R}))$ quantum group. This can be explained directly at the Lagrangian level by the above dilaton potential \eqref{sinhdilpot} in the Poisson-sigma model framework of Sect.~\ref{s:firstorder} \cite{Fan:2021bwt,Mertens:2022aou}.
    \item For the particular case where the matter sector is the ($p,q$) minimal model, the resulting string model is called the \textbf{($p,q)$ minimal string}, which famously was conjectured to be equivalent to a random matrix integral, stemming from ideas around random triangulations of a 2d surface \cite{Brezin:1990rb,Douglas:1989ve,Gross:1989vs}. The fact that JT gravity is a matrix integral is hence embedded into this older statement.
    \item Multi-boundary and higher genus effects can be studied within the random matrix framework of the minimal string, leading to a deformation of the decomposition of amplitudes given in Sect.~\ref{s:sss} \cite{Mertens:2020hbs}.
\end{itemize}

\subsection{Universe field theory and quantum chaos}
The non-perturbative completion of JT gravity into a matrix integral is very useful, but it has its downsides. For instance, a matrix integral is not an ordinary quantum mechanical system, instead representing a maximally disordered average over models.

What one would want is an all-encompassing quantum mechanical framework that correctly describes multi-universes as quantum mechanical operator insertions, and allows for quantum processes that capture the emission and absorption of entire universes.\footnote{Notice that there is no tension with Sect.~\ref{s:discrete}, since we are not working with a fixed number of boundaries, and do not require the interpretation of the boundaries as Euclidean time directions of this quantum mechanical system.} This is a third quantization of sorts, similar to string field theory. Such a quantum picture was already developed in the past by \cite{Cole,GiStInc}, and was reinvestigated in the current context in \cite{Marolf:2020xie, Giddings:2020yes}, where basic principles were formulated.

A concrete description in which such multi-universe phenomena are captured in JT gravity is in terms of Kodaira--Spencer theory. This observation was made in  \cite{Post:2022dfi}, and we refer to that work for all the details of this correspondence.
The Kodaira-Spencer model is defined on a 2d surface \cite{Dijkgraaf:2007sx} (the spectral curve of the associated matrix model):\footnote{It starts its life as a 6d model on the topological closed string B-model that is reduced to 2d.}
\begin{equation}
\label{KS}
    S_{\text{KS}} = \int d^2z \left(\frac{1}{2} \partial \Phi \bar{\partial} \Phi - (\mathcal{J}-2zy(z)) \bar{\partial} \Phi \right) + \frac{\lambda}{2} \oint_{z=0} \frac{dz}{2zy(z)}\Phi \mathcal{J}^2,
\end{equation}
in terms of two fields $\Phi$ (unrelated to the dilaton!) and $\mathcal{J}$, which can be thought of as $\mathbb{Z}_2$-twisted boson fields. The free theory has the propagators $\left\langle \mathcal{J}(z_0) \Phi(z) \right\rangle = \frac{1}{z_0-z} - \frac{1}{z_0+z}$ and $\left\langle \mathcal{J}(z_i)\mathcal{J}(z) \right\rangle = \frac{1}{(z_i-z)^2} + \frac{1}{(z_i+z)^2}$.
The cubic interaction vertex in \eqref{KS} is to be identified with the elementary three-holed sphere in hyperbolic geometry. The full hyperbolic surface is then created as a Feynman graph that coincides with the ``skeleton graph'' one draws on the surface. This theory is a string field theory model for which the perturbative string worldsheets are reinterpreted as spacetime universes, hence the name \textbf{universe field theory}. \\

\noindent \textbf{Example}: \\
Let us consider the genus one correction to the disk amplitude, given in 1$^{st}$ order perturbation theory by bringing down a single instance of the interaction vertex \eqref{KS} as
\begin{align}
\label{onelo}
  \left\langle \mathcal{J}(z_0)\right\rangle_1^{c} &= \frac{\lambda}{2} \oint_{z=0} \frac{dz}{2zy(z)} \left\langle \mathcal{J}(z_0) \Phi(z) \right\rangle \left\langle \mathcal{J}(z)\mathcal{J}(z) \right\rangle,
\end{align}
corresponding to a one-loop Feynman diagram. Evaluating gives
\begin{align}
 \left\langle \mathcal{J}(z_0)\right\rangle_1^{c} &= \frac{\lambda}{4}\text{Res}_{z=0}\frac{2}{z_0^2-z^2}\frac{2\pi}{\sin 2\pi z} \frac{1}{4z^2} = \lambda\frac{3+2\pi^2z_0^2}{24z_0^4},
\end{align}
which is indeed the correct contribution $W_{1,1}$.
\begin{center}
---o---
\end{center}
\vspace{0.5cm}

Feynman diagrammatics with the cubic vertex in \eqref{KS} allows one to immediately write down recursive Dyson--Schwinger relations that can be shown to match with Eynard--Orantin's topological recursion relations \cite{Dijkgraaf:2007sx}.\footnote{A subtlety with this expression is that naively not all contractions are taken into account: the external $\mathcal{J}$ lines only contract with the $\Phi$-leg of an internal vertex. This is related to the fact that the boson fields are chiral \cite{Dijkgraaf:2007sx}. This is also present in \eqref{onelo} above.} 

FZZT brane (and anti-brane) operators can be inserted into the model as $e^{\Phi(z)}$ (and $e^{-\Phi(z)}$), and can be utilized to get to the same non-perturbative physics as the matrix model completion of JT gravity reviewed in Sect.~\ref{s:nonpt}. The model with a fixed number of branes can be rewritten in terms of a Kontsevich-like matrix model in terms of flavor degrees of freedom. The resulting flavor matrix theory can then be shown to describe universal chaotic dynamics \cite{Altland:2022xqx}. This hence provides a direct connection between the universal ergodic behavior in chaotic systems, and the universe field theory of a gravity model.

\setlength{\leftskip}{0cm}


\backmatter

\bmhead{Acknowledgments}

We would like to thank all our collaborators throughout the past years, without whom we would not have been able to write the current review. We thank Roberto Emparan, Juan Maldacena, Francesca Mariani and Subir Sachdev for comments on the draft. We would like to specially thank Herman Verlinde for introducing us to this subject. TM acknowledges financial support from Research Foundation Flanders (FWO Vlaanderen) and the European Research Council (grant BHHQG-101040024). Funded by the European Union. Views and opinions expressed are however those of the author(s) only and do not necessarily reflect those of the European Union or the European Research Council. Neither the European Union nor the granting authority can be held responsible for them. GJT is supported by the Institute for Advanced Study and the National Science Foundation under Grant No. PHY-2207584, and by the Sivian Fund. GJT thanks the University of Warsaw for hospitality during final stages of writing this review.



\phantomsection
\addcontentsline{toc}{section}{References}
{\small \bibliography{sn-bibliographyv2}}


\end{document}